\documentclass[11pt, a4paper]{article}

\usepackage{geometry}
\usepackage[multidot]{grffile}
\usepackage{enumitem}
\usepackage{setspace}
\usepackage{amsmath}
\usepackage[single]{accents}
\usepackage[nodayofweek]{datetime}
\usepackage{natbib, url}
\usepackage{graphicx}
\graphicspath{{Figures/}}
\usepackage[dvipsnames]{xcolor}
\usepackage{threeparttable}
\usepackage{rotating}
\usepackage{xr-hyper} 
\usepackage{booktabs} 
\usepackage[bookmarks,bookmarksnumbered=true,colorlinks=true,
     linkcolor=blue,urlcolor=blue,citecolor=blue,breaklinks = true,backref=page]{hyperref}
\usepackage{color}
\usepackage{scalerel,amssymb}

\usepackage{doi}
\newif\ifempty
\providecommand{\noopsort}[1]{} 

\newcommand{\betahat}{\widehat{\beta}}

\newcommand{\btheta}{\boldsymbol{\theta}}

\newcommand{\DATA}{\textrm{DATA}}
\newcommand{\data}{\textrm{data}}
\newcommand{\dd}{\mathrm{d}}

\newcommand{\dfreedom}{r_{d}}
\newcommand{\dlrtlst}{\textrm{dlrtlst}}
\newcommand{\dlrtnorm}{\textrm{dlrtnorm}}
\newcommand{\dltlst}{\textrm{dltlst}}
\newcommand{\dltnorm}{\textrm{dltnorm}}
\newcommand{\dtnorm}{\texttt{dtnorm}}

\newcommand{\etahat}{\widehat{\eta}}
\newcommand{\E}{\textrm{E}}
\newcommand{\fail}{\textrm{fail}}
\newcommand{\Fhat}{\widehat{F}}

\newcommand{\intervspace}{\;\;}

\newcommand{\lev}{\textrm{lev}}

\newcommand{\like}{{L}}

\newcommand{\logis}{\textrm{logis}}
\newcommand{\loglike}{{\cal L}}
\newcommand{\logLST}[1]{\textrm{$<$LLST#1$>$}}
\newcommand{\logNormalPrior}{\textrm{$<$LNORM$>$}}
\newcommand{\lst}{\textrm{lst}}

\newcommand{\NORM}{\textrm{NORM}}

\newcommand{\norm}{\textrm{norm}}

\newcommand{\sev}{\textrm{sev}}

\newcommand{\thetadummyvec}{{\boldsymbol{\theta}}}
\newcommand{\thetavec}{{\boldsymbol{\theta}}}

\newcommand{\TNormalPrior}{\textrm{$<$TNORM$+>$}}
\newcommand{\TNORM}{\textrm{TNORM}}

\newcommand{\typeII}{\textrm{Type~2}}
\newcommand{\typeI}{\textrm{Type~1}}

\newcommand{\weibscale}{\eta}

\newcommand{\footremember}[2]{%
  \footnote{#2}
  \newcounter{#1}
  \setcounter{#1}{\value{footnote}}%
}
\newcommand{\footrecall}[1]{%
  \footnotemark[\value{#1}]%
} 
\def\msquare{\mathord{\scalebox{1.0}[1]{\scalerel*{\Box}{X}}}}

\newcommand{\rsplidafiguresize}[2]{\includegraphics[keepaspectratio=true,width=#2,angle=0]{#1}}
\newcommand{\rsplidafiguresizetwo}[2]{\includegraphics[keepaspectratio=true,width=#2,angle=-90]{#1}}

\newdateformat{mydate}{\twodigit{\THEDAY}{  }\monthname[\THEMONTH] \THEYEAR}
\title{Specifying Prior Distributions in Reliability Applications}
\author{%
  Qinglong Tian\footremember{waterloo}{Department of Statistics and Actuarial Science, University of Waterloo}%
  \and Colin Lewis-Beck\footremember{amazon}{Amazon.com Inc.}%
  \and Jarad B. Niemi\footremember{isu}{Department of Statistics, Iowa State University}%
  \and William Q. Meeker\footrecall{isu}
}
\mydate

\begin{document}
\maketitle

\begin{abstract}
Especially when facing reliability data with limited information (e.g.,
a small number of failures), there
are strong motivations for using Bayesian inference methods.
These include the option to use information
from physics-of-failure or previous experience with a failure mode
in a particular material to specify an informative
prior distribution. Another advantage is the ability
to make statistical inferences without
having to rely on specious (when the number of failures is small)
asymptotic theory needed to justify
non-Bayesian methods. Users of non-Bayesian methods are faced with
multiple methods of constructing uncertainty intervals (Wald,
likelihood, and various bootstrap methods) that can give
substantially different answers when there is little information in
the data. For Bayesian inference, there is only one method of
constructing equal-tail credible intervals---but
it is necessary to provide a prior
distribution to fully specify the model.
Much work has been done to find default prior distributions
that will provide inference methods with good (and in some cases
exact) frequentist coverage properties. This paper reviews some of
this work and provides, evaluates, and illustrates principled
extensions and adaptations of these methods to the practical
realities of reliability data (e.g., non-trivial censoring).
\end{abstract}


%

\newpage

\section{Background and Motivating Examples}
\subsection{Bayesian Methods in Reliability Applications}
The use of probability plotting and maximum likelihood (ML) methods
for the analysis of censored reliability data has matured over the
past 30 years. These methods appear in numerous textbooks and are
readily available in several widely-used commercial statistical software
packages (e.g., JMP, MINITAB, and SAS). More recently,
commercial statistical software that provides capabilities for doing Bayesian
estimation has become available (e.g., JMP and SAS). Bayesian
estimation requires the specification of a joint prior distribution
for the model parameters. The purpose of this paper is to provide
guidance on how prior specification should be done in reliability
applications. We focus on applications requiring a single
distribution (e.g., Weibull or lognormal). The basic ideas, however, can
be applied to more complicated models, as described in our
concluding remarks section.

\subsection{Motivating Examples}
\label{section:motivating.examples}
\subsubsection{Bearing cage field data}
\begin{figure}[!htbp]
\begin{tabular}{cc}
(a) & (b) \\[-3.2ex]
\rsplidafiguresize{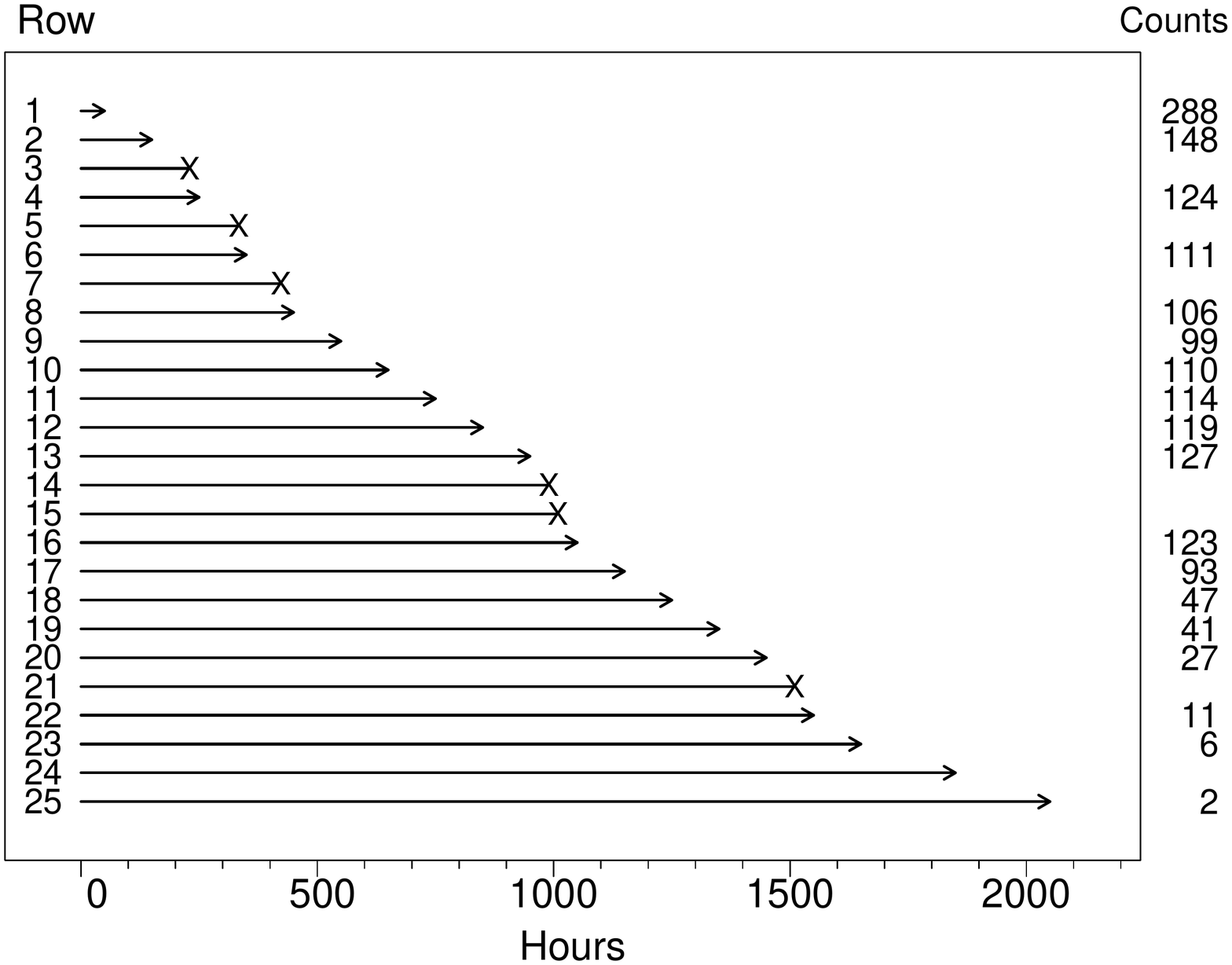}{3.35in} &
\rsplidafiguresize{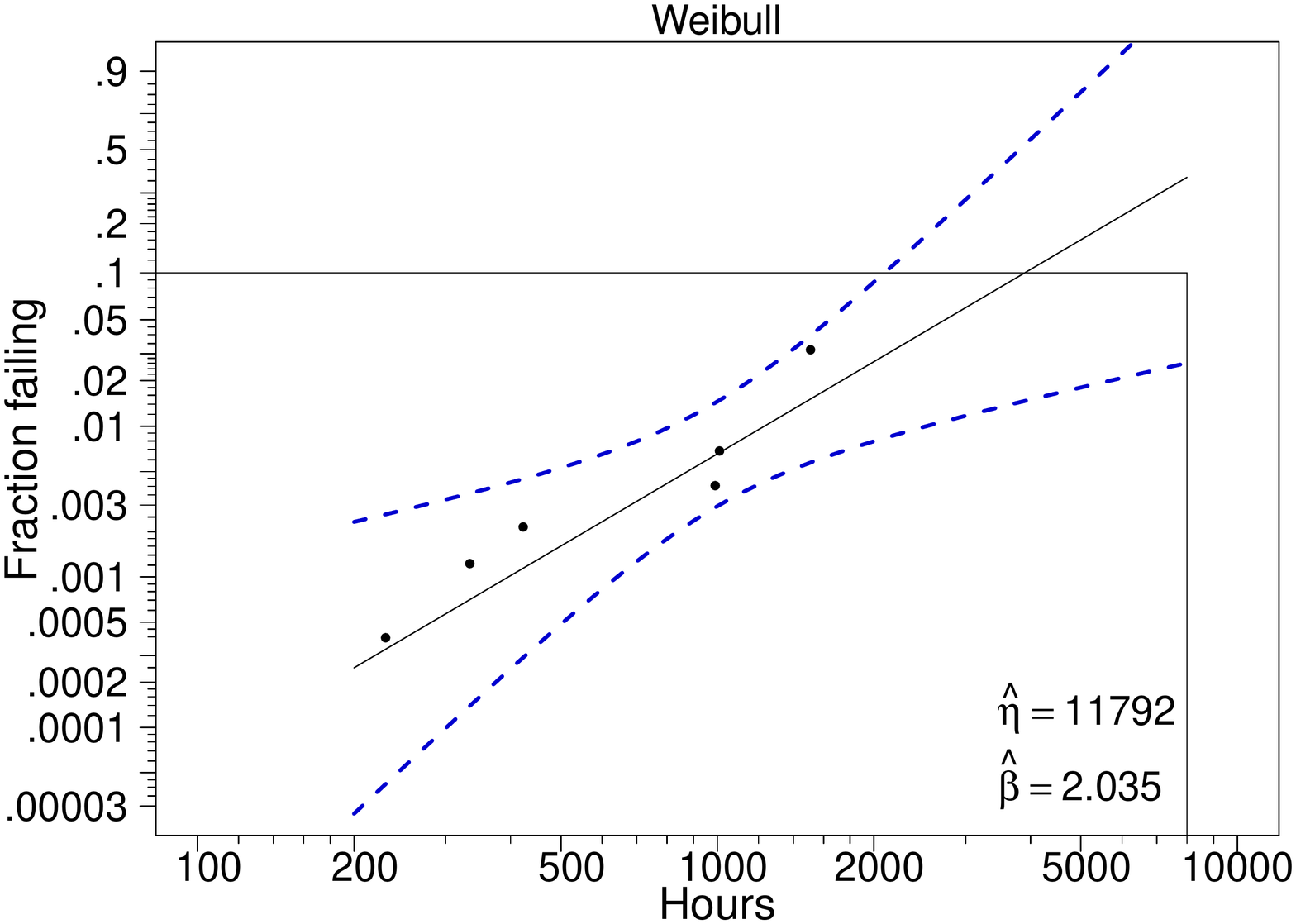}{3.35in}
\end{tabular}
\caption{Bearing cage event plot~(a) and Weibull probability plot~(b).}
\label{figure:BearingCage.plots}
\end{figure}

Figure~\ref{figure:BearingCage.plots}(a) is an event plot of
bearing-cage fracture times for six failed units as well as running
times for 1,697 units that had accumulated various amounts of service
time without failing.  The data and an analysis appear in
\citet{Abernethyetl1983}.
These data represent a population of units that had been introduced
into service over time and the data are multiply censored
(censoring at multiple points in time). There
were concerns about the adequacy of the bearing-cage design.
Analysts wanted to use these initial data to decide if a redesign
would be needed to meet the design-life specification.
This requirement was that the 0.10 quantile of bearing life
(sometimes referred to as
B10) be at least 8,000 hours.
Figure~\ref{figure:BearingCage.plots}(b) is a Weibull probability
plot. Because of the small number of
failures, the confidence interval for the fraction failing at 8,000
hours is wide and deciding whether the reliability
goal is being met is difficult. A likelihood ratio confidence
interval for the fraction failing at 8,000 hours is $[0.026, \,\,
  0.9999]$, which is not useful.

\subsubsection{Rocket motor field data}
This example was first presented in \citet{OlwellSorell2001}.  The
US Navy had an inventory of approximately 20,000 missiles. Each
included a rocket motor---one of five critical components. These
missiles were subject, over time, to unmonitored thermal cycling due
to environmental variability during, storage, transit, and
take-off-landing cycles.
Only 1,940
of the missiles had actually been put into use over a period of
time up to 18 years subsequent to their manufacture. At their time
of use, 1,937 of these motors performed satisfactorily; but there
were three catastrophic launch failures.  Responsible scientists and
engineers believed that these failures were caused by the thermal
cycling. In particular, it was believed that the thermal cycling
resulted in failed bonds between the solid propellant and the
missile casing.

The failures raised concern about the previously unanticipated
possibility of a sharply increasing failure rate over time (i.e.,
rapid wearout) as the motors aged and were subjected to thermal
cycling while in storage. If this were indeed the case, a
possible---but costly---remedial strategy might be to
replace aged rocket motors with new ones.  Thus, to assess the
magnitude of the problem, it was desired to quantify the rocket-motor
failure probability as a function of the amount of thermal cycling
to which a motor was exposed and to obtain appropriate confidence
bounds around such estimates based on the results for the 1,940
rocket motors---assuming these to be a random sample from the larger
population (at least concerning their failure-time
distribution).

Because no information was directly available on the thermal cycling
history of the individual motors, the age of the motor (i.e.,  time
since manufacture) at launch was used as a surrogate.  This was not
an ideal replacement because the thermal cycling rate, or rate of
accumulation of other damage mechanisms, varied across the
population of motors, depending on an individual missile's
age and environmental storage history.
Compared to a scale based on the number of thermal cycles, the
effect of using time since manufacture
is to increase the variability in the observed failure-time
response, as described in \citet{MeekerEscobarHong2009}. The failure
probability 20 years after manufacture was of particular interest.

\begin{figure}[!htbp]
\begin{tabular}{cc}
(a) & (b) \\[-3.2ex]
\rsplidafiguresizetwo{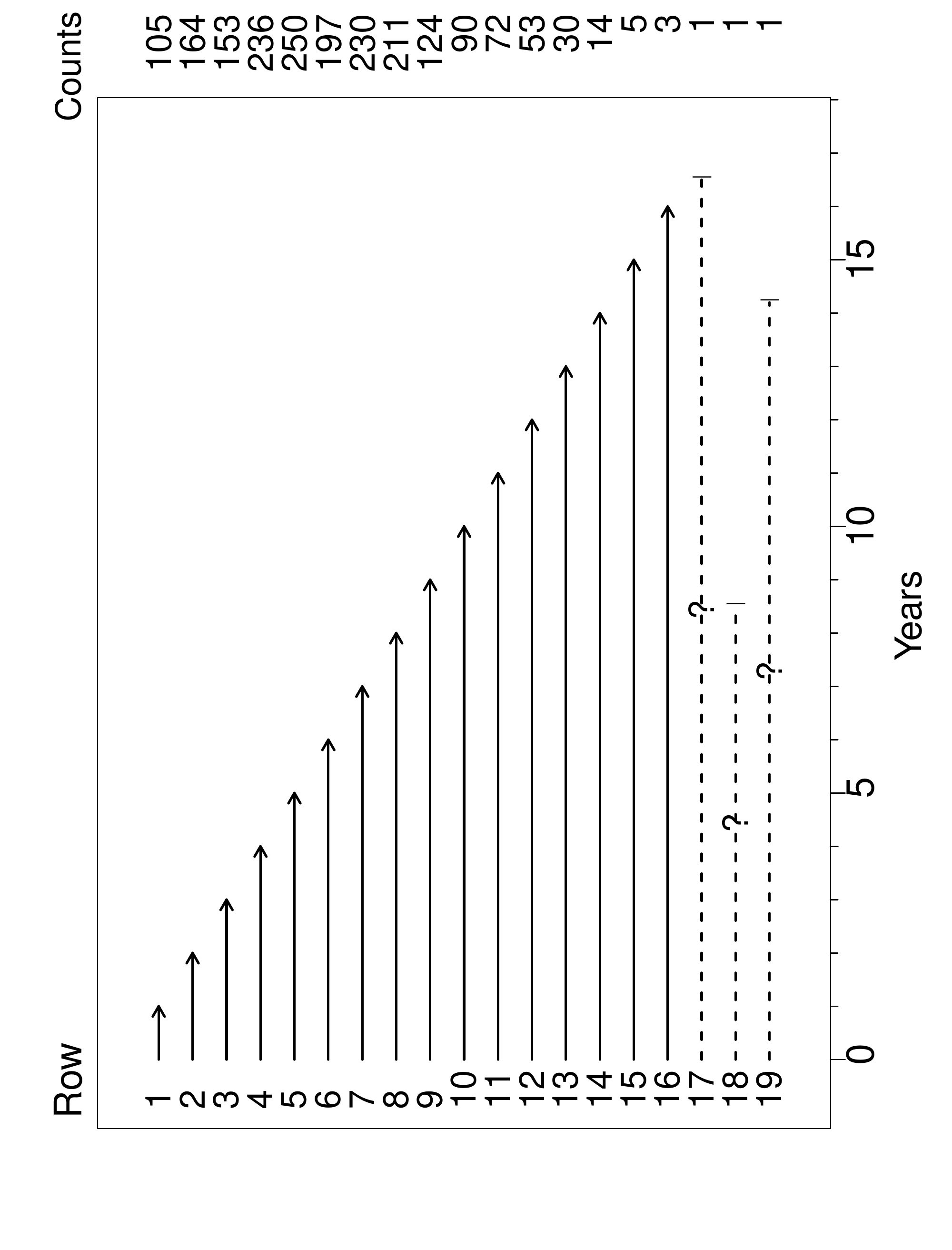}{2.5in} &
\rsplidafiguresizetwo{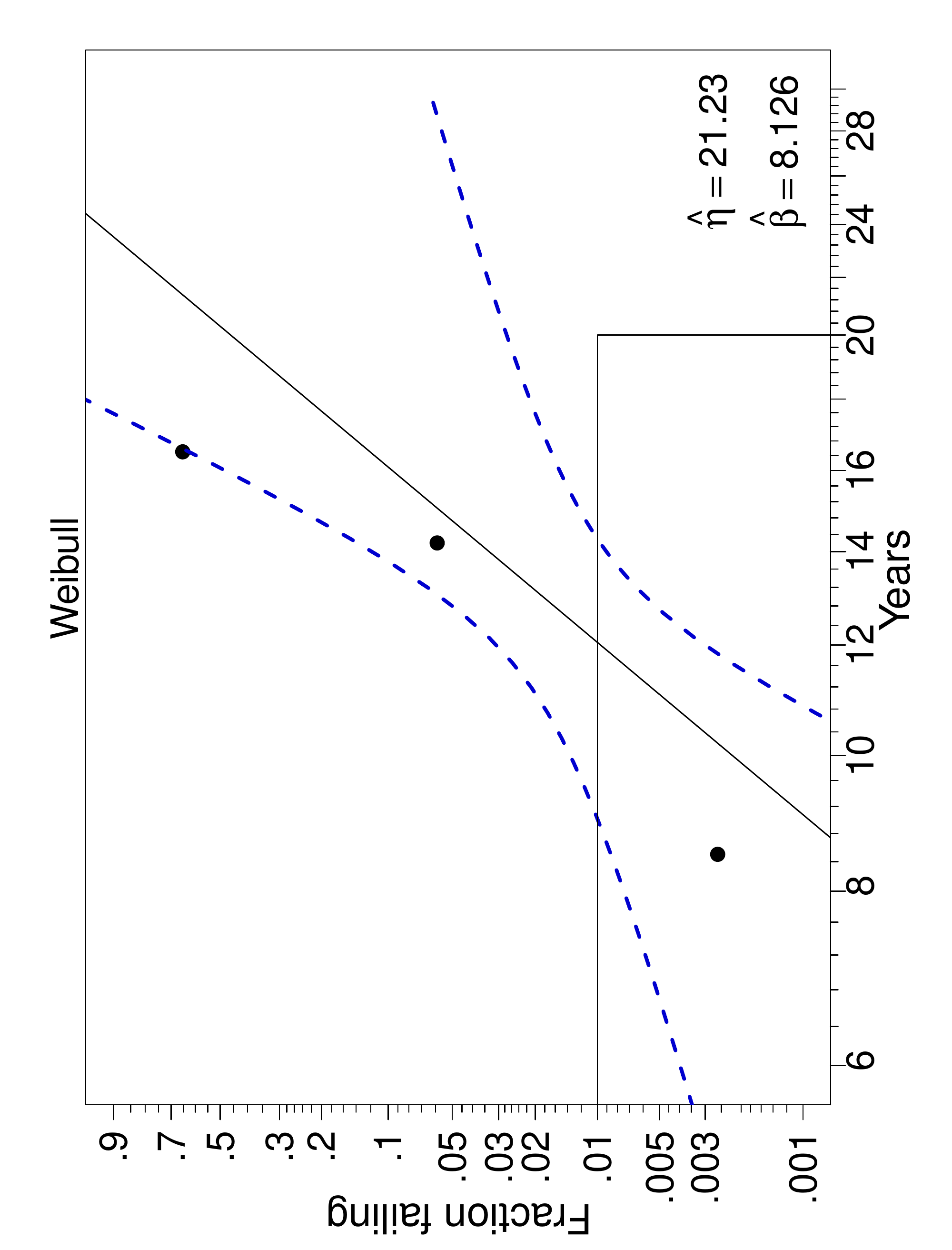}{2.5in}
\end{tabular}
\caption{Rocket motor event plot~(a) and Weibull probability plot~(b).}
\label{figure:RocketMotor.plots.ps}
\end{figure}

The specific age at failure of each of the three failed motors was
not known---all that was known was that failure, in each case, had
occurred sometime before the time of launch---thus making the time
since manufacture at launch left-censored observations of the actual
failure times.  Similarly, the information of (eventual) failure age
for the 1,937 successful motors is right-censored---all that is
known is that the time to the yet-to-occur failure exceeds the
calendar age at the time of launch.  Thus, the available
rocket-motor field-performance data, contained only left- and
right-censored observations---but \textit{no} known exact failure
times (such data are known as ``current-status data'').
Figure~\ref{figure:RocketMotor.plots.ps}(a) is an event
plot that further illustrates the structure of the data.

Because failure times are only bounded (no exact failure times)
and because of the
very small number of known failures, the amount of information in
the data is severely limited. Nevertheless, it is possible to
estimate the Weibull parameters from these
data. Figure~\ref{figure:RocketMotor.plots.ps}(b) is a Weibull
probability plot of the data. The ML
estimates of the Weibull parameters are $\etahat=21.23$
years and $\betahat=8.126$.
A likelihood ratio confidence interval for the
fraction failing at 20 years is $[0.023, \,\, 0.9999]$, which,
again, is not useful.

For most failure mechanisms operating in the field, the Weibull
shape parameter $\beta$ will be less than 4. Using the surrogate
years since manufacturer in place of the unknown number of thermal
cycles will further increase the relative variability in the data
which would make $\beta$ even smaller.  Thus the estimate
$\betahat=8.126$ was surprisingly large.

\subsection{Literature Review}
Books that focus on Bayesian methods for reliability data analysis
include~\citet{MartzWaller1982}, \citet{Hamadaetal2008}, and
\citet{LiuAbeyratne2019}.  \citet{SanderBadoux1991} contains six contributions
that describe early work on the application of Bayesian methods
in different reliability applications.
Numerous papers describing particular applications of
Bayesian methods have appeared in the engineering and statistical
literature over the past 30 years.

This paper, primarily, focuses on finding noninformative or other
default prior distributions for reliability applications using a
single distribution. As described more fully in
Section~\ref{section:noninformative.prior}, in many reliability
applications that we have encountered there is a need to be
minimally informative so that the prior choice can be easily
defended.  Section~\ref{section:noninformative.prior.literature}
reviews some of the literature on \textit{noninformative} prior
distributions. 

There has, however, been much work done on eliciting informative
priors for reliability applications based on expert judgment. Here
we review some of this work. \citet{Lindley1983} describes
analytical methods for combining information (either location or
both location and scale) from several different sources that might
include expert opinion or other information sources. The paper
demonstrates the advantages of using $t$ distributions instead of
normal distributions for the underlying model being used to combine
the information into one probability distribution. As we mention
in Section~\ref{section:motivation.for.partially.informative.prior},
we have found location-scale-$t$ distribution to be useful for this
purpose. \citet{LindleySingpurwalla1986} use the conceptual ideas
from \citet{Lindley1983} in the application of combining expert
opinions to quantify the reliability of a multicomponent parallel
redundant system.  \citet{VanNoortwijk_etal_1992} describe
procedures for combining expert information, based on discretized
life distributions at the elicitation
stage. \citet{CampodonicoSingpurwalla1995} describe how to use prior
information for the intensity function of a nonhomogeneous Poisson
point processes and illustrate the approach using two reliability
applications.  \citet{WallsQuigley2001} describe an elicitation
process that allows for the identification and management of expert
bias. They apply the methods to reliability growth modeling.
\citet{BedfordQuigleyWalls2006} review how expert knowledge is used
to make needed assessments of reliability in applications involving
engineering system design.

\citet{Gutierrez-PulidoAguirre-TorresAndres2005} suggest specifying
fully informative prior distributions for a two-parameter
distribution by specifying intervals for the mean and standard
deviation or two quantiles for the failure-time distribution.
\citet{KaminskiyKrivtsov2005} suggest, for a Weibull distribution,
specifying fully informative joint prior distributions for the
parameters by specifying priors for two points on the cdf of
the Weibull distribution. \citet{KrivtsovFrankstein2017} extend the
method to other failure-time distributions.

\citet{MeyerBooker2001} provides a book-length guide to methods and
procedures for eliciting information from experts without focus on
any particular area of application. \citet{OHagan_etal2006} provide
a more technical book-length guide for eliciting information from
experts.

\subsection{Overview}
The remainder of this paper is organized as follows.
Section~\ref{section:single.distribution.reliability.models}
provides a brief review and defines
notation for reliability models, censoring, and likelihood.
Section~\ref{section:bayesian.methods.prior.information} briefly
introduces the basic concepts of Bayesian inference and explains
motivation for and mechanics of needed reparameterization.  The
commonly used noninformative prior distributions are derived from
the Fisher information matrix (FIM). Section~\ref{section:lls.fim}
describes how to obtain the elements of the FIM for different kinds
of censoring.  Section~\ref{section:noninformative.prior} reviews
the commonly used noninformative prior distributions and describes
extensions to \typeI{} and \typeII{} censoring.
Section~\ref{section:random.censoring.ij.prior} explains how the
results in Section~\ref{section:noninformative.prior} can be applied
in situations where there is random censoring.
Section~\ref{section:weakly.informative.prior} explains the
importance of weakly informative prior distributions for some
applications and illustrates the use of noninformative and weakly
informative priors in the examples.  In most reliability
applications, engineers will have strong prior information on only
one parameter (e.g., the Weibull shape parameter).
Section~\ref{section:combining.informative.with.noninformative}
shows how to combine prior information for one parameter with a
noninformative or weakly informative prior for the other parameter
and applies the ideas to the examples.
Section~\ref{section:weibull.type.one.simulation} describes a
simulation that was conducted to study the coverage probabilities of
credible intervals computed under different noninformative priors.
Section~\ref{section:prior.sensitivity.analysis} suggests and
illustrates methods of doing prior distribution sensitivity
analysis.  Section~\ref{section:concluding.remarks} gives
concluding remarks and suggests extensions and areas for future research.

\section{Review of Single Distribution Reliability Models, Censoring, and Likelihood}
\label{section:single.distribution.reliability.models}
This section briefly reviews the commonly used models, censoring, and methods
for fitting a single distribution to reliability data.
\subsection{Log-Location-Scale Distributions}
The most frequently used distributions for failure-time data are in
the log-location-scale family of distributions. A random variable
$T>0$ belongs to the log-location-scale family if
$Y=\log(T)$ is a member of the location-scale family.  The cdf
for a log-location-scale distribution is
\begin{align}
\label{equation:lls.cdf}
 F(t;\mu, \sigma)&=\Phi\left [\frac{\log(t)-\mu}{\sigma}
\right ], \,\, t > 0
 \end{align}
and the corresponding pdf is
\begin{align}
\label{equation:lls.pdf}
 f(t;\mu, \sigma)&=\frac{1}{ \sigma t} \phi
\left [
\frac{\log(t)-\mu}{\sigma} \right],
 \end{align}
where $\Phi(z)$ and $\phi(z)=d\Phi(z)/dz$ are, respectively,
the cdf and pdf for the
particular standard
location-scale distribution.
The most common log-location-scale distributions are the
lognormal ($\Phi(z)=\Phi_{\norm}(z)$ is the standard normal cdf),
Weibull ($\Phi(z)=\Phi_{\sev}(z)=1-\exp[-\exp(z)]$) , Fr\'{e}chet
($\Phi(z)=\Phi_{\lev}(z)=\exp[-\exp(-z)]$), and loglogistic
distributions ($\Phi(z)=\Phi_{\logis}(z)=1/[1+\exp(-z)]$).

The hazard function is important in reliability theory and
applications and is defined by
\begin{align*}
h(t)&=\lim_{\Delta t \rightarrow 0} \frac{\Pr(
t < T\le
t+\Delta t  \mid T > t)}{\Delta t}
= \frac{f(t)}{1-F(t)}.
\end{align*}
The hazard function is proportional to the probability of failing in the next small
interval of time, conditional on having survived to the beginning of
that interval. That is, for
small $\Delta t$,
\begin{align*}
\Pr(t < T \le t+\Delta t \mid
T > t) \approx   h(t) \times \Delta t .
\end{align*}

\subsubsection{Example 1: The lognormal distribution}
If $T$ has a lognormal distribution, then $Y=\log(T) \sim \NORM(\mu,
\sigma)$, where $\mu$ is the mean and $\sigma$ is the standard deviation of
the underlying normal distribution. For the lognormal distribution,
$\sigma$ is the shape parameter and $\exp(\mu)$ is the median (and a
scale parameter).

\subsubsection{Example 2: The Weibull distribution}
The Weibull cdf is often given as
\begin{align}
\label{equation:bayes.weibull.cdf}
\Pr(T \leq t;\weibscale,\beta ) &= F(t;\weibscale,\beta)=1-
\exp \left [-\left (\frac{t}{\weibscale} \right )^{\beta}
\right ], \,\,  t > 0,
\end{align}
where $\beta$ is a shape parameter and $\eta$ is a scale
parameter---sometimes called ``characteristic life'' and is
approximately the 0.63 quantile of the distribution.

The Weibull cdf can also be expressed by (\ref{equation:lls.cdf})
with the parameters $\mu=\log(\eta)$ and $\sigma=1/\beta$. Although
results (e.g., from software) are typically presented in the more
familiar $(\eta, \beta)$ parameterization, it is common practice to
use the $(\mu, \sigma)$ parameterization for the development of
theory and software for the entire (log-)location-scale families and
this is especially true for regression models like those used for
accelerated testing \citep[e.g.,][Chapters 18 and
  19]{MeekerEscobarPascual2022}.

\subsection{Quantities of Interest in Reliability Data Analysis}
In reliability applications, the usual distribution parameters are not
of primary interest. Instead, there is generally a need to estimate
failure probabilities (computed using (\ref{equation:lls.cdf})) at a
specified time or a
failure-time distribution quantile. The $t_{p}$
quantile for a distribution in the
log-location-scale family can be expressed as
$ t_{p}=\exp[\mu + \Phi^{-1}(p)\sigma]$.
These quantiles also play an important role in prior elicitation, as
described in Section~\ref{section:reparameterization}.

\subsection{Censoring}
\label{section:censoring.types}
Censoring is ubiquitous in the analysis of reliability and other
time-to-event data.
There are different kinds of censoring that arise in applications.
\begin{itemize}[itemsep=1mm, parsep=0pt]
\item
Right censoring arises when one or more units have not failed when the
data are analyzed and occur for different reasons.
\begin{itemize}[itemsep=1mm, parsep=0pt]
\item
Time (\typeI{}) censoring arises in life tests where the test ends
at a specified time. The number of failures is random (and could be zero).
\item
Failure (\typeII{}) censoring arises in life tests where the test ends
after a specified number of units have failed. The length of the test
is random. Such tests are not common in practice because of the need to adhere
to schedules.
\item
Multiple right censoring (many different censoring times) is common in
field data. Differing censoring times arise from some combination of
staggered entry, differing use rates (when time is measured in
amount of use since entering service), and competing risks (e.g., failure
modes unrelated to the one of primary interest).
\end{itemize}
\item
Interval censoring arises when failures are found at inspection
times. All that is known is that a failure occurred between the
most recent previous and the current inspection.
\item
Left censored observations arise when a failure has already occurred
at the first time a unit is observed and is equivalent to an
interval-censored observation that has zero as its beginning time.
\end{itemize}
An assumption of noninformative censoring \citep[e.g.,][pages
  59--60]{Lawless2003} is generally required to use the common
methods for analyzing censored data.

\subsection{Log-likelihood}
\label{section:lls.likelihood}
For
data consisting of $n$  independent and identically distributed
(iid) exact failure times and right-censored observations,
with no explanatory variables, the likelihood is
\begin{align}
\nonumber
\like( \DATA | \mu,\sigma) &= \prod_{i=1}^{n}
\like(\data_{i} | \mu,\sigma)
\\
\label{equation:log.location.scale.likelihood}
 &= \prod_{i=1}^{n}
\left\{ \frac{1}{\sigma t_{i}} \, \phi
\left[\frac{ \log(t_{i}) -\mu}{\sigma}
\right]
\right\}^{\delta_{i}}
\times
\left\{1-\Phi \left[\frac{ \log(t_{i}) -\mu}{\sigma}
\right] \right\}^{1-\delta_{i}},
\end{align}
where, for observation $i$, $\data_{i}=(t_{i}, \delta_{i})$, $t_{i}$
is either a failure time or a right-censored time, $\delta_{i}=1$
for an exact failure time, and $\delta_{i}=0$ for a right-censored
observation. For interval-censored observations with lower and
upper interval endpoints $t_{L,i}$ and $t_{U,i}$, the factor on the
left in (\ref{equation:log.location.scale.likelihood}) is replaced
by
\begin{align}
\label{equation:interval.censoring.contribution}
\like( \DATA | \mu,\sigma) &=  \prod_{i=1}^{n}
\left\{\Phi \left[\frac{ \log(t_{U,i}) -\mu}{\sigma}\right] - \Phi
\left[\frac{ \log(t_{L,i}) -\mu}{\sigma}\right] \right\}^{\delta_{i}}
\times
\left\{1-\Phi \left[\frac{ \log(t_{i}) -\mu}{\sigma}
\right] \right\}^{1-\delta_{i}}.
\end{align}
 For a left-censored observation (an interval that starts at zero),
$\Phi[(\log(t_{L,i}) -\mu)/\sigma]$  in
 (\ref{equation:interval.censoring.contribution}) is replaced by zero. For
 more information about likelihoods for censored data, see
 Chapters 2, 7, and 8 in \citet{MeekerEscobarPascual2022}.  For many
 purposes (e.g., computational and for the development of theory),
 it is convenient to use the log-likelihood $\loglike( \DATA | \mu,\sigma) =
 \log[\like( \DATA | \mu,\sigma)]$.

\section{Using Bayesian Methods and Prior Information in Reliability
  Applications}
\label{section:bayesian.methods.prior.information}
As we saw in the motivating examples in
Section~\ref{section:motivating.examples}, the data in reliability
applications often have few failures and thus contain little
information. Engineers, however, often have additional useful, but
imprecise information that can be combined with the limited
data. For example, if failures are caused by a wearout mechanism,
the hazard function would be increasing and thus the Weibull shape
parameter would be greater than one. Previous experience with
a particular failure mechanism in the same material may allow
bounding a Weibull or lognormal shape parameter more precisely.
On the other hand,
there may not be information that can be used to set an informative
prior distribution on a scale parameter. In such cases, the
informative prior for the shape parameter needs to be used in
conjunction with a noninformative or weakly information prior for
the scale parameter.

\subsection{Bayes' Theorem}

 Bayes' theorem for continuous parameters in $\thetavec$ can be
 written as
\begin{equation}
\label{equation:bayes.theorem}
\pi(\thetavec | \DATA)=
\frac{\like(\DATA | \thetavec ) \pi(\thetavec) }
{ \int \like(\DATA | \thetadummyvec^{\prime} ) \pi(\thetadummyvec^{\prime} ) d \thetadummyvec^{\prime} }
\end{equation}
where the joint prior distribution $\pi(\thetavec)$ quantifies the
available prior information about the unknown parameters in
$\thetavec$. The output of (\ref{equation:bayes.theorem}) is
$\pi(\thetavec|\DATA)$, the joint posterior distribution for
$\thetavec$, reflecting knowledge of $\thetavec$ after the
information in the data and the prior distribution have been
combined. For (log-)location-scale distributions used here,
$\thetavec=(\mu, \sigma)$.

\subsection{Reparameterization}
\label{section:reparameterization}
In many situations, it is important to replace the traditional
parameters (e.g., $\mu$ and $\sigma$ for log-location-scale distributions)
with alternative parameters. For a
log-location-scale distribution, it is useful to replace the usual scale
parameter $\exp(\mu)$ with a particular quantile $t_{p_{r}}$ as an
alternative scale parameter for a value of $p_{r}$ that is chosen in
a purposeful manner. Doing this has important advantages for the
following reasons.
\begin{enumerate}[itemsep=1mm, parsep=0pt]
\item
Elicitation of a prior distribution is facilitated because the parameters have
practical interpretations and are familiar to practitioners.
\item
When prior knowledge is accumulated based on experience involving
heavy censoring the traditional parameters $\mu$ and $\sigma$ would
have strong correlation. Using $t_{p_{r}}$ and $\sigma$ for a
suitably specified value of $p_{r}$ would allow specifying the joint prior
density as a product of marginal densities.
\item
The numerical performance of MCMC algorithms is generally better when
strong correlation is avoided.
\end{enumerate}

A useful reparameterization for the Weibull distribution replaces
$\eta$ with a particular distribution quantile that could be
suggested by available data. For
heavily right-censored data from a high-reliability
product, this would be a lower-tail quantile of the failure-time
distribution. For example, if in certain applications one typically
sees 10\% of a population of units failing, then something like the
$0.05$ quantile would be a more appropriate scale parameter. Also,
it is easier to elicit prior information for such a small quantile
compared to the time at which a proportion 0.63 would fail.

The $p_{r}$ quantile of the Weibull
distribution is $t_{p_{r}}=\weibscale \left
[-\log(1-p_{r})\right ]^{1/\beta}$.  Replacing $\weibscale$ with the
equivalent expression $\eta=t_{p_{r}}/[-\log(1-p_{r})]^{1/\beta}$ in
(\ref{equation:bayes.weibull.cdf}) provides a reparameterized version of
the Weibull distribution:
\begin{align}
\nonumber
\Pr(T \leq t;t_{p_{r}},\beta ) = F(t;t_{p_{r}},\beta)&=1-
\exp \left [-\left (\frac{t}{t_{p_{r}}/[-\log(1-p_{r})]^{1/\beta}} \right )^{\beta}
\right ]\\[0.5ex]
\label{equation:reparameterized.weibull.cdf}
&=1-\exp\left[\log(1-p_{r})\left(\frac{t}{t_{p_{r}}}\right)^{\beta}\right],
\,\, t > 0.
\end{align}
The latter expression shows that $t_{p_{r}}$ is an alternative scale parameter.

Especially when there is heavy censoring (i.e., only a small
fraction failing), estimation of $(t_{p_{r}}, \beta)$ will be more
stable than estimating $(\eta, \beta)$, for some appropriately
chosen value of $p_{r}$. Thus one could choose $p_{r}$, based on the
data, by taking the largest value of the nonparametric estimate of
$F(t)$ and dividing it by two, assuring that the parameter $t_{p_{r}}$
is within the data.  Another alternative is to choose $p_{r}$ based
on engineering knowledge that would allow elicitation of an
informative or weakly informative prior distribution for
$t_{p_{r}}$.  Note that it is possible to have two separate
reparameterizations (one for estimation and one for elicitation), as
a prior with one parameterization can be easily translated into a
prior for another parameterization. Usually, however, the
elicitation-motivated reparameterization will be sufficient for both
purposes.

\begin{figure}[!htbp]
\begin{tabular}{cc}
\rsplidafiguresize{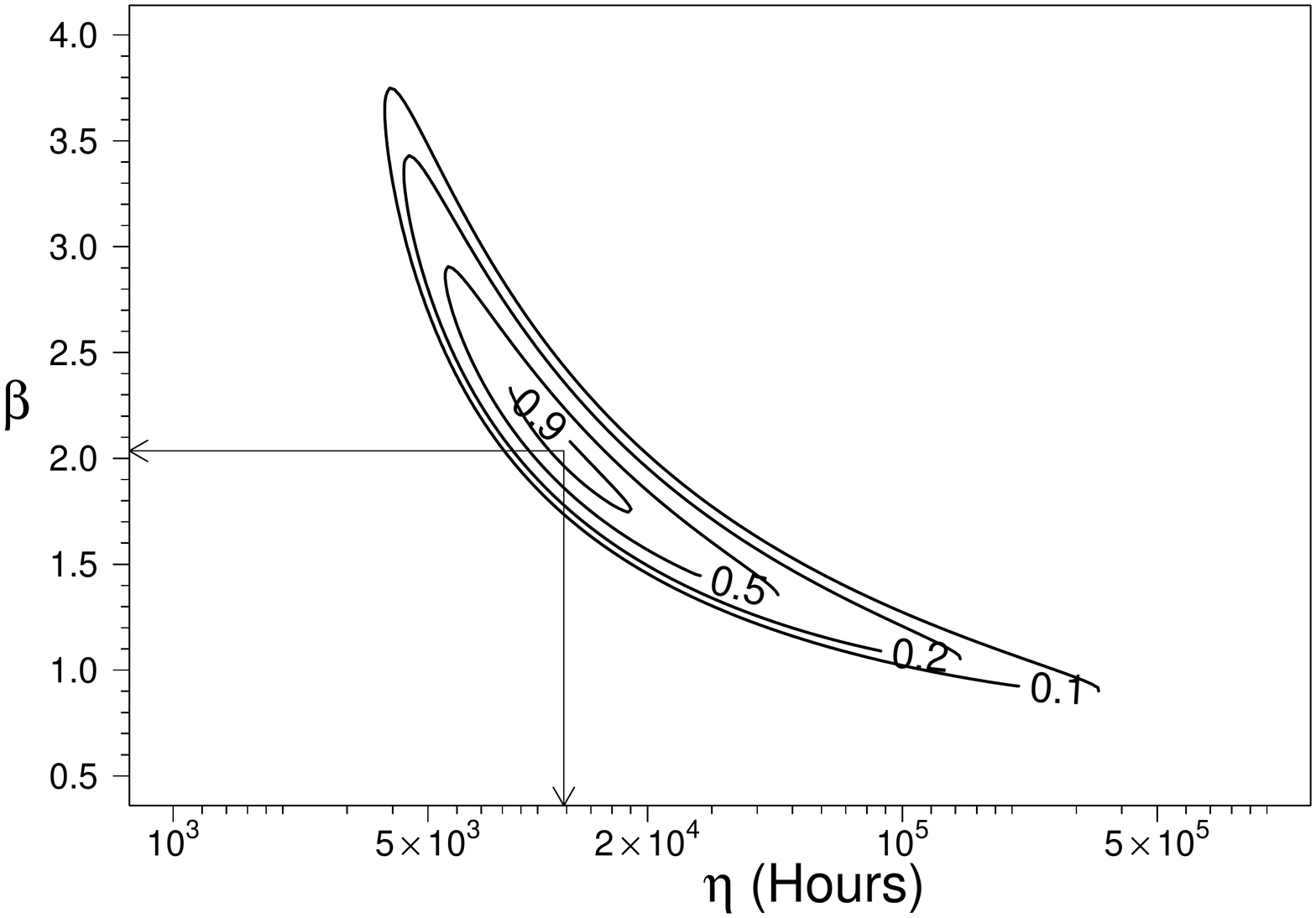}{3.2in}
&
\rsplidafiguresize{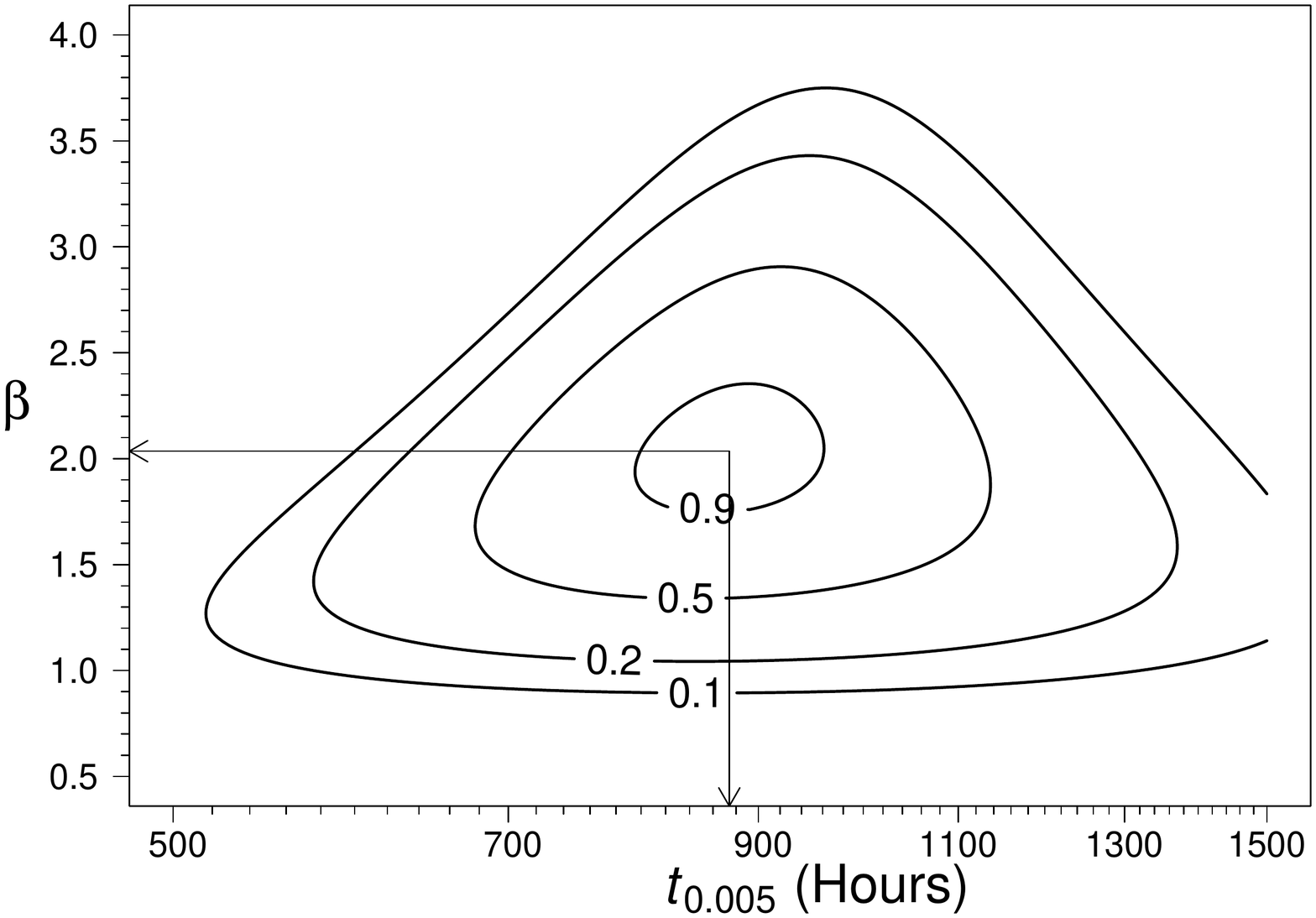}{3.2in}\\[-4ex]
\rsplidafiguresize{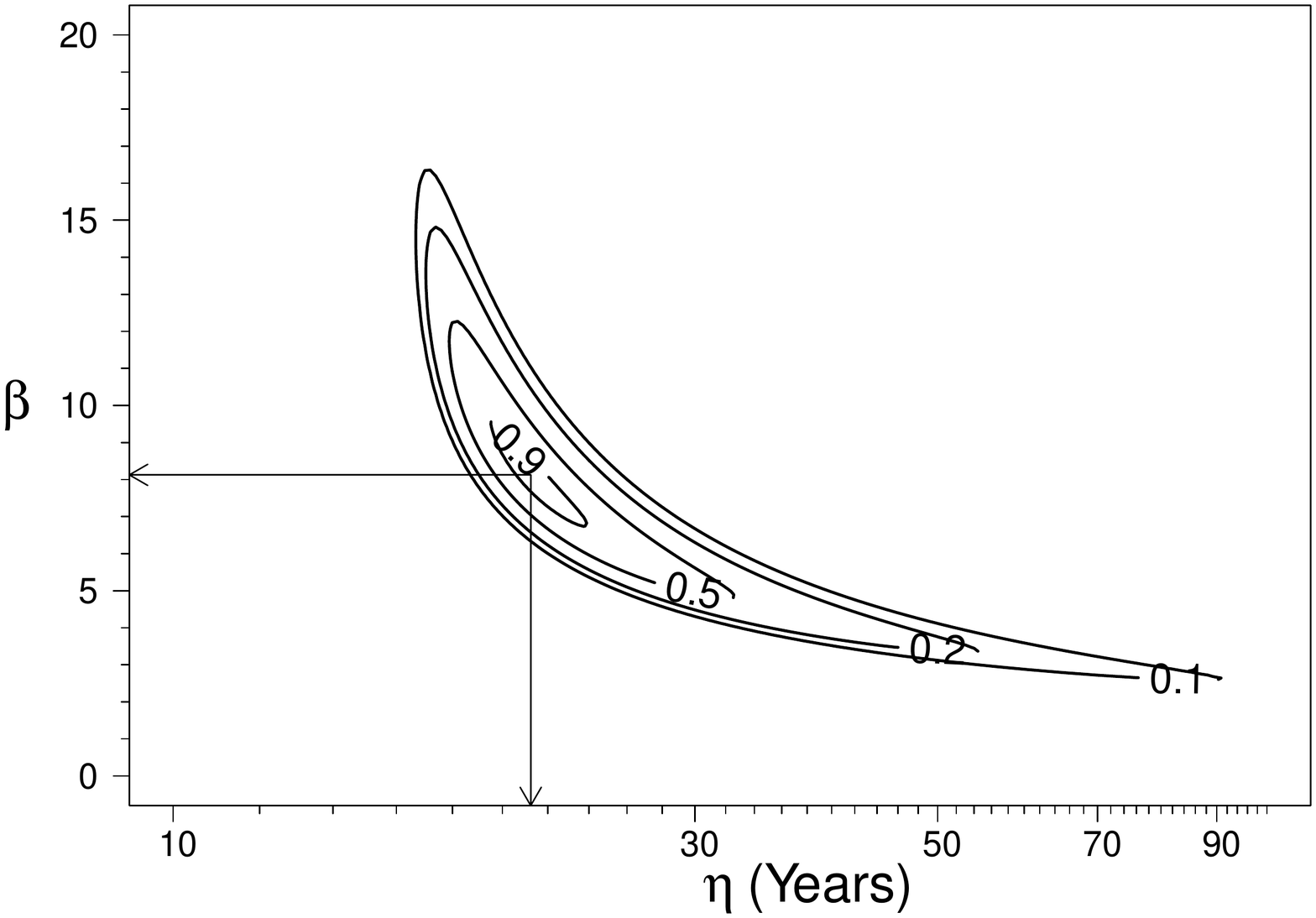}{3.2in}
&
\rsplidafiguresize{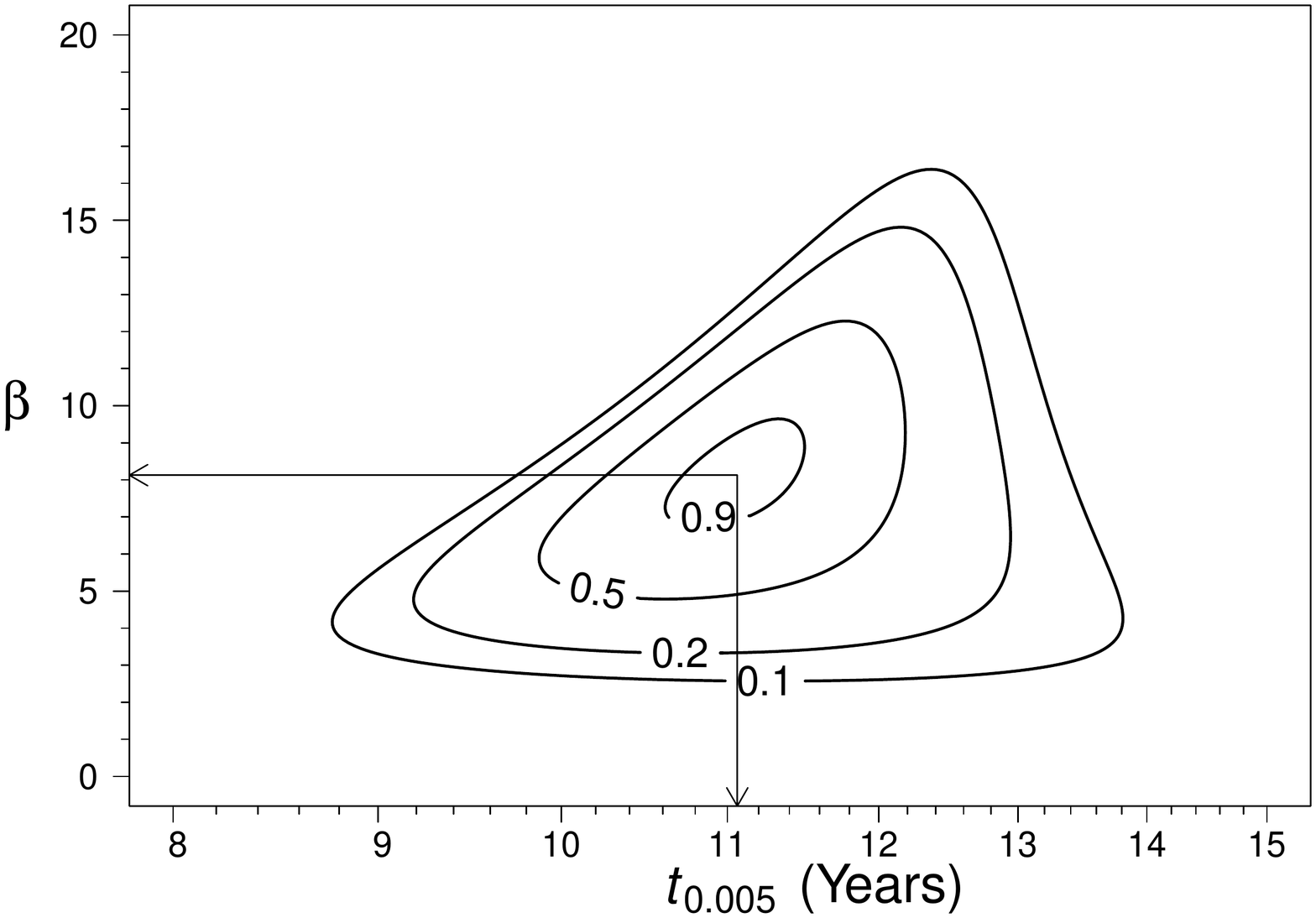}{3.2in}
\end{tabular}
\caption{Likelihood contour plots for the bearing-cage (top) and
  rocket motor data (bottom) in the traditional
  $(\eta, \beta)$ parameterization (left) and in the
  $(t_{0.005}, \beta)$ reparameterization (right).}
\label{figure:likelihood.contour.plots.reparameterization}
\end{figure}

Figure~\ref{figure:likelihood.contour.plots.reparameterization}
illustrates such reparameterizations for the bearing-cage and
rocket-motor data. For these examples, $p_{r}$ was chosen to provide a
well-behaved likelihood surface. The lognormal and other
log-location-scale distributions can be similarly reparameterized.

\section{The Fisher Information Matrix}
\label{section:lls.fim}
As described in Section~\ref{section:noninformative.prior} and
Section~\ref{S.section:derivations.noninformative.priors} of the
appendix, the Fisher information matrix (FIM)
is used to define certain noninformative prior
distributions. This section shows how to compute the FIM for
(log-)location-scale distributions and different kinds of censoring.
\subsection{Scaled Fisher Information Matrix Elements}
\label{section:scaled.fim.elements}
The \textit{scaled} FIM elements are:
\begin{align}
  \begin{split}
\label{equation:lls.fim}
f_{11}(z_{c}) &= \frac{\sigma^2}{n} E\left [-\frac{\partial^2 \log \like(\mu,\sigma) }{\partial\mu^2} \right]
\\[2ex]
f_{12}(z_{c}) &= \frac{\sigma^2}{n} E\left
[-\frac{\partial^2\log \like(\mu,\sigma) }{\partial\mu\partial\sigma} \right]
\\[2ex]
f_{22}(z_{c}) &= \frac{\sigma^2}{n} E\left
[-\frac{\partial^2 \log \like(\mu,\sigma) }{\partial \sigma^2}\right ].
    \end{split}
\end{align}
The $\sigma^2/n$ term cancels with a term $n/\sigma^2$ arising from
the expectations in (\ref{equation:lls.fim}) and thus these scaled
FIM elements depend only on $z_{c}$ and the assumed distribution (but
not on $n$ or $\sigma$).

For \typeI{} (time) censoring, $z_{c} = [\log(t_{c})-\mu]/\sigma$ is a
standardized censoring time where $t_{c}$ is the censoring time and
$p_{c}=\Phi(z_{c})$ is the expected fraction failing. For \typeII{} (failure)
censoring $z_{c} = \Phi^{-1}(r/n)$ where $r/n$ is the given fraction
failing.  An algorithm to compute these elements is given by
\citet{EscobarMeeker1994} and implemented in the
function \texttt{lsinf} in the R package
\texttt{lsinf} \citep{Meeker2022}.

\subsection{Fisher Information for \typeI{} and \typeII{} Censored Samples}
\label{section:fisher.typeI.typeII.censored}
For a sample of $n$ iid observations, singly censored (i.e., all
censoring is at one time point) at time
$t_{c}$, the FIM for $(\mu, \sigma)$ is
\begin{align}
\label{equation:fisher-mu-sigma}
I_{(\mu, \sigma)} &= \frac{n}{\sigma^{2}}
\left[\begin{array}{lr}
f_{11}(z_{c})  & f_{12}(z_{c})\\[1ex]
\textrm{symmetric} & f_{22}(z_{c})
\end{array}\right].
\end{align}
With \typeII{} censoring, the scaled FIM $(\sigma^{2}/n)\, I_{(\mu,
  \sigma)}$ depends on $r$, the known number of failures, and has no
unknown parameters and thus does not depend on any unknown parameters.  With
\typeI{} censoring, the scaled FIM depends on $p_{c}=F(t_{c}; \mu, \sigma)$, the
unknown expected fraction failing at time $t_{c}$. 

To simplify the presentation in the remainder of this
paper, when it is possible, we will suppress the dependency of the
$f_{ij}$ elements on $z_{c}$.  That is, for example, we write
$f_{11}$ instead of $f_{11}(z_{c})$.

\subsection{Fisher Information Matrix for Randomly Censored Samples}
\label{section:fisher.random.censored}
Random censoring arises for different reasons such as staggered
entry, differing use rates in the population, and competing
risks. The competing risk model provides a convenient model to
describe or characterize random censoring. Suppose that $T$ is a
random failure time having a log-location-scale distribution with
parameters $(\mu, \sigma)$ and $C$ is a random censoring time for a
unit. Then the unit fails if $T \leq C$ and is censored if $T >
C$. If $\log(C)$ has a pdf $h(x)$, then, using conditional
expectation, as outlined in \citet{EscobarMeeker1998}, the FIM for a
sample of size $n$ is
\begin{align}
\label{equation.fim.random.censoring}
I_{(\mu, \sigma)} &= \frac{n}{\sigma^{2}}
\left[\begin{array}{lr}
\int_{-\infty}^{\infty}f_{11}(w)h(x)dx  & \int_{-\infty}^{\infty}f_{12}(w)h(x)dx\\[1ex]
\textrm{symmetric} &  \int_{-\infty}^{\infty}f_{22}(w)h(x)dx
\end{array}\right],
\end{align}
where $w=(x-\mu)/\sigma$.

\section{Noninformative Prior Distributions for Log-Location-Scale Distributions}
\label{section:noninformative.prior}
\subsection{Motivation}
In many (if not most) applications of Bayesian methods, there is a
desire to analyze data without using any information that might be
available to specify a prior distribution (sometimes known as
objective-Bayesian analysis). This would be the case, for example,
when interested parties might not agree on a subjective prior (e.g.,
engineers and managers or manufacturers and consumers) or when there
is a need to avoid having to defend an assumed prior distribution
(e.g., in legal or regulatory proceedings). 
Practitioners may be unable to specify their belief using a
probability distribution due to lack of statistical expertise thus
requiring the use of a default prior.
In these situations, an
alternative is to use what is generically called a noninformative
prior distribution.  Several different methods for specifying a
noninformative prior distribution have been suggested.

\subsection{Previous Work on Noninformative Prior Distributions}
\label{section:noninformative.prior.literature}
There is extensive literature concerning noninformative prior
distributions. Here we mention work most closely related to ours.
\citet{Bernardo1979} reviews the early literature and describes
criteria and methods for constructing reference prior
distributions. \citet{BergerBernardo1992a} extend previous work on
reference prior distributions, focusing on multiparameter models and
recommend a specification of order, based on parameter
importance. \citet{BergerBernardo1992b} provide an updated
literature review of this area and describe a general algorithm for
finding a reference prior for continuous multiparameter models with
a given parameter ordering.  \citet{Sun1997} reviews earlier work
showing various conditions for second-order and third-order
probability matching
priors for two-parameter distributions and applies these results to
the Weibull distribution demonstrating that certain reference priors
meet the conditions.  \citet{SunBerger1998} consider informative
priors when there is external information (e.g., for only one
parameter), with the idea of using a reference prior distribution
for other parameters, similar to what we suggest in
Section~\ref{section:combining.informative.with.noninformative}.
\citet{AbbasTang2015,AbbasTang2016} describe reference prior
distributions for the Fr\'{e}chet and loglogistic distributions.
\citet[Chapter 5]{GhoshDelampadyTapas2006} provide a summary of
methods and operational details for obtaining noninformative prior
distributions.

\subsection{Parameterization for Prior Distributions}
\label{subsection:param.for.priors}
In our log-location-scale distribution examples,
when specifying a prior distribution we will use the
parameterization $(t_{p_{r}}, \sigma)$ (or $(t_{p_{r}}, \beta=1/\sigma)$ for
the Weibull distribution) because these are the parameters that have
a practical interpretation, allowing elicitation of informative or
specifying weakly informative prior distributions, when needed. When using
algorithms to compute MCMC draws, however, we use the
$(y_{p_{r}}=\log(t_{p_{r}}), \log(\sigma))$ parameterization
because the expressions for the priors
tend to be simpler, MCMC algorithms work better in the unconstrained
parameter space, and plots of the posterior draws tend to be easier to
interpret on the log scales. Prior distributions for $(t_{p_{r}},
\sigma)$ are easily translated into priors for $(\log(t_{p_{r}}),
\log(\sigma))$.  Examples are given in
Section~\ref{S.section:derivations.noninformative.priors} of the
appendix.
\subsection{A Fundamental Principle for Specifying Noninformative
  Prior Distributions}
\label{section:noninformative.prior.fundamental.principle}
There is an important fundamental principle for specifying
noninformative prior distributions in situations where there is only a
small amount of information in the data corresponding to the desired
inference(s). The prior should put negligible
density in parts of the parameter space that are impossible or
clearly implausible and that would otherwise lead to nonnegligible posterior
probability in such parts of the parameter space. This is the
often-stated justification for the use of weakly informative priors
(Section~\ref{section:weakly.informative.prior}). For example, using
a prior $\pi[\log(t_{p_{r}}), \log(\sigma)]\propto 1$ (also known as
``flat'') in
situations with a small number of failures can result (because of
the diffuseness of the likelihood) in non-negligible posterior
probabilities in nonsensical regions of the parameter space.
We have developed our recommendations (summarized in
Section~\ref{section:recommended.lls.prior})
to be consistent with this principle.

\subsection{Jeffreys Prior Distributions}
\label{section:jeffreys.prior.distribution}
The Jeffreys prior can be derived as being proportional to
the square root of the determinant of the FIM
(defined in Section~\ref{section:fisher.typeI.typeII.censored}).
For models with one parameter (e.g., the exponential distribution or
the normal distribution with known standard deviation), the Jeffreys
prior distribution has been shown to provide results (e.g.,
credible, tolerance, and prediction intervals) that are the same as
or close to classical non-Bayesian methods for certain models.

The Jeffreys prior, even if there is more than one parameter, is
invariant to reparameterization. This means if you use
the square root of the determinant of the FIM
definition for a given parameterization you will get a Jeffreys
prior. For any other parameterization, the Jeffreys prior can be
obtained by finding the square root of the determinant
of the FIM in the new parameterization or by
using the transformation of a random variable method.
That is, either method will lead to the same Jeffreys prior distribution.

From (\ref{equation:lls.fim}),
the Jeffreys prior is defined as
\begin{equation}
\label{equation:jeffreys-mu-sigma}
\pi(\mu,\sigma)\propto\sqrt{\left|\text{I}_n(\mu,\sigma)\right|}=\frac{n}{\sigma^2}\sqrt{f_{11}f_{22}-f_{12}^2}.
\end{equation}
Here the $f_{ij}$ values are scaled elements of the FIM defined in
Section~\ref{section:scaled.fim.elements}, which depend on the
standardized censoring time $z_{c}$.  For \typeII{} censoring or
complete data, $z_{c}=\Phi^{-1}(r/n)$ is a known constant, so that
the Jeffreys prior is $\pi(\mu,\sigma)\propto{1}/{\sigma^2}.$ Then,
using transformation of random variables, the Jeffreys prior for
the unrestricted parameterization is $\pi[\mu,\log(\sigma)]\propto1/\sigma$
for \typeII{} censoring (or complete data).

The priors for \typeI{} censoring are more complicated because the
$f_{ij}$ elements depend on the unknown parameters through
$z_{c}=(\log(t_{c})-\mu)/\sigma$ (more specifically, they depend on
$p_{\fail}=\Phi(z_{c})$, the unknown expected fraction failing at
the censoring time $t_{c}$). For \typeI{} censoring, the Jeffreys
prior is
\[
\pi(\mu,\sigma)\propto\frac{1}{\sigma^2}\sqrt{f_{11}f_{22}-f_{12}^2}.
\]
Again, for the unrestricted parameterization, the \typeI{} censoring
Jeffreys prior is
\begin{align*}
\pi[\mu, \log(\sigma)] & \propto
\frac{1}{\sigma}\sqrt{f_{11}f_{22}-f_{12}^2}.
\end{align*}

For (log-)location-scale distributions (and other distributions with
more than one parameter), the Jeffreys priors have well-known
deficiencies \citep[][page 182]{Jeffreys1961}. Also, in models with
more than one parameter, Jeffreys priors 
may not have the desirable classical properties of agreeing with reference
priors that are probability matching \citep[e.g.,][]{Sun1997}.

\subsection{Independence Jeffreys Prior Distributions}
\label{section:ij.prior}
 The independence Jeffreys (IJ) prior  (also
known as the modified Jeffreys prior) is obtained by finding the
Conditional Jeffreys (CJ) prior for each parameter, assuming that it is the only
unknown parameter, and then using the product of these conditional
priors as the joint prior, as if the parameters were independent
random variables (but notably, they are not independent).
In contrast to the Jeffreys prior, for
 (log-)location-scale distributions,
the IJ prior distribution has an appealing
property. In particular, it provides,
for complete and \typeII{} censored data, the same exact inferences
(i.e., the credible/confidence intervals procedures with
coverage probabilities that are
the same as the nominal credible/confidence level) as
non-Bayesian pivotal-based methods (and approximately the same for
other kinds of censoring). This result is given for complete data in
\citet{DiCiccioKuffnerYoungAlastair2017} but is also true for
\typeII{} censored data \citep[e.g., page 565 in][]{Lawless2003}.

\subsubsection{\typeII{} censoring IJ prior distributions}
  For
\typeII{} censoring or complete data, because $f_{11}$ and $f_{22}$
are known constants, the CJ prior for $\mu$ is
$\pi(\mu|\sigma)\propto(1/\sigma)\sqrt{f_{11}}\propto1$.  The
CJ prior for $\sigma$ is
$\pi(\sigma|\mu)\propto(1/\sigma)\sqrt{f_{22}}\propto1/\sigma$.  So,
the IJ prior for $(\mu,\sigma)$  under \typeII{} censoring is
$\pi(\mu,\sigma)\propto\pi(\mu|\sigma)\pi(\sigma|\mu)\propto1/\sigma$.

Again, using transformation of random variables, the CJ prior for
$\log(t_{p_{r}})$ given $\log(\sigma)$ is
$\pi[\log(t_{p_{r}})|\log(\sigma)]\propto 1$.  The CJ prior for
$\log(\sigma)$ given $\log(t_{p_{r}})$ is $\pi[\log(\sigma)
  |\log(t_{p_{r}})] \propto 1$. Thus the IJ prior is
$\pi[\log(t_{p_{r}}),\log(\sigma)]\propto 1$, which we (following
common usage) call ``flat.''

\subsubsection{\typeI{} censoring IJ prior distributions}
For \typeI{} censoring, the CJ prior for $\mu$, when $\sigma$ is a known
constant, is
$\pi(\mu|\sigma)\propto(1/\sigma)\sqrt{f_{11}}\propto\sqrt{f_{11}}$
and the CJ prior for $\sigma$ is
$\pi(\sigma|\mu)\propto(1/\sigma)\sqrt{f_{22}}$.  So, the
IJ prior for $(\mu,\sigma)$ is
$$
\pi(\mu,\sigma)\propto\frac{1}{\sigma}\sqrt{f_{11}f_{22}}.
$$
Then,
using transformation of random variables, the
CJ prior for $\mu$ given $\log(\sigma)$ is
$\pi[\mu|\log(\sigma)] \propto \sqrt{f_{11}}$. The CJ
prior for $\log(\sigma)$ given $\mu$ is
\begin{align*}
\pi[\log(\sigma) | \mu] &\propto
\sqrt{f_{22}}.
\end{align*}
So, the IJ prior is
$\pi[\mu, \log(\sigma)] \propto \sqrt{f_{11}f_{22}}$.

\subsubsection{Results for other parameterizations}
The FIM for the parameterization $(\log(t_{p_{r}}),\log(\sigma))$ is
easily obtained by using the delta method on the inverse of the FIM
in the $(\mu, \sigma)$ parameterization given in
(\ref{section:fisher.typeI.typeII.censored}). Details are given in
Appendix
Section~\ref{section:priors.y.log.sigam.parameterization}. Then
expressions for the Jeffreys, CJ, and IJ priors are obtained by
following the same steps described earlier in this section and
leads, for example, to the IJ prior
\begin{align}
\nonumber
\pi[\log(t_{p_{r}}),\log(\sigma)] &\propto
\pi[\log(t_{p_{r}})|\log(\sigma)] \times \pi[\log(\sigma)|\log(t_{p_{r}})]\\[1ex]
\label{equation:IJ.typeI.yp.log.sigma}
& \propto
\sqrt{f_{11}\left\{f_{11}[\Phi^{-1}(p_{r})]^2-2f_{12}\Phi^{-1}(p_{r})+f_{22}\right\}},
\end{align}
where, again, the $f_{ij}$ values are scaled elements of the FIM
defined in Section~\ref{section:scaled.fim.elements}, which depend
on the standardized censoring time $z_{c}=[\log(t_{c})-\mu]/\sigma$
or $p_{c}=\Phi(z_{c})$.

Relative to flat priors, these IJ priors tend to be more
consistent with the fundamental principle
described in
Section~\ref{section:noninformative.prior.fundamental.principle},
and provide advantages over traditional noninformative priors. This is
explained in Section~\ref{section:implementing.ij.priors} and
demonstrated in Section~\ref{simulation.results.conclusions}.

\subsection{Reference Prior Distributions}
\label{section:reference.prior.distributions}
Another well-studied approach to finding a noninformative prior
distribution is to use a reference prior (e.g., \citet{Bernardo1979}
and \citet{BergerBernardo1992a}).  Generally, a reference
prior is the prior distribution that maximizes the
Kullback-Leibler divergence
between the prior and the
expected posterior distribution. 
An \textit{ordered reference prior} specifies the order of
importance of the parameters, and different reference priors can
arise, depending on the specified order of
importance of the parameters (or functions of the
parameters). 
Interestingly, certain definitions of reference priors
lead exactly to the IJ priors (and others lead to the Jeffreys
prior).

As an example, with \typeII{} censoring or complete data, the IJ
prior for the log-location-scale distribution parameters $(t_{p_{r}},
\sigma)$ is (proportional to) $1/(t_{p_{r}}\sigma)$ and for
the parameterization $(\log(t_{p_{r}}), \log(\sigma))$, the IJ prior is flat (i.e.,
uniform over the entire $(\log(t_{p_{r}}), \log(\sigma))$ plane for any
$p_{r}$). As shown in
Section~\ref{S.section.parameterization.tp.sigma} of the
appendix, these IJ priors are also reference priors when
either $t_{p_{r}}$ or $\sigma$ is the parameter of first importance.

As described in
Section~\ref{S.section:reference.priors.for.typeI.censoring} of the
appendix, expressions for computing reference priors
are not readily available for \typeI{} censoring. However, given the
equivalence of IJ and ordered reference priors described above for
\typeII{} censoring, we expect that the IJ priors for \typeI{}
censoring (Section~\ref{section:ij.prior}) would provide a good
approximation to the corresponding reference priors.

\subsection{Improper Priors and Posteriors}
\label{section:improper.with.few.failures}
Noninformative priors are generally improper (i.e. their integrals
over the parameter space are not finite). Thus, when using such prior
distributions (e.g., flat, IJ, or CJ combined with a proper marginal
for the other parameter), it is important to assure that there is a
sufficient amount of information in the data that the posterior will
be proper. \citet{RamosRamosLouzada2020} give conditions under which
a Weibull posterior distribution will be proper for some improper
priors, suggesting that having two failures is sufficient to result
in a proper posterior.  We did extensive numerical experiments with
flat and IJ prior distributions and simulated \typeI-censored
samples with different numbers of failures. These experiments
indicated that the posterior is proper and reasonably well behaved
when there are at least three failures.  With only two failures and
a flat prior, however, even when an appropriate parameterization was
used, we were not able to find a stable sampler.  For example, the
sampler would often return large numbers or infinite values of the
parameters, perhaps due to limitations in computer floating-point
representation of real numbers (we also encountered infinite values
in our experiments with three failures, but their occurrence was
relatively rare).
In summary, even if a proper posterior is assured theoretically with
two failures, there may be practical difficulties.

\subsection{A Summary of Noninformative Priors for
  (Log-)Location-Scale Distributions}
\label{section:summary.of.noninformative.priors}

Table~\ref{table:summary.noninformative.priors} summarizes Jeffreys,
IJ, and reference noninformative prior distributions for
(log-)location-scale distributions using different
parameterizations.
Section~\ref{S.section:derivations.noninformative.priors} of the
appendix gives derivations of these priors.
The relationships and correspondences (some of which have been noted
previously in the literature) are interesting. For example,
\begin{itemize}[itemsep=1mm, parsep=0pt]
\item
For all parameterizations, the usual Jeffreys prior is the same as
the reference prior when no parameter-importance is specified
\citep[as noted by][page 1,350]{KassWasserman1996}.
\item
When the distribution parameters are defined as the shape parameter
$\sigma$ and a scale parameter (e.g., the traditional $\exp(\mu)$ or
the $p$ quantile $t_{p}$ for any $p$), the ordered reference prior
is the same as the the IJ prior, irrespective of the ordering. This
equivalence property also holds for one-to-one functions of the
individual parameters (e.g., when $\sigma$ is replaced by
$\log(\sigma)$).
\item
Related to the previous point, although IJ priors for
log-location-scale distributions are not in general invariant to
reparameterization (in the sense described in
Section~\ref{section:jeffreys.prior.distribution}), they are
invariant to one-to-one monotone reparameterizations of either or
both of the $(t_{p},\sigma)$ parameters. The proof of this result is
given in
Section~\ref{S.section:invariant.one.one.reparameterization} of the
appendix.
\item
This IJ/ordered-reference equivalence property does \emph{not} hold
when the distribution parameters are defined as the shape parameter
$\sigma$ and $\zeta_{e}=[\log(t_{e})-\mu)/\sigma]$ and $\sigma$ is
in second order.
\end{itemize}

\begin{sidewaystable}
\caption{Improper noninformative joint prior distributions
  $\pi(\theta_{1},\theta_{2})$ for different parameterizations for
  log-location-scale distributions based on  \typeII{} and  \typeI{}
  censored data
\label{table:summary.noninformative.priors}}
\begin{tabular}{lcc}
\toprule
& \multicolumn{2}{c}{Type of censoring} \\
\cmidrule(l{3em}r{5.5em}){2-3}
Prior type and parameterization & \typeII{} & \typeI{} \\
\midrule
{\small Jeffreys ($\mu, \sigma$)} & {\small $1/\sigma^{2}$}  & {\small $(1/\sigma^2)\sqrt{f_{11}f_{22}-f_{12}^2}$} \\[0.5ex]
{\small Jeffreys ($\log(t_{p_{r}}), \sigma$)}  & {\small $1/\sigma^{2}$}  & {\small $(1/\sigma^2)\sqrt{f_{11}f_{22}-f_{12}^2}$} \\[0.5ex]
{\small Jeffreys ($t_{p_{r}}, \sigma$)} & {\small $1/(t_{p_{r}}\sigma^{2})$ } &  {\small  $1/(t_{p_{r}}\sigma^2)\sqrt{f_{11}f_{22}-f_{12}^2}$ } \\[0.5ex]
{\small Jeffreys ($\log(t_{p}), \log(\sigma)$)} &  {\small $1/\sigma$}   &  {\small $(1/\sigma)\sqrt{f_{11}f_{22}-f_{12}^2}$} \\[0.5ex]
{\small Jeffreys ($\zeta_{e}, \sigma$)} &{\small $1/\sigma$} &  {\small  $(1/\sigma)\sqrt{f_{11}f_{22}-f_{12}^2}$} \\[0.5ex]
{\small Jeffreys ($\zeta_{e}, \log(\sigma)$)} &{\small 1} &  {\small  $\sqrt{f_{11}f_{22}-f_{12}^2}$} \\[0.5ex]
{\small IJ  ($\mu, \sigma$)} &   {\small $1/\sigma$}  & {\small $(1/\sigma)\sqrt{f_{11}f_{22}}$} \\[0.5ex]
{\small IJ  ($\log(t_{p_{r}}), \sigma$)} &   {\small $1/\sigma$}  & {\small $(1/\sigma)\sqrt{f_{11}\left\{f_{11}[\Phi^{-1}(p_{r})]^2-2f_{12}\Phi^{-1}(p_{r})+f_{22}\right\}}$} \\[0.5ex]
{\small IJ  ($t_{p_{r}}, \sigma$)} & {\small $1/(t_{p_{r}}\sigma)$ }  &  {\small  $1/(t_{p_{r}}\sigma)\sqrt{f_{11}\left\{f_{11}[\Phi^{-1}(p_{r})]^2-2f_{12}\Phi^{-1}(p_{r})+f_{22}\right\}}$ } \\[0.5ex]
{\small IJ ($\log(t_{p_{r}}), \log(\sigma)$)} &  {\small 1}   &  {\small $\sqrt{f_{11}\left\{f_{11}[\Phi^{-1}(p_{r})]^2-2f_{12}\Phi^{-1}(p_{r})+f_{22}\right\}}$}  \\[0.5ex]
{\small IJ  ($\zeta_{e}, \sigma$)} &  {\small $1/\sigma$}  &  {\small  $(1/\sigma)\sqrt{f_{11}[f_{11}\zeta_{e}^2-2f_{12}\zeta_{e}+f_{22}]}$} \\[0.5ex]
{\small IJ  ($\zeta_{e}, \log(\sigma)$)} &  {\small 1}  &  {\small  $\sqrt{f_{11}[f_{11}\zeta_{e}^2-2f_{12}\zeta_{e}+f_{22}]}$} \\[0.5ex]
{\small Reference $(\log(t_{p_{r}}), \sigma)$} & {\small  $1/\sigma^{2}$}  &  \\[0.5ex]
{\small Reference $(\{\log(t_{p_{r}}), \sigma\})$} &  {\small $1/\sigma$}  &  \\[0.5ex]
{\small Reference $(\{\sigma, \log(t_{p_{r}}) \})$} &  {\small $1/\sigma$}  & \\[0.5ex]
{\small Reference $(t_{p_{r}}, \sigma)$} & {\small $1/(t_{p_{r}}\sigma^{2})$ } & \\[0.5ex]
{\small Reference $(\{t_{p_{r}}, \sigma\})$} & {\small  $1/(t_{p_{r}}\sigma)$ } &  \\[0.5ex]
{\small Reference $(\{\sigma, t_{p_{r}}\})$} & {\small  $1/(t_{p_{r}}\sigma)$ } &  \\[0.5ex]
{\small Reference $(\log(t_{p_{r}}), \log(\sigma)) $}  & {\small $1/\sigma$ } &  \\[0.5ex]
{\small Reference $( \{\log(t_{p_{r}}), \log(\sigma) \})$}  & {\small 1} &  \\[0.5ex]
{\small Reference $(\{\log(\sigma), \log(t_{p_{r}}) \} )$}  & {\small 1} &  \\[0.5ex]
{\small Reference $(\zeta_{e}, \sigma)$} &{\small $1/\sigma$}  &  \\[0.5ex]
{\small Reference $(\{\zeta_{e}, \sigma\})$} &{\small    $\left[\sigma\sqrt{f_{11}\zeta_{e}^2-2f_{12}\zeta_{e}+f_{22}} \right]^{-1}$} &   \\[0.5ex]
{\small Reference $(\{\sigma, \zeta_{e}\})$} &{\small $1/\sigma$}&   \\[0.5ex]
{\small Reference $(\zeta_{e}, \log(\sigma))$} & {\small  1} & \\[0.5ex]
{\small Reference $(\{\zeta_{e}, \log(\sigma)\})$} & {\small  $\left[\sqrt{f_{11}\zeta_{e}^2-2f_{12}\zeta_{e}+f_{22}} \right]^{-1}$} &  \\[0.5ex]
{\small Reference $(\{\log(\sigma), \zeta_{e}\})$} & {\small  1} &  \\[0.5ex]
\bottomrule
\end{tabular}
\begin{tablenotes}
\item[1] The scaled FIM elements $f_{11}$,
  $f_{12}$, and $f_{22}$ depend on the standardized censoring time
  $z_{c}=[\log(t_{c})-\mu]/\sigma$. Parameters shown within $\{\dots\}$
  indicate parameter-importance order, if there is one. The parameter
  $\zeta_{e}=[\log(t_{e})-\mu)/\sigma]$ where $t_{e}$ is the time at which a
  failure probability is to be estimated. The three reference
  priors for $(\mu, \sigma)$ are exactly the same as for
  $(\log(t_{p_{r}}), \sigma)$ and are thus not presented here.
\end{tablenotes}
\end{sidewaystable}

\subsection{Implementing and Interpreting the IJ Prior Distributions}
\label{section:implementing.ij.priors}
Implementing the IJ priors (e.g.,
$\pi(\log(t_{p_{r}}), \log(\sigma)) \propto 1$ or
$\pi(\log(t_{p_{r}}), \sigma) \propto 1/\sigma$) that arise with
complete data or \typeII{} censoring is straightforward.
For \typeI{} censoring we describe
the IJ prior for $(\log(t_{p_{r}}),\log(\sigma))$
in (\ref{equation:IJ.typeI.yp.log.sigma}) as an example.
The prior is well defined over the entire
parameter space (e.g., the real plane for
$(\log(t_{p_{r}}),\log(\sigma))$. For the given $p_{r}$ and censoring
time $t_{c}$ (the only inputs needed),
$z_{c}=[\log(t_{c})-\mu]/\sigma$ (the argument for the
$f_{ij}$ values) is computed as a function of
$(\log(t_{p_{r}}),\log(\sigma))$ (i.e., by using $\mu =
\log(t_{p_{r}})- \Phi^{-1}(p_{r})\sigma$). 

When specifying a reparameterization that replaces the usual scale
parameter $\exp(\mu)$ of a log-location-scale distribution with a
specific quantile (i.e., $t_{p_{r}}$), the specific value of $p_{r}$
is not critical and is usually chosen so that the likelihood
contours are well behaved (as illustrated in
Figure~\ref{figure:likelihood.contour.plots.reparameterization}). Due
to the invariance property of ML estimators, ML estimates of
$\sigma$, quantiles, and cdf values will not be affected by the
choice of $p_{r}$. In the case of Bayesian estimation, with a
noninformative prior distribution (e.g., the priors outlined in
Table~\ref{table:summary.noninformative.priors}) the invariance
property for the choice of $p_{r}$ will hold for \typeII{} censoring
and otherwise approximately. Extensive numerical experiments
suggest that the approximation is excellent with the IJ prior and
\typeI{} censoring.

\begin{figure}[!htbp]
\begin{tabular}{cc}
\rsplidafiguresize{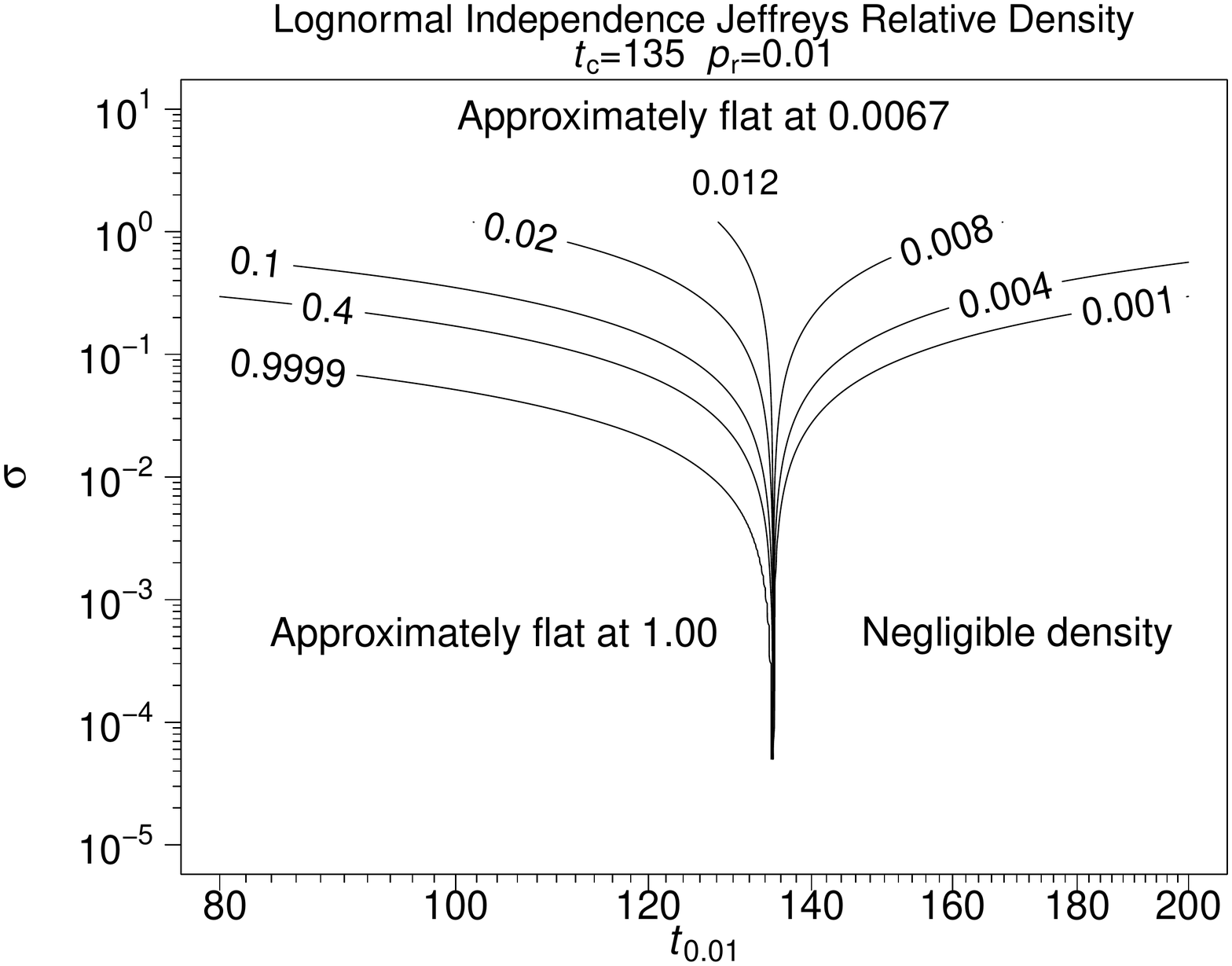}{3.3in}&
\rsplidafiguresize{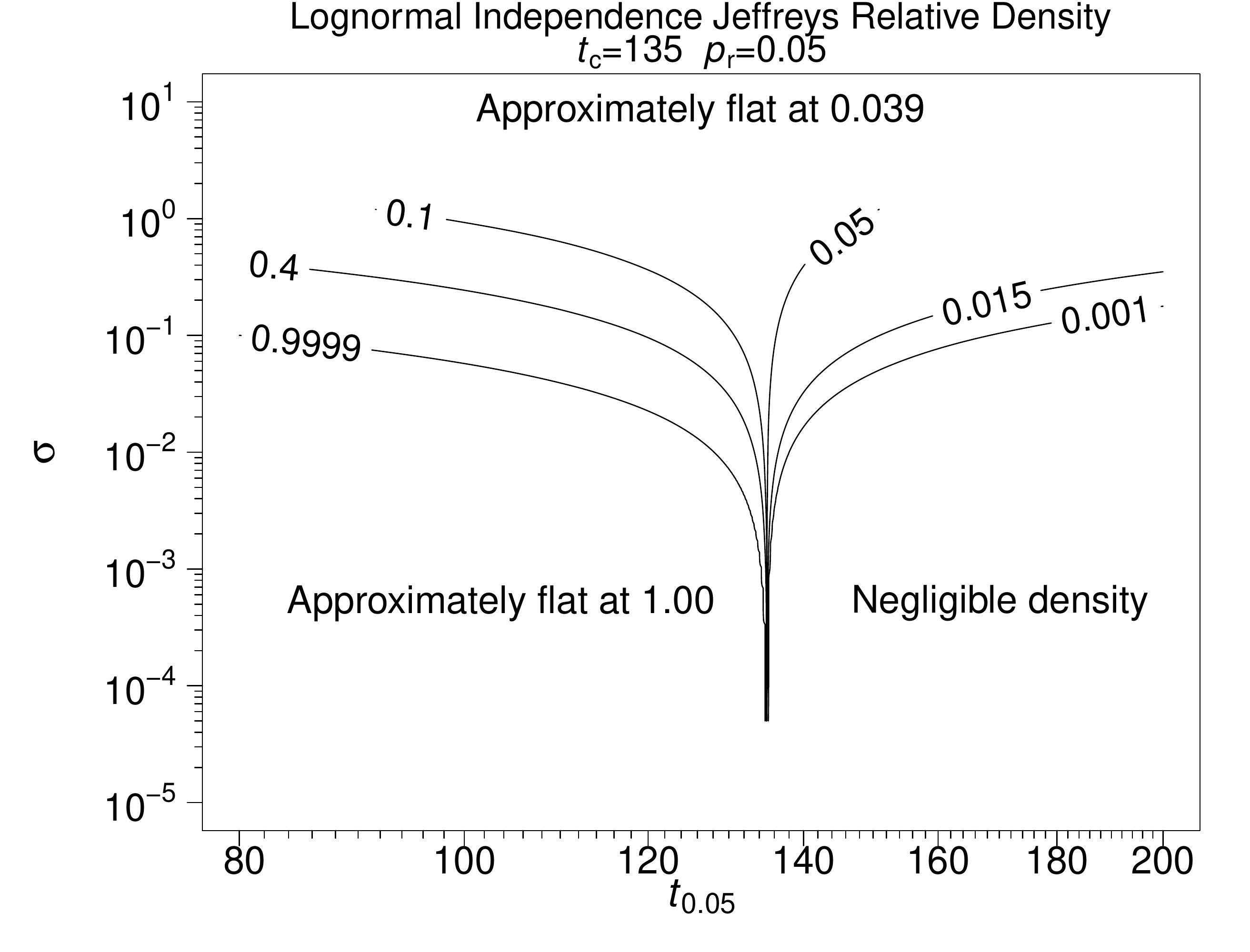}{3.3in}\\
\rsplidafiguresize{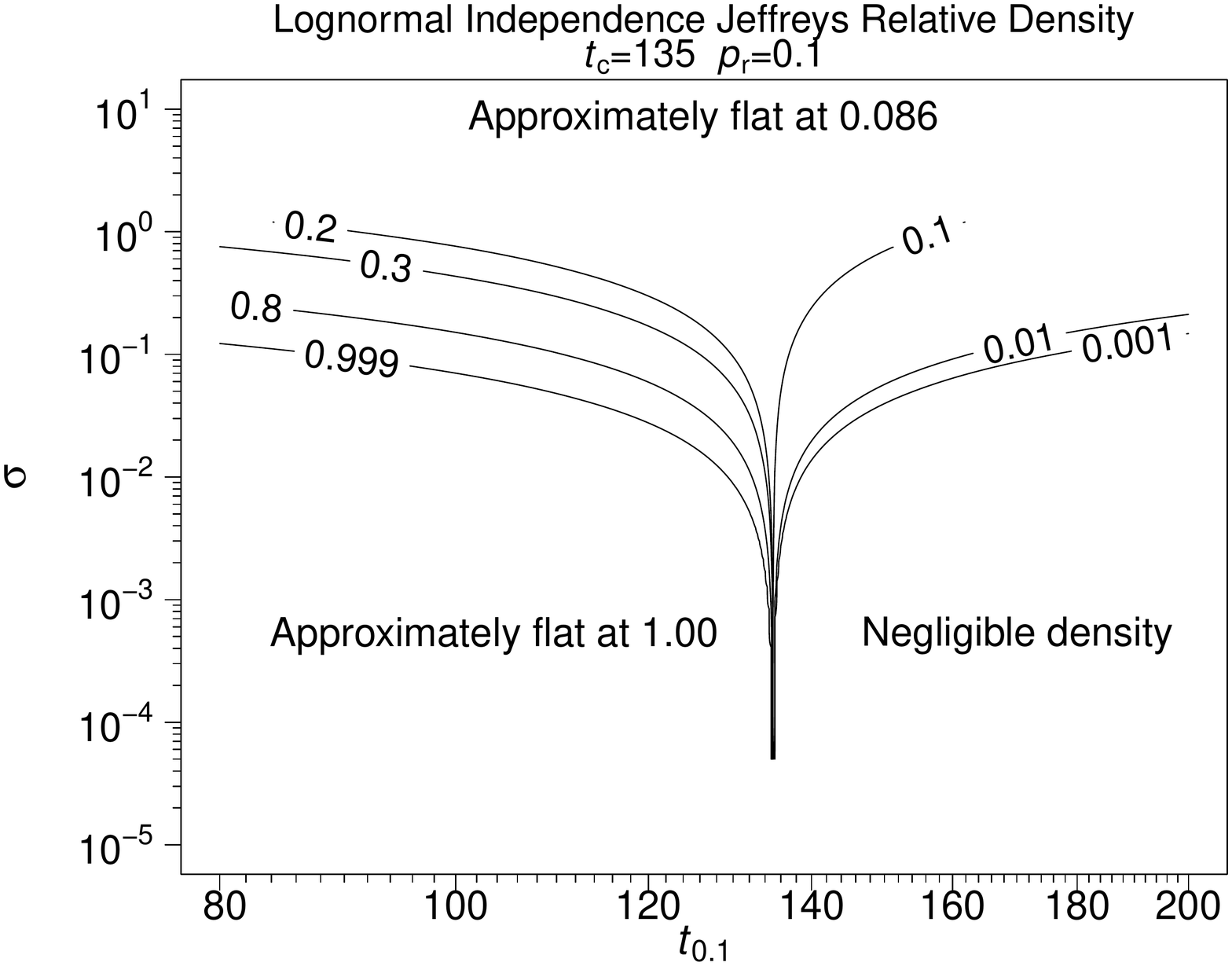}{3.3in}&
\rsplidafiguresize{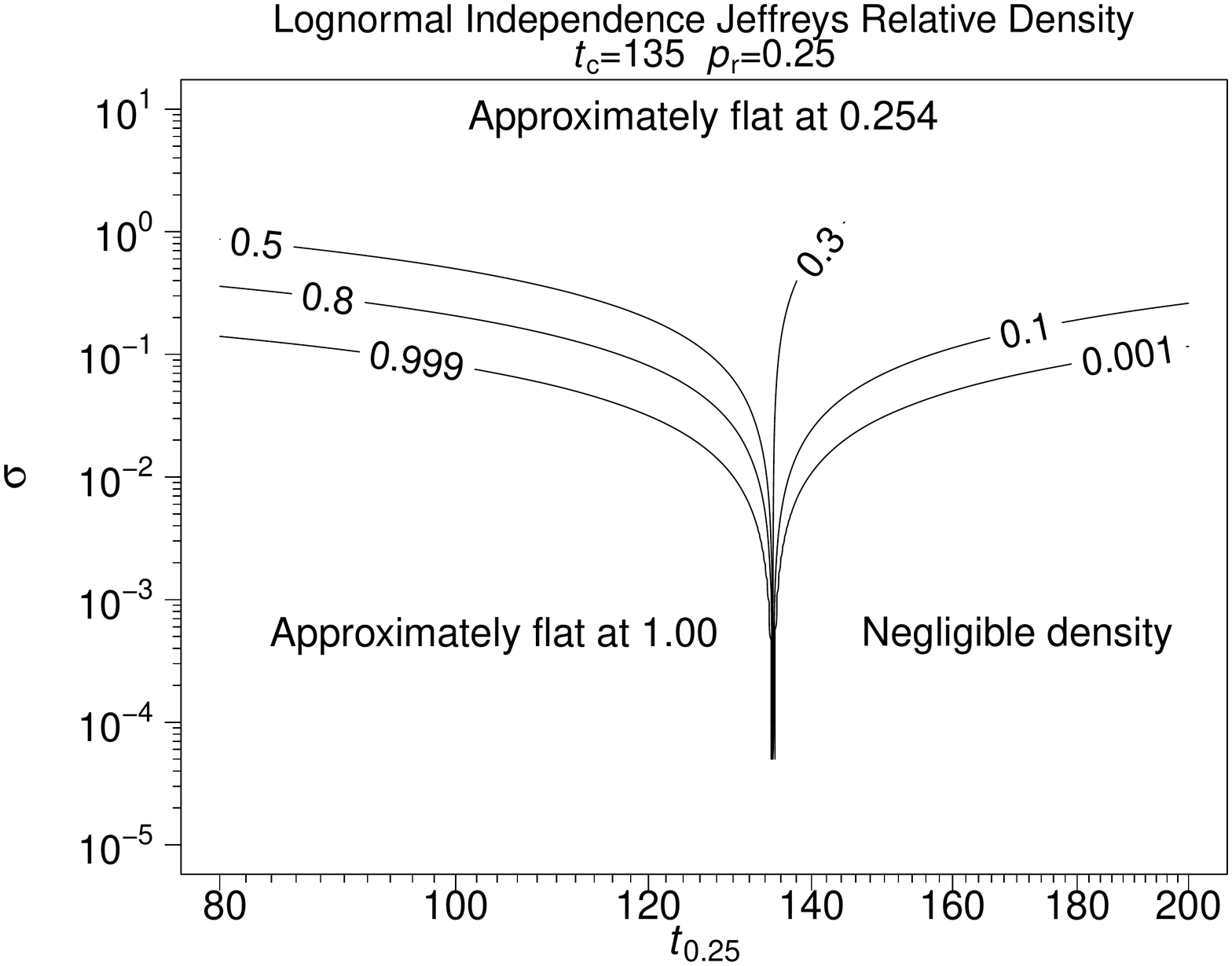}{3.3in}\\
\rsplidafiguresize{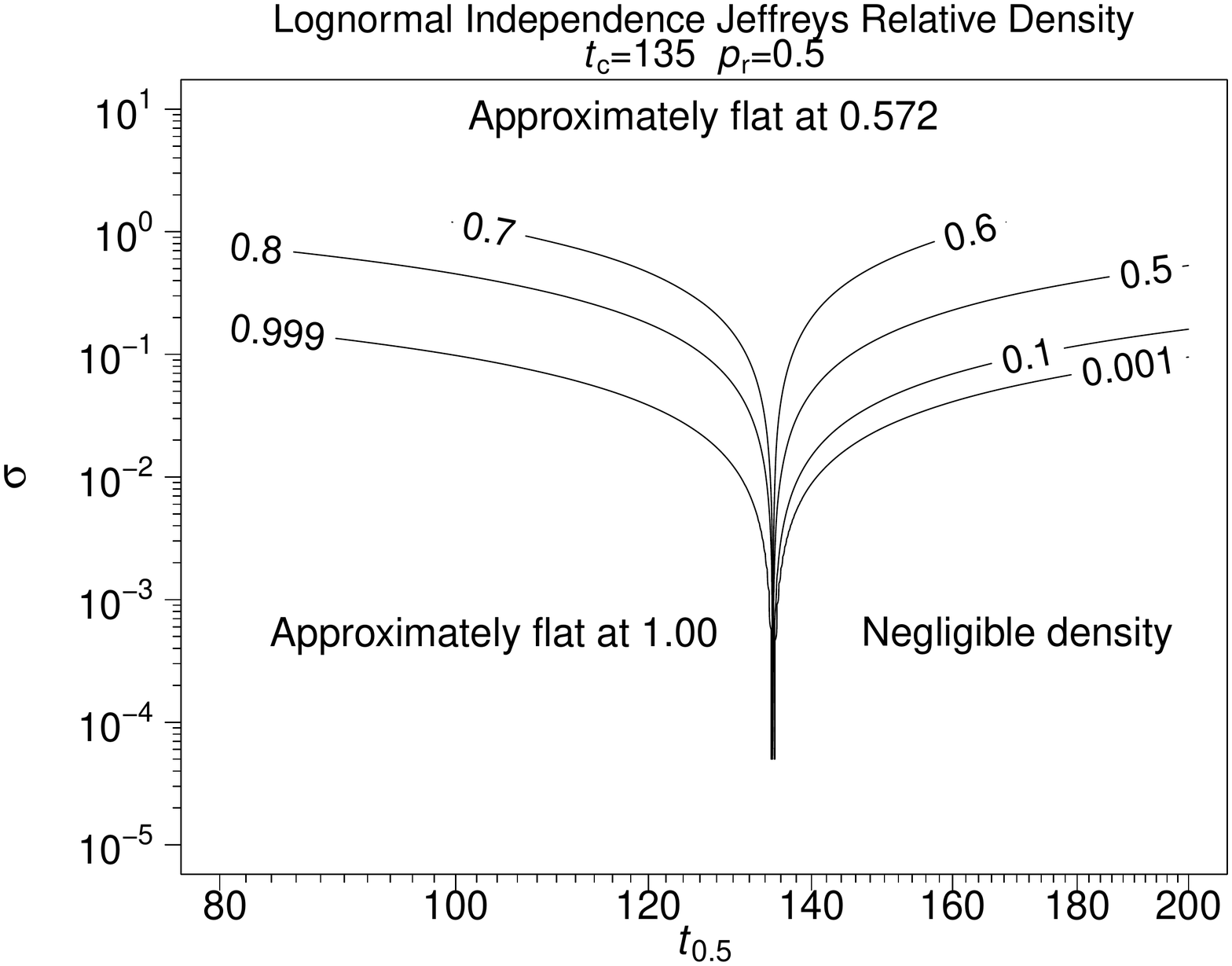}{3.3in}&
\rsplidafiguresize{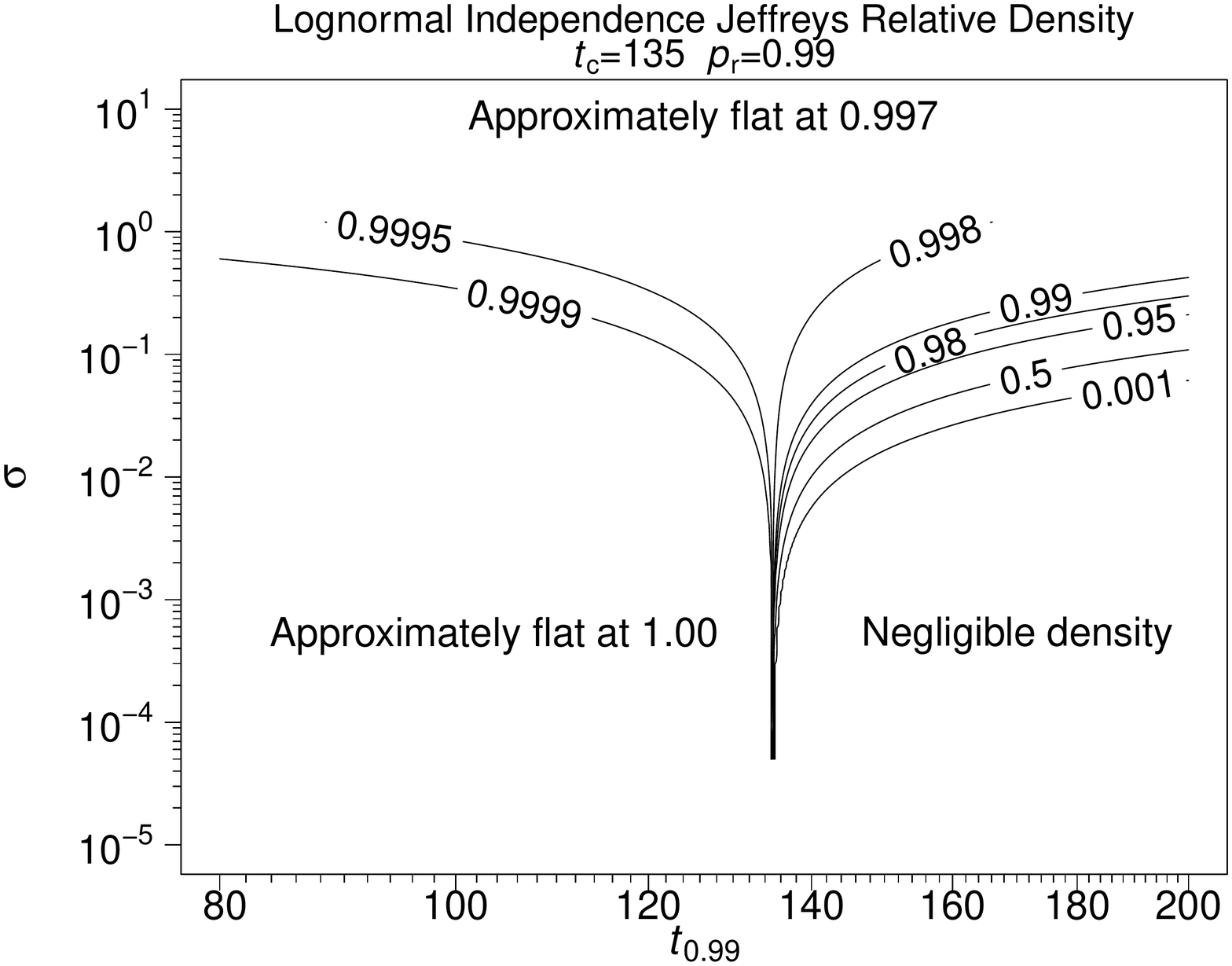}{3.3in}
\end{tabular}
\caption{Independence Jeffreys prior densities for $p_{r}=0.01$ (top
  left), $p_{r}=0.05$ (top right), $p_{r}=0.10$ (middle left), 
$p_{r}=0.25$ (middle right), $p_{r}=0.50$ (bottom left), and
  $p_{r}=0.99$ (bottom right).}
\label{figure:ijprior.density.examples}
\end{figure}

Section~\ref{S.section:understanding.ij.prior.features} in the appendix provides a detailed description of the \typeI{}
censoring IJ prior distribution features and the reasons those
features arise. Here we provide a brief summary of that material.
Figure~\ref{figure:ijprior.density.examples} illustrates the general
shapes of the IJ priors for different values of $p_{r}$.
These improper IJ priors have been scaled to have a maximum
of 1.0 and thus we refer to them as relative densities.
Generally, the IJ prior for small $\sigma$ and $t_{p_{r}} < t_{c}$
is flat at a level of 1.0. 
For any small $\sigma$, as the value of $t_{p_{r}}$ crosses
$t_{c}$ from the left, there is a steep cliff, with the level of the density
dropping to near 0. Note that this is because, in bottom-right part
of the parameter space, the probability of getting even one failure
is negligible. Detailed examples are given in
Section~\ref{S.section:reasons.for.ij.prior.features} of the
appendix.
The importance of this result is that,
relative to other noninformative priors, the
IJ priors (and CJ priors when used in conjunction with an
informative or weakly information marginal prior for the
other parameter)
follow the fundamental principle described in
Section~\ref{section:noninformative.prior.fundamental.principle}.
For large values of $\sigma$, the IJ prior is approximately flat
at a level close to $p_{r}$.

\section{Random Censoring IJ Prior Distributions}
\label{section:random.censoring.ij.prior}
As described in Section~\ref{section:censoring.types}, field
reliability data
almost always result in multiply time-censored data (e.g.,
Figures~\ref{figure:BearingCage.plots}(a)
and~\ref{figure:RocketMotor.plots.ps}(a)).  The random
censoring IJ prior, based on a competing risk model, can be obtained
from the random censoring FIM described in
Section~\ref{section:fisher.random.censored}. Suppose that $T$ is
the failure time for a unit but that it will not be observed if the
random censoring time $C < T$. There are two
situations to consider, described in the following subsections.

\subsection{IJ Prior Distributions for Limited-Time Random Censoring}
\label{section:limited.time.random.censoring}
High-reliability products or any product that has been in the field
for a small amount of time (e.g., a newly released smart phone model) will have a largest value of a
nonparametric estimate (e.g., Kaplan--Meier) of the marginal cdf of
$T$ that is considerably less than 1. This situation is
common and arises when data are analyzed at a particular data-freeze
date when only a small fraction of units in the field has failed. This is
similar to \typeI{} (time) censoring, except that there will be
some additional right-censored observations (e.g., due to staggered entry
into service) before the censoring time $t_{c}$.
The bearing cage field data (Section~\ref{section:motivating.examples})
provides an example of such data.
The rocket motor field data (also
Section~\ref{section:motivating.examples}) is different because all
three failures were left censored. The nonparametric ML
estimate of $F(t)$ jumps to 1.0 at the left-censored
observation at 16.5 years because this is larger than 16 years, the largest
right-censored observation (there is additional discussion of this
point at the end of Section~\ref{section:comparisons.bayes.noninformative}).
However, the similarity of the shapes of the
bearing cage and rocket motor likelihoods
(see Figure~\ref{figure:likelihood.contour.plots.reparameterization}),
suggests a much smaller effective $t_{c}$ for the rocket motor
data---perhaps 11 years.

It would be possible to define and compute an IJ prior for a
competing risk model describing multiple censoring
by using the scaled FIM elements inside the square
brackets in (\ref{equation.fim.random.censoring}) to replace the
$f_{ij}$ values in (\ref{equation:IJ.typeI.yp.log.sigma}). The pdf
of $\log(C)$, $h(x)$ would describe the pattern up to the censoring
time $t_{c}$, where all of the remaining mass would be
concentrated. The shape of the resulting IJ prior in this situation
will, however, be similar to the IJ prior for \typeI{} censoring,
described in Sections~\ref{section:ij.prior}
and~\ref{section:implementing.ij.priors} and thus those can be
used instead. We use this approach for our examples in
Sections~\ref{section:comparisons.ml.and.bayes}
and~\ref{section:comparisons.bayes.noninformative}.

\subsection{IJ Prior Distributions for Unlimited-Time Random Censoring}
For products that have a substantial amount of field experience and
are run until failure, after which they are replaced (e.g.,
single-use batteries), the largest value of the nonparametric estimate of
the marginal cdf of $T$ will typically be close to 1. An example of this kind
of data is given in the mechanical switch example in \citet{Nair1984},
where he focused on estimating the marginal distribution of failure mode A
and the occurrence of failure mode B resulted in the random censoring.

Again, it would be possible to define and compute an IJ prior for
this situation by using the scaled FIM elements inside the square
brackets in (\ref{equation.fim.random.censoring}) to replace the
$f_{ij}$ values in (\ref{equation:IJ.typeI.yp.log.sigma}). The pdf
of $\log(C)$, $h(x)$ could be obtained by looking at the
distribution of censoring times.  The shape of the resulting IJ
prior in this situation will be approximately flat because there is
not a single censoring time that ends the failure-observation process.

\section{Motivation for Weakly Informative Prior Distributions}
\label{section:weakly.informative.prior}
\subsection{Potential difficulties with noninformative prior distributions}
The noninformative prior distributions described in
Section~\ref{section:noninformative.prior} have
appealing theoretical properties under certain \textit{specified conditions}
(e.g. (log-)location-scale distributions with complete data or
\typeII{} censoring). In most applications, such conditions will not
be met exactly.  Noninformative prior distributions can put large
amounts of relative density at unreasonable (e.g., impossible) parts
of the parameter space. Then, with limited information in the data
(e.g., a small number of failures), a noninformative prior
distribution can strongly influence inferences and possibly
result in misleading conclusions. In such situations, especially, it is
better to use a prior distribution that rules out combinations of
parameters that are impossible or nonsensical.

\subsection{Previous Work on Weakly Informative Prior Distributions}
While there is a vast literature on noninformative or reference
prior distributions, less work has been done on weakly informative
prior distributions.  Weakly informative priors are constructed to
be diffuse relative to the likelihood and known scale of the data.
However, as opposed to noninformative priors,
weakly informative priors put density
on reasonable values of the parameters while down weighting
nonsensical values.  When there is a small amount of information in the
data (e.g., few failures) or if
fitting a complex model with many parameters, weakly informative
priors can help stabilize estimation whereas a noninformative prior
(e.g., a flat prior) can result in dispersed posterior distributions
with probability mass on extreme parameter values.
\cite{GelmanJakulinPittauSugelman2008}
recommend weakly informative priors for the
parameters of logistic and other regression models.  More recently,
\cite{GelmanSimpsonBetancourt2017} provide a historical overview of the
different classes of Bayesian priors such as, uniform, Jeffreys,
reference, and weakly informative priors. The authors illustrate
the dangers of using flat or default priors and recommend weakly
informative prior distributions that are selected based on the data
and subject-specific domain. \cite{Lemoine2019} uses
simulation to demonstrate how noninformative priors can produce
spurious parameter estimates, and advocates for weakly informative
priors as the new default choice for Bayesian estimation.

\subsection{Weakly Informative Prior Distributions for
  Log-Location-Scale Distributions}
\label{section:weakly.informative.priors.lls.distributions}
When there is little or no prior information about certain
parameters or when there is need to present an analysis
where results do not depend on subjective prior information, a commonly-used
alternative is to specify weakly informative marginal priors for
those parameters.  Commonly-used weakly informative priors include a
normal distribution with a large variance for parameters that are
unrestricted in sign or a lognormal distribution
with a large value of the 
shape parameter (log-standard-deviation) for parameters that
must be positive.

\begin{figure}[!htbp]
\begin{tabular}{cc}
(a) & (b) \\[-3.2ex]
\rsplidafiguresize{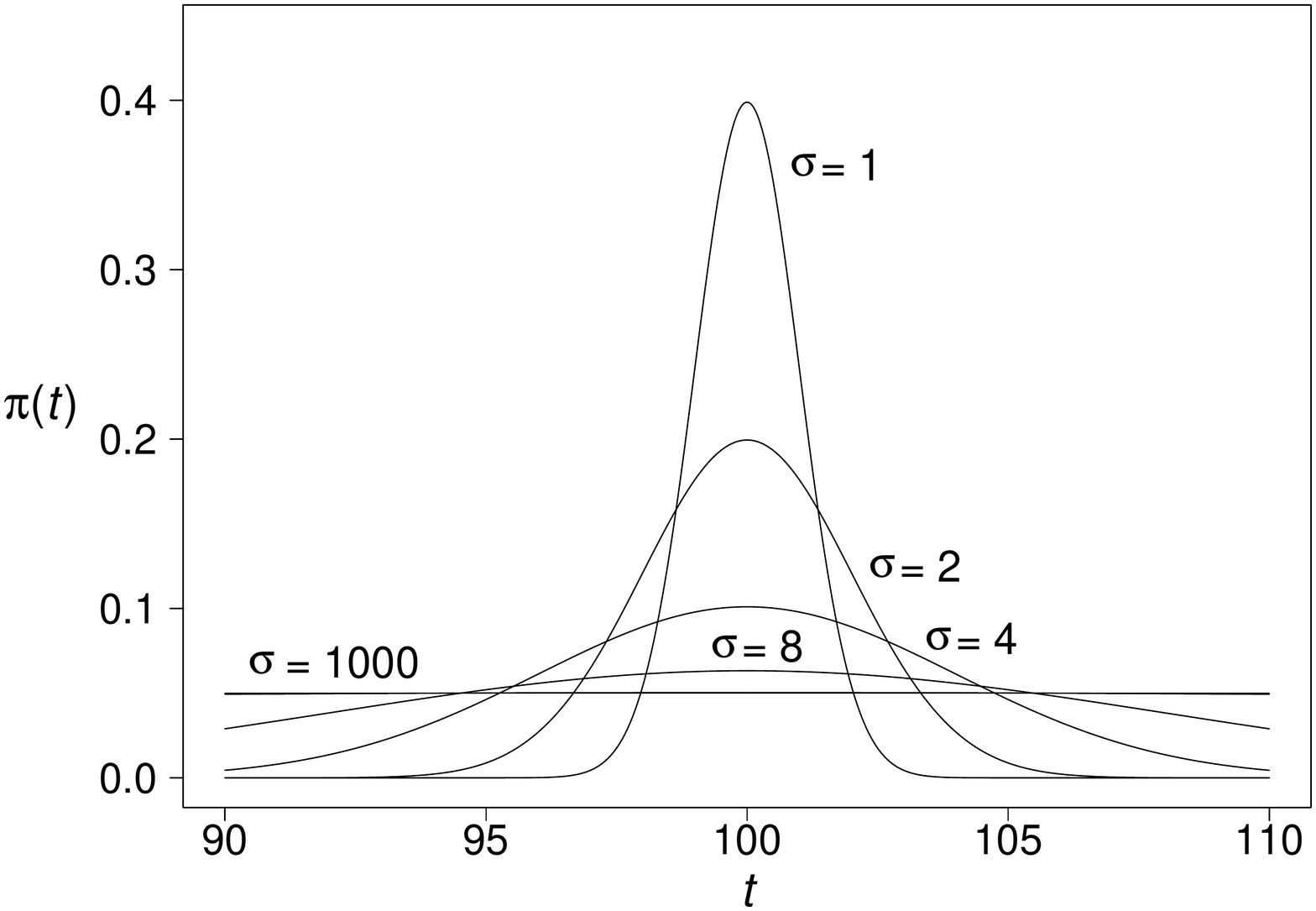}{3.3in} & \rsplidafiguresize{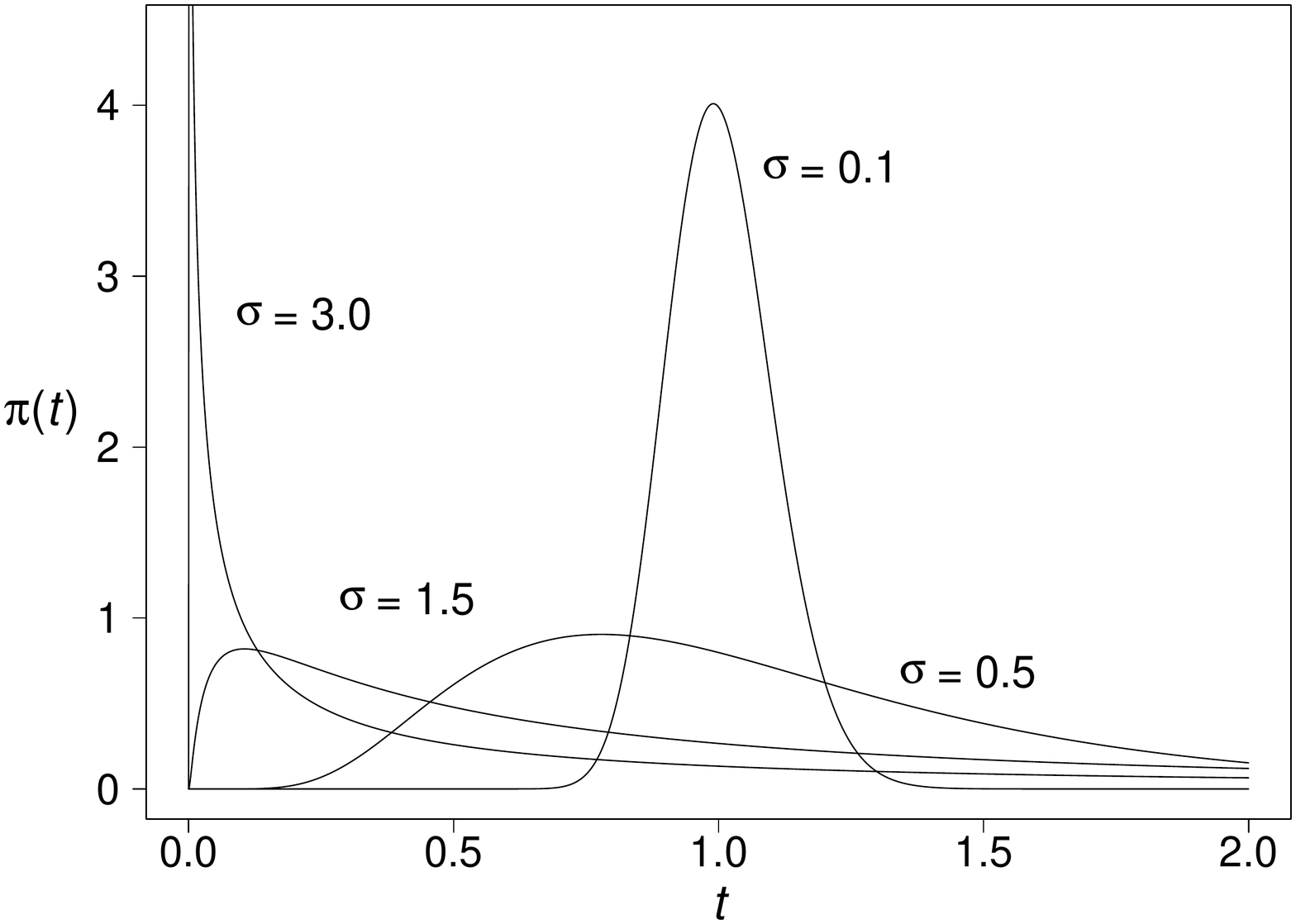}{3.3in}
\end{tabular}
\caption{Illustration that a normal distribution  density approaches a
  uniform (flat prior) distribution as $\sigma$ increases~(a) and that a lognormal
  density in $t$
  is proportional to $1/t$ for large values of $\sigma$~(b).}
\label{figure:limiting.weakly.informative.plots}
\end{figure}

These choices can be motivated by the fact that a normal
distribution prior density with any mean will approach a flat prior
as the standard deviation of the normal distribution increases.
Correspondingly, a lognormal distribution prior $\pi(t)$
with any log-mean will be proportional to $1/t$ as the log-standard-deviation
increases. These results are illustrated in
Figure~\ref{figure:limiting.weakly.informative.plots} and proofs are
in
Section~\ref{S.section:limiting.results.weakly.informative.prior.distributions}
of the appendix.

The parameters for these weakly informative marginal prior
distributions can be chosen using knowledge about the scale of the
response (i.e., depending on the units of the response), what
values of the parameters are physically possible,
and knowledge based on previous experience and engineering
knowledge.
The center of these
distributions might be chosen in a conservative manner.

Instead of specifying the prior distribution parameters (especially
for the lognormal distribution), it is generally better (and much
easier for prior elicitation) to specify a
range that contains a large proportion of the probability
distribution. Here we use a 0.99 probability range which is defined
as the 0.005 and the 0.995 quantiles of the prior distribution. For
example, to
specify that the Weibull shape parameter has a lognormal prior distribution
with probability 0.99 between 1.5 and 5, we write $\beta \sim \logNormalPrior(1.5, 5)$.

The use of quantiles to elicit/specify prior distributions has been
discussed previously \citep[e.g.,
  in][]{DeyLiu2007}. \citet{MeyerBooker2001} point out that
individuals tend to underestimate uncertainty. Thus, unless the
prior interval is based on quantitative information (e.g., interval
estimates on the same parameter from a previous study or studies) one might
want to ask for a 99\% interval and treat it as if it were a 95\%
interval. \citet{Mikkola.et.al2021} describe the current state of the art of
prior elicitation and provide an extensive literature review, including other
papers that use quantiles in elicitation.

\subsection{Comparisons of ML and Bayesian Estimation using
  Noninformative or Weakly Informative
Prior Distributions}
\label{section:comparisons.ml.and.bayes}
\subsubsection{Bearing cage field data}
\begin{figure}[!htbp]
\begin{tabular}{cc}
(a) & (b) \\[-3.2ex]
\rsplidafiguresize{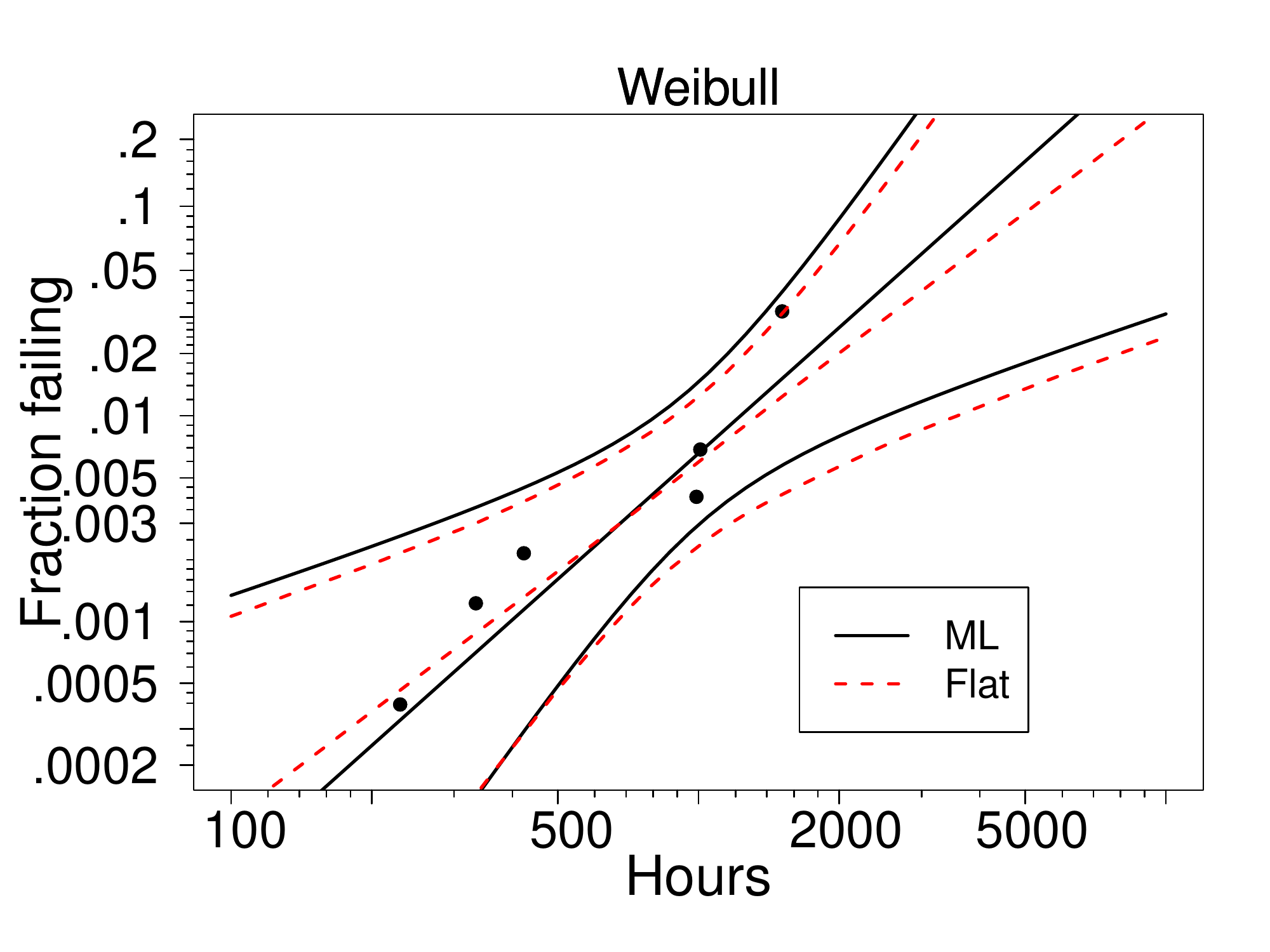}{3.35in} &
\rsplidafiguresize{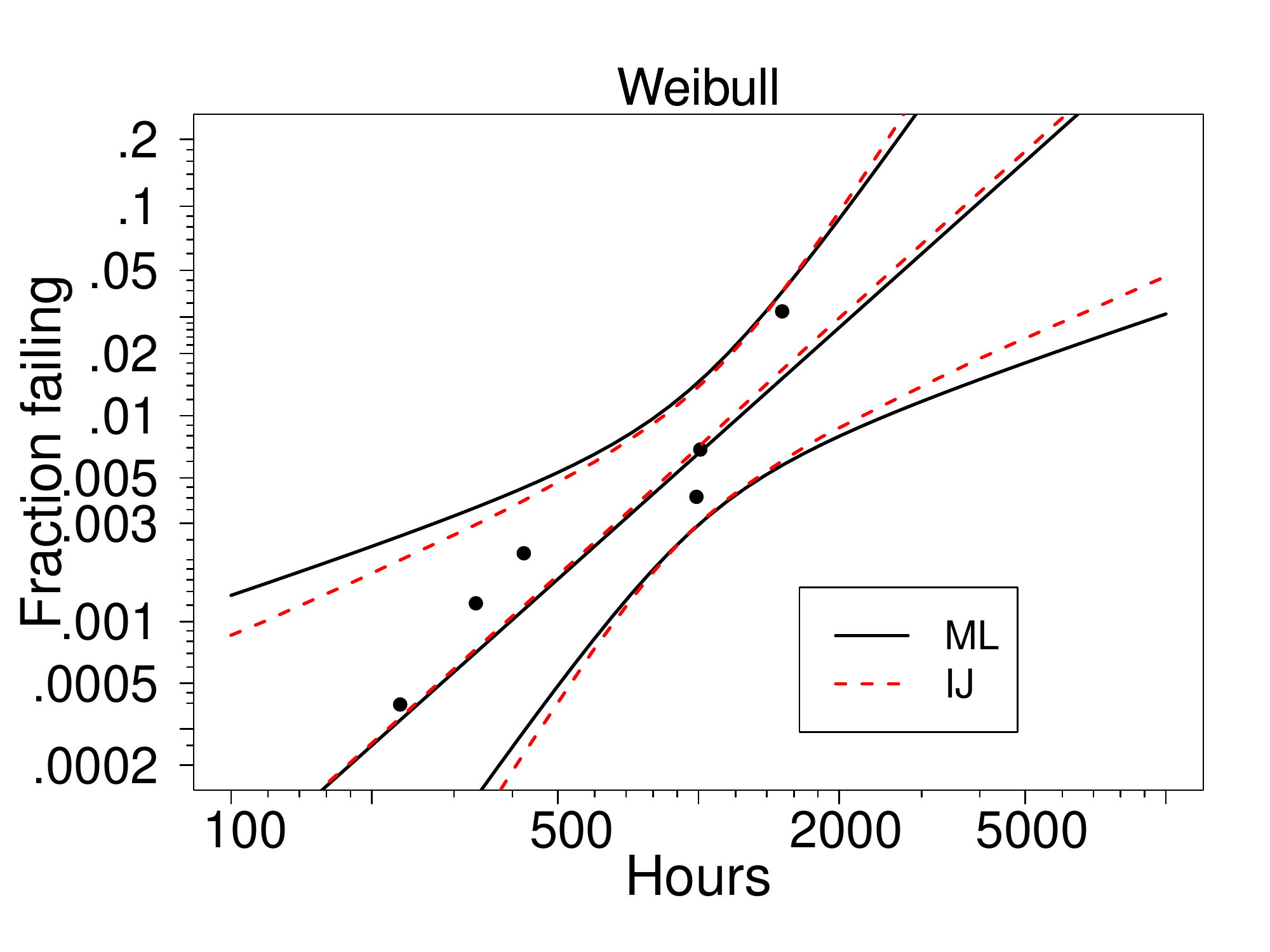}{3.35in}
\end{tabular}
\caption{Bearing cage comparison of estimation results for ML and
  flat prior~(a) and ML and IJ prior~(b).}
\label{figure:BearingCage.ml.and.noninformative}
\end{figure}

This is a continuation of the bearing cage examples in
Sections~\ref{section:motivating.examples}
and~\ref{section:reparameterization} where we compare ML and Bayes
estimation using alternative noninformative
priors. Figure~\ref{figure:BearingCage.ml.and.noninformative}
compares ML and Bayes estimation with a flat (on $\log(t_{0.10})$ and
$\log(\sigma)$)~(a) and ML and IJ priors~(b).
The IJ priors agree well with the ML results except for a little
deviation in the upper confidence bounds on $F(t)$ in the lower tail
of the distribution and the lower confidence bounds on $F(t)$ in the
upper tail of the distribution. The flat prior results are more
optimistic (smaller failure probabilities) in the upper tail of the
distribution (the region of interest).

\subsubsection{Rocket motor field data}
This is a continuation of the rocket motor examples in
Sections~\ref{section:motivating.examples}
and~\ref{section:reparameterization}.
\begin{figure}[!htbp]
\begin{tabular}{cc}
(a) & (b) \\[-3.2ex]
\rsplidafiguresize{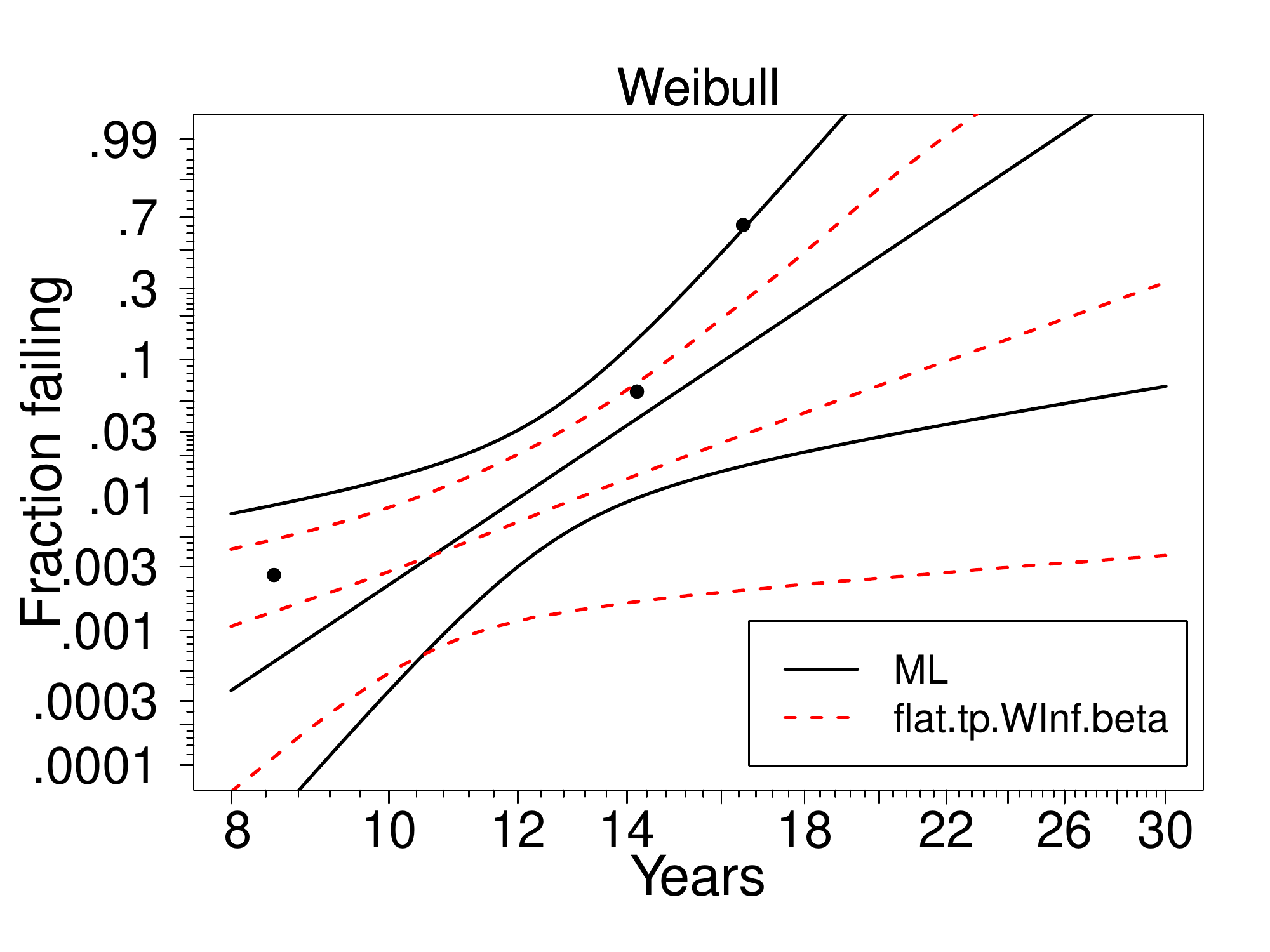}{3.35in} &
\rsplidafiguresize{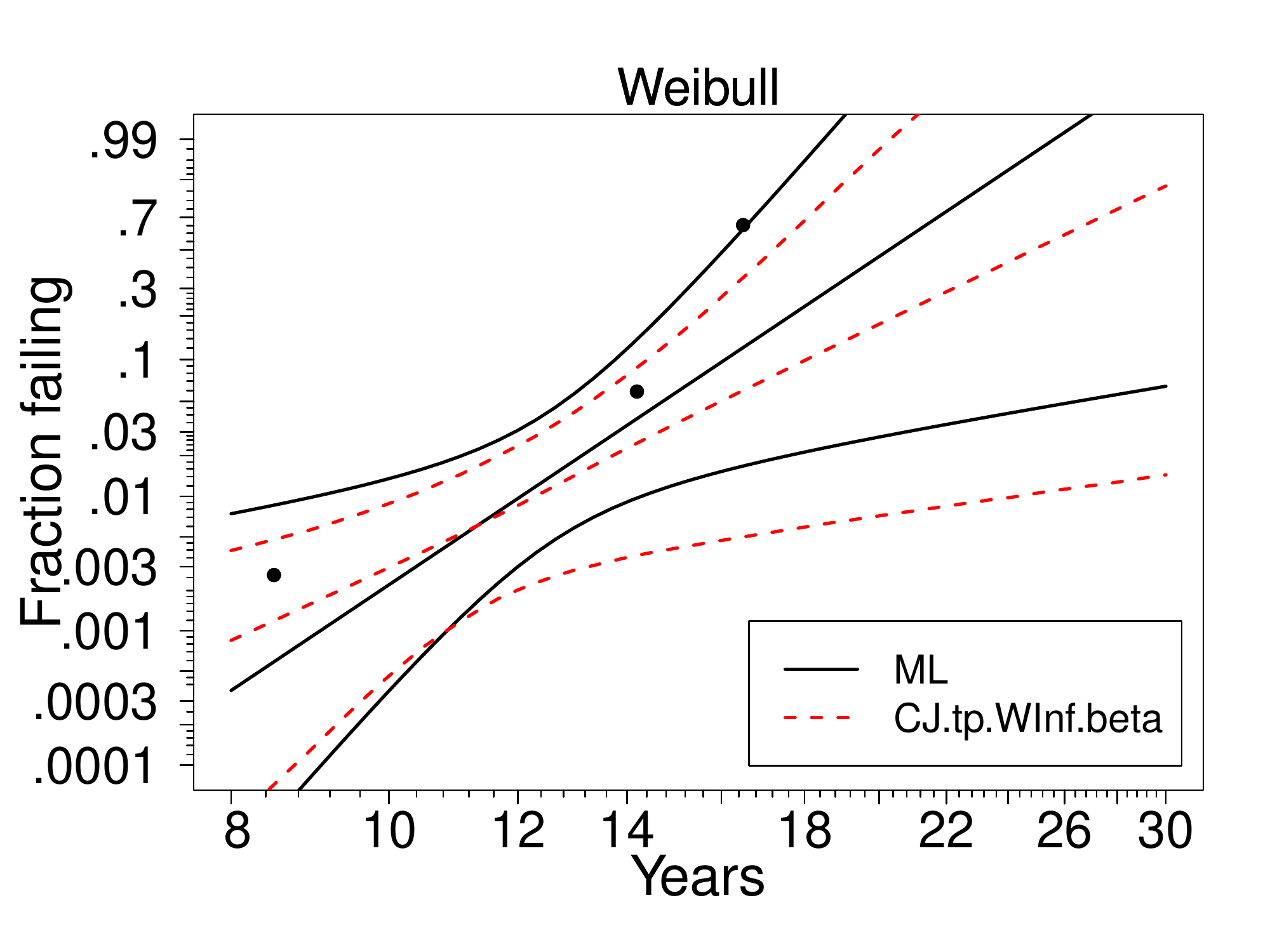}{3.35in}
\end{tabular}
\caption{Rocket motor comparison of estimation results comparing ML
and Bayes with flat  prior for
$t_{p_{r}}$~(a) and  a CJ prior for $t_{p_{r}}$~(b) using, in both
cases a weakly informative priors on $\beta$ .}
\label{figure:RocketMotor.ml.weaklyinformative.cj}
\end{figure}
Although ML estimates and associated confidence intervals exist for
the rocket motor data, because the failures are left-censored
observations, using Bayes estimation with flat or IJ priors
apparently results in an improper posterior distribution. We avoid
this problem by using a weakly informative prior $\beta \sim
\logNormalPrior(0.2, 25)$ for the comparisons in this example (note
that this is an extremely wide range relative to values of $\beta$
seen in typical applications) that is approximately flat for the
$\log(\sigma)$ parameterization.
Figure~\ref{figure:RocketMotor.ml.weaklyinformative.cj}(a) compares
estimation results for ML and Bayes with a flat (uniform) prior for
$\log(t_{p_{r}})$. Figure~\ref{figure:RocketMotor.ml.weaklyinformative.cj}(b)
is similar but replaces the flat prior for $\log(t_{p_{r}})$ with a
CJ prior. The estimates using the CJ prior $\log(t_{p_{r}})$ in
Figure~\ref{figure:RocketMotor.ml.weaklyinformative.cj}(b) are
considerably closer to the ML estimates and less optimistic when
compared to the flat prior in
Figure~\ref{figure:RocketMotor.ml.weaklyinformative.cj}(a).

One might ask why the ML and Bayesian estimate of $F(t)$ in
Figure~\ref{figure:RocketMotor.ml.weaklyinformative.cj} differ so
much from the nonparametric estimate represented by the three
plotted points.  The slope of the ML estimate of $F(t)$ is the same
as the ML estimate of $\beta$, as shown in
\citet[][Section~6.2.4]{MeekerEscobarPascual2022} and this is
approximately true for the Bayesian estimate.  There were only three
failures and all were left censored.  The nonparametric ML estimate
(NPMLE) of $F(t)$ for current-status data can be obtained by using
the ``Pool Adjacent Violators'' algorithm for the observed fraction
failing as a function of years. For the rocket motor data, the NPMLE
is a step function that jumps from 0 to $1/384=0.026$ at 8.5 years,
to $1/9=0.111$ at 14.2 years, and to 1.0 at 16.5 years. Then the
points are plotted at half of the jump height, as suggested in
\citet[][Section~3.3.1.2]{Lawless2003}. The left censoring implies
that these jumps occur after the (unknown) failure times, which
would result in a kind of upward bias in the plotted points,
relative to what would be plotted if the exact failure times had
been known.

\section{Combining  Informative with Noninformative or Weakly Informative Prior Distributions}
\label{section:combining.informative.with.noninformative}
\subsection{Motivation for Partially Informative Prior Distributions}
\label{section:motivation.for.partially.informative.prior}
An important reason for using Bayesian methods is that they
provide a formal mechanism for including prior information (i.e.,
knowledge beyond that provided by the data) into the analysis.
When combining a CJ prior with an informative or
weakly informative prior our approach is similar to that suggested
in \citet{SunBerger1998}. For example, as described in
Section~\ref{section:ij.prior}, the  CJ prior
$\pi[\log(t_{p_{r}}) \, |\log(\sigma)]$ can be combined with an
informative or weakly informative marginal prior $\pi(\log(\sigma))$
to give the joint prior $\pi[\log(t_{p_{r}}),\log(\sigma)]$. In our
software, for convenience, the prior is specified
in terms of the familiar Weibull shape
parameter $\beta=1/\sigma$ (or the lognormal shape parameter $\sigma$)
and then transformed into the marginal prior for $\log(\sigma)$.

Section~\ref{section:reparameterization} discussed the need for
reparameterization.
If there is prior information for one or more of the model
parameters and if the definition of the parameters (i.e., the particular
parameterization) has been chosen such that the information about
parameters is approximately mutually independent, then one can specify a
joint prior density as the product of marginal densities for
each parameter.

Informative marginal prior distributions can be used for those
parameters for which there is appreciable prior
information. As mentioned by \citet{MeyerBooker2001}, individuals
providing an informative marginal prior
distribution will typically feel more comfortable expressing their
knowledge about a parameter by using a symmetric distribution.
The normal distribution (truncated
below zero for a positive parameter) is a reasonable (approximately symmetric) choice.  Then
noninformative (e.g., flat or CJ) or weakly
informative (e.g., normal with a large 99\% range) marginal prior
distributions can be specified for the other unrestricted parameters
(e.g., $\log(t_{p_{r}})$ or $\log(\sigma)$).

A useful generalization of the normal distribution is
the location-scale-$t$ (LST) distribution. That is, for a specified
degrees-of-freedom parameter $\dfreedom>0$, there is a symmetric
LST distribution having tails that are heavier (or
much heavier) than the normal distribution. For large values of
$\dfreedom$ (e.g., greater than 60), the LST distribution is
approximately the same as a normal distribution. For $\dfreedom=1$,
the LST distribution is a Cauchy distribution.

\subsection{A Summary of Recommended Log-Location-Scale Prior Distributions}
\label{section:recommended.lls.prior}
Table~\ref{table:summary.llc.priors} provides a summary of the
recommended prior distributions for use with log-location-scale
distributions.
\begin{table}[!htbp]
\caption{Summary of recommended prior distributions for
  log-location-scale distribution parameters
\label{table:summary.llc.priors}
}
\centering
\begin{tabular}{llc}
\toprule
Prior distributions for $t_{p_{r}}$ & Type of prior &Prior distribution inputs\\
\midrule
Lognormal for $t_{p_{r}}$ & \parbox{15em}{Informative\\Weakly informative}
&$(\mu_{\log(t_{p_{r}})}, \sigma_{\log(t_{p_{r}})} )$\\[2.5ex]
Truncated $(>0)$ normal for $t_{p_{r}}$ & \parbox{15em}{Informative}
&$(\mu_{t_{p_{r}}}, \sigma_{t_{p_{r}}} )$\\[2.5ex]
Log-location-scale-$t$ for $t_{p_{r}}$ & \parbox{15em}{Informative\\Weakly informative}
&$(\mu_{\log(t_{p_{r}})}, \sigma_{\log(t_{p_{r}})}, \dfreedom  )$\\[2.5ex]
Truncated $(>0)$ location-scale-$t$ for $t_{p_{r}}$ & \parbox{15em}{Informative}
&$(\mu_{t_{p_{r}}}, \sigma_{t_{p_{r}}}, \dfreedom )$\\[2.5ex]
Flat for  $\log(t_{p_{r}})$ & Noninformative &None\\[2.5ex]
\parbox{15em}{Conditional Jeffreys\\  for $\log(t_{p_{r}})|\log(1/\beta)$}
&Noninformative &  $t_{c}$ and $p_{r}$\\[5ex]
\toprule
Prior distributions for $\beta=1/\sigma$ (or $\sigma$)&
Type of prior &Prior distribution inputs\\
\midrule
Lognormal for $\beta$ & \parbox{15em}{Informative\\Weakly informative}
& $(\mu_{\log(\beta)}, \sigma_{\log(\beta)})$\\[2.5ex]
Truncated $(>0)$ normal for $\beta$ &\parbox{15em}{Informative}
  &$(\mu_{\beta}, \sigma_{\beta} )$\\[2.5ex]
Log-location-scale-$t$ for  $\beta$ & \parbox{15em}{Informative\\Weakly informative}
& $(\mu_{\log(\beta)}, \sigma_{\log(\beta)}, \dfreedom )$\\[2.5ex]
Truncated $(>0)$ location-scale-$t$ for $\beta$ &\parbox{15em}{Informative}
  &$(\mu_{\beta}, \sigma_{\beta}, \dfreedom )$\\[2.5ex]
Flat for  $\log(1/\beta)$ & Noninformative &None\\[2.5ex]
\parbox{15em}{Conditional Jeffreys\\ for $\log(1/\beta)|\log(t_{p_{r}})$}
& Noninformative & $t_{c}$ and $p_{r}$\\
\bottomrule
\end{tabular}
\end{table}
Some comments on these prior distributions and Table~\ref{table:summary.llc.priors} are:
\begin{itemize}[itemsep=1mm, parsep=0pt]
\item
The CJ priors referenced in
Table~\ref{table:summary.llc.priors} are for \typeI{} censoring
(Section~\ref{section:ij.prior}) and limited-time random censoring
(Section~\ref{section:limited.time.random.censoring}).
\item
Table~\ref{table:summary.llc.priors} gives priors for the Weibull
failure-time distribution shape parameter $\beta=1/\sigma$.
The recommendations
are similar for the lognormal failure-time distribution with
shape parameter $\sigma$.
\item
As described in
Section~\ref{section:weakly.informative.priors.lls.distributions},
lognormal, truncated normal, log-location-scale-$t$, and truncated
location-scale-$t$ prior distributions for the interpretable
Weibull parameters $t_{p_{r}}$ and $\beta=1/\sigma$ are initially specified
by a 0.99 probability range. This range is then translated into
parameters for the marginal prior distributions for both $t_{p_{r}}$
and $\beta=1/\sigma$.
\item
Subsequently, the marginal priors for $t_{p_{r}}$ and
$\beta=1/\sigma$ are used to obtain (by standard methods for
obtaining the distribution of a transformation of random variables),
the marginal priors for $\log(t_{p_{r}})$ and
$\log(1/\beta)=\log(\sigma)$ that are used in the MCMC
computations. Details are given in
Section~\ref{S.section:log.truncated.normal.distributions} of the
appendix.
\item
For the CJ priors, the censoring time $t_{c}$ is
not a parameter, but it is a necessary input.
\item
The noninformative flat and CJ priors do not
require specification of any parameters and are thus given directly
in terms of the $\log(t_{p_{r}})$ or $\log(1/\beta)$
parameterization.
\item
When the CJ priors for
$\log(t_{p_{r}})|\log(1/\beta)$ and for
$\log(1/\beta)|\log(t_{p_{r}})$ are used together, the result is a
joint IJ prior for $(\log(t_{p_{r}}), \log(1/\beta))$. Note that
when these two densities are used together, the result is not a
joint distribution of independent random variables. This is the
reason that we use the name Independence Jeffreys (IJ).
\end{itemize}

\subsection{Comparisons of Bayesian Estimation using
  Noninformative and Partially Informative
Prior Distributions}
\label{section:comparisons.bayes.noninformative}
This section returns to the two motivating examples, comparing
noninformative (or weakly informative) priors with partially
informative priors.
\subsubsection{Bearing cage field data}
\begin{figure}[!ht]
\begin{tabular}{cc}
\multicolumn{2}{c}{Weibull posterior cdf estimates and 95\% credible
  intervals based on the }\\
\phantom{xxxxx}(a) noninformative prior & \phantom{xxxxx}(b) partially informative prior\\[-3.2ex]
\rsplidafiguresize{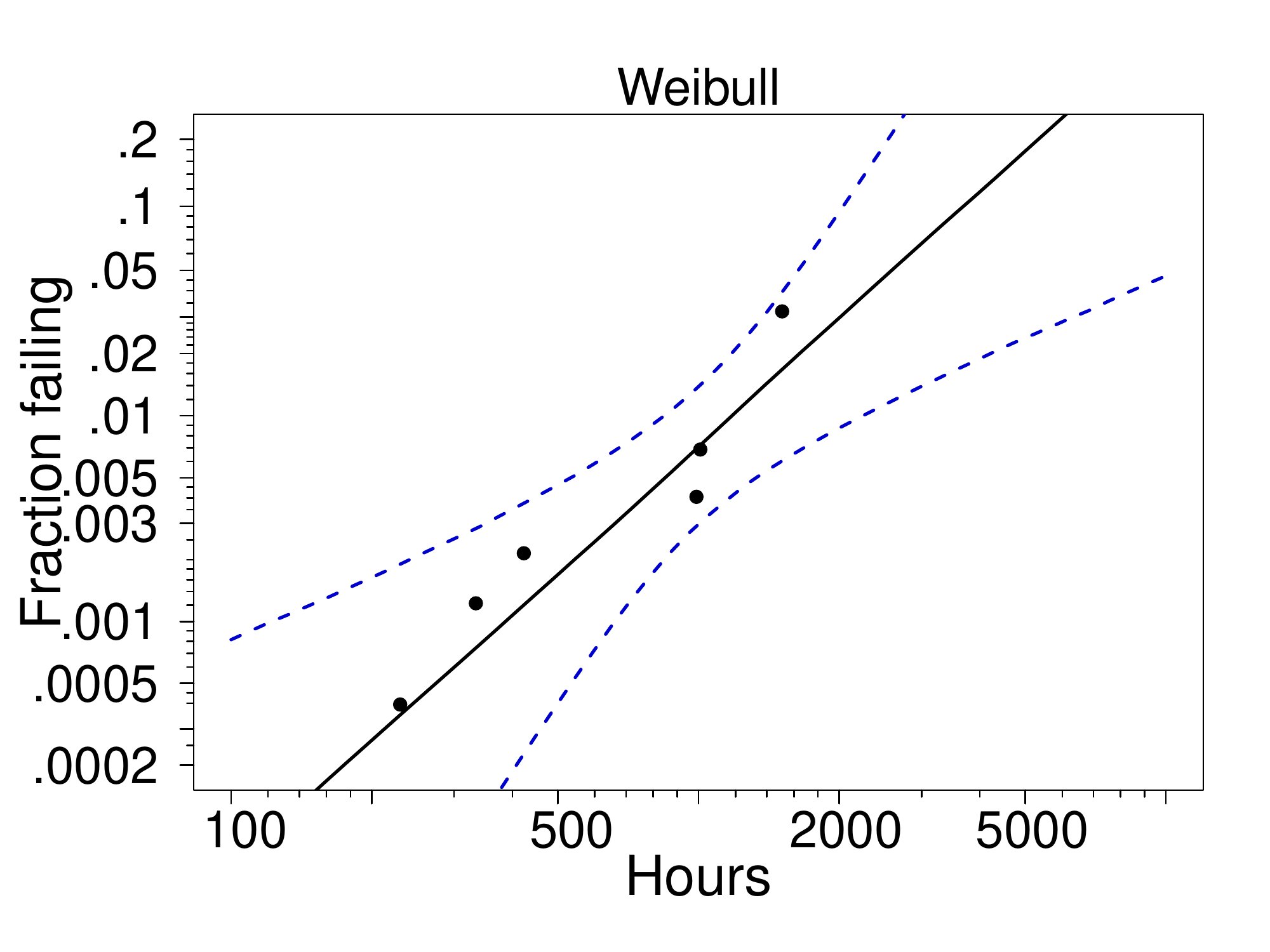}{3.25in} &
\rsplidafiguresize{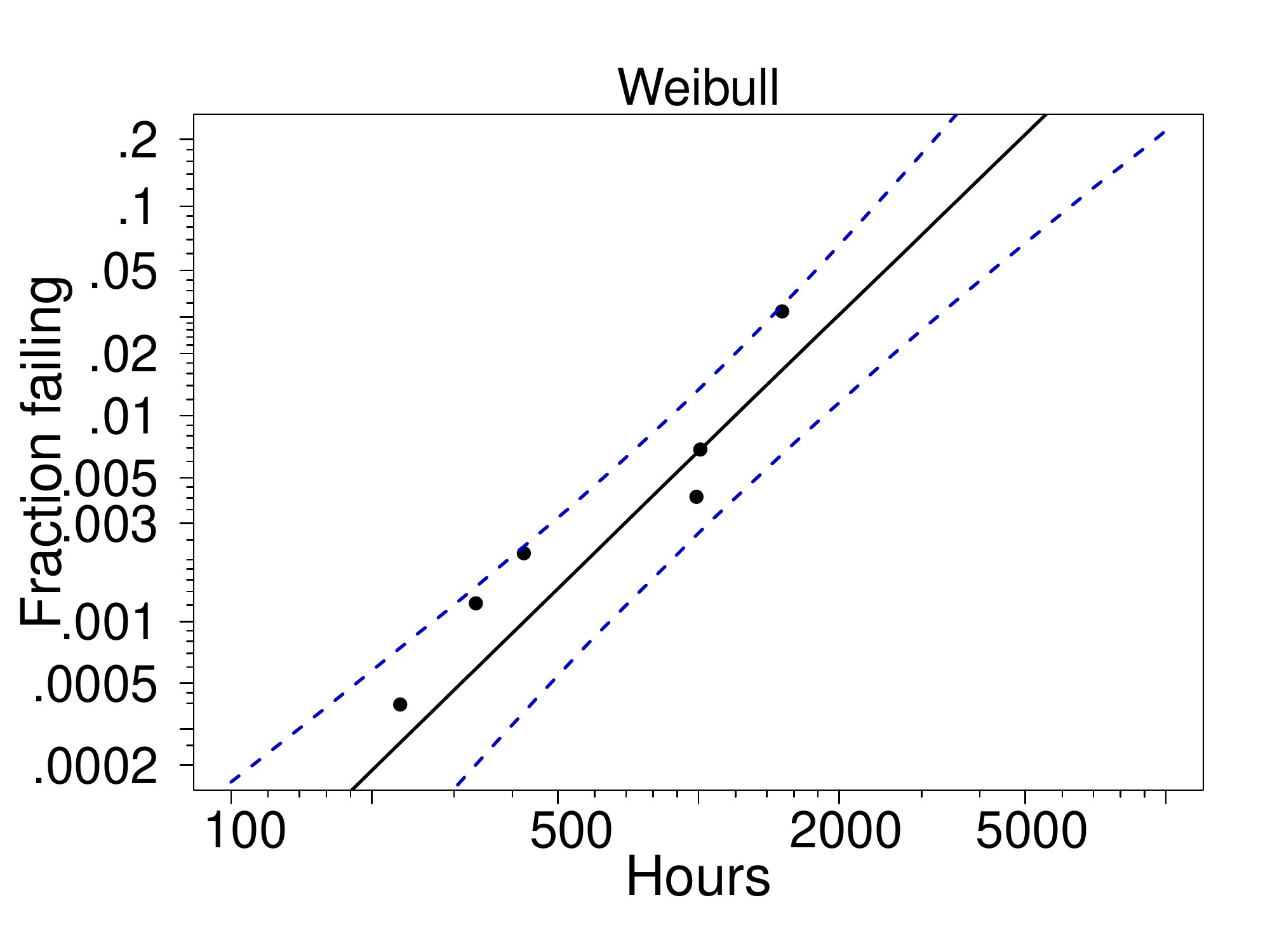}{3.25in}\\
\multicolumn{2}{c}{Weibull likelihood contours and  draws from the }\\
\phantom{xxxxx}(c) noninformative prior & \phantom{xxxxx}(d) partially informative prior\\[-3.2ex]
\rsplidafiguresize{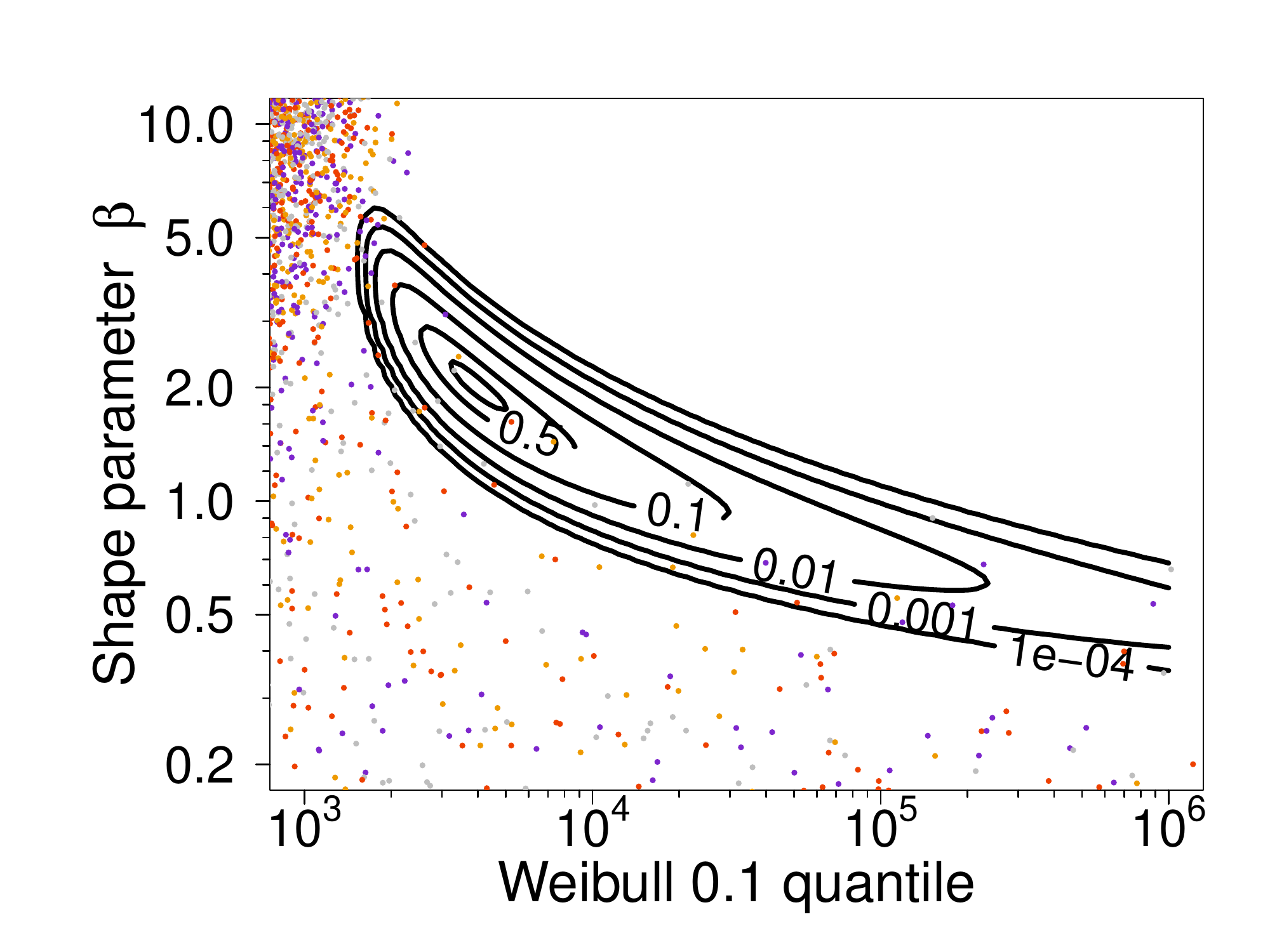}{3.25in} &
\rsplidafiguresize{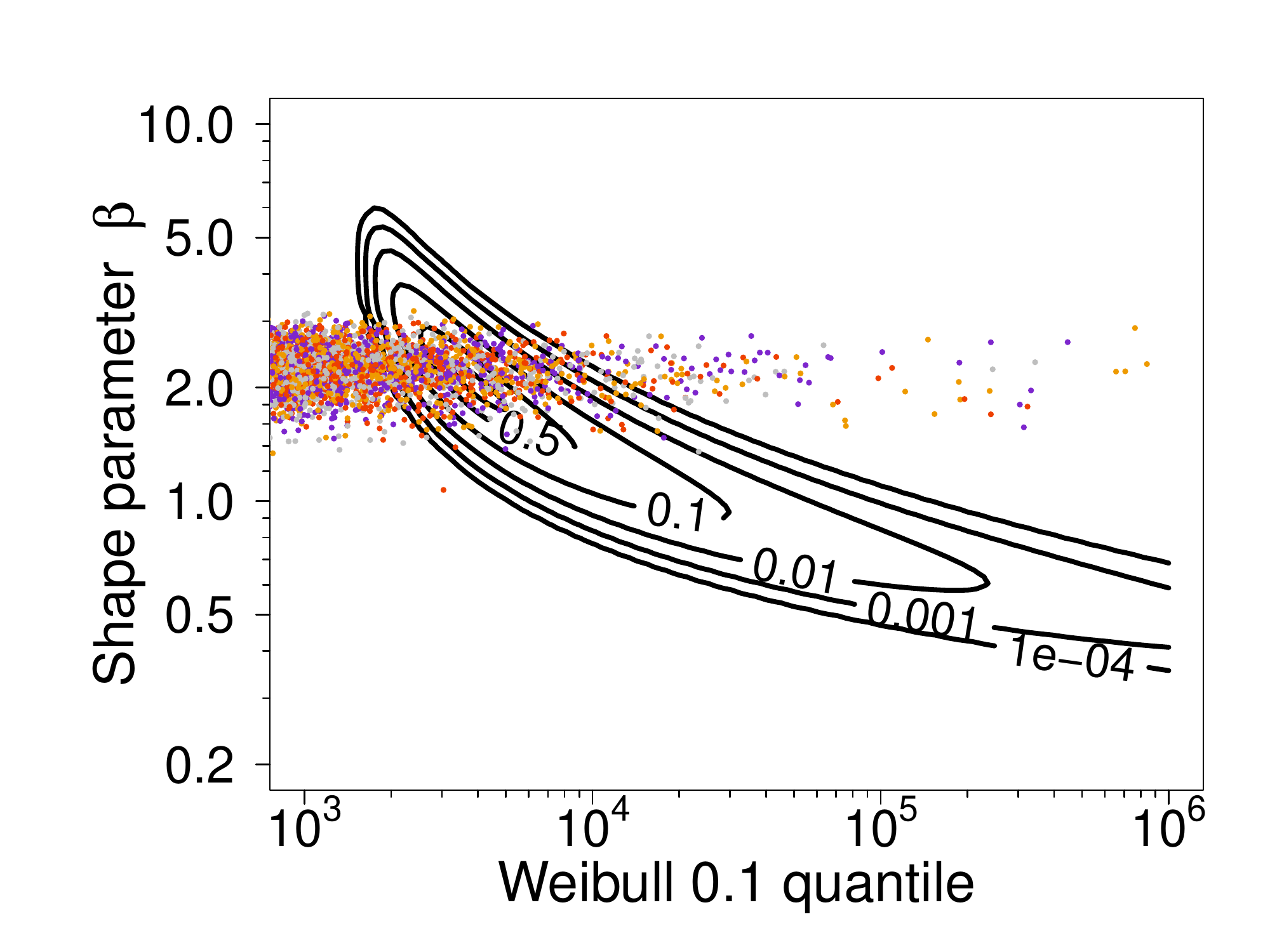}{3.25in}\\
\multicolumn{2}{c}{Weibull likelihood contours and posterior draws based on the}\\
\phantom{xxxxx}(e) noninformative prior & \phantom{xxxxx}(f) partially informative prior\\[-3.2ex]
\rsplidafiguresize{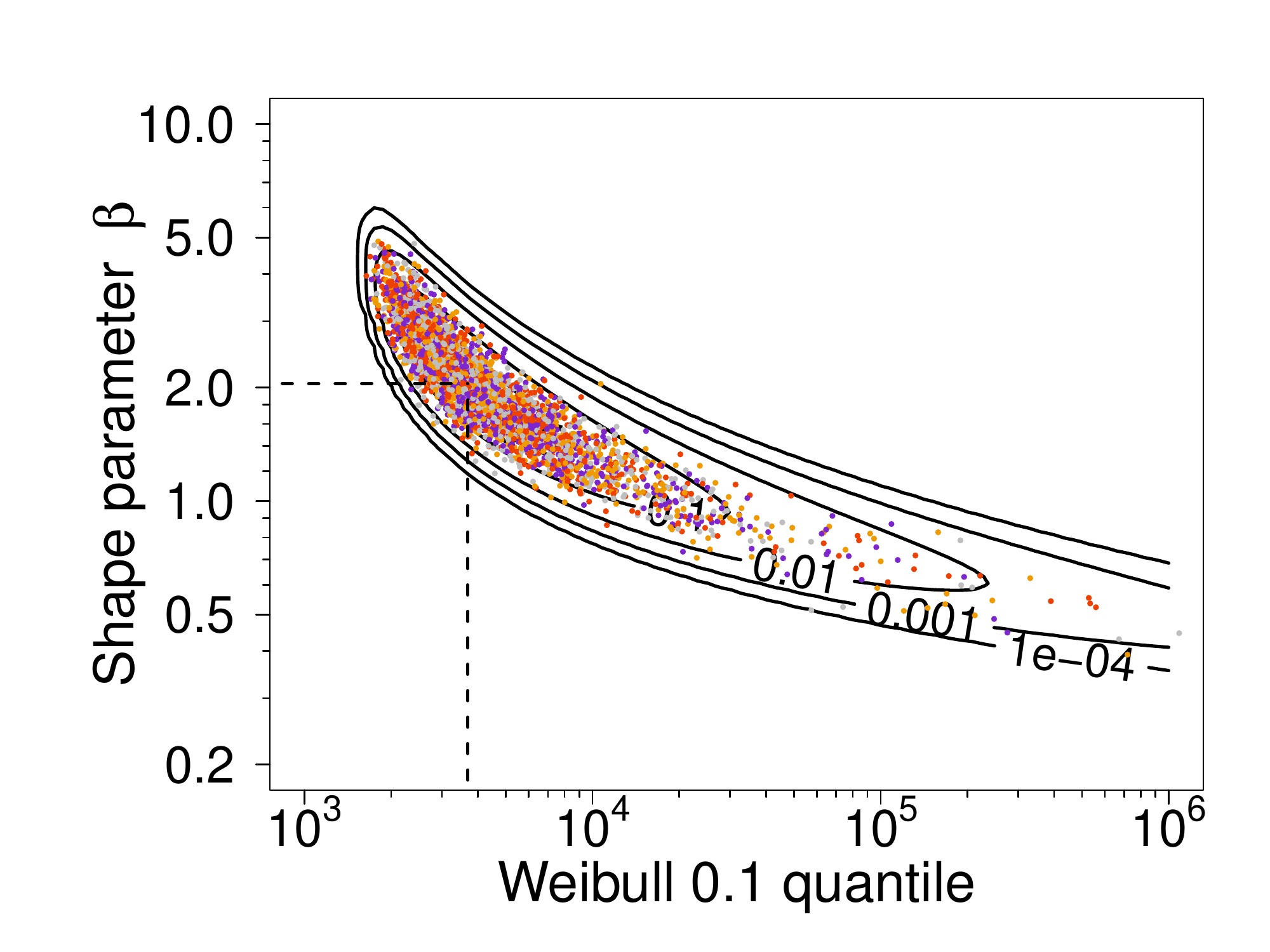}{3.25in} &
\rsplidafiguresize{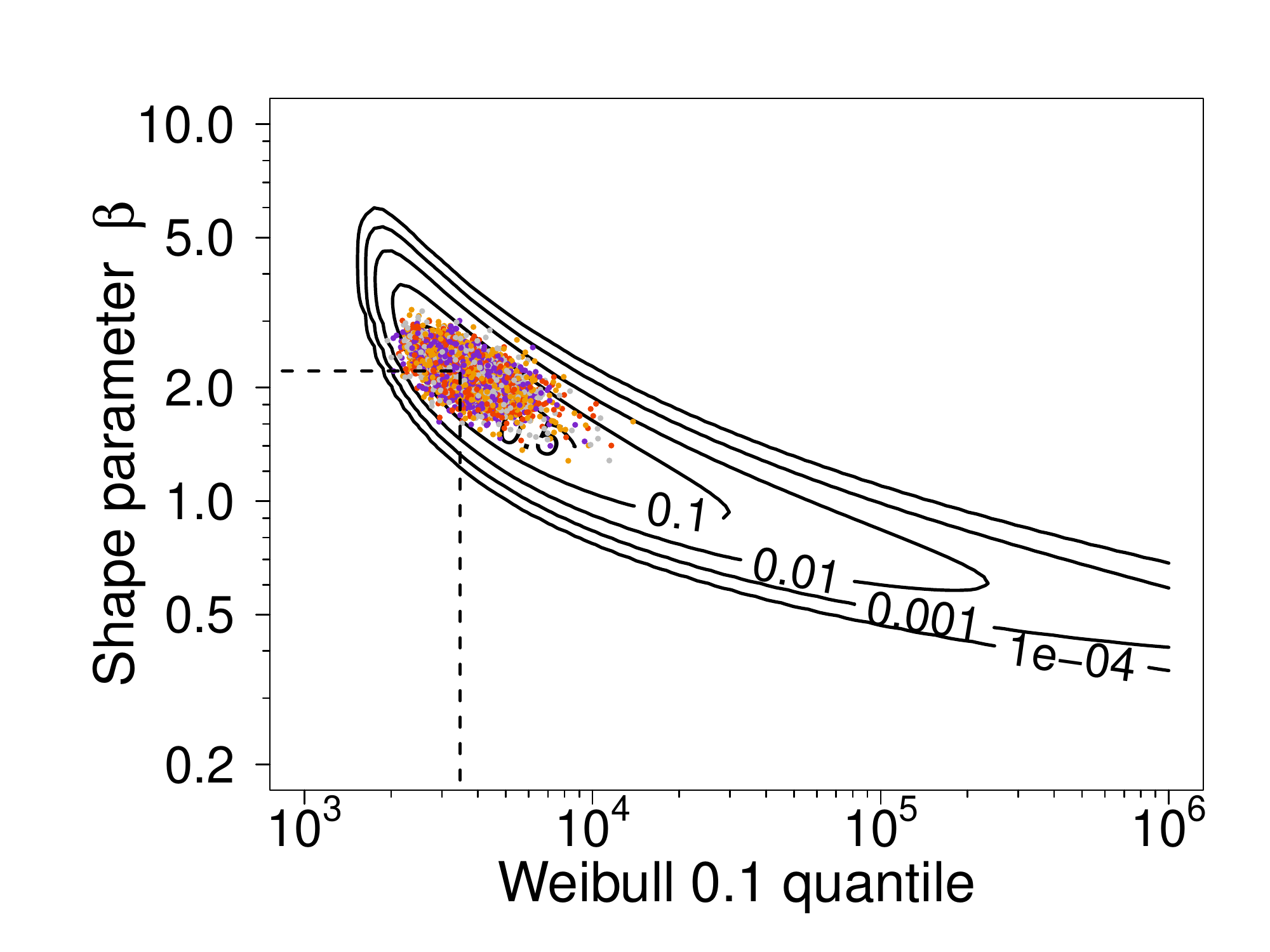}{3.25in}
\end{tabular}
\caption{Bearing cage Bayesian estimation results comparing
  a noninformative prior (on the left) with a partially informative
  prior (on the right). For each prior there is a Weibull
  probability plot showing the point estimate and credible intervals
for $F(t)$ (top), draws from the bounded joint prior and likelihood contours
(middle), and draws from the joint posterior and likelihood contours
(bottom).}
\label{figure:BearingCage.noninformative.informative}
\end{figure}

This is a continuation of the bearing cage example in
Section~\ref{section:comparisons.ml.and.bayes} where ML estimates
were compared with Bayesian estimates based on
noninformative prior distributions.
For the bearing cage field data,
Figure~\ref{figure:BearingCage.noninformative.informative}
compares Bayesian estimation results for a noninformative prior (on
the left) and a partially informative prior (on the right).
For the noninformative prior, we used an IJ prior. For
the partially informative prior, we combined the CJ prior for $y_p$
given $\log(\sigma)$ (which, from
Section~\ref{subsubsection:ij.prior.yp.log.sigma} is
$\pi(y_p|\tau)\propto\sqrt{f_{11}}$) with an informative truncated
(to be positive) normal distribution prior
$\beta \sim \TNormalPrior(1.5, 3)$.

The probability plot on the top-left in
Figure~\ref{figure:BearingCage.noninformative.informative} gives
Bayesian estimation results with the noninformative IJ prior
and they are similar to the ML results given on the right in
Figure~\ref{figure:BearingCage.plots}. The credible interval for
$F(8000)$ is $[0.03, \intervspace 0.99992]$, which is not useful for
assessing whether the goal of fraction failing less than 0.10 has
been met.

The probability plot on the top-right gives Bayesian estimation
results with the partially informative prior, showing the improved
precision. A 95\% credible interval for $F(8000)$ is
$[0.15, \intervspace 0.92]$ indicating clearly that the goal of
fraction failing of less than 0.10 has not been met.

The middle and bottom rows of plots show likelihood contours along
with prior and posterior draws, respectively.
As described more fully in
Section~\ref{S.section:implementation.using.bayes.without.tears} of
the appendix, prior draws were obtained by sampling
from versions of the noninformative priors that were bounded by a
large rectangle (much larger than the boundaries of the contour
plots where the draws are plotted) so that the resulting prior
distributions are proper.
These plots provide a
visualization of how constraining the values of the Weibull shape
parameter $\beta$ to those that are consistent with engineering
knowledge importantly improves precision for estimation.

\subsubsection{Rocket motor field data}
\begin{figure}[!ht]
\begin{tabular}{cc}
\multicolumn{2}{c}{Weibull posterior cdf estimates and 95\% credible
  intervals based on the }\\
(a) weakly/noninformative prior & \phantom{xxxxx}(b) partially informative prior\\[-3.2ex]
\rsplidafiguresize{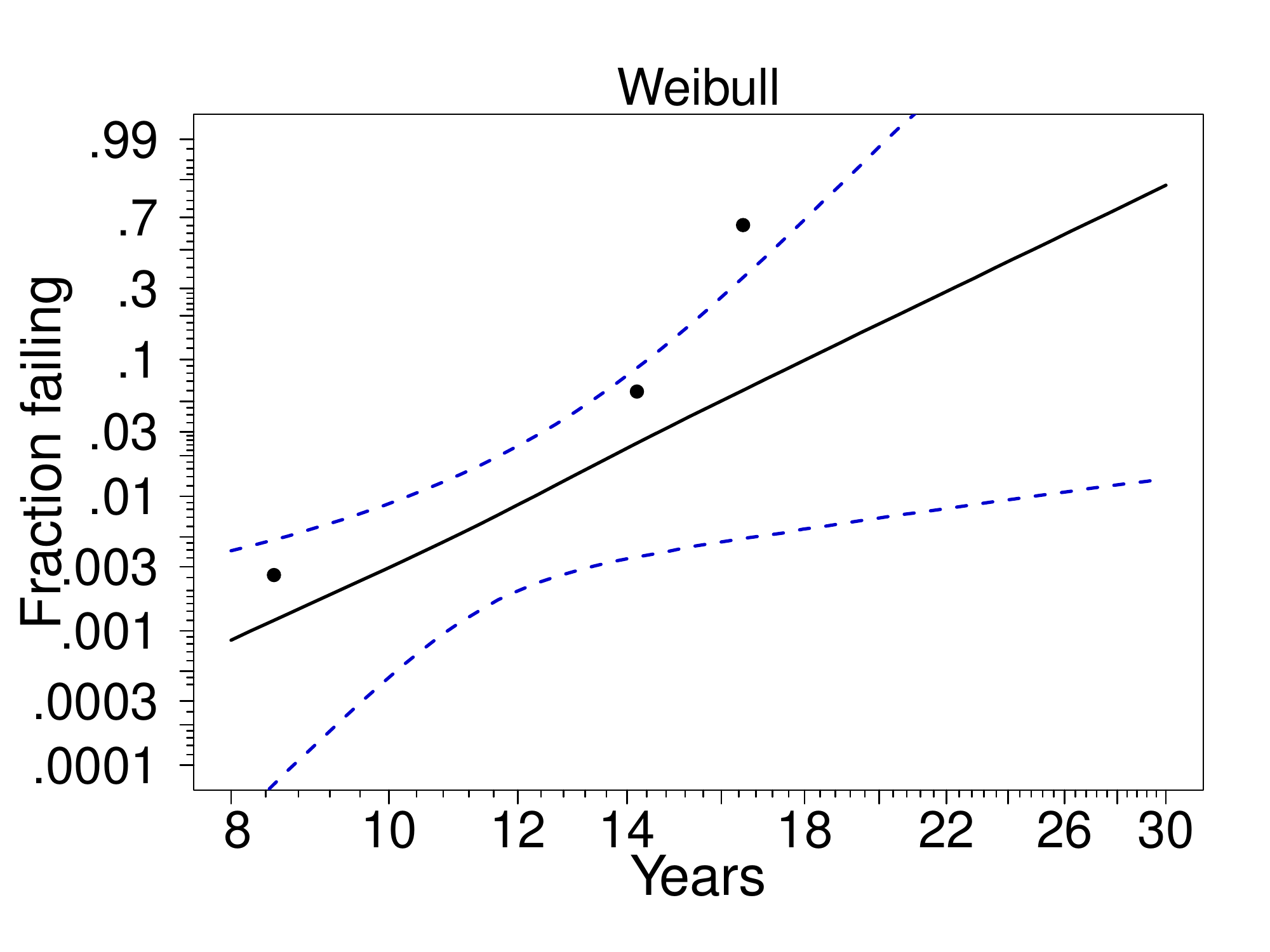}{3.25in} &
\rsplidafiguresize{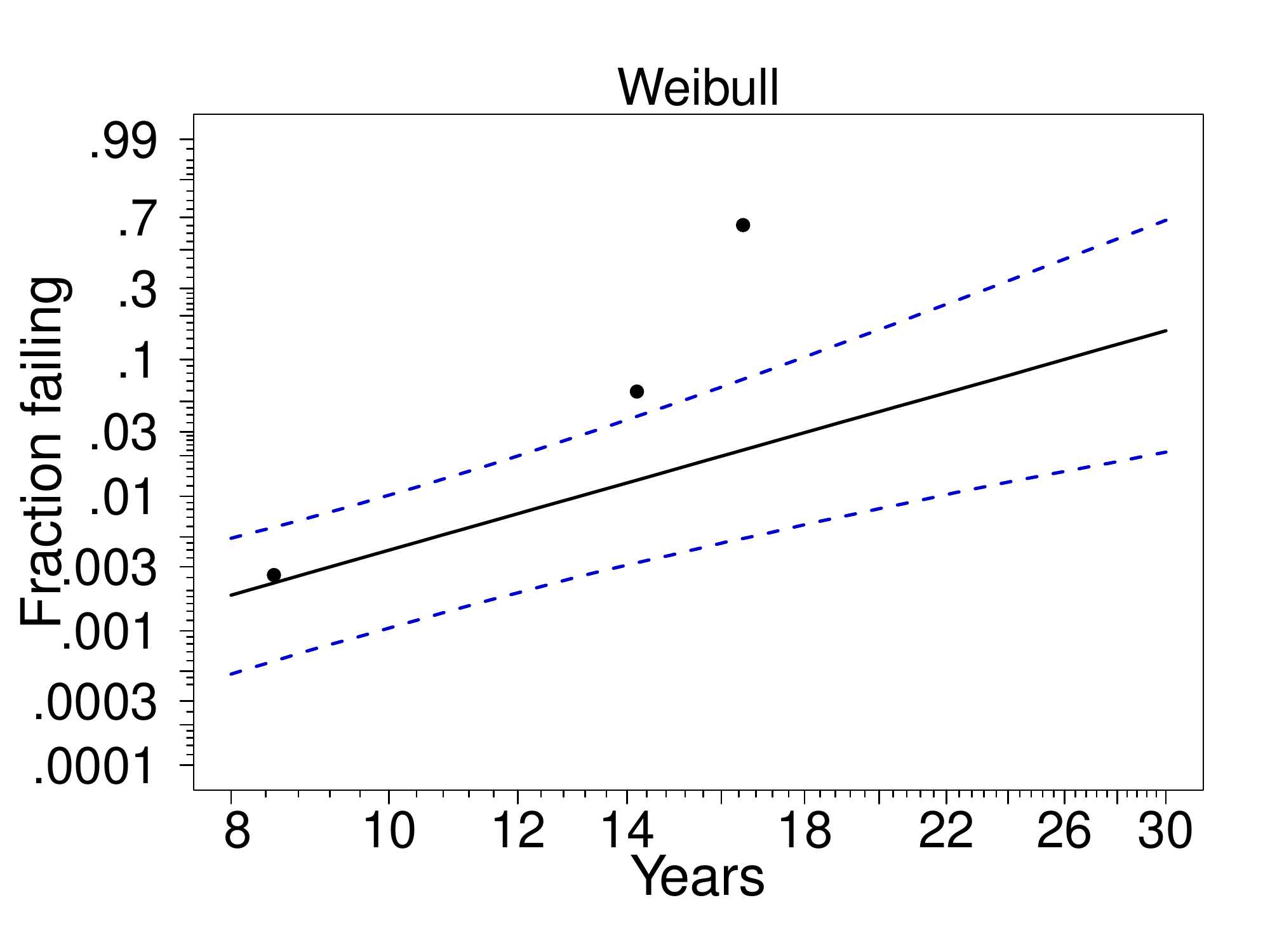}{3.25in}\\
\multicolumn{2}{c}{Weibull likelihood contours and  draws from the }\\
(c) weakly/noninformative prior & \phantom{xxxxx}(d) partially informative prior\\[-3.2ex]
\rsplidafiguresize{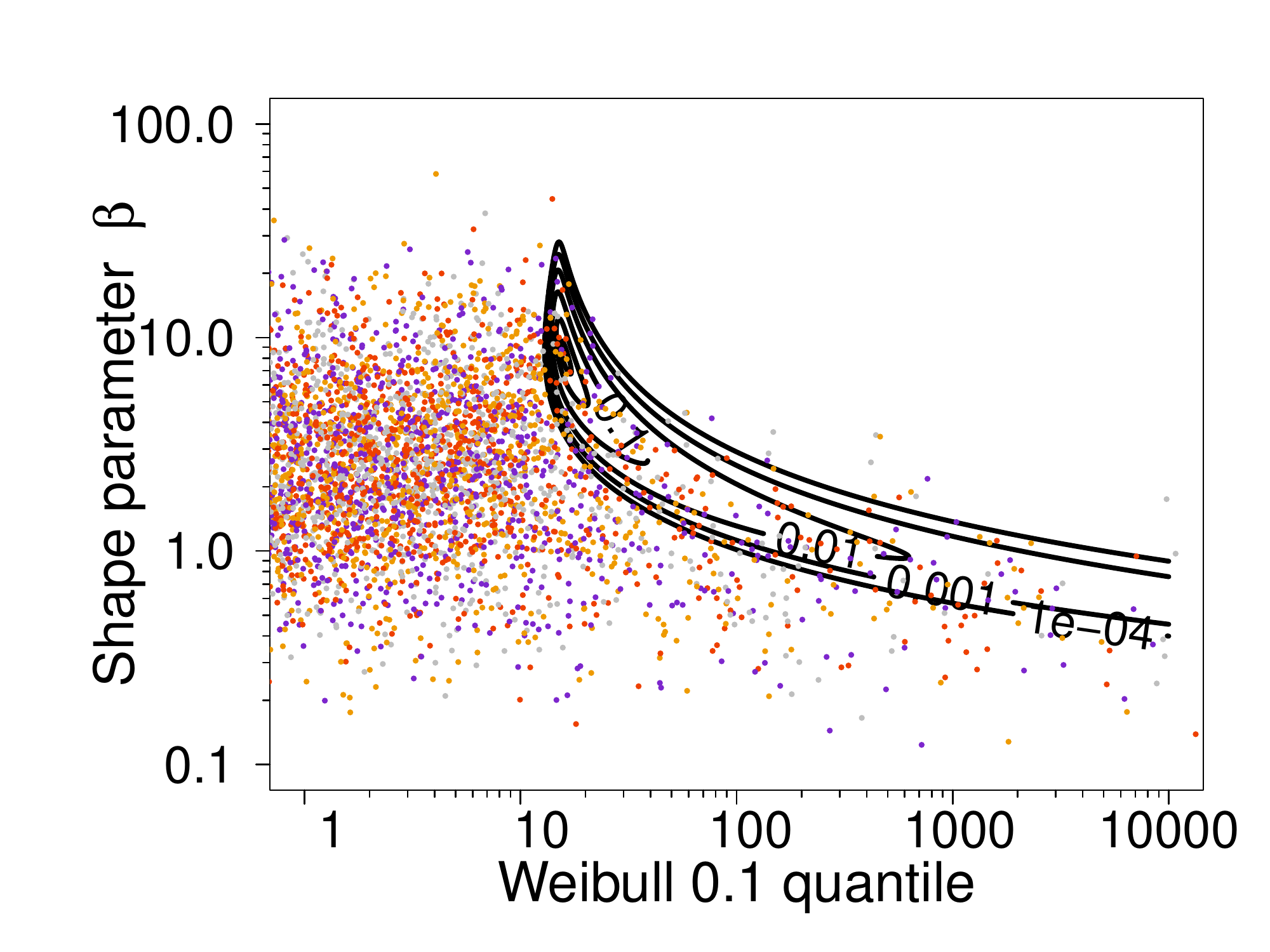}{3.25in} &
\rsplidafiguresize{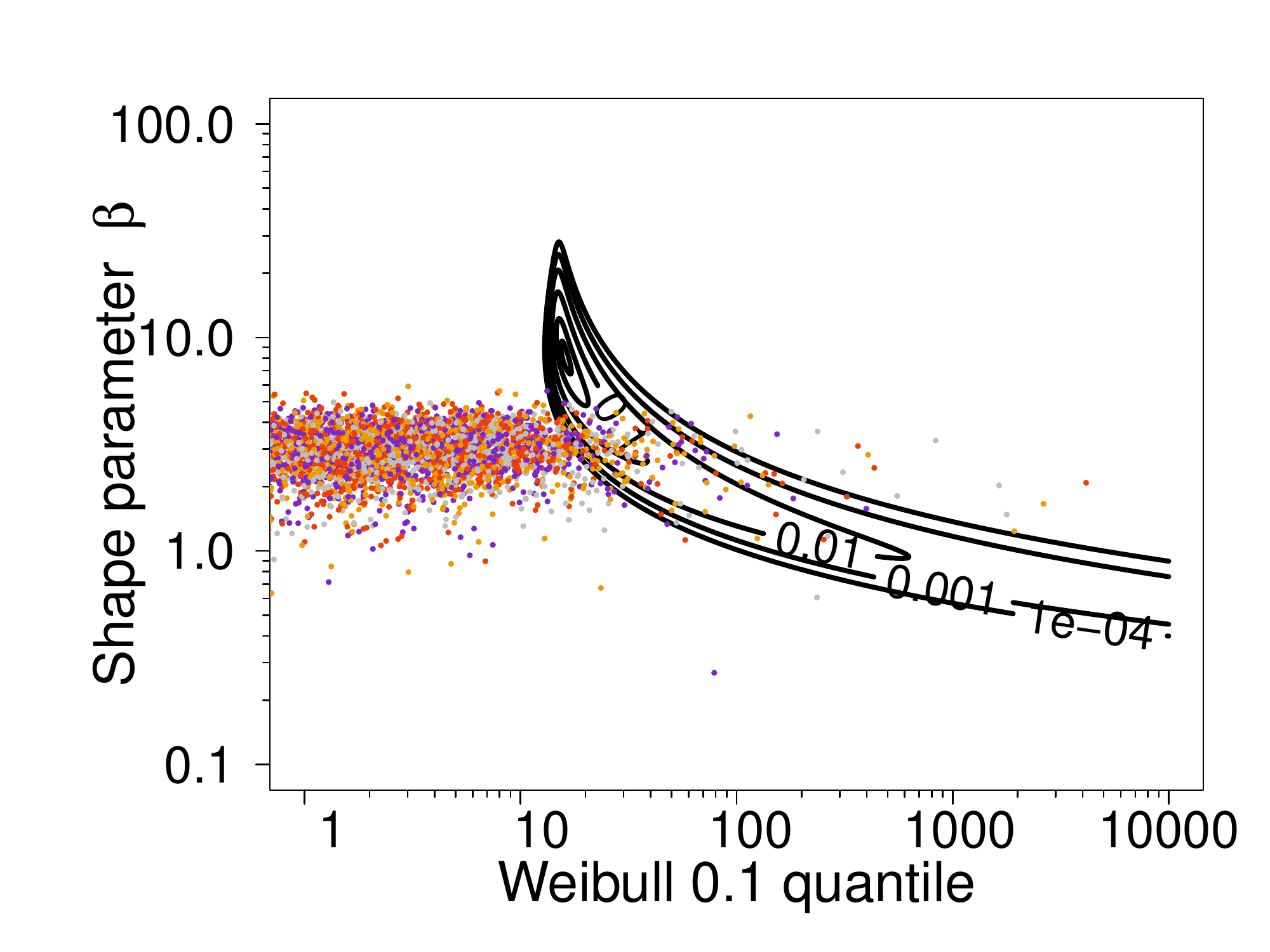}{3.25in}\\
\multicolumn{2}{c}{Weibull likelihood contours and posterior draws based on the}\\
(e) weakly/noninformative prior & \phantom{xxxxx}(f) partially informative prior\\[-3.2ex]
\rsplidafiguresize{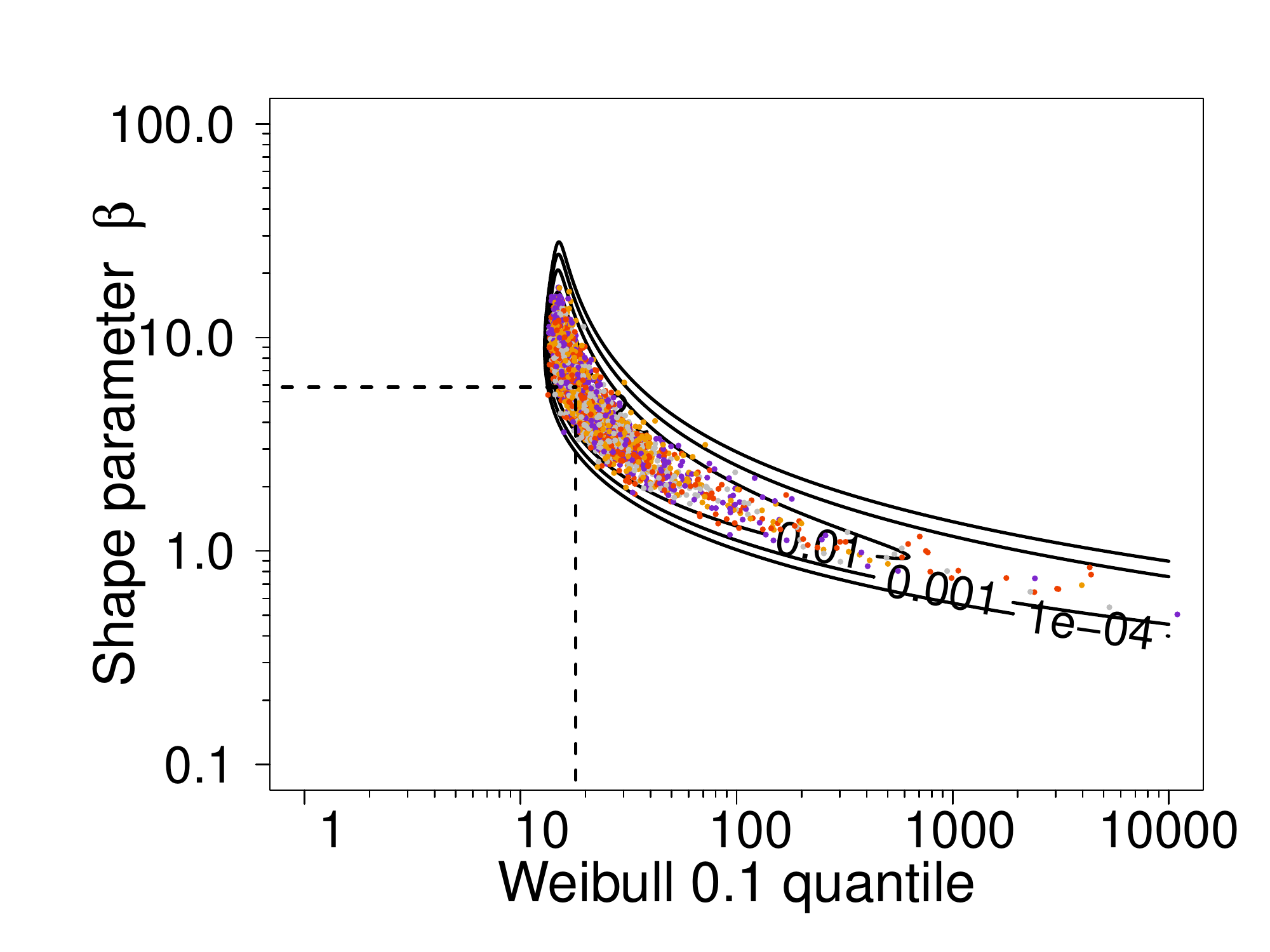}{3.25in} &
\rsplidafiguresize{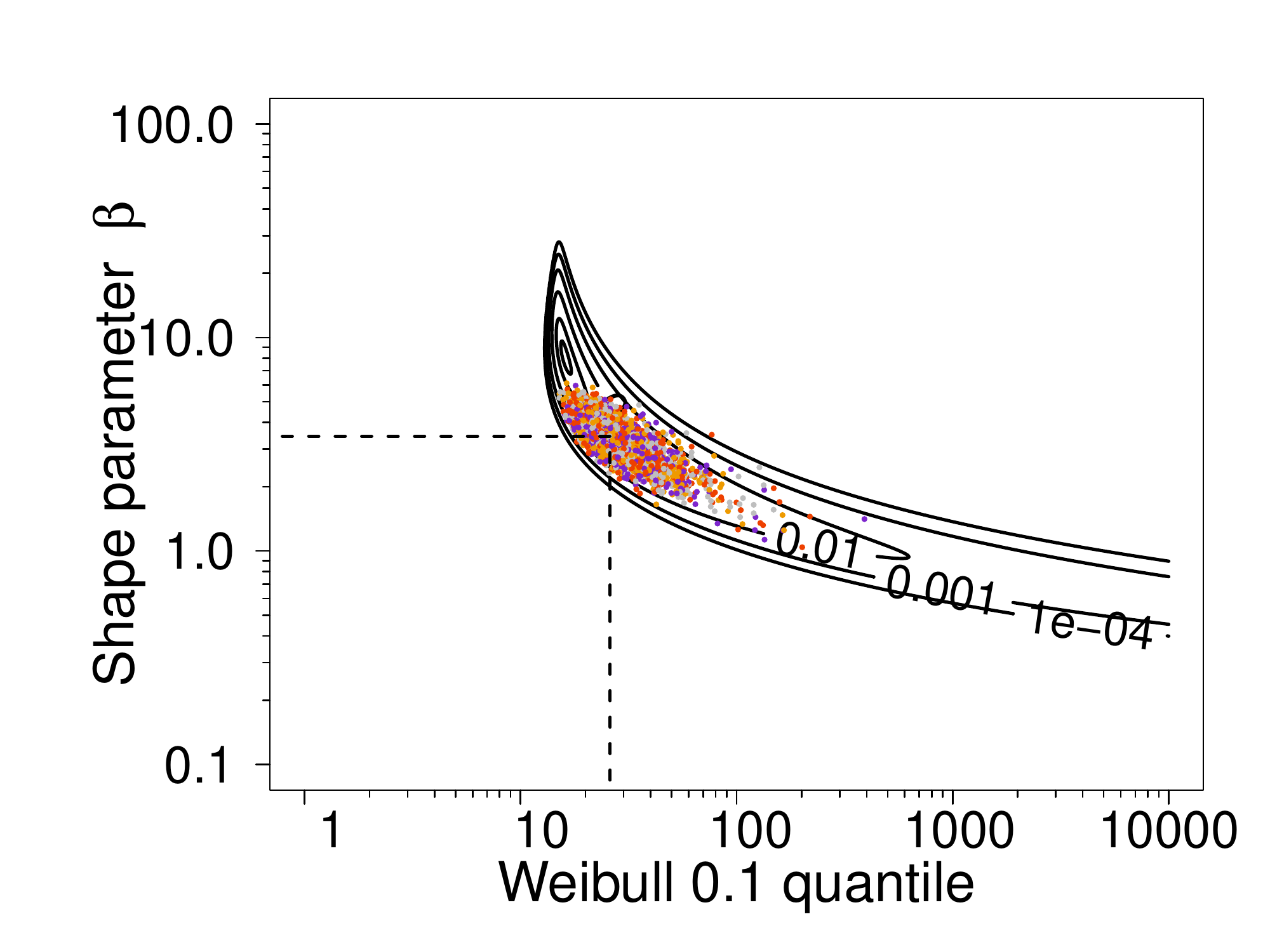}{3.25in}
\end{tabular}
\caption{Rocket motor Bayesian estimation results comparing
  a weakly informative/noninformative prior (on the left) with a
  partially informative prior (on the right). For each prior there is a Weibull
  probability plot showing the point estimate and credible intervals
for $F(t)$ (top), draws from the bounded joint prior and likelihood contours
(middle), and draws from the joint posterior and likelihood contours
(bottom).}
\label{figure:RocketMotor.noninformative.informative}
\end{figure}

This is a continuation of the rocket motor example in
Section~\ref{section:comparisons.ml.and.bayes} where ML estimates
were compared with Bayesian estimates based on
weakly informative prior distributions.
For the rocket motor field data,
Figure~\ref{figure:RocketMotor.noninformative.informative} compares
Bayesian estimation results for a noninformative/weakly informative
prior (on the left) and a partially informative prior (on the right).
As described in Section~\ref{section:comparisons.ml.and.bayes},
because of the limited amount of information in the three
left-censored observations, a weakly informative $\beta \sim
\logNormalPrior(0.2, 25)$ marginal prior was used for the
noninformative part of the comparison and, as in
Figure~\ref{figure:RocketMotor.ml.weaklyinformative.cj}(b), this was
paired with a noninformative CJ prior. \citet{OlwellSorell2001}
stated, ``It is unusual in Weibull analysis to get shape parameters
greater than 5.'' Also, because thermal cycling is a wearout mode,
$\beta>1$. Thus for the informative part of the comparison, the
weakly informative prior was replaced by a somewhat informative
$\beta \sim \logNormalPrior(1, 5)$.

The probability plot on the top-left in
Figure~\ref{figure:RocketMotor.noninformative.informative} gives
Bayesian estimation results with the noninformative/weakly
informative prior, and they are similar to the ML results given on
the right in Figure~\ref{figure:RocketMotor.plots.ps}. The credible
interval for $F(20)$ is $[0.007, \intervspace 0.98]$, which does not
help assess whether there is a serious problem or not.
The probability plot on the top-right gives Bayesian estimation
results with the partially informative prior, showing considerably
better precision for estimating $F(20)$. A 95\% credible interval for
$F(20)$ is $[0.008, \intervspace 0.16]$ which would help assess the
need for corrective action.
Similar to Figure~\ref{figure:BearingCage.noninformative.informative}
for the bearing cage example, the plots in the middle and bottom rows
of Figure~\ref{figure:RocketMotor.noninformative.informative} show
likelihood contours along with prior and posterior draws,
respectively. Again, these plots provide a visualization of how the
partially informative prior for the Weibull shape parameter $\beta$
improves estimation precision.

As mentioned in
Section~\ref{section:motivating.examples}, the ML estimate
$\betahat=8.126$ which, as suggested by \citet{OlwellSorell2001}, is
physically unreasonable for Weibull field data. 
Additionally, using years after manufacture as the surrogate
response for the unknown number and range of the thermal cycles has
the effect increasing variability (causing $\beta$ to be smaller).
\citet{OlwellSorell2001} also mention that it is well known
that the Weibull ML estimator for $\beta$ has serious upward bias
when there are few failures. The Bayesian estimate with the
noninformative/weakly informative prior $\betahat=5.9$, is an
improvement but still physically unreasonable. Using the partially
informative prior, described above, gives a much more reasonable
$\betahat=3.4$, but the change in slope results in the deviation
between the Bayesian and the nonparametric estimates of $F(t)$ seen
in Figure~\ref{figure:RocketMotor.noninformative.informative}(b).

\section{Simulation Study to Evaluate Alternative Noninformative Prior Distributions}
\label{section:weibull.type.one.simulation}
\subsection{Goals of the Simulation}
\label{section:goals}
As mentioned in Section~\ref{section:ij.prior}, for complete and
\typeII{} censored data from a (log-)location-scale distribution,
when using an IJ prior distribution, Bayesian credible intervals
have coverage probabilities that are the same as the nominal
credible/confidence level (i.e., giving ``exact'' interval procedures).
For \typeI{} and random censoring, this
result is approximate.
This section describes a simulation study to compare coverage
probabilities of credible intervals for the
flat and IJ noninformative prior distributions for the
most commonly used failure-time distributions (Weibull and
lognormal) under \typeI{} censoring. Here we focus on the Weibull
distribution simulation. Results for the lognormal distribution
simulation were similar and are given in
Section~\ref{S.section:lognormal.simulation.results.conclusions} of
the appendix. The rest of this section describes the
design of the study and the results.

\subsection{Simulation Factors}
\label{section:factors}
We simulated a life test where all units are put on test
simultaneously and are observed until a fixed censoring time,
$t_c$. Simulated failure times larger than $t_c$ are right-censored
at $t_c$ (i.e., \typeI{} censoring).  Without loss of generality, we
use the Weibull parameters $\mu=0$ and $\beta=1/\sigma=1$. The
experimental factors for the simulation were:
\begin{itemize}[itemsep=1mm, parsep=0pt]
\item $\E\big( r \big)$: the expected number of failures before $t_c$,
\item $p_{\fail}$: the expected proportion of failures before time $t_c$, and
\item $\pi[\log(t_{p_{r}}), \sigma]$: the joint prior distribution for the
  unknown parameters, where $p_{r}$ is the reparameterization
  quantile and is always chosen to be $r/(2n)$, resulting in a
  well-behaved likelihood (and posterior).
\end{itemize}
We use the expected number of failures instead of the sample size as
an experimental factor because it is a better measure of the
expected amount
of information in a data set and avoids a strong interaction that
arises when sample size and expected proportion censored are used as
factors.
This simulation is similar to that used in
\citet{JengMeeker2000} except that instead of comparing ten different
non-Bayesian confidence interval methods we compare \emph{the} Bayesian
method under different noninformative priors.

\subsection{Simulation Factor Levels}
\label{section:simulation.factor.levels}
In order to cover a range of situations and joint prior
specifications, we use the following levels of the factors:
\begin{itemize}[itemsep=1mm, parsep=0pt]
\item $\E\big( r \big)$ = 10, 25, 35, 50, 75, 100,
\item $p_{\fail}$ = 0.01, 0.05, 0.10, 0.50, and
\item $\pi[\log(t_{p_{r}}), \sigma]$ = flat and IJ.
\end{itemize}
For each factor-level combination, the sample size is computed as
$n=\E(r)/p_{\fail}$.  The censoring time is computed as $t_c =
\text{exp}\left(\mu + \Phi^{-1}_{\sev}(p_{\fail}) \sigma
\right)$.
We simulate $t_i$, $i=1 \dots n$.  Observations with $t_i >
t_c$ are coded as being censored at $t_c$.
As discussed in Section~\ref{section:improper.with.few.failures},
with fewer than three failures, the posterior is either improper
or poorly behaved. Thus we condition on at least 3 failures.

For each factor-level combination, we simulated 5,000 data sets.
As outlined in Section \ref{section:implementing.ij.priors},
the IJ prior $\pi[\log(t_{p_r}), \log(\sigma)]$ 
is specified in terms of the reparameterization
$(\log(t_{p_r}), \log(\sigma))$.  We obtain posterior draws
for each simulated data set, for both flat
and IJ noninformative prior distributions.

\subsection{Estimation}
\label{section:simulation.estimation}
Similar to the other examples in this paper,
the Weibull distribution models were fit using codes based on the
{\tt R} package {\tt rstan} \citep{rstan}.  {\tt R} package
\texttt{lsinf} was used to compute the scaled FIM needed for the IJ
priors.  Implementation details are described in
Section~\ref{S.section:experiences.using.stan.with.ij.prior} of the
appendix.
Four chains were run for each estimation run, resulting in
10,000 draws after warmup and thinning. Before doing the production
simulation runs, extensive experiments were conducted to investigate
the performance of the {\tt rstan} NUTS sampler for our different
factor-level combinations. 
Initial values for the NUTS sampler were obtained by generating
values from 99\% confidence intervals based on maximum likelihood estimates.
The Gelman-Rubin potential scale
reduction factor and close examination of select trace plots were
used to check for adequate mixing of the four chains
\citep{GelmanRubin1992}.
Performance of the prior distributions was evaluated in terms of
the error probabilities for each interval end point,
computed as the proportion of times the
computed 95\% credible interval lower (upper)
endpoint was greater than (was less than) the true value of the
quantiles being estimated.  Evaluations were done for credible intervals
for the Weibull quantiles $t_{0.01}$, $t_{0.05}$,
$t_{0.10}$, and $t_{0.50}$.

\subsection{Weibull Distribution Simulation Results and Conclusions}
\label{simulation.results.conclusions}
\begin{figure}
\centering
\begin{tabular}{cc}
\rsplidafiguresize{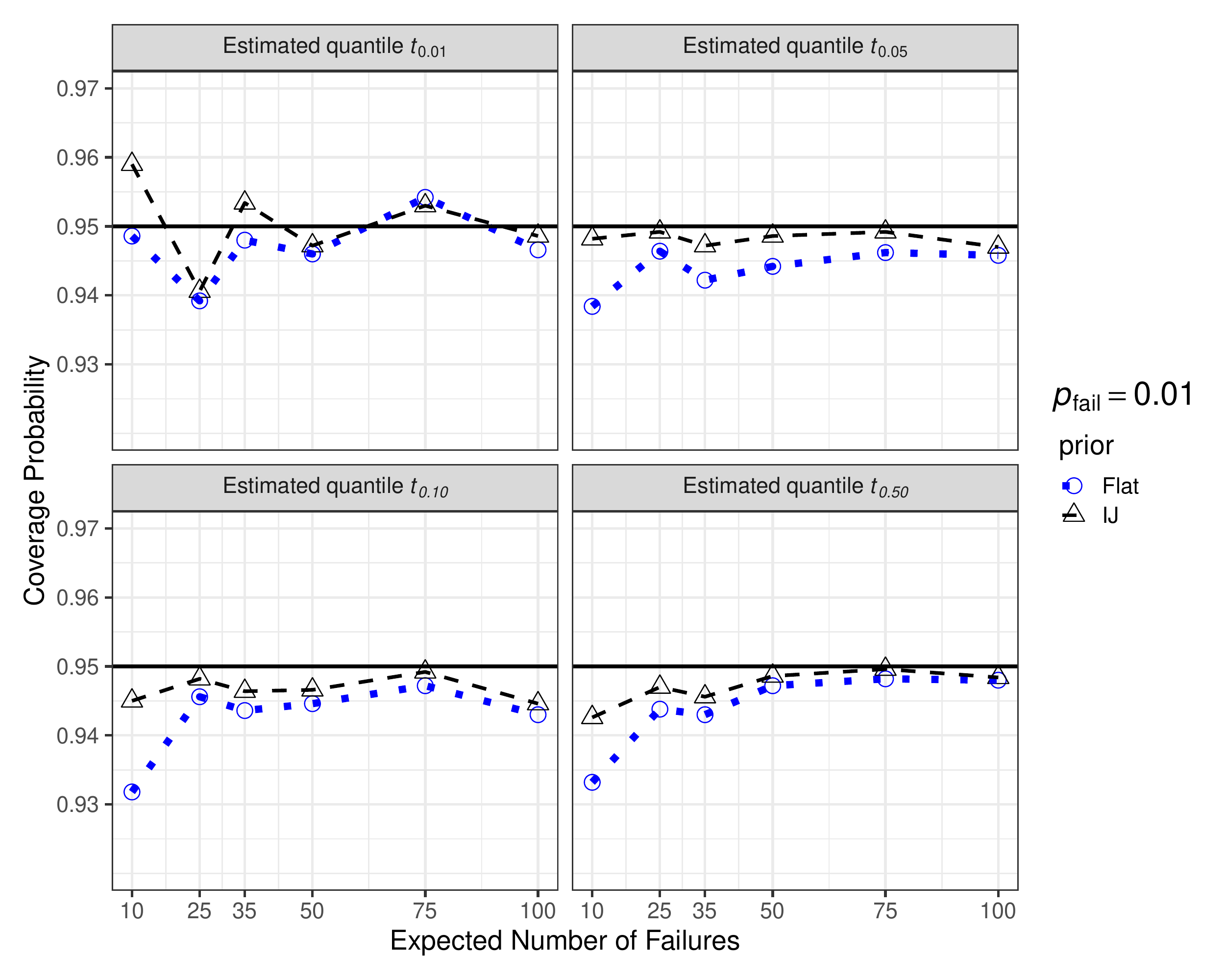}{6.0in}\\
\rsplidafiguresize{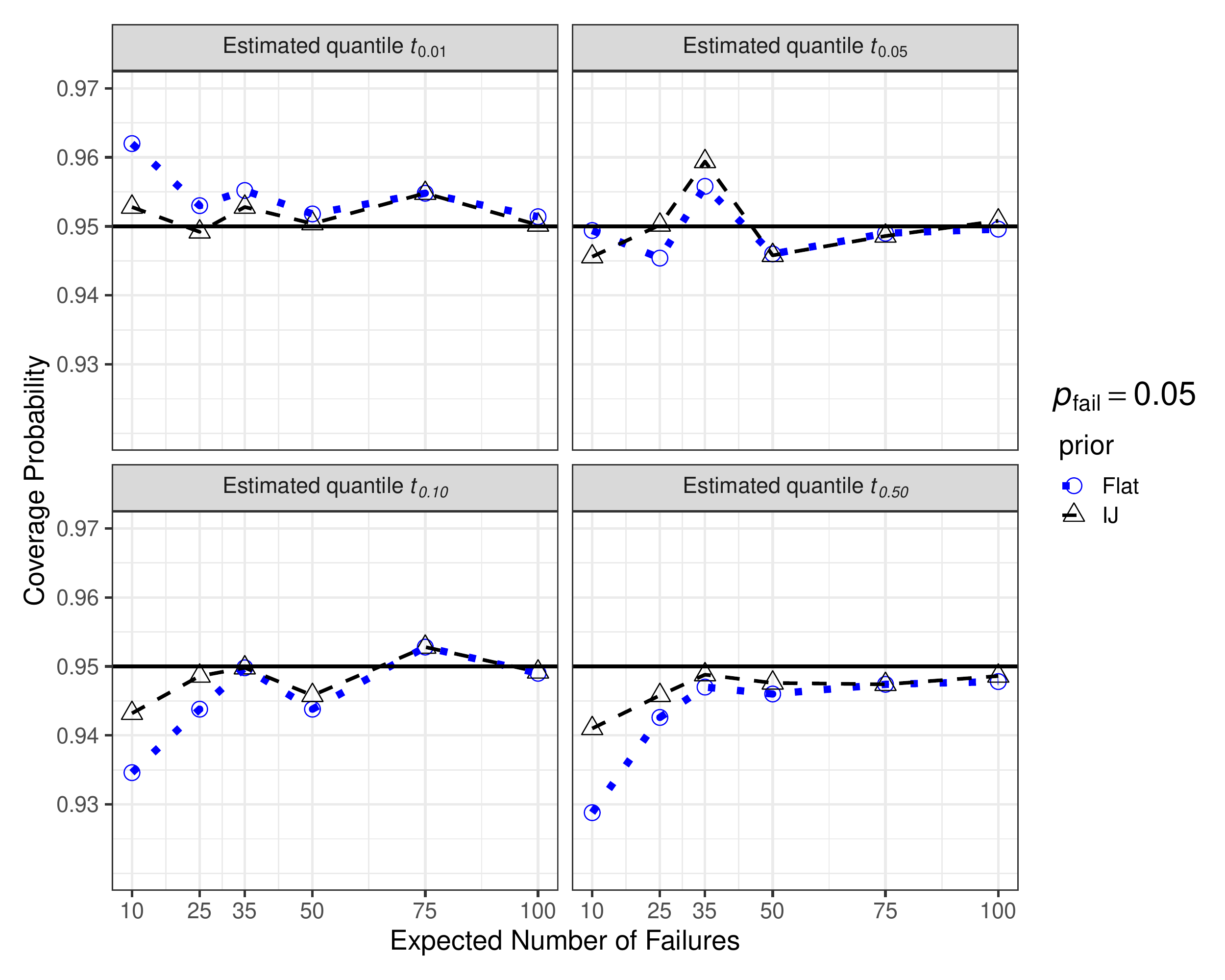}{6.0in}
\end{tabular}
\caption{Weibull distribution two-sided estimated coverage probabilities for $p_{\fail}=0.01$
  on the top and $p_{\fail}=0.05$ on the bottom.}
\label{figure:WeibullSimulationTwoSidedResults.small.pfail}
\end{figure}

\begin{figure}
\centering
\begin{tabular}{cc}
\rsplidafiguresize{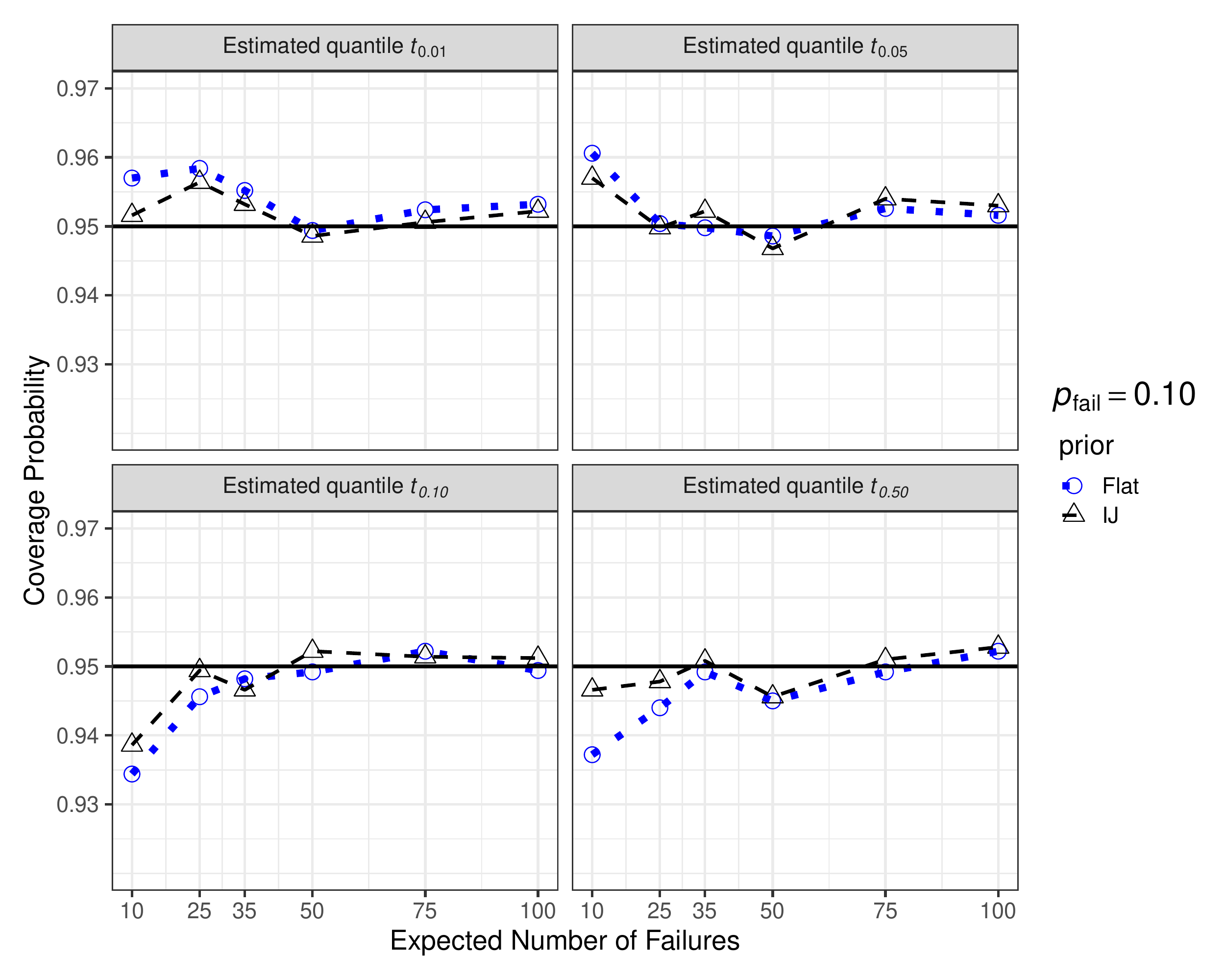}{6.0in}\\
\rsplidafiguresize{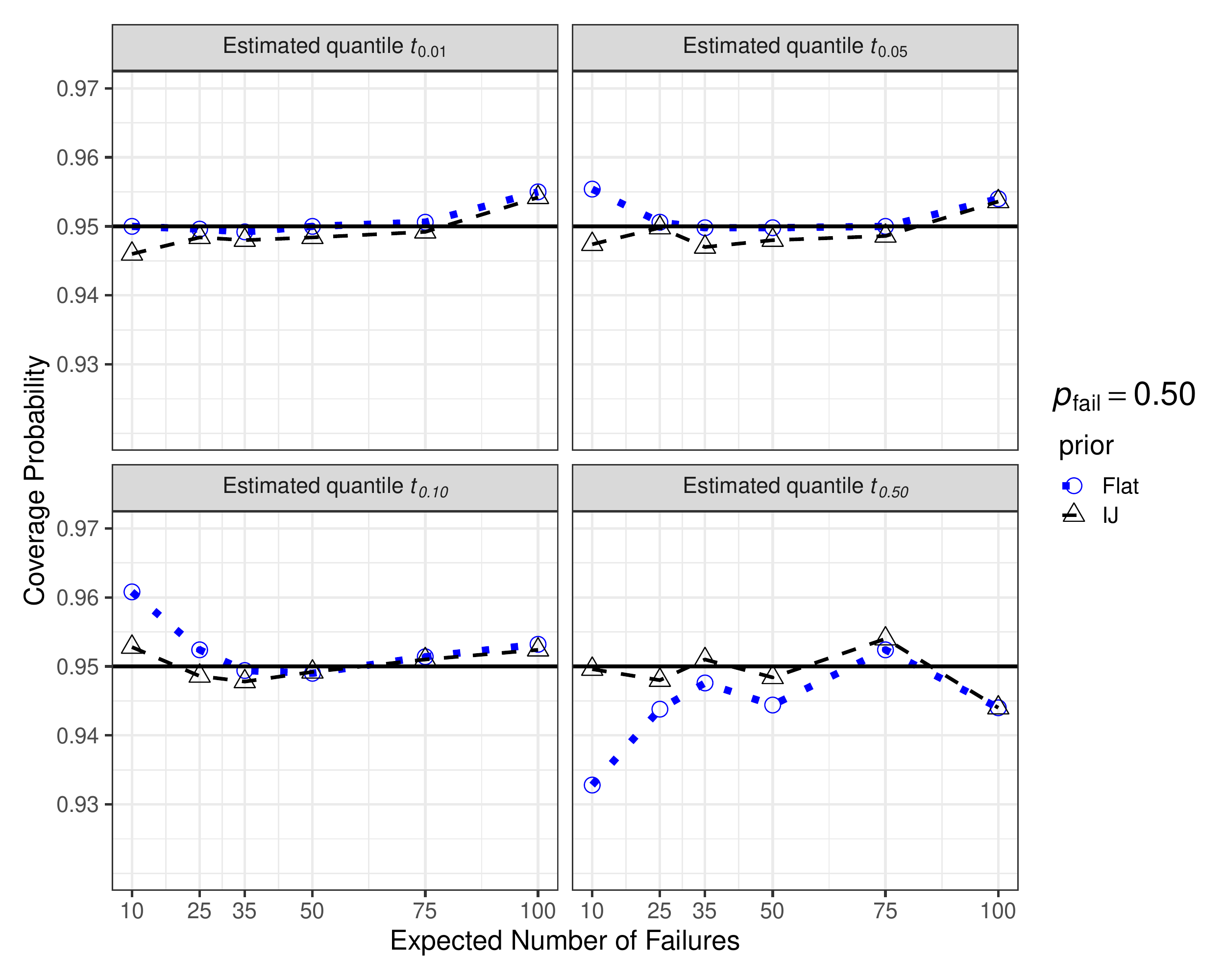}{6.0in}
\end{tabular}
\caption{Weibull distribution two-sided estimated coverage probabilities for $p_{\fail}=0.10$
  on the top and $p_{\fail}=0.50$ on the bottom.}
\label{figure:WeibullSimulationTwoSidedResults.large.pfail}
\end{figure}

\begin{figure}
\centering
\begin{tabular}{cc}
\rsplidafiguresize{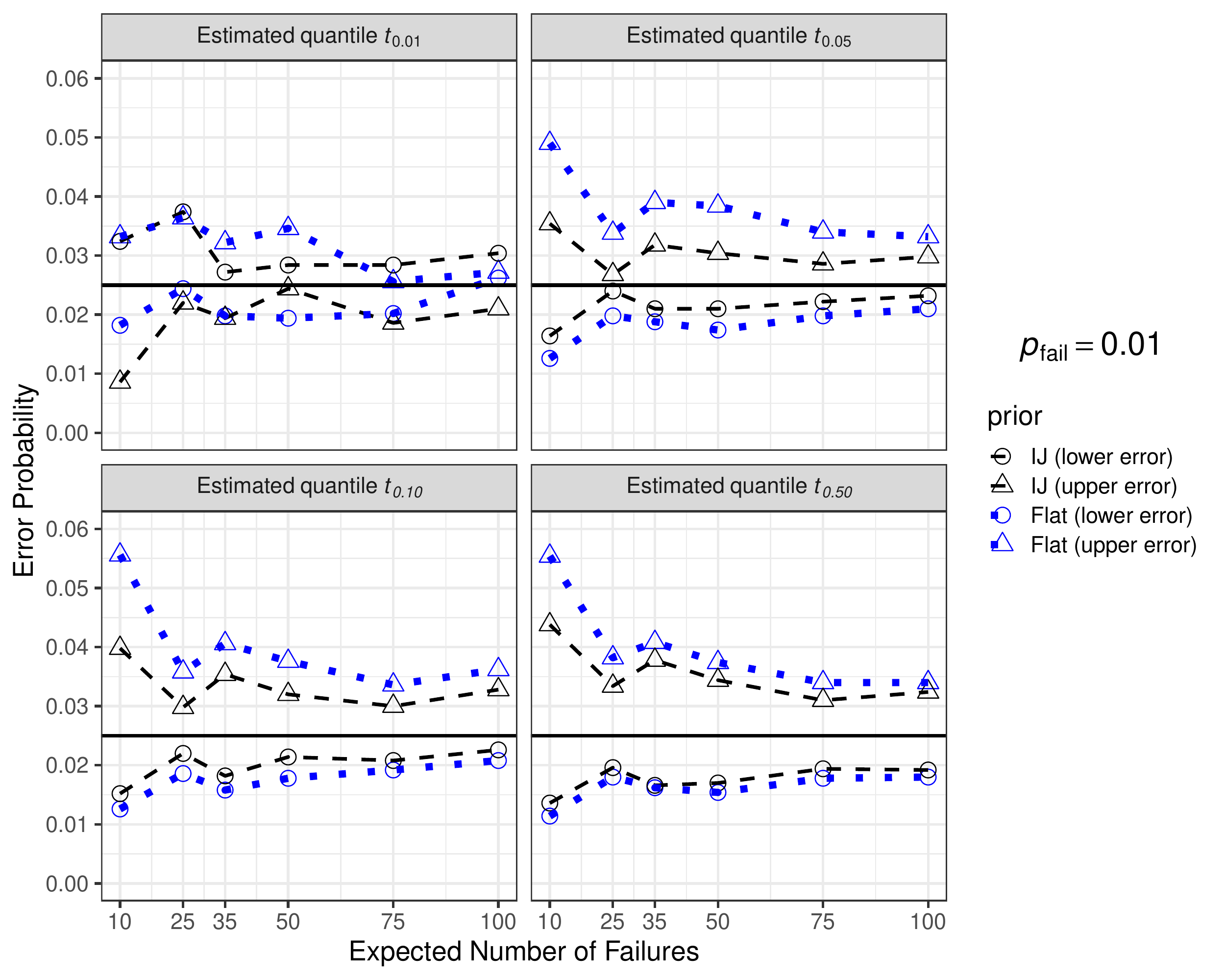}{6.0in}\\
\rsplidafiguresize{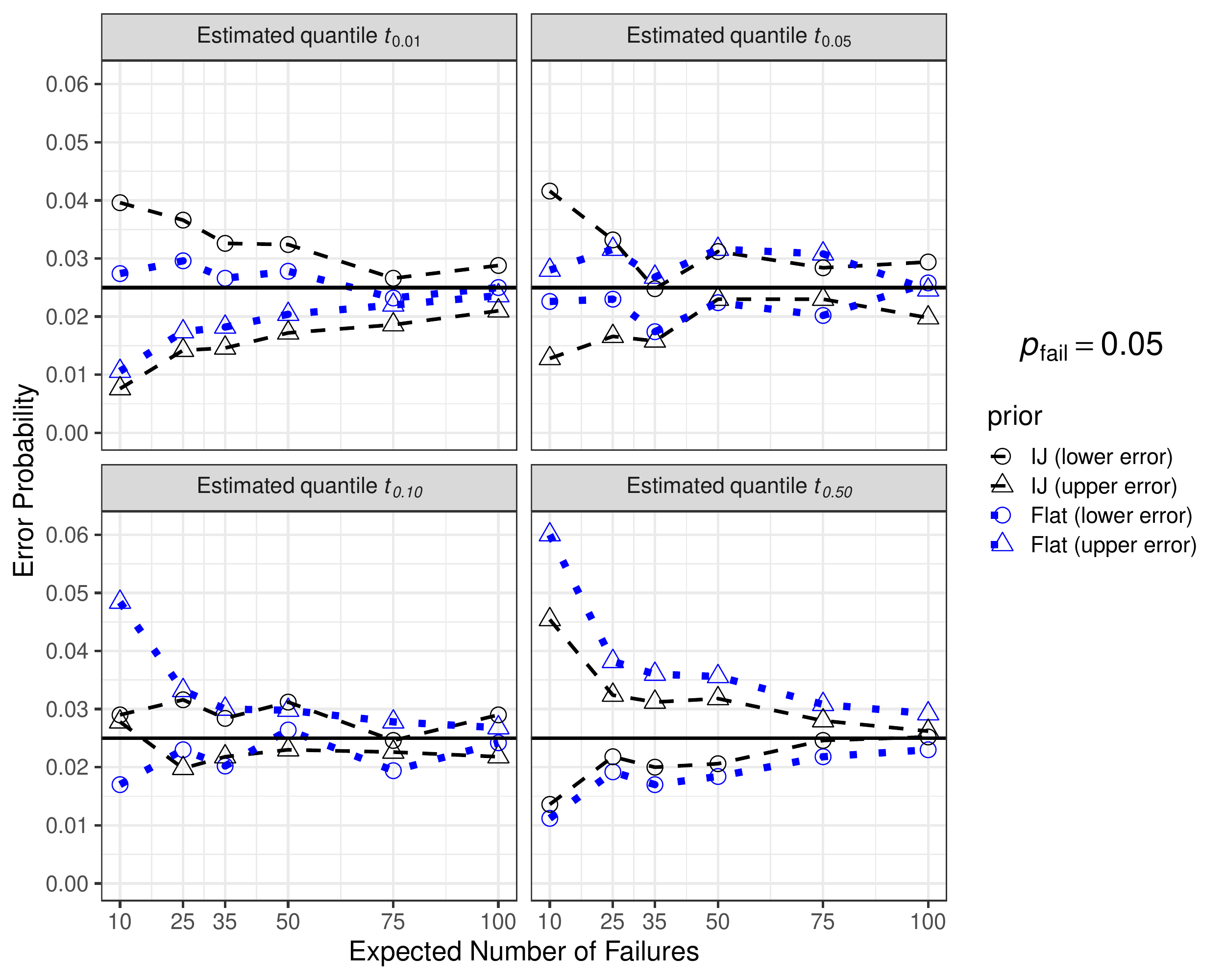}{6.0in}
\end{tabular}
\caption{Weibull distribution one-sided
estimated error probabilities for $p_{\fail}=0.01$
  on the top and $p_{\fail}=0.05$ on the bottom.}
\label{figure:WeibullSimulationOneSidedResults.small.pfail}
\end{figure}

\begin{figure}
\centering
\begin{tabular}{cc}
\rsplidafiguresize{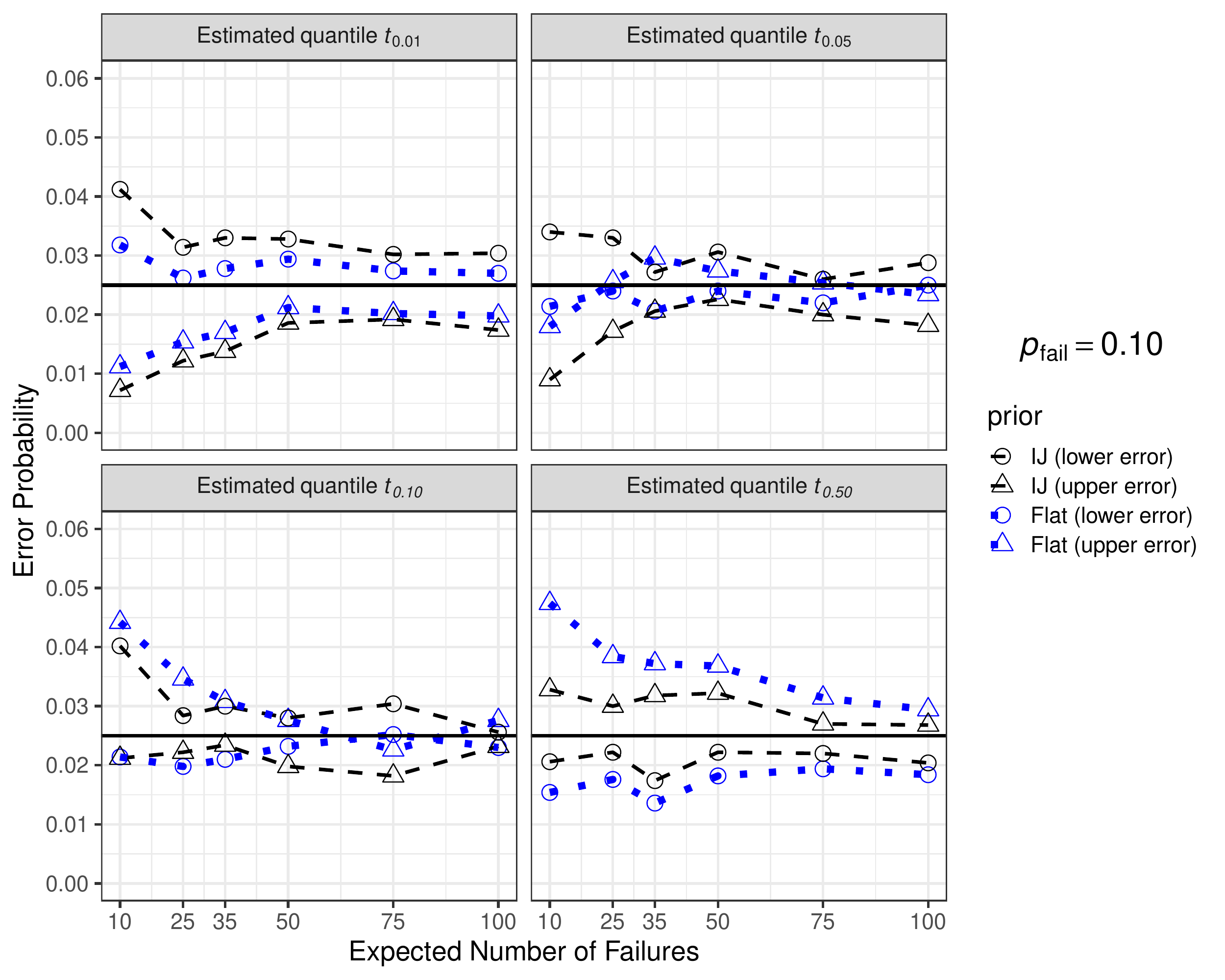}{6.0in}\\
\rsplidafiguresize{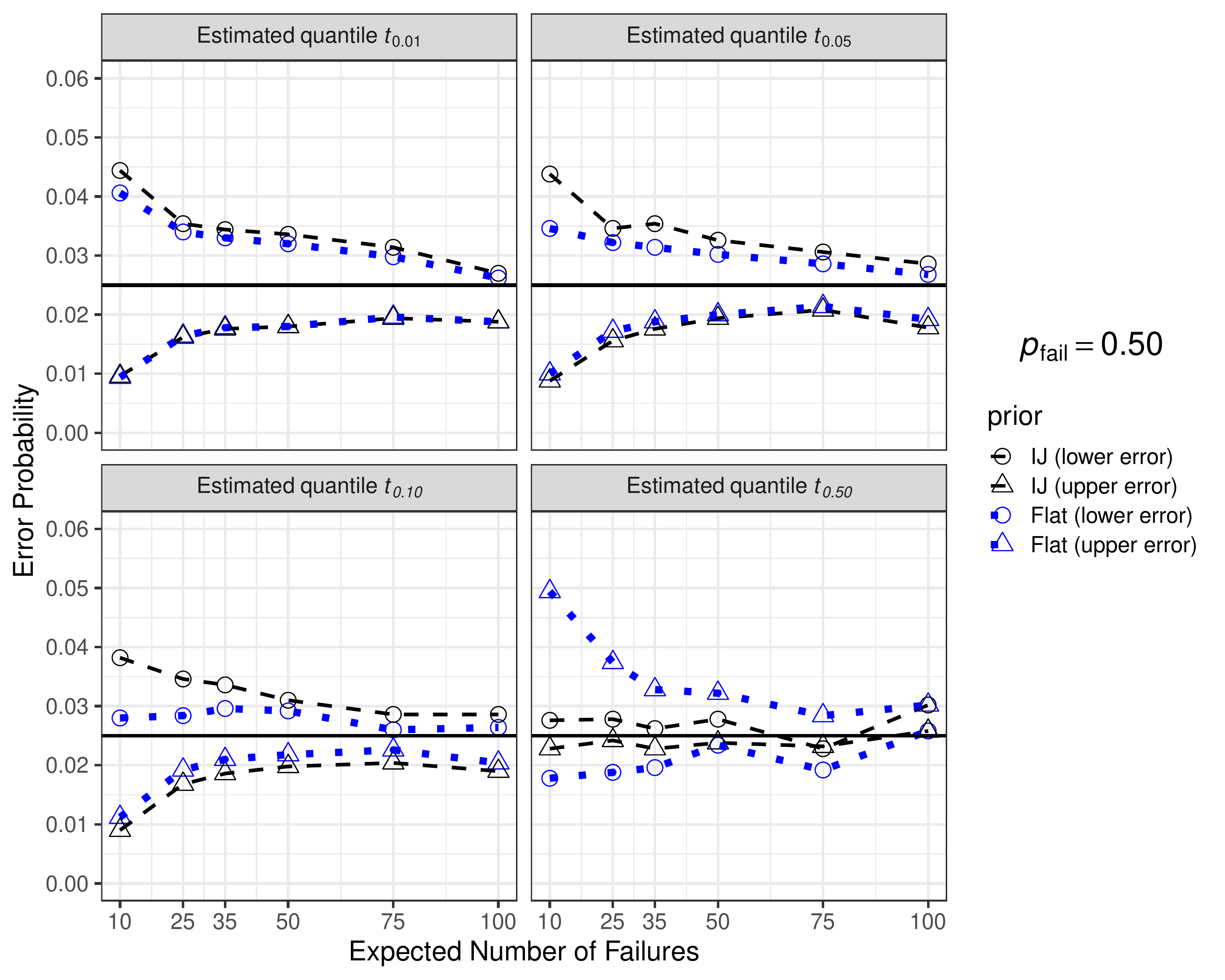}{6.0in}
\end{tabular}
\caption{Weibull distribution  one-sided
estimated error probabilities for $p_{\fail}=0.10$
  on the top and $p_{\fail}=0.50$ on the bottom.}
\label{figure:WeibullSimulationOneSidedResults.large.pfail}
\end{figure}

Figures~\ref{figure:WeibullSimulationTwoSidedResults.small.pfail}
($p_{\fail}=0.01$ and $p_{\fail}=0.05$) and~\ref{figure:WeibullSimulationTwoSidedResults.large.pfail} ($p_{\fail}=0.10$ and $p_{\fail}=0.50$) summarize
the Weibull distribution simulation results with two-sided coverage probabilities.
Figures~\ref{figure:WeibullSimulationOneSidedResults.small.pfail}
($p_{\fail}=0.01$ and $p_{\fail}=0.05$)
and~\ref{figure:WeibullSimulationOneSidedResults.large.pfail}
($p_{\fail}=0.10$ and $p_{\fail}=0.50$) summarize the Weibull
distribution simulation results with one-sided error probabilities.
Section~\ref{S.section:lognormal.simulation.results.conclusions} of
the appendix gives similar results for the lognormal
distribution.

Note that if $\alpha_{L}$ is the error probability for the lower
endpoint of a credible interval and $\alpha_{U}$ is the error
probability for the upper endpoint, then the two-sided coverage
probability is $1 - \alpha_{L} -\alpha_{U}$.
When interpreting the results of the simulation, it is important to
keep in mind that, as mentioned in
Section~\ref{section:simulation.factor.levels}, within each of the
eight plots in
Figures~\ref{figure:WeibullSimulationTwoSidedResults.small.pfail}--\ref{figure:WeibullSimulationOneSidedResults.large.pfail}, the results
are based on the same set of 5,000 simulated data sets. Thus, for
example, both points in a plot for a particular value of
$\E(r)$  and estimated quantiles tend to move together. Given the
nominal credible level of 0.95 for the intervals, the nominal error
probabilities for each tail are 0.025. 
The standard error of the estimated coverage probabilities is
approximately $\sqrt{0.95(1-0.95)/5000}=0.003$.
The standard error
of the estimated error probabilities is approximately
$\sqrt{0.025(1-0.025)/5000}=0.002$. Also, recall that the simulation
results are conditional on observing at least three failures. The
probabilities of fewer than three failures when $\E(r) = 10$ are
0.0027, 0.0023, 0.0019, and  0.0002, respectively, for
$p_{\fail}=0.01$, 0.05 0.10, and 0.50.

Some observations from the two-sided estimated coverage
probabilities in Figures~\ref{figure:WeibullSimulationTwoSidedResults.small.pfail}
and~\ref{figure:WeibullSimulationTwoSidedResults.large.pfail} are:
\begin{itemize}[itemsep=1mm, parsep=0pt]
\item
The IJ coverage probabilities tend to be better or the same as flat almost everywhere.
\item
There can be a large departure from the nominal $0.95$ when
$\E(r)=10$, perhaps partially because of the conditioning on $r \ge 3$.
\item
Both flat and IJ priors are close to nominal for $\E(r) \ge 35$.
\end{itemize}

\noindent
Some observations from the one-sided estimated error
probabilities in Figures~\ref{figure:WeibullSimulationOneSidedResults.small.pfail}
and~\ref{figure:WeibullSimulationOneSidedResults.large.pfail} are:
\begin{itemize}[itemsep=1mm, parsep=0pt]
\item
As is common for approximate interval procedures the lower and upper
bound error probabilities are not the same. When one error probability (either lower or
upper) is above nominal, the other is below nominal resulting in
some cancellation that makes the two-sided coverage close to nominal.
\item
In most cases, the IJ performs better than flat (i.e., smaller error probabilities), but there are some
exceptions. Generally, the differences are not large.
\item
For the $p_{\fail}=0.01$, the IJ priors have smaller error
probabilities for all levels of $\E(r)$.
\item
The error
probabilities for the IJ and flat priors tend to get closer together
as $p_{\fail}$ increases and especially for $p_{\fail}=0.50$.
\end{itemize}

The results for the lognormal distribution
Section~\ref{S.section:lognormal.simulation.results.conclusions} of
the appendix are similar. From these simulation results, and
consistent with the comments at the end of
Section~\ref{section:implementing.ij.priors}, we conclude that the
IJ prior can provide useful improvement in performance in \typeI{}
(and similar) applications: particularly when there are  few failures or a small fraction of failures.

\section{Prior Distribution Sensitivity Analysis for the Motivating Examples}
\label{section:prior.sensitivity.analysis}
As mentioned previously, with limited information in the data (e.g.,
a small number of failures due to censoring in reliability
applications), the choice of a prior distribution could have a strong
influence on inferences. When attempting to use weakly informative
prior distributions, it is useful to experiment with different
specifications to assess sensitivity.

\begin{figure}
\begin{tabular}{cc}
\rsplidafiguresize{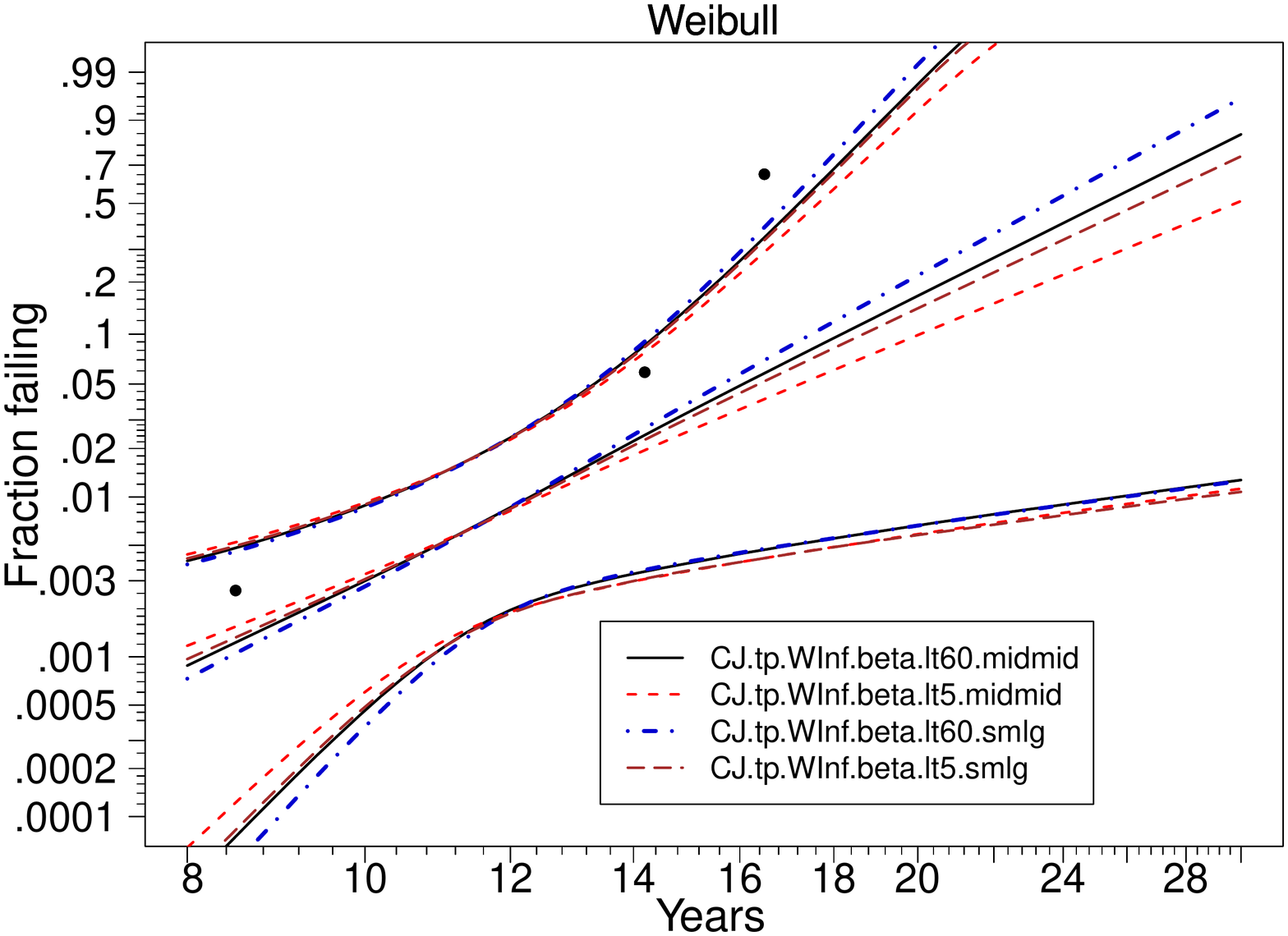}{3.65in}&
\rsplidafiguresize{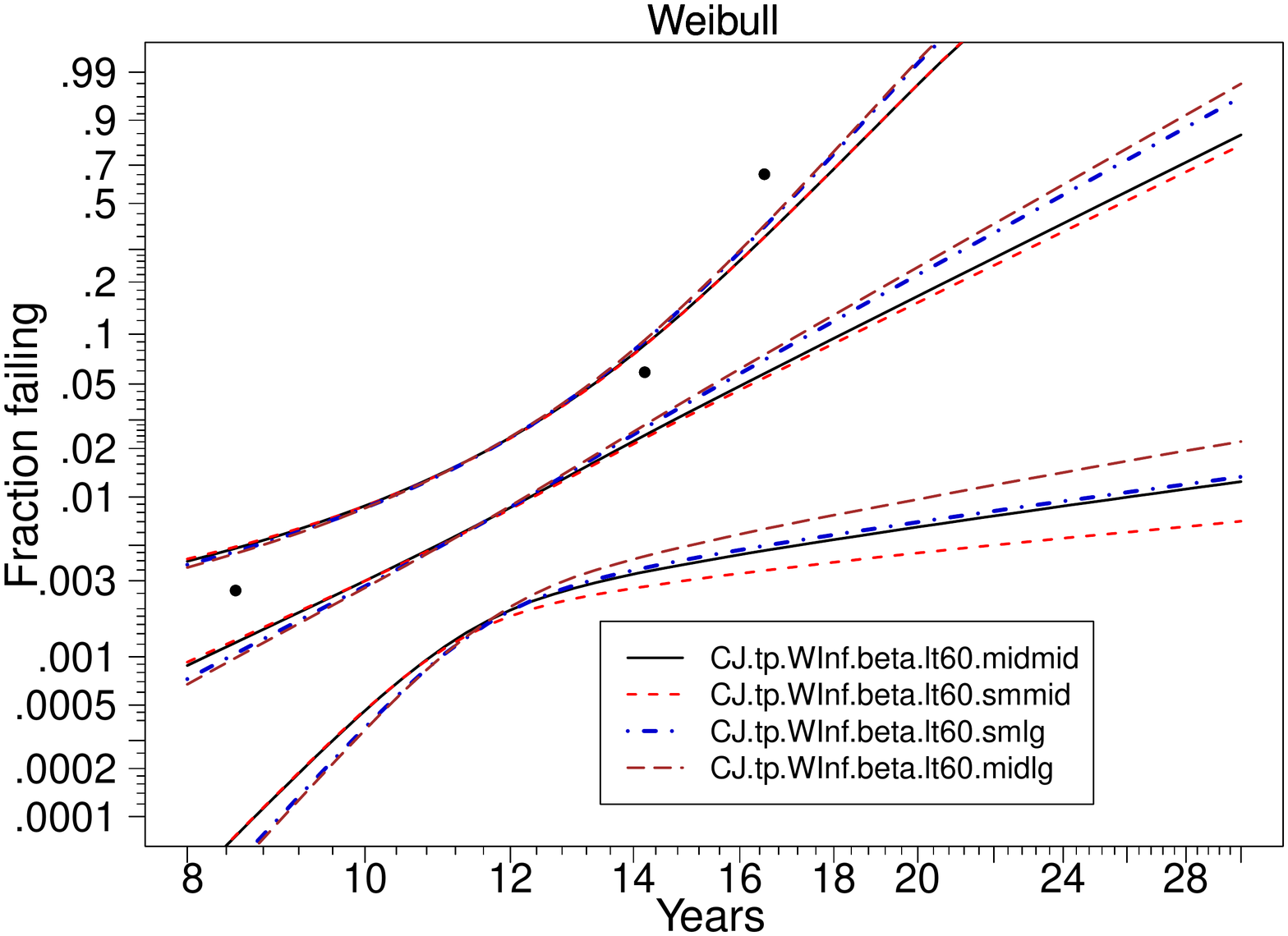}{3.65in}
\end{tabular}
\caption{Sensitivity analysis comparing different weakly informative
  priors for the Weibull shape parameter $\beta$.}
\label{figure:RocketMotorWinfSensitivity}
\end{figure}

The plots in Figure~\ref{figure:RocketMotorWinfSensitivity} show two
different $2 \times 2$ factorial comparisons.
In both cases, the baseline prior is the log-location-scale-$t$
distribution with 60 degrees of freedom, denoted by $\beta \sim
\logLST{60}(0.20, 25)$, which is essentially equivalent to the $\beta
\sim \logNormalPrior(0.20, 25)$ used in the rocket motor example in
Section~\ref{section:comparisons.ml.and.bayes}.
The plot on the left varies the degrees of freedom between 5 and 60
(heavy versus lighter tails) and the 99\% quantile range for $\beta$
between $(0.20,
25)$ and $(0.10, 50)$ (smaller to larger range for the Weibull shape parameter).
The plot on the right varies the lower endpoint of the 99\% quantile
range for $\beta$ between
range between $0.10$ and $0.20$ and the upper endpoint between 25 and 50
(changing both the log-location-scale-$t$ prior median and shape
parameter). Note that the plots have two common factor-level
combinations.

There is not much difference among the different priors between years
11 and 14 where there is more information. Not surprisingly, when
extrapolating beyond 16 years, the point estimates and the upper
uncertainty bounds have what might be considered to be important
differences. In practice, one would consider the plausible ranges for
a parameter (even the narrower 99\% quantile range $(0.20, 25)$ for $\beta$ is
extremely wide relative to $\beta$ values typically encountered in
practice) and then perhaps choose a weakly informative prior that is
a compromise or somewhat conservative.

\section{Concluding Remarks and Areas for Future Research}
\label{section:concluding.remarks}
This paper provides guidance for setting prior distributions for
log-location-scale distributions used in reliability
applications. We applied our recommendations to
field data applications with complicated censoring
and derived new noninformative
priors that have good coverage-probability properties
that will be especially useful in small-information (e.g., few
failures) applications
where the use of methods based on
asymptotic theory can give misleading results.

The general
principles and prior distributions we suggest are applicable in many
other reliability models and in domains outside of reliability. For
example, in accelerated testing, engineers often have useful prior
information about the activation energy of a temperature-accelerated
failure mode or regression coefficients in other kinds of
acceleration models (but not the other parameters in the model).
\citet{XuFuTangGuan2015} describe the use of
noninformative prior distributions for accelerated test models, and
their methods could be extended to the more commonly used \typeI{}
censoring.  Bayesian methods are also being used for other types of
reliability data, including degradation data,
fatigue S-N data, and recurrent events
data. Noninformative priors are needed for most, if not all, of the
model parameters in such applications.

The further development and implementation (e.g., in widely
available software) of default noninformative and partially
informative prior distributions is important for making the
advantages of Bayesian methods (described in our abstract) more
accessible to practitioners. For example, after selecting a failure-time
distribution, an analyst could be presented with options
to override default noninformative priors and specify
a weakly informative or an informative prior for each of the failure-time
distribution parameters.  Such functionality would
allow users to easily perform a sensitivity analysis to see the
effect that various noninformative/weakly informative
priors have on the estimation results.
Once this type of software is available, other than additional
computational effort, there would be little justification for
recommending non-Bayesian model fitting in reliability applications.

\section{Acknowledgments}
Luis Escobar provided assistance in refining the proofs in
Section~\ref{S.section:limiting.results.weakly.informative.prior.distributions} of
the appendix and for suggesting simplifications for
the computation of the log-truncated-normal and
log-reciprocal-truncated-normal probability density functions
described in
Section~\ref{S.section:log.truncated.normal.distributions} of the
appendix.
Necip Doganaksoy, Luis Escobar,
Mike Hamada, Yili Hong, Qingpei Hu, Yew Meng Koh,
Larry Leemis, Peng Liu, Lu Lu, Jave Pascual, and Grant Reinman provided helpful
suggestions that improved the paper.
We would also like to thank the associate editor and the
anonymous reviewers who made many
comments and suggestions that helped us improve the paper.

\newpage

\appendix

\part*{Appendix}

\section{Overview of the Appendix}
This appendix provides derivations, proofs,
additional detailed descriptions, and additional simulation results.
Section~\ref{S.section:derivations.noninformative.priors} shows how
to derive Jeffreys, Independence Jeffreys (IJ),
and reference priors for different parameterizations.
Section~\ref{S.section:understanding.ij.prior.features} describes
features of the \typeI{} censoring IJ priors
and the reasons that those features arise.
Section~\ref{S.section:experiences.using.stan.with.ij.prior}
describes some details of implementing a Conditional Jeffreys (CJ) or
IJ prior using Stan.
Section~\ref{section:weibull.type.one.simulation} in the main paper
gives results for a simulation using the Weibull
distribution. Simulations were also done for the lognormal
distribution and they are described in
Section~\ref{S.section:lognormal.simulation.results.conclusions}.
Section~\ref{S.section:careful.look.three.failures} looks carefully
at estimation results under different noninformative priors for two
data sets with only three failures,
based on two of the simulation
(Section~\ref{section:weibull.type.one.simulation}) factor-level combinations.
Section~\ref{S.section:limiting.results.weakly.informative.prior.distributions}
provides proofs for the limiting results for normal/lognormal weakly
informative prior distributions that are described in
Section~\ref{section:weakly.informative.priors.lls.distributions}.
Section~\ref{S.section:log.truncated.normal.distributions} gives
expressions for the log-truncated and log-reciprocal-truncated pdfs
that are used in our Stan implementation of truncated normal and truncated 
location-scale-$t$ informative prior distributions.

\section{Derivations of the Noninformative Prior Distributions}
\label{S.section:derivations.noninformative.priors}
\subsection{Motivation for and Overview of the Derivations}
Section~\ref{section:summary.of.noninformative.priors} and
Table~\ref{table:summary.noninformative.priors} in the paper
provide a summary of noninformative prior distributions (Jeffreys,
IJ, and Reference) for different parameterizations of
(log-)location-scale distributions. The different parameterizations
are needed because
\begin{itemize}
\item
In reliability applications,
inferences are generally needed on alternative ``parameters'' such
as quantiles or failure probabilities.
\item
Priors may be elicited for parameters that differ from the
traditional parameters.
\item 
MCMC computations may be conducted using a parameterization
that differs from the parameterization where the prior is
elicited/specified.
\end{itemize} 
This section provides technical details and
derivations of these prior distributions.
Section~\ref{S.subsec:mu-sigma} shows how to compute the elements of
the Fisher information matrix (FIM) for the $(\mu, \sigma)$
parameterization.  Section~\ref{S.section:jeffreys.priors} defines
and gives general methods for obtaining the Jeffreys, CJ, and IJ
priors for the $(\mu, \sigma)$ parameterization for \typeII{} and
\typeI{} censoring.  Section~\ref{S.section-ref-prior} defines and
gives general methods for obtaining reference priors for the $(\mu,
\sigma)$ parameterization for \typeII{} censoring.
Sections~\ref{S.subsec:use-y_p-sigma}--\ref{S.section-zeta-log.sigma}
derive the noninformative priors for other parameterizations.

\subsection{The Fisher Information Matrix for $(\mu, \sigma)$}
\label{S.subsec:mu-sigma}
The FIM plays an important role in
computing noninformative priors such as Jeffreys, IJ, and reference priors.
Suppose the random variable $Y$ has a location-scale
distribution with cdf
\begin{equation*}
    G(y;\mu,\sigma)=\Phi(z),
\end{equation*}
and pdf $g(y;\mu,\sigma)=\phi(z)/\sigma$, where $z=(y-\mu)/\sigma$.
$\Phi(\cdot)$ and $\phi(\cdot)$ are the standard cdf and pdf for the
particular distribution.  We use the results in
\citet{EscobarMeeker1994}.  For \typeII{} (failure) censoring after
$r$ out of $n$ failures, the scaled elements of the
FIM for the parameters $\mu$ and $\sigma$ are
\begin{equation}
  \label{S.eq:fim-1}
  \begin{split}
    f_{11}(z_{c}) =&  \ \frac{\sigma^2}{n}\E\left[-\frac{\partial^2\log(\mathcal{L})}{\partial\mu^2}\right]=\Psi_0(z_{c}),\\
    f_{12}(z_{c}) =& \  \frac{\sigma^2}{n}\E\left[-\frac{\partial^2\log\mathcal{L}}{\partial\mu\partial\sigma}\right]=\Psi_1(z_{c}),\\
    f_{22}(z_{c}) =& \  \frac{\sigma^2}{n}\E\left[-\frac{\partial^2\log\mathcal{L}}{\partial\sigma^2}\right]=\Psi_2(z_{c}),
  \end{split}
\end{equation}
 where the largest
$n-r$ sample values are censored, $\Phi^{-1}(p)$ is the $p$-quantile
of $\Phi(\cdot)$, and $z_{c}=\Phi^{-1}(r/n)$.
Here $\mathcal{L}\equiv\mathcal{L}(\mu,\sigma)$ is the likelihood function
and
\[
\Psi_i(a)=\int_{-\infty}^{a}\left[1+xH(x)\right]^iH(x)^{2-i}\phi(x)dx,\quad(i=0,1,2)
\]
where
\[
H(x)=\frac{\phi^\prime(x)}{\phi(x)}+\frac{\phi(x)}{1-\Phi(x)},
\]
$\phi^\prime(x)$ is the derivative of $\phi(x)$ and $\E$ is the
expectation with respect to $Y$.  

For \typeI{} (time) censored data
with fixed right censoring time $y_c$, the $f_{jk}$ $(jk=11,12,22)$
are still given by (\ref{S.eq:fim-1}) but $z_{c}=(y_c-\mu)/\sigma$ and
the number of failures $0\leq r\leq n$ is random having a binomial
distribution with probability $p=\Phi(z_c)$.  Then for either
\typeI{} or \typeII{} censoring, the FIM for
$(\mu,\sigma)$ is
\begin{equation}
\label{S.equation:fisher-mu-sigma}
\text{I}_n(\mu,\sigma)=\frac{n}{\sigma^2}
\begin{bmatrix}
\Psi_0(z_{c}) & \Psi_1(z_{c})\\
\textrm{symmetric} & \Psi_2(z_{c})
\end{bmatrix}=\frac{n}{\sigma^2}
\begin{bmatrix}
f_{11}(z_{c}) & f_{12}(z_{c})\\
\textrm{symmetric} & f_{22}(z_{c})
\end{bmatrix}
.
\end{equation}
To simplify the presentation in the remainder of this section we
suppress the dependency of the $f_{ij}$ elements on $z_{c}$.  That
is, for example, we write $f_{11}$ instead of $f_{11}(z_{c})$.

\subsection{Jeffreys Priors and Independence Jeffreys Priors}
\label{S.section:jeffreys.priors}
\subsubsection{Jeffreys prior}
The Jeffreys priors for the $(\mu, \sigma)$ parameterization for
\typeI{} and \typeII{} censoring are derived in
Section~\ref{section:jeffreys.prior.distribution} of the main paper
and will not be repeated here.

\subsubsection{Independence Jeffreys priors}
\label{S.mu-sigma-ind-jeff-prior}
The Independence Jeffreys priors for the $(\mu, \sigma)$
parameterization for \typeI{} and \typeII{} censoring are derived in
Section~\ref{section:ij.prior} of the main paper and will not be
repeated here.

\subsection{Reference Priors}
\label{S.section-ref-prior}
\subsubsection{General description}
\label{S.subsubsec-ref-prior}
It is well known that Jeffreys prior has desirable properties for
one-parameter models, but
deficiencies for multi-parameter models.  For example, for the
$\mathrm{Norm}(\mu,\sigma)$ distribution, the Jeffreys prior is
$\pi(\mu,\sigma)\propto1/\sigma^2$, which does not have the
desirable properties of the right invariant prior
$\pi(\mu,\sigma)\propto1/\sigma$, as described in \citet[][Chapter
  5]{GhoshDelampadyTapas2006}. Also see \citet[][pages 182--184]{Jeffreys1961}.
\citet{Bernardo1979}
proposed an alternative to the Jeffreys prior, called a reference
prior.  The basic idea is to find the prior that maximizes the
Kullback-Leibler (KL) divergence between the prior and the
expected posterior.  Reference priors have been further studied and extended by
\citet{BergerBernardo1989,BergerBernardo1992a,
  BergerBernardo1992b}.  For a scalar parameter, the reference prior
is the same as the Jeffreys prior.

For multiple parameters, in the case where all parameters are of
same importance, the reference prior again leads to the Jeffreys
prior (cf.~\citealt{KassWasserman1996}, page 1350).  But one can also allow
parameters to have different importance.  For example, for a two
parameter vector $(\theta_1,\theta_2)$, $\theta_1$ might be of
primary importance.

In this section, we employ the general methods in
\citet{GhoshDelampadyTapas2006} to derive reference priors for \typeII{}
censored data.  We first obtain the conditional prior for
$\theta_2$ given $\theta_1$, which is denoted by
$\pi(\theta_2|\theta_1)\propto\sqrt{\text{I}_{22}(\boldsymbol{\theta})}$,
where
$\text{I}_{22}(\boldsymbol{\theta})=\text{E}\left[-\partial^2\log\mathcal{L}(\boldsymbol{\theta})/\partial\theta_2^2\right]$.
The marginal prior for $\theta_1$, denoted by $\pi(\theta_1)$ is
defined as the maximizer of the KL-divergence between the prior
$\pi(\boldsymbol{\theta})=\pi(\theta_1)\pi(\theta_2|\theta_1)$ and
the expected posterior distribution.  It should be noted that the
computing of a reference prior, including the KL-divergence, is done
over a sequence of compact sets $K_i$ that increase in size, where
the union $\bigcup\limits_{i=1}^{\infty}K_i$ is the parameter space.
A reference prior is computed on each compact set $K_i$, followed by
a limiting operation.  This approach is used to avoid improper prior
distributions that would otherwise arise after the limiting operation.  With two
parameters, we use increasing rectangles $K_i$.

\subsubsection{Reference priors for \typeII{} censoring}
For \typeII{} censoring (or complete data), we first consider the
reference priors for parameter ordering $\theta_{(1)}=\mu$ and
$\theta_{(2)}=\sigma$, where $\theta_{(1)}$ is more important than
$\theta_{(2)}$ (later this ordering is denoted by $\{\mu,\sigma\}$).
The conditional prior is
$\pi_i(\sigma|\mu)\propto\sqrt{f_{22}}/\sigma$ on the rectangle
$K_i=K_{1i}\times K_{2i}$.  Because $\sqrt{f_{22}}$ is a constant,
the conditional prior becomes
$\pi_i(\sigma|\mu)=c_i/\sigma$, where $c_i$ is generic notation for
a normalizing constant.  The marginal prior for $\mu$ on $K_i$ that
maximizes the KL-divergence is
\begin{equation}\label{S.eq:marginal-mu-max}
\pi_i(\mu)\propto\exp\left\{\int_{K_{2i}}\frac{1}{2}\log\left[h_1(\btheta)\right]\pi_i(\sigma|\mu)d\sigma\right\},
\end{equation}
where
\begin{equation}\label{S.eq:h1-function}
h_1(\btheta)=\frac{|\text{I}_n(\mu,\sigma)|}{\text{I}_{22}(\mu,\sigma)}=\frac{1}{\sigma^2}\frac{n(f_{11}f_{22}-f_{12}^2)}{f_{22}}.
\end{equation}
So, the marginal prior for $\mu$ on rectangle $K_i$
\begin{equation}\label{S.eq:marginal-simple-form}
\pi_i(\mu)\propto\exp\left\{\int_{K_{2i}}\frac{1}{2}\log\left(\frac{\textrm{const}}{\sigma^2}\right)\frac{c_i}{\sigma}d\sigma\right\}
\end{equation}
is a constant.  Thus, $\pi_i(\mu)$ is flat on $K_i$ and the
reference prior for parameter ordering $\theta_{(1)}=\mu$,
$\theta_{(2)}=\sigma$ is
$\pi_i(\mu,\sigma)=\pi_i(\sigma|\mu)\pi_i(\mu)\propto1/\sigma$.
Therefore, the reference prior for parameter ordering $\{\mu,\sigma\}$ is
$\pi(\mu,\sigma)\propto1/\sigma$.
We denote such a prior by $\pi(\{\mu,\sigma\})$
to indicate the order of parameters.

For parameter ordering $\theta_{(1)}=\sigma$ and $\theta_{(2)}=\mu$,
the conditional prior is
$\pi_i(\mu|\sigma)\propto\sqrt{nf_{11}}/\sigma$.  Given the value of
$\sigma$, the conditional prior for $\mu$ on $K_i$, which is denoted
by $\pi_i(\mu|\sigma)=c_i$, is flat as $f_{11}$ is a known constant.
The marginal prior for $\sigma$ on $K_{i}$ is
\[
\pi_i(\sigma)\propto\exp\left\{\int_{K_{1i}}\frac{1}{2}\log\left[h_2(\btheta)\right]\pi_i(\mu|\sigma)d\mu\right\},
\]
where
$$
h_2(\btheta)=\frac{|\text{I}_n(\mu,\sigma)|}{\text{I}_{11}(\mu,\sigma)}=\frac{1}{\sigma^2}\frac{n(f_{11}f_{22}-f_{12}^2)}{f_{11}}.
$$
So, the marginal prior for $\sigma$ on rectangle $K_i$
\[
\pi_i(\sigma)\propto\exp\left\{\int_{K_{1i}}\frac{1}{2}\log\left(\frac{\textrm{const}}{\sigma^2}\right)c_id\mu\right\}=\frac{c_i}{\sigma}.
\]
Thus, the joint prior on the rectangle $K_i$ is $\pi_i(\mu,\sigma)=c_i/\sigma$.
Then, as a limit of $\pi_i(\mu,\sigma)$, the reference prior for $\{\sigma,\mu\}$ is $\pi(\{\sigma,\mu\})\propto1/\sigma$.

\subsubsection{Reference priors for \typeI{} censoring}
\label{S.section:reference.priors.for.typeI.censoring}
Although we could use the same procedure to compute reference priors
for \typeI{} censoring data, the computations are usually intractable
and do not have closed forms.  For \typeI{} censoring, $h_1(\btheta)$
is no longer proportional to $1/\sigma^2$ because $f_{11}$,
$f_{12}$, and $f_{22}$ are no longer constant.  Thus, the integral
in (\ref{S.eq:marginal-mu-max}) will not reduce to the simple integral
in (\ref{S.eq:marginal-simple-form}).  Obtaining the analytical form
of the integral in~(\ref{S.eq:marginal-mu-max}) under \typeI{} censoring
is difficult and in the rest of this work, we only provide reference
priors for \typeII{} censoring (or complete) data. We will, however,
provide results for Jeffreys, CJ and IJ priors for \typeI{} censoring.

\subsection{Using the Parameterization $(y_p,\sigma)$}
\label{S.subsec:use-y_p-sigma}
\subsubsection{FIM for $(y_p,\sigma)$}
The $p$ quantile of a location-scale random variable
$Y$ is $y_p=\mu+\sigma\Phi^{-1}(p)$ so that
$z_{c}=\Phi^{-1}(r/n)$ for \typeII{} censoring and
$z_{c}\equiv(y_c-\mu)/\sigma=[y_c-y_p-\sigma\Phi^{-1}(p)]/\sigma$
for \typeI{} censoring.
The large-sample approximate covariance matrices of ML estimators
$(\widehat{\mu},\widehat{\sigma})$ and ML estimators
$(\widehat{y}_p,\widehat{\sigma})$ are, respectively, the inverse of
the FIMs $\text{I}_n^{-1}(\mu,\sigma)$ and
$\text{I}_n^{-1}(y_p,\sigma)$. 
The large-sample approximate covariance matrix for
$(\widehat{y}_p,\widehat{\sigma})$ is $\text{I}^{-1}_n(y_p,\sigma)$
and can be computed from the large-sample approximate covariance matrix for
$(\widehat{\mu},\widehat{\sigma})$ using the delta method:
\[
\begin{split}
&\text{I}_n^{-1}[y_p(\mu,\sigma),\sigma(\mu,\sigma)]=\mathcal{J} \, \text{I}_n^{-1}(\mu,\sigma)\mathcal{J}^\prime\\[1ex]=&
  \frac{\sigma^2}{n(f_{12}^2-f_{11}f_{22})}\times\begin{bmatrix}
  \Phi^{-1}(p)\left[f_{12}-\Phi^{-1}(p)f_{11}\right]-f_{22}+\Phi^{-1}(p)f_{12}
  & f_{12}-\Phi^{-1}(p)f_{11}\\ \textrm{symmetric} & -f_{11}
\end{bmatrix},
\end{split}
\]
where the Jacobian $\mathcal{J}$ is
\[
\mathcal{J}=
\begin{bmatrix}
\dfrac{\partial y_p}{\partial\mu} & \dfrac{\partial y_p}{\partial\sigma}\\
\dfrac{\partial\sigma}{\partial\mu} & \dfrac{\partial\sigma}{\partial\sigma}
\end{bmatrix}=
\begin{bmatrix}
1 & \Phi^{-1}(p) \\
0 & 1 
\end{bmatrix}.
\]
Then the FIM for $(y_p,\sigma)$ is
\begin{equation}
\label{S.equation-page-5}
\text{I}_n(y_p,\sigma)=\frac{n}{\sigma^2}
\begin{bmatrix}
f_{11} & f_{12}-\Phi^{-1}(p)f_{11} \\
\textrm{symmetric} & f_{11}[\Phi^{-1}(p)]^2-2f_{12}\Phi^{-1}(p)+f_{22}
\end{bmatrix}.
\end{equation}

\subsubsection{Jeffreys prior}
\label{S.subsec-yp-logsigma}
The determinant of the FIM (\ref{S.equation-page-5}) is
\[
\left|\text{I}_n(y_p,\sigma)\right|=\frac{n^2}{\sigma^4}\left(f_{11}f_{22}-f_{12}^2\right).
\]
Because the determinant is the same as in
(\ref{equation:jeffreys-mu-sigma}), the Jeffreys prior for $(y_p,\sigma)$
is the same as $(\mu,\sigma)$ for both \typeII{} and \typeI{} censoring.

\subsubsection{IJ prior}

For \typeII{} censored or complete data, the elements of the FIM are
known constants and thus as $\pi(y_p|\sigma)\propto1$ and
$\pi(\sigma|y_p)\propto1/\sigma$.  Then the IJ
prior is $\pi(y_p,\sigma)\propto{1}/{\sigma}$.  
For \typeI{}
censoring, the CJ prior for $y_p$ is
\[
\pi(y_p|\sigma)\propto\sqrt{f_{11}},
\]
and the CJ prior for $\sigma$ is
\[
\pi(\sigma|y_p)\propto\frac{1}{\sigma}\sqrt{f_{11}[\Phi^{-1}(p)]^2-2f_{12}\Phi^{-1}(p)+f_{22}},
\]
which is the same as $\pi(\sigma|\mu)$
in Section~\ref{S.mu-sigma-ind-jeff-prior} when $p$ is chosen such
that $\Phi^{-1}(p)=0$.
Then the IJ prior for $(y_p,\sigma)$ is
\[
\begin{split}
\pi(y_p, \sigma)\propto&\,\pi(y_p|\sigma)\pi(\sigma|y_p)\\
\propto&\,\frac{1}{\sigma}\sqrt{f_{11}\left\{f_{11}[\Phi^{-1}(p)]^2-2f_{12}\Phi^{-1}(p)+f_{22}\right\}}.
\end{split}
\]

\subsubsection{Reference prior}
We consider \typeII{} censored or complete data.
For parameter ordering $\{y_p, \sigma\}$, the conditional Jeffreys prior
for $\sigma$ is $\pi(\sigma|y_p)\propto1/\sigma$.
The marginal prior for $y_p$ is
\[
\pi(y_p)\propto\exp\left\{\int\frac{1}{2}\log\left[\frac{|\textrm{I}_n(y_p,\sigma)|}{\mathrm{const}/\sigma^2}\right]\frac{1}{\sigma}d\sigma\right\}\propto 1.
\]
Thus the reference prior for $\{y_p,\sigma\}$ is
\[
\pi(\{y_p,\sigma\})\propto\frac{1}{\sigma}.
\]
For parameter ordering $\{\sigma, y_p\}$,
the conditional Jeffreys prior is $\pi(y_p|\sigma)\propto 1$.
The marginal prior for $\sigma$ is given by
\[
\pi(\sigma)\propto\exp\left\{\int\frac{1}{2}\log\left[\frac{\mathrm{I}_n(y_p,\sigma)}{\mathrm{const}/\sigma^2}\right]dy_p\right\}\propto\frac{1}{\sigma}.
\]
Thus the reference prior for $\{\sigma,y_p\}$ is
\[
\pi(\{\sigma,y_p\})\propto\frac{1}{\sigma}.
\]

\subsection{Using the Parameterization $(t_p, \sigma)$}
\label{S.section.parameterization.tp.sigma}
The $p$ quantile of a log-location-scale random variable $T$ is 
$t_p=\exp(y_p)=\exp\left[\mu+\sigma\Phi^{-1}(p)\right]$.
\subsubsection{FIM for $(t_p, \sigma)$}
\label{S.section:fim.tp.sigma}
Using the delta method,  the inverse of the FIM for $(t_p, \sigma)$ is
\[
\begin{split}
  \text{I}^{-1}_n\left(t_p,\sigma\right)=&\mathcal{J} \, \text{I}^{-1}_n(\mu,\sigma)\mathcal{J}^\prime\\ =&\frac{\sigma^2}{n(f_{12}^2-f_{11}f_{22})}\times\begin{bmatrix}
        t_p^2\left\{\Phi^{-1}(p)\left[f_{12}-\Phi^{-1}(p)f_{11}\right]-f_{22}+\Phi^{-1}(p)f_{12}\right\}
        & t_p\left[f_{12}-\Phi^{-1}(p)f_{11}\right]\\ \textrm{symmetric} &
        -f_{11}
  \end{bmatrix},
\end{split}
\]
where the Jacobian $\mathcal{J}$ is
\[
\mathcal{J}=
\begin{bmatrix}
  \dfrac{\partial t_p}{\partial\mu} & \dfrac{\partial
          t_p}{\partial\sigma}\\ \dfrac{\partial\sigma}{\partial\mu}
        & \dfrac{\partial\sigma}{\partial\sigma}
\end{bmatrix}=
\begin{bmatrix}
  t_p & t_p\Phi^{-1}(p) \\
  0 & 1 
\end{bmatrix}.
\]
Then the FIM for $(t_p, \sigma)$ is
\[
\mathrm{I}_n(t_p,\sigma)=\frac{n}{\sigma^2}\begin{bmatrix}
  \begin{array}{lr}
    {f_{11}}/{t_p^2} &
                {[f_{12}-f_{11}\Phi^{-1}(p)]}/{t_p}
                \\ \textrm{symmetric} &
                   {f_{11}\Phi^{-1}(p)-2f_{12}\Phi^{-1}(p)+f_{22}}
  \end{array}
\end{bmatrix}.
\]

\subsubsection{Jeffreys prior}

The determinant of the FIM is
\[
\mathrm{det}\left[\mathrm{I}_n(t_p,\sigma)\right]=n^2\frac{(f_{11}f_{12}-f_{12}^2)}{t_p^2\sigma^4}.
\]
For \typeII{} censoring, the Jeffreys prior is
\[
\pi(t_p,\sigma)\propto\frac{1}{t_p\sigma^2}.
\]
For \typeI{} censoring, the Jeffreys prior is
\[
\pi(t_p,\sigma)\propto\frac{1}{t_p\sigma^2}\sqrt{f_{11}f_{12}-f_{12}^2}.
\]

\subsubsection{IJ prior}
\label{S.section:ij.prior.tp.sigma}
The CJ prior for $t_p$ given $\sigma$ is
$\pi(t_p|\sigma)\propto\sqrt{f_{11}}/(t_p\sigma )$ and that for
$\sigma$ given $t_p$ is
\begin{align*}
\pi(\sigma|t_p) & \propto \frac{1}{\sigma}\sqrt{f_{11}\Phi^{-1}(p)-2f_{12}\Phi^{-1}(p)+f_{22}}.
\end{align*}
For \typeII{} censoring, we have $\pi(t_p|\sigma)\propto1/t_p$ and
$\pi(\sigma|t_p)\propto1/\sigma$. Then the IJ
prior is
\[
\pi(t_p,\sigma)\propto\frac{1}{t_p\sigma}.
\]
For \typeI{} censoring, we have
$\pi(t_p|\sigma)\propto\sqrt{f_{11}}/t_p$ and
\begin{align*}
\pi(\sigma|t_p) & \propto \frac{1}{\sigma} \sqrt{f_{11}\Phi^{-1}(p)-2f_{12}\Phi^{-1}(p)+f_{22}}.
\end{align*}
Then the IJ prior is
\[
\pi(t_p,\sigma)\propto\frac{1}{t_p\sigma}\sqrt{f_{11}[f_{11}\Phi^{-1}(p)-2f_{12}\Phi^{-1}(p)+f_{22}]}.
\]

\subsubsection{Reference prior}

We only consider complete or \typeII{} censored data.
For $\{t_p,\sigma\}$, the conditional Jeffreys prior for $\sigma$ is $\pi(\sigma|t_p)\propto1/\sigma$.
Then the marginal prior for $t_p$ is
\[
\pi(t_p)=\exp\left\{\frac{1}{2}\int\log\left[\frac{1/(t_p^2\sigma^4)}{1/\sigma^2}\right]\frac{1}{\sigma}d\sigma\right\}\propto\frac{1}{t_p}.
\]
The reference prior for $\{t_p,\sigma\}$ is
\[
\pi(\{t_p,\sigma\})\propto\frac{1}{t_p\sigma}.
\]
For $\{\sigma,t_p\}$, the conditional Jeffreys prior for $t_p$ is $\pi(t_p|\sigma)\propto1/t_p$.
The marginal prior for $\pi(\sigma)$ is
\[
\pi(\sigma)\propto\exp\left\{\frac{1}{2}\int\log\left[\frac{1/(t_p^2\sigma^4)}{1/(t_p^2\sigma^2)}\right]dy_p\right\}\propto\frac{1}{\sigma}.
\]
The reference prior for
$$
\pi(\{\sigma,t_p\})\propto\frac{1}{t_p\sigma}.
$$

\subsection{Using the Parameterization $(\log(t_{p_{r}}),\log(\sigma))$}
\label{section:priors.y.log.sigam.parameterization}
As described in Section~\ref{subsection:param.for.priors} of the
paper, $(\log(t_{p_{r}}),\log(\sigma))$ is the parameterization that we used for
our MCMC computations.
To simplify the notation in the derivation,
let $\tau\equiv\log(\sigma)$ and $y_p=\log(t_{p_{r}})$. 
\subsubsection{FIM for $(y_p,\log(\sigma))$}
Then the inverse
of the FIM for $(y_p,\tau)$ is
\[
\begin{split}
&\text{I}_n^{-1}(y_p,\tau)=\mathcal{J} \, \text{I}_n^{-1}(\mu,\sigma)\mathcal{J}^\prime\\
=&\frac{1}{n\left[f_{12}^2-f_{11}f_{22}\right]}\times\\
&\begin{bmatrix}
\sigma^2\left\{\Phi^{-1}(p)\left[f_{12}-\Phi^{-1}(p)f_{11}\right]-f_{22}+\Phi^{-1}(p)f_{12}\right\} &  \sigma\left[f_{12}-\Phi^{-1}(p)f_{11}\right]\\
\textrm{symmetric} & -f_{11}
\end{bmatrix},
\end{split}
\]
where the Jacobian is
\[
\mathcal{J}=
\begin{bmatrix}
  \dfrac{\partial y_p}{\partial\mu} & \dfrac{\partial y_p}{\partial\sigma}\\
  \dfrac{\partial\tau}{\partial\mu} & \dfrac{\partial\tau}{\partial\sigma}
\end{bmatrix}
=
\begin{bmatrix}
  1 & \Phi^{-1}(p) \\
  0 & 1/\sigma
\end{bmatrix}.
\]
Thus, the FIM for $(y_p,\tau)$ is
\[
\begin{split}
  \text{I}_n(y_p,\tau)&=n
  \begin{bmatrix}
    f_{11}/{\sigma^2} & [f_{12}-\Phi^{-1}(p)f_{11}]/{\sigma}\\
    \textrm{symmetric} & f_{11}[\Phi^{-1}(p)]^2-2f_{12}\Phi^{-1}(p)+f_{22} \\
  \end{bmatrix}\\[1.5ex]
&=n\begin{bmatrix}
    f_{11}/{\exp(2\tau)} & [f_{12}-\Phi^{-1}(p)f_{11}]/{\exp(\tau)}\\
    \textrm{symmetric} & f_{11}[\Phi^{-1}(p)]^2-2f_{12}\Phi^{-1}(p)+f_{22} \\
\end{bmatrix}.
\end{split}
\]

\subsubsection{Jeffreys prior}
The determinant of the FIM is
\[
\left|\text{I}_n(y_p,\tau)\right|=\frac{n^2}{\exp(2\tau)}\left(f_{11}f_{22}-f_{12}^2\right)=\frac{n^{2}}{\sigma^2}\left(f_{11}f_{22}-f_{12}^2\right).
\]
Then, for \typeII{} censoring,
the Jeffreys prior is $\pi(y_p,\tau)\propto1/\sigma$
because the $f_{ij}$ elements are fixed constants.
This result also can be obtained directly
from Section~\ref{S.subsec-yp-logsigma} using the fact that the
Jeffreys prior is invariant to parameter transformation.  For
example, we already know that $\pi(y_p,\sigma)\propto1/\sigma^2$.
Then the Jeffreys prior for parameters $(y_p,\tau)$ can be computed
as
\[
\pi(y_p,\tau)\propto\frac{1}{\exp(2\tau)}\left|
\begin{bmatrix}
  \dfrac{\partial y_p}{\partial y_p} & \dfrac{\partial y_p}{\partial \tau} \\
  \dfrac{\partial \sigma}{\partial y_p}& \dfrac{\partial \sigma}{\partial \tau} \\
\end{bmatrix}
\right|=\frac{1}{\exp(2\tau)}\left|\begin{bmatrix}
  1 & 0\\
  0 & \exp(\tau)
\end{bmatrix}\right|=\frac{1}{\exp(\tau)}=\frac{1}{\sigma}.
\]
For \typeI{} censoring, the Jeffreys prior is
$$
\pi(y_p,\tau)\propto\frac{1}{\exp(\tau)}\sqrt{f_{11}f_{22}-f_{12}^2}=\frac{1}{\sigma}\sqrt{f_{11}f_{22}-f_{12}^2}.
$$

\subsubsection{IJ prior}
\label{subsubsection:ij.prior.yp.log.sigma}
For \typeII{} censoring, the CJ prior for $y_p$
given $\tau$ is $\pi(y_p|\tau)\propto1$; the CJ
prior for $\tau$ given $y_p$ is $\pi(\tau|y_p)\propto 1$. Thus thee
IJ prior is $\pi(y_p,\tau)\propto 1$.  
For \typeI{}
censoring, the CJ prior for $y_p$ given $\tau$ is
$\pi(y_p|\tau)\propto\sqrt{f_{11}}$; the CJ prior
for $\tau$ given $y_p$ is
$$
\pi(\tau|y_p)\propto\sqrt{f_{11}[\Phi^{-1}(p)]^2-2f_{12}\Phi^{-1}(p)+f_{22}}.
$$
So, the IJ prior is
\[
\begin{split}
\pi(y_p,\tau)\propto&\pi(y_p|\tau)\pi(\tau|y_p)\\[1ex]
\propto&\sqrt{f_{11}\left\{f_{11}[\Phi^{-1}(p)]^2-2f_{12}\Phi^{-1}(p)+f_{22}\right\}}.
\end{split}
\]

\subsubsection{Reference prior}

For \typeII{} censoring or complete data, the reference prior for
parameter ordering $\{y_p,\tau\}$
is computed by first deriving the conditional Jeffreys prior $\tau$
\[
\pi(\tau|y_p)\propto1;
\]
the marginal prior for $y_p$ is
\[
\pi(y_p)\propto\exp\left\{\frac{1}{2} \int \log\left[\frac{1/\exp(2\tau)}{1}\right]d\tau\right\}\propto1.
\]
Thus the reference prior for $\{y_p,\tau\}$ is given by
\[
\pi(\{y_p,\tau\})\propto1.
\]
For parameter ordering $\{\tau,y_p\}$, the conditional Jeffreys prior for $y_p$ is $\pi(y_p|\tau)\propto1$ and the marginal prior for $\tau$ is
\[
\pi(\tau)\propto\exp\left\{\frac{1}{2} \int \log\left[\frac{1/\exp(2\tau)}{1/\exp(2\tau)}\right]
\pi(y_p|\tau)   dy_p\right\}\propto
1.
\]
The reference prior for $\{\tau,y_p\}$ is
\[
\pi(\{\tau, y_p\})\propto1.
\]

\subsection{Using the Parameterization $(\zeta_{e},\sigma)$}
\subsubsection{FIM for $(\zeta_{e},\sigma)$}
Let $\zeta_{e}=(y_e-\mu)/\sigma$, where $y_e>0$ is a given value.  The
parameter $\zeta_{e}$ is important because 
inferences are often needed for $\Pr(Y\leq y_e)=\Phi(\zeta_{e})$.
For \typeII{} censoring, $z_{c}=\Phi^{-1}(r/n)$; for \typeI{} censoring,
$z_{c}=(y-y_e+\sigma\zeta_{e})/\sigma$. The inverse of the FIM for
$(\zeta_{e},\sigma)$ is
\[
\begin{split}
\text{I}_n^{-1}[\zeta_{e}(\mu,\sigma),\sigma(\mu,\sigma)]&=\mathcal{J} \,
\text{I}_n^{-1}(\mu,\sigma)\mathcal{J}^\prime\\
&=\frac{1}{n(f_{12}^2-f_{11}f_{22})}\times
\begin{bmatrix}
\zeta_{e}[f_12-\zeta_{e} f_{11}]-f_{22}+\zeta_{e} f_{11} & \sigma[\zeta_{e} f_{11}-f_{12}] \\
\textrm{symmetric} & -\sigma^2f_{11}
\end{bmatrix}
\end{split}
\]
where the Jacobian $\mathcal{J}$ is
\[
\mathcal{J}=
\begin{bmatrix}
\dfrac{\partial\zeta_{e}}{\partial\mu} & \dfrac{\partial\zeta_{e}}{\partial\sigma} \\
\dfrac{\partial\sigma}{\partial\mu} & \dfrac{\partial\sigma}{\partial\sigma}
\end{bmatrix}=
\begin{bmatrix}
-\dfrac{1}{\sigma} & -\dfrac{\zeta_{e}}{\sigma}\\
0 & 1
\end{bmatrix}.
\]
By replacing $(\mu,\sigma)$ with $(\zeta_{e},\sigma)$ in
$\text{I}_n[\zeta_{e}(\mu,\sigma),\sigma(\mu,\sigma)]$, the FIM for
$(\zeta_{e},\sigma)$ is
\begin{equation}
\label{S.equation-psi}
\text{I}_n(\zeta_{e},\sigma)=\frac{n}{\sigma^2}
\begin{bmatrix}
\sigma^2f_{11} & \sigma(f_{11}\zeta_{e}-f_{12})\\
\textrm{symmetric} & f_{11}\zeta_{e}^2-2f_{12}\zeta_{e}+f_{22}
\end{bmatrix}.
\end{equation}

\subsubsection{Jeffreys prior}
The determinant of the FIM (\ref{S.equation-psi}) is
\[
|\text{I}_n(\zeta_{e},\sigma)|=\frac{n^2}{\sigma^2}\left(f_{22}f_{11}-f_{12}^2\right).
\]
For \typeII{} censoring or complete data, the Jeffreys prior is
$
\pi(\zeta_{e},\sigma)\propto{1}/{\sigma}.
$
For \typeI{} censoring, the Jeffreys prior is
\[
\pi(\zeta_{e},\sigma)\propto\frac{1}{\sigma}{\sqrt{f_{22}f_{11}-f_{12}^2}}.
\]

\subsubsection{IJ prior}
For \typeII{} censoring using the appropriate elements of
(\ref{S.equation-psi}), the CJ prior for $\zeta_{e}$ given $\sigma$ is
$\pi(\zeta_{e}|\sigma)\propto1$; the prior for $\sigma$ given $\zeta_{e}$ is
$\pi(\sigma|\zeta_{e})\propto1/\sigma$. So, the IJ prior for
$(\zeta_{e}, \sigma)$ is
$\pi(\zeta_{e},\sigma)\propto{1}/{\sigma}.  $ 
For \typeI{} censoring, the
CJ prior for $\zeta_{e}$  given $\sigma$
is $\pi(\zeta_{e}|\sigma)\propto\sqrt{f_{11}}$ and the
CJ prior for $\sigma$ given $\zeta_{e}$ is
$\pi(\sigma|\zeta_{e})\propto(1/\sigma)\sqrt{f_{11}\zeta_{e}^2-2f_{12}\zeta_{e}+f_{22}}$.
So, the IJ prior for $(\zeta_{e},\sigma)$ is
\[
\pi(\zeta_{e},\sigma)\propto\frac{1}{\sigma}{\sqrt{f_{11}(f_{11}\zeta_{e}^2-2f_{12}\zeta_{e}+f_{22})}}.
\]

\subsubsection{Reference prior}
We consider the reference prior for \typeII{} censoring and complete
data.
For parameter ordering $\theta_{(1)}=\zeta_{e}$ and
$\theta_{(2)}=\sigma$, the conditional prior on
$K_i=K_{1i}\times K_{2i}$ is $\pi(\sigma|\zeta_{e})\propto1/\sigma$ and
it denoted by $\pi_i(\sigma|\zeta_{e})=c_i/\sigma$. Then the marginal
prior for $\zeta_{e}$ is
\begin{equation*}
\begin{split}
\pi_i(\zeta_{e})=&\,c_i\left(\int_{K_{2i}}\frac{1}{2}\log\left\{\frac{|\text{I}_n(\zeta_{e},\sigma)|}{[f_{11}\zeta_{e}^2-2f_{12}\zeta_{e}+f_{22}]/\sigma^2}\right\}\frac{c_i}{\sigma}d\sigma\right)\\
=&\frac{c_i}{\sqrt{f_{11}\zeta_{e}^2-2f_{12}\zeta_{e}+f_{22}}}.
\end{split}
\end{equation*}
Here $c_i$ is the generic notation for a normalizing constant.
Thus, the reference prior for $\{\zeta_{e},\sigma\}$ is
\begin{equation*}
\pi(\{\zeta_{e},\sigma\})\propto\frac{1}{\sigma\sqrt{f_{11}\zeta_{e}^2-2f_{12}\zeta_{e}+f_{22}}}.
\end{equation*}
For parameter ordering  $\{\sigma,\zeta_{e}\}$, the conditional
prior on $K_i=K_{1i}\times K_{2i}$ is $\pi(\zeta_{e}|\sigma)\propto
c_i$.  The marginal prior for $\zeta_{e}$ is
\[
\pi_i(\sigma)\propto\exp\left\{\int_{K_{1i}}\frac{1}{2}\log\left[\frac{|\text{I}_n(\zeta_{e},\sigma)|}{f_{11}}\right]{c_i}d\zeta_{e}\right\}\propto\frac{1}{\sigma}.
\]
So, the corresponding reference prior is
$
\pi(\{\sigma,\zeta_{e}\})\propto{1}/{\sigma}
$.

\subsection{Using the Parameterization $(\zeta_{e},\log(\sigma))$}
\label{S.section-zeta-log.sigma} 
\subsubsection{FIM for $(\zeta_{e},\log(\sigma))$}
We define $\tau=\log(\sigma)$; then the inverse of the FIM for
$(\zeta_{e},\tau)$ is
\[
\begin{split}
\text{I}^{-1}_n\left[\zeta_{e},\tau(\sigma)\right]=&\mathcal{J} \, \text{I}^{-1}_n(\zeta_{e},\sigma)\mathcal{J}^\prime\\
=&\frac{1}{n[f_{12}^2-f_{11}f_{22}]}\times
\begin{bmatrix}
  \zeta_{e}[f_{12}-\zeta_{e} f_{11}]-f_{22}+\zeta_{e} f_{11} & \zeta_{e} f_{11}-f_{12} \\
  \textrm{symmetric} & -f_{11}
\end{bmatrix},
\end{split}
\]
where
\[
\mathcal{J}=\begin{bmatrix}
  1 & 0\\
  0 & \dfrac{1}{\sigma}
\end{bmatrix}.
\]
Then the FIM for $(\zeta_{e},\tau)$  is
\begin{equation}\label{S.fim-psi-tau}
\text{I}_n(\zeta_{e},\tau)=n\begin{bmatrix}
  f_{11} & \zeta_{e} f_{11}-f_{12}\\
  \textrm{symmetric} & f_{11}\zeta_{e}^2-2f_{12}\zeta_{e}+f_{22}
\end{bmatrix}.
\end{equation}

\subsubsection{Jeffreys prior}
The determinant of the FIM in (\ref{S.fim-psi-tau}) is
\[
\left|\text{I}_n(\zeta_{e},\tau)\right|=n^2\left[f_{11}f_{22}-f_{12}^2\right].
\]
For \typeII{} censoring or complete data, the Jeffreys prior is
$\pi(\zeta_{e},\tau)\propto1$.
For \typeI{} censoring data, the Jeffreys prior is
$$
\pi(\zeta_{e},\tau)\propto\sqrt{f_{11}f_{22}-f_{12}^2}.
$$

\subsubsection{IJ prior}
For \typeII{} censoring or complete data using the appropriate
elements of (\ref{S.fim-psi-tau}), the CJ prior for $\zeta_{e}$ is
$\pi(\zeta_{e}|\tau)\propto1$ and the CJ prior for $\tau$ is
$\pi(\tau|\zeta_{e})\propto1$; thus the IJ prior is
$\pi(\zeta_{e},\tau)\propto1$.  
For \typeI{} censoring, the prior for
$\zeta_{e}$ is $\pi(\zeta_{e}|\tau)\propto\sqrt{f_{11}}$ and the prior for
$\tau$ is
$\pi(\tau|\zeta_{e})\propto\sqrt{f_{11}\zeta_{e}^2-2f_{12}\zeta_{e}+f_{22}}$.
Then the IJ prior is
\[
\pi(\zeta_{e},\tau)\propto{\sqrt{f_{11}[f_{11}\zeta_{e}^2-2f_{12}\zeta_{e}+f_{22}]}}.
\]

\subsubsection{Reference prior}

We consider the reference prior for \typeII{} or complete data.
For parameter ordering  $\{\zeta_e,\tau\}$, the CJ prior for $\tau$ is $\pi(\tau|\zeta_e)\propto1$ and the marginal distribution for $\zeta_e$ is
\[
\pi(\zeta_e)\propto\exp\left\{\frac{1}{2} \int \log\left[\frac{1}{f_{11}\zeta_e^2-2f_{12}\zeta_e+f_{22}}\right]d\tau\right\}\propto\frac{1}{\sqrt{f_{11}\zeta_e^2-2f_{12}\zeta_e+f_{22}}}.
\]
Thus the reference prior for parameter ordering $\{\zeta_e,\tau\}$ is
$$
\pi(\{\zeta_e,\tau\})\propto\frac{1}{\sqrt{f_{11}\zeta_e^2-2f_{12}\zeta_e+f_{22}}}.
$$
For $\{\tau,\zeta_e\}$, the CJ prior for $\zeta_e$ is $\pi(\zeta_e|\tau)\propto1$ and the marginal prior for $\tau$ is given by
\[
\pi(\tau)\propto\exp\left\{\frac{1}{2} \int \log\left[\frac{|\mathrm{I}|}{nf_{11}}\right]d\zeta_e\right\}\propto1.
\]
Then the reference prior for $\{\tau,\zeta_e\}$ is $\pi(\{\tau,\zeta_e\})\propto1$.

\subsection{\,Proof of invariance of IJ priors to one-to-one reparameterizations}
\label{S.section:invariant.one.one.reparameterization}
 Here we show that IJ priors for log-location-scale distributions
 are invariant (in the sense described in
 Section~\ref{section:jeffreys.prior.distribution} of the paper) to
 one-to-one monotone reparameterizations of either or both of the
 $(t_{p},\sigma)$ parameters.

Suppose that the original parameters are $(t_p,\sigma)$. The
new parameters are
$\theta_1=A(t_p)$ and
$\theta_2=B(\sigma)$ where $A$ and $B$ are one-to-one monotone
transformations. Denote the derivatives by
\[
a=\frac{\dd \theta_1}{\dd t_p}\quad\text{and}\quad b=\frac{\dd \theta_2}{\dd\sigma}.
\]
Here $a$ is written as a function of $\theta_1$ as $a=a(\theta_1)$
and $b$ is written as a function of $\theta_2$ as $b=b(\theta_2)$.
The inverse of the FIM (large-sample approximate covariance matrix) under
parameterization $(t_p,\sigma)$ is (from
Section~\ref{S.section:fim.tp.sigma})
$$
\mathrm{I}^{-1}(t_p,\sigma)=\frac{\sigma^{2}}
{n\left(f_{12}^{2}-f_{11} f_{22}\right)} \times\left[\begin{array}{cc}
  t_{p}^{2}\left\{\Phi^{-1}(p)
\left[f_{12}-\Phi^{-1}(p) f_{11}\right]-f_{22}+\Phi^{-1}(p)
f_{12}\right\} & 
t_{p}\left[f_{12}-\Phi^{-1}(p) f_{11}\right] \\
  \text { symmetric } & -f_{11}
\end{array}\right].
$$

Using the delta method, the inverse FIM (large-sample approximate covariance
matrix) for the $(\theta_1,\theta_2)$ parameterization
\[
\begin{split}
  \mathrm{I}^{-1}(\theta_1,\theta_2)=&
\frac{\left[\sigma(\theta_2)\right]^2}
{n\left(f_{12}^{2}-f_{11} f_{22}\right)}\times\\
  &\mathcal{J}\left[\begin{array}{cc}
    \left[t_p(\theta_1)\right]^{2}
\left\{\Phi^{-1}(p)\left[f_{12}-\Phi^{-1}(p) f_{11}\right]-f_{22}+\Phi^{-1}(p) f_{12}\right\} & t_p(\theta_1)\left[f_{12}-\Phi^{-1}(p) f_{11}\right] \\
    \text { symmetric } & -f_{11}
  \end{array}\right]\mathcal{J}^{\prime}
\end{split},
\]
where
\[
\mathcal{J}=\begin{bmatrix}
  a & 0\\
  0 & b
\end{bmatrix}.
\]
Then the FIM for the $(\theta_1,\theta_2)$ parameterization is
\[
\begin{split}
  \mathrm{I}(\theta_1,\theta_2)=&
\frac{n}{\left[t_p(\theta_1)\right]^2
\left[\sigma(\theta_2)\right]^2a^2b^2}\times\\
  &\begin{bmatrix}
    b^2f_{11} & ab\sigma(\theta_2)
\left[f_{12}-\Phi^{-1}(p)f_{11}\right] \\
    \text{ symmetric } & -a^2
\left[t_p(\theta_1)\right]^2\left\{\Phi^{-1}(p)
\left[f_{12}-\Phi^{-1}(p)f_{11}\right]-f_{22}+\Phi^{-1}(p)f_{12}\right\} \\
  \end{bmatrix}
\end{split}
\]
For the CJ priors, we have
\[
\pi(\theta_1|\theta_2)\propto
\frac{1}{at_p(\theta_1)}\quad\text{and}\quad\pi(\theta_2|\theta_1)
\propto\frac{1}{b\sigma(\theta_2)}.
\]
So, the IJ prior is
\[
\pi_1(\theta_1,\theta_2)\propto
\frac{1}{t_p(\theta_1)\sigma(\theta_2)}\frac{1}{ab}.
\]

To prove that the IJ prior is invariant to the transformations, we can
first perform the variable transformation on the IJ prior using
$(t_p,\sigma)$ and show that the prior after transformation is the
same as $\pi_1(\theta_1,\theta_2)$.  From
Section~\ref{S.section:ij.prior.tp.sigma}, the IJ prior using
$(t_p,\sigma)$ is  $1/(t_p\sigma)$.  First replace $t_p$
and $\sigma$ with $t_p(\theta_1)$ and $\sigma(\theta_2)$, then
finish the transformation with the Jacobian
\[
\pi_2(\theta_1,\theta_2)\propto\frac{1}{t_p(\theta_1)\sigma(\theta_2)}|\mathcal{J}|,
\]
where
\[
|\mathcal{J}|=\left|\begin{bmatrix}
  \dfrac{\partial t_p(\theta_1)}{\partial \theta_1} & \dfrac{\partial t_p(\theta_1)}{\partial \theta_2}\\
  \dfrac{\partial \sigma(\theta_2)}{\partial \theta_1} & \dfrac{\partial \sigma(\theta_2)}{\partial \theta_2}
\end{bmatrix}\right|=\dfrac{1}{|\mathcal{J}^\ast|},
\]
and
\[
\mathcal{J}^\ast=\begin{bmatrix}
  \dfrac{\partial \theta_1}{\partial t_p} & \dfrac{\partial \theta_1}{\partial \sigma}\\
  \dfrac{\partial \theta_2}{\partial t_p} & \dfrac{\partial \theta_2}{\partial \sigma}
\end{bmatrix}=
\begin{bmatrix}
  a & 0\\
  0 & b\\
\end{bmatrix}
\quad\text{and}\quad|\mathcal{J}^\ast|=ab.
\]
Thus
\begin{align*}
\pi_2(\theta_1,\theta_2)\propto
\frac{1}{t_p(\theta_1)\sigma(\theta_2)}\frac{1}{ab},
\end{align*}
 giving the desired result.

\section{Understanding the \typeI{} Censoring Independence Jeffreys Prior Distribution}
\label{S.section:understanding.ij.prior.features}
\subsection{Features of the \typeI{} Censoring Independence Jeffreys Prior
  Distribution}
\label{S.section:ij.prior.features}
Figure~\ref{figure:ijprior.density.examples} shows contour plots of the
\typeI{} censoring IJ prior densities in
(\ref{equation:IJ.typeI.yp.log.sigma}) of the main paper for
various values of $p_{r}$. That is,
\begin{align}
\label{S.equation:IJ.typeI.yp.log.sigma}
\pi(\log(t_{p_{r}}),\log(\sigma)) &\propto
\sqrt{f_{11}\left\{f_{11}[\Phi^{-1}(p_{r})]^2-2f_{12}\Phi^{-1}(p_{r})+f_{22}\right\}}
\end{align}
where the $f_{ij}$ values are scaled elements of the FIM defined in
Section~\ref{section:scaled.fim.elements} of the main paper. These
scaled elements depend on the standardized censoring time
$z_{c}=[\log(t_{c})-\mu]/\sigma$ or or the expected fraction failing
$p_{c}=\Phi(z_{c})$. As described in
Section~\ref{section:implementing.ij.priors},
$\pi(\log(t_{p_{r}}),\log(\sigma))$ in
(\ref{section:implementing.ij.priors}) can be computed for any values
of $\log(t_{p_{r}}$ and $\log(\sigma)$ and
the only inputs needed are $t_{c}$ and $p_{r}$.
To make it easier to compare across different
inputs, the densities in
Figure~\ref{figure:ijprior.density.examples} were scaled to have a maximum
of 1.0 and thus we refer to them as relative densities.

We use the lognormal distribution for the
examples (because of the well-known interpretation of the usual
$(\mu, \sigma)$ parameters and the median $t_{0.50}=\exp(\mu)$. The
results are largely similar for the Weibull distribution when
presented in the $(t_{p_{r}}, \sigma)$ parameterization (where the
Weibull shape parameter is $\beta=1/\sigma$).

Without any important loss of generality, all of these examples
given here used $t_{c}=135$. Note that the density is computed as a
function of $\log(t_{p_{r}})$ and $\log(\sigma)$ but presented in
the plots (for
the sake of easier interpretation) using log axes for
$t_{p_{r}}$ and $\sigma$.

The prior joint relative densities in
Figure~\ref{figure:ijprior.density.examples} have some common
features. In particular,
\begin{itemize}
\item
For any given value of $\sigma$, $\pi(\log(t_{p_{r}}),\log(\sigma))$
is a decreasing function of $t_{p_{r}}$.
\item
For small $\sigma$ and $t_{p_{r}} < t_{c}$, the relative density is
approximately flat at a level of 1.0.
\item
Following from the previous point, the joint densities are improper
in the sense that the area under the density is infinite.
\item
For small $\sigma$ and $t_{p_{r}} > t_{c}$, the level of the density
is, relative to the flat region, negligible, for all values of
$p_{r}$.
\item
Following from the features in the previous two points, there is a
steep cliff as one crosses $t_{c}$ for small values of $\sigma$.
\item
For large values of $\sigma$ the level of the density is
approximately flat, but at a level that is approximately equal to $p_{r}$.
\item
With $p_{r}=0.99$ (or larger) the density is approximately flat
except in the ``Negligible density'' region of the parameter space.
\end{itemize}
The reasons for these features are given in
Section~\ref{S.section:reasons.for.ij.prior.features}.

\subsection{Reasons for the \typeI{} Censoring Independence Jeffreys Prior
  Distribution Features}
\label{S.section:reasons.for.ij.prior.features}

\begin{figure}
\begin{tabular}{cc}
\rsplidafiguresize{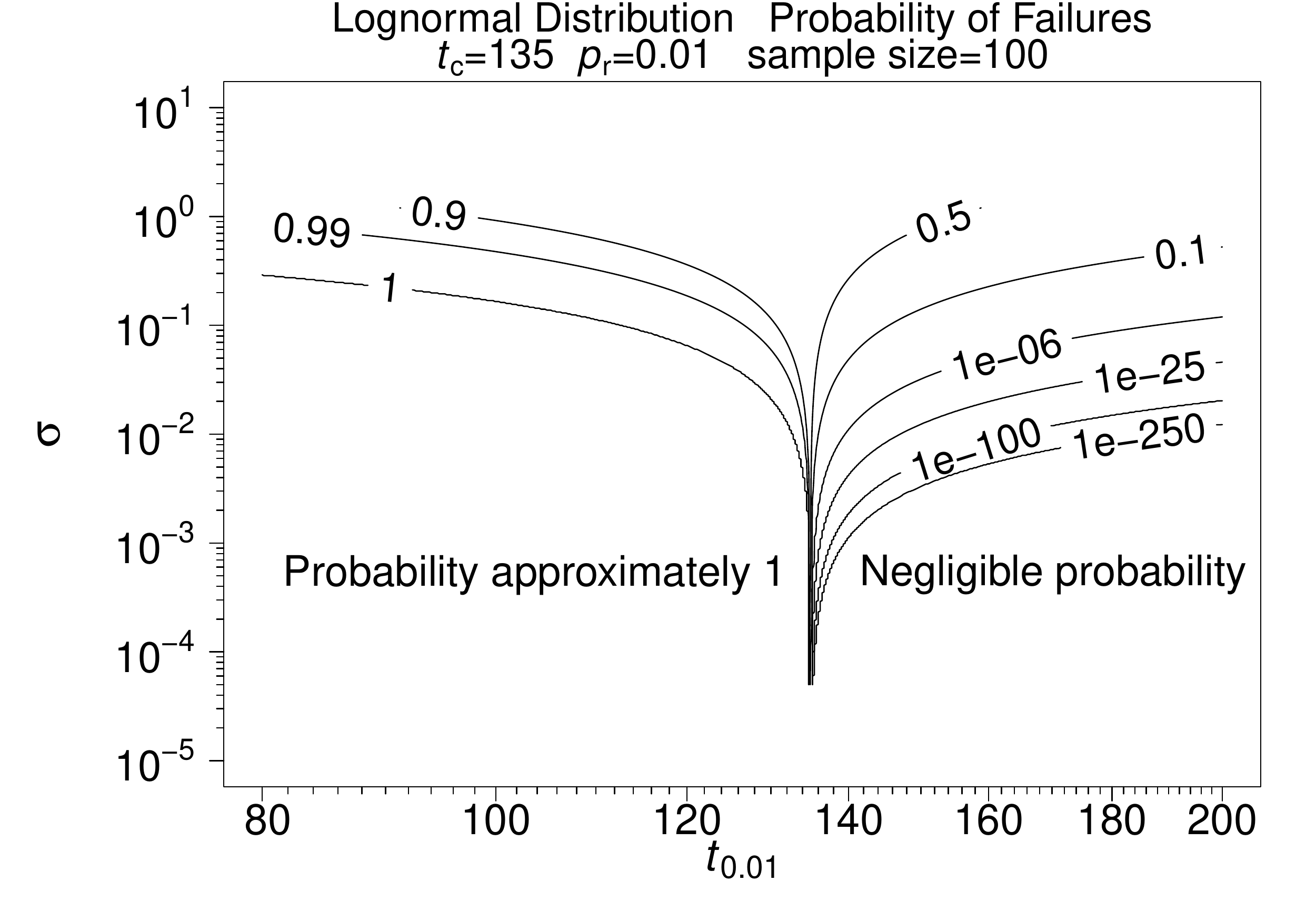}{3.5in}&
\rsplidafiguresize{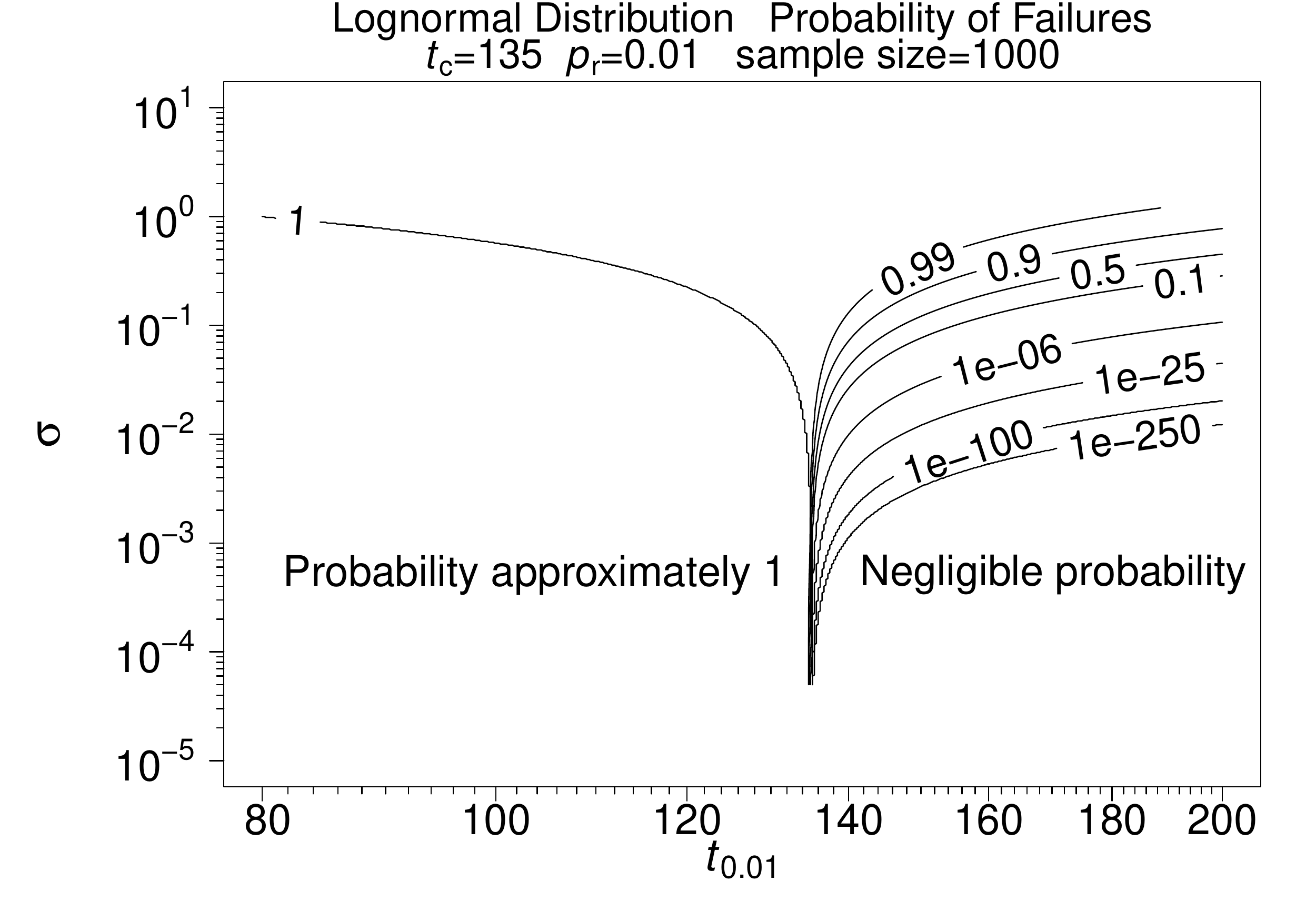}{3.5in}\\
\rsplidafiguresize{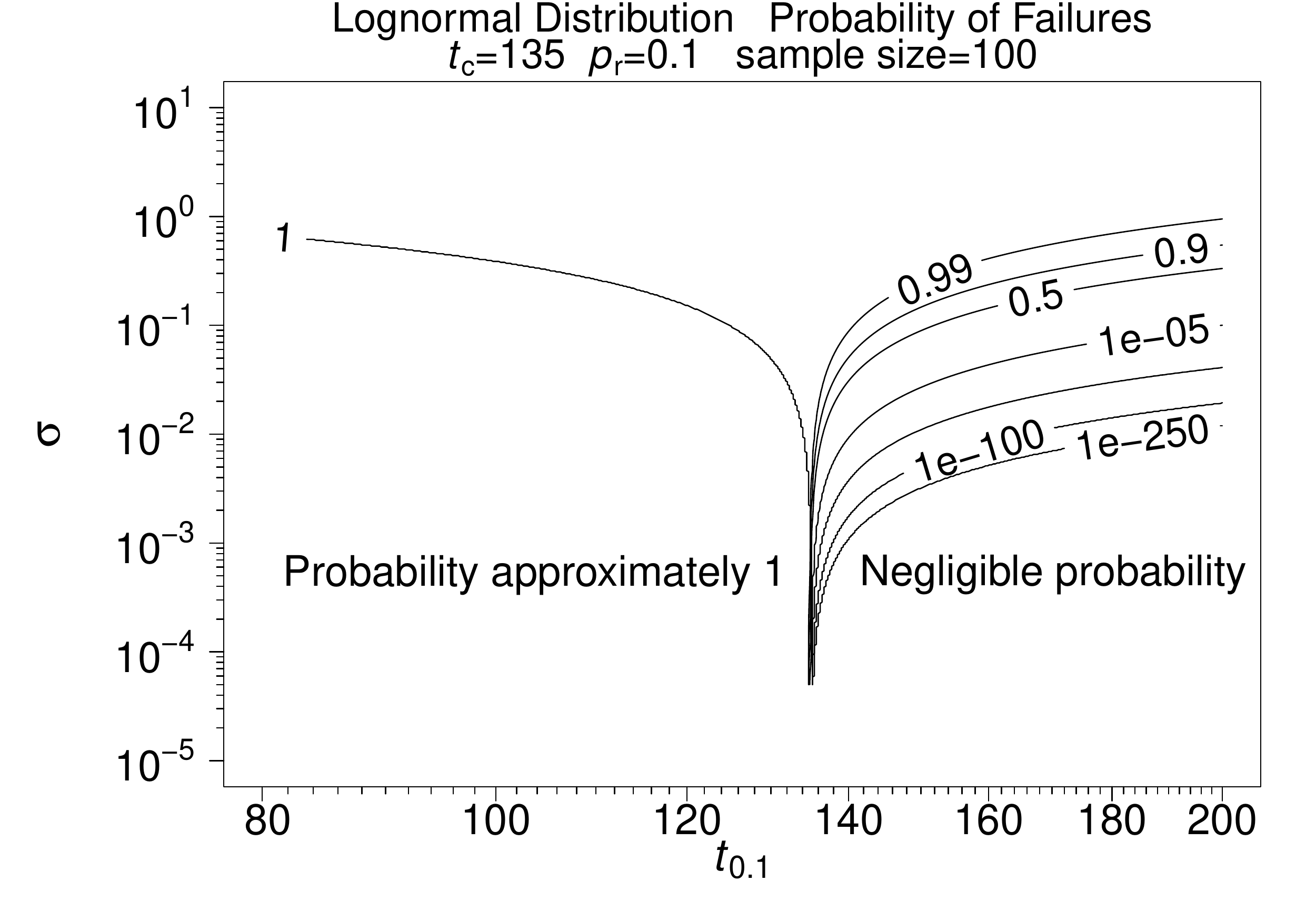}{3.5in}&
\rsplidafiguresize{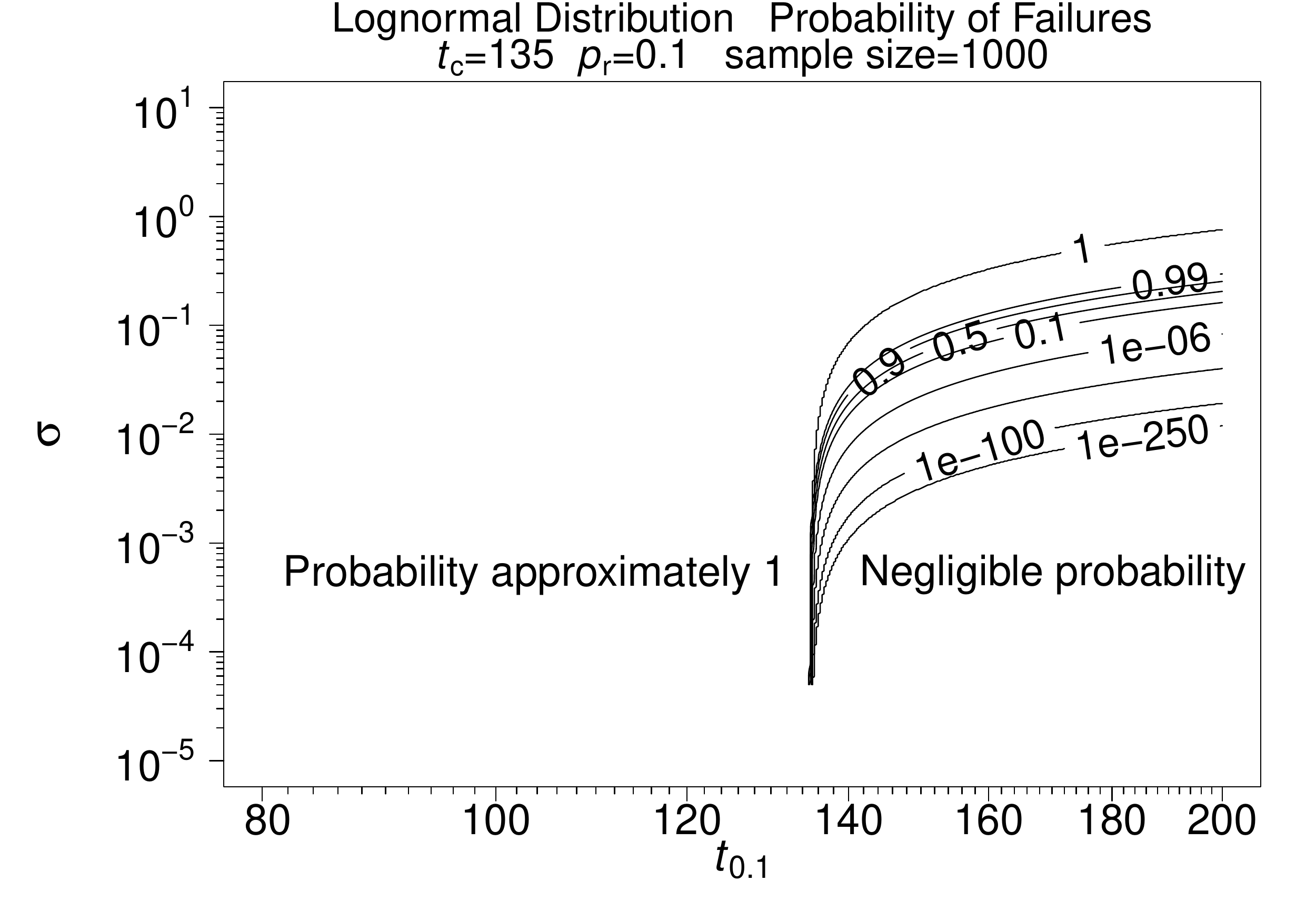}{3.5in}\\
\rsplidafiguresize{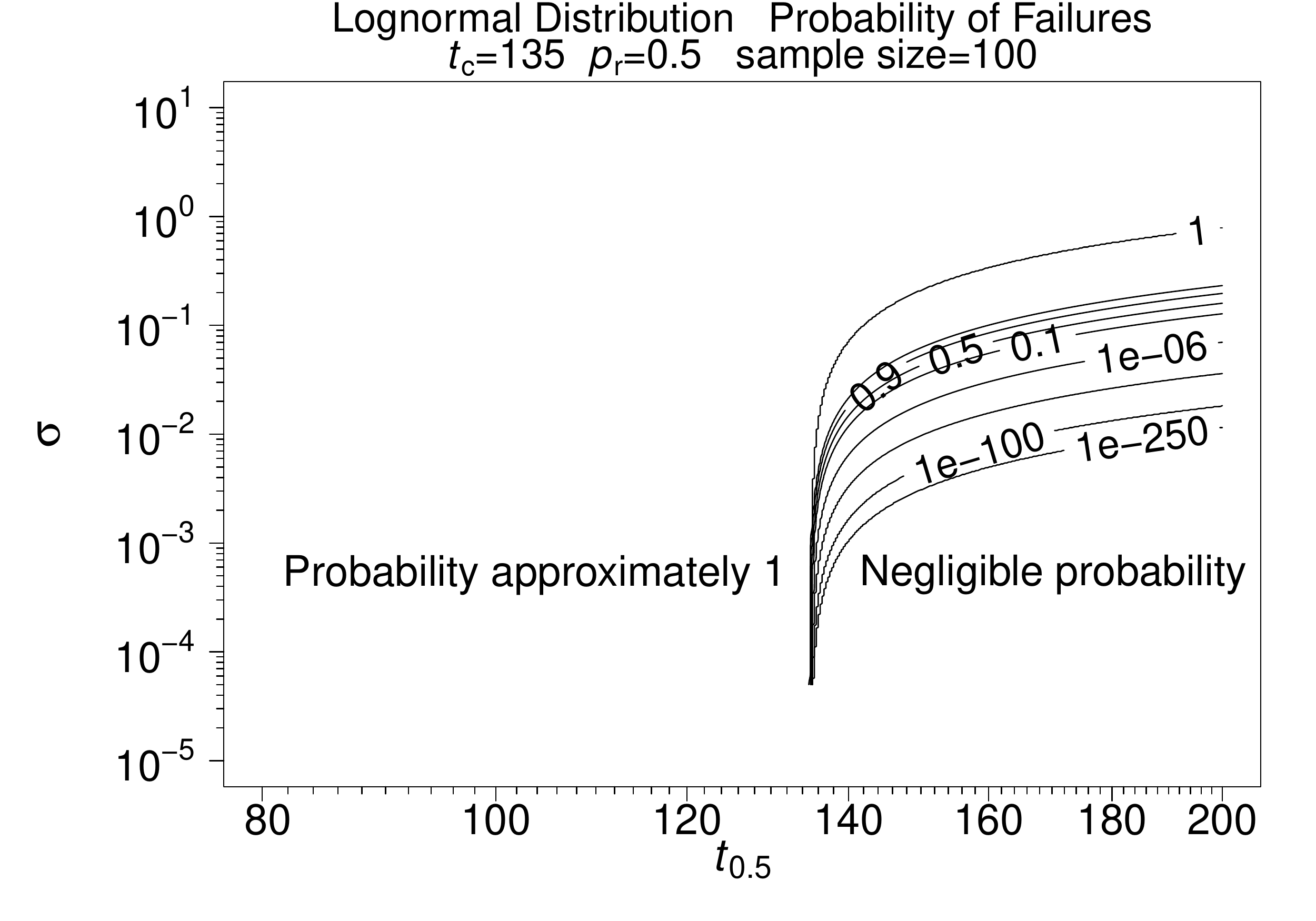}{3.5in}&
\rsplidafiguresize{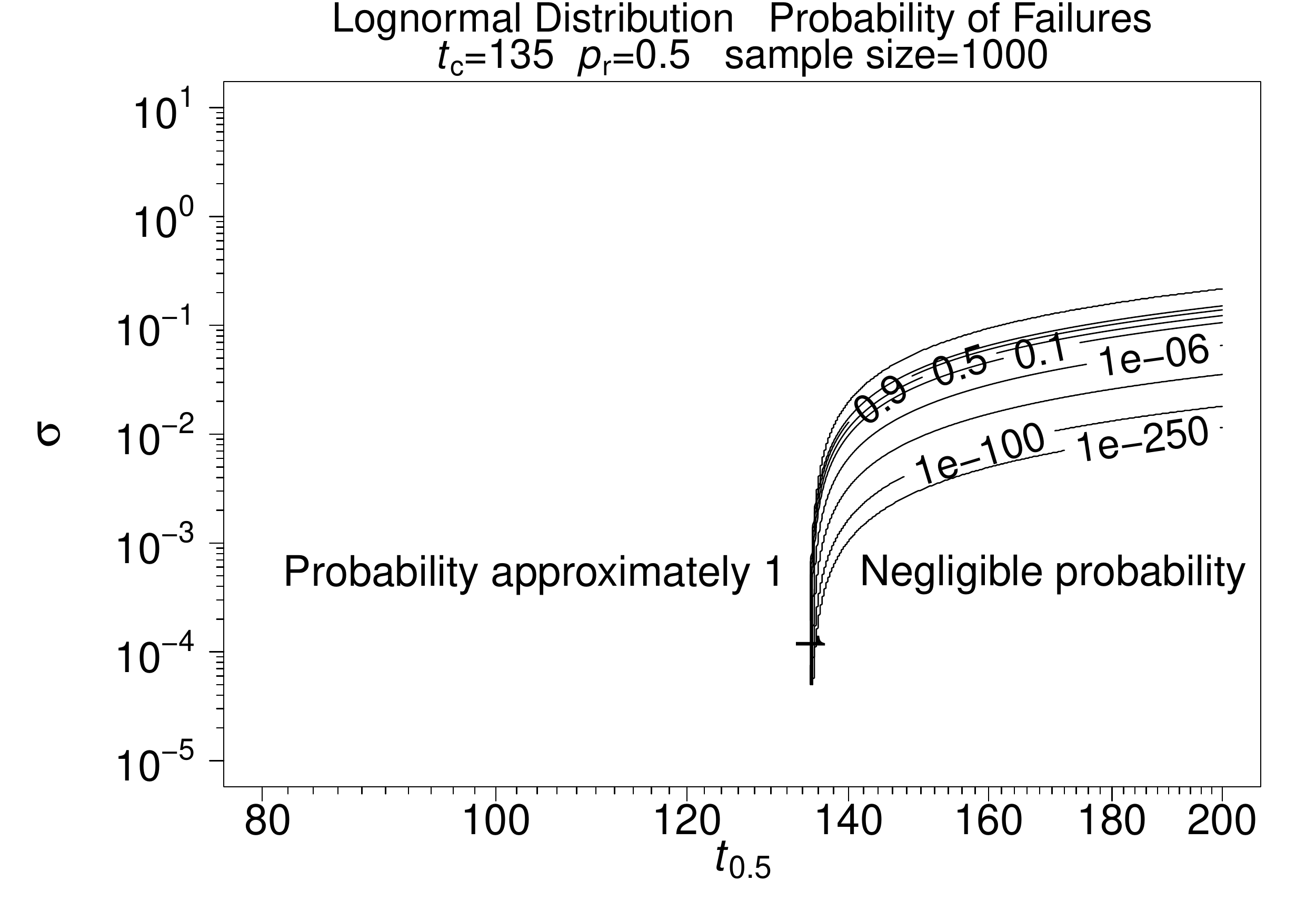}{3.5in}\\
\end{tabular}
\caption{Probability of one or more failures before the the
  censoring time $t_{c}=135$ for sample sizes $n=100$ (on the left)
  and $n=1000$ (on the right) for reparameterization quantile
  $p_{r}=$ 0.01 (top), 0.10 (middle), and 0.50 (bottom).  }
\label{S.figure:probability.of.failures}
\end{figure}

Here we explore the reasons for the different IJ prior density shapes and
features in Figure~\ref{figure:ijprior.density.examples}.
Figure~\ref{S.figure:probability.of.failures} shows the probability of
having one or more failures in a \typeI-censored life test as a
function the parameters $t_{p_{r}}$ and $\sigma$ and sample size
$n$. This binomial probability can be computed as
\begin{align*}
\Pr(r>0) = 1 - \Pr(r=0) = 1 - [1 - \Pr(T < t_{c})]^{n}
\end{align*}
where
\begin{align}
\nonumber
\Pr(T < t_{c}) &= \Phi_{\norm}\left [\frac{\log(t_{c}) - \mu}{\sigma}
\right ]\\
\label{S.equation:pr.t.le.tc}
              &= \Phi_{\norm}\left [\frac{\log(t_{c}) -
        \log(t_{p_{r}})}{\sigma} + \Phi_{\norm}^{-1}(p_{r})
\right ].
\end{align}
Figure~\ref{S.figure:probability.of.failures} shows that the
negligible-density region of the IJ prior densities in
Figure~\ref{figure:ijprior.density.examples} corresponds to that part of the
parameter space where the probability of having a failure is
negligible.

\begin{figure}
\begin{tabular}{cc}
\rsplidafiguresize{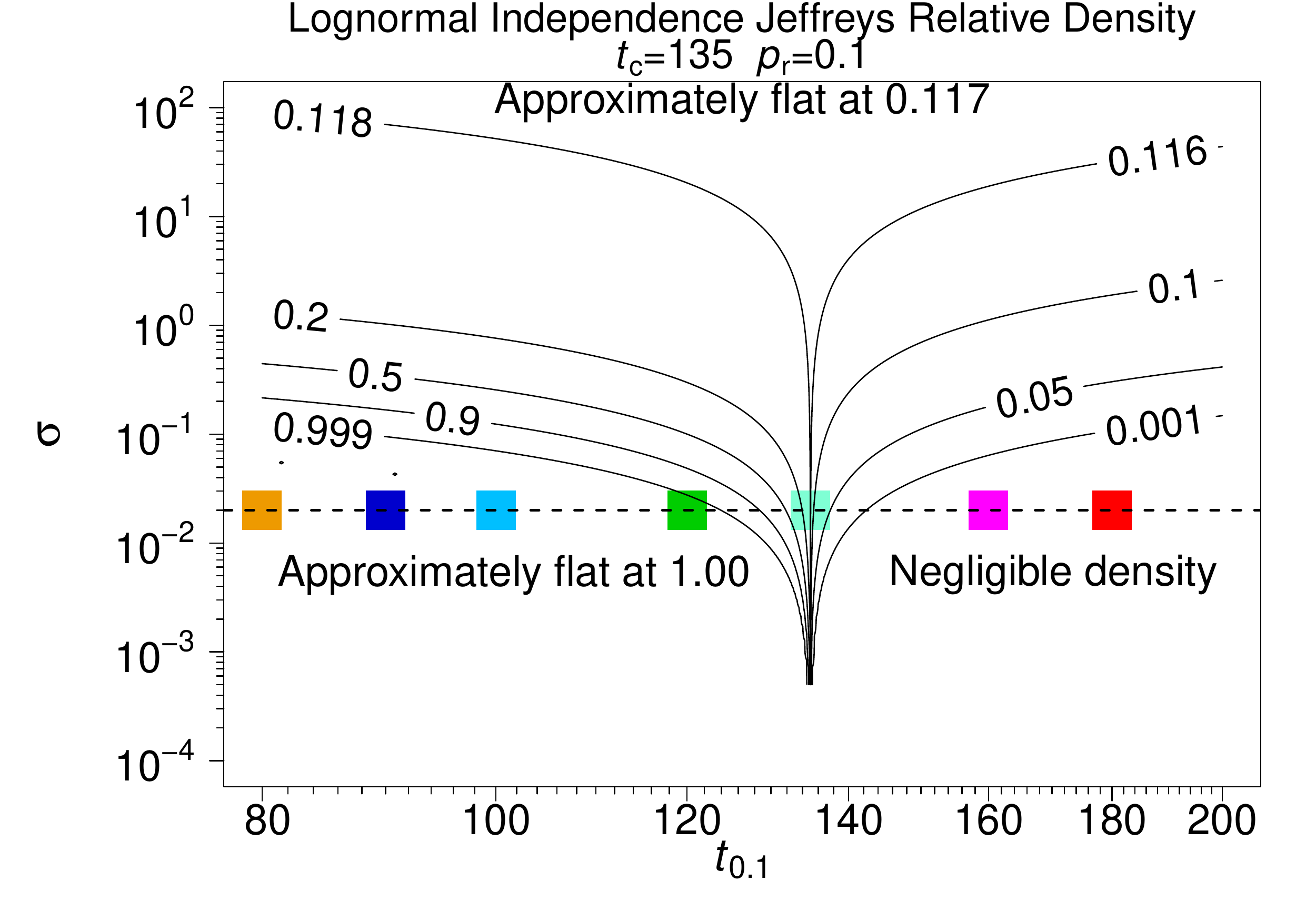}{3.5in}&
\rsplidafiguresize{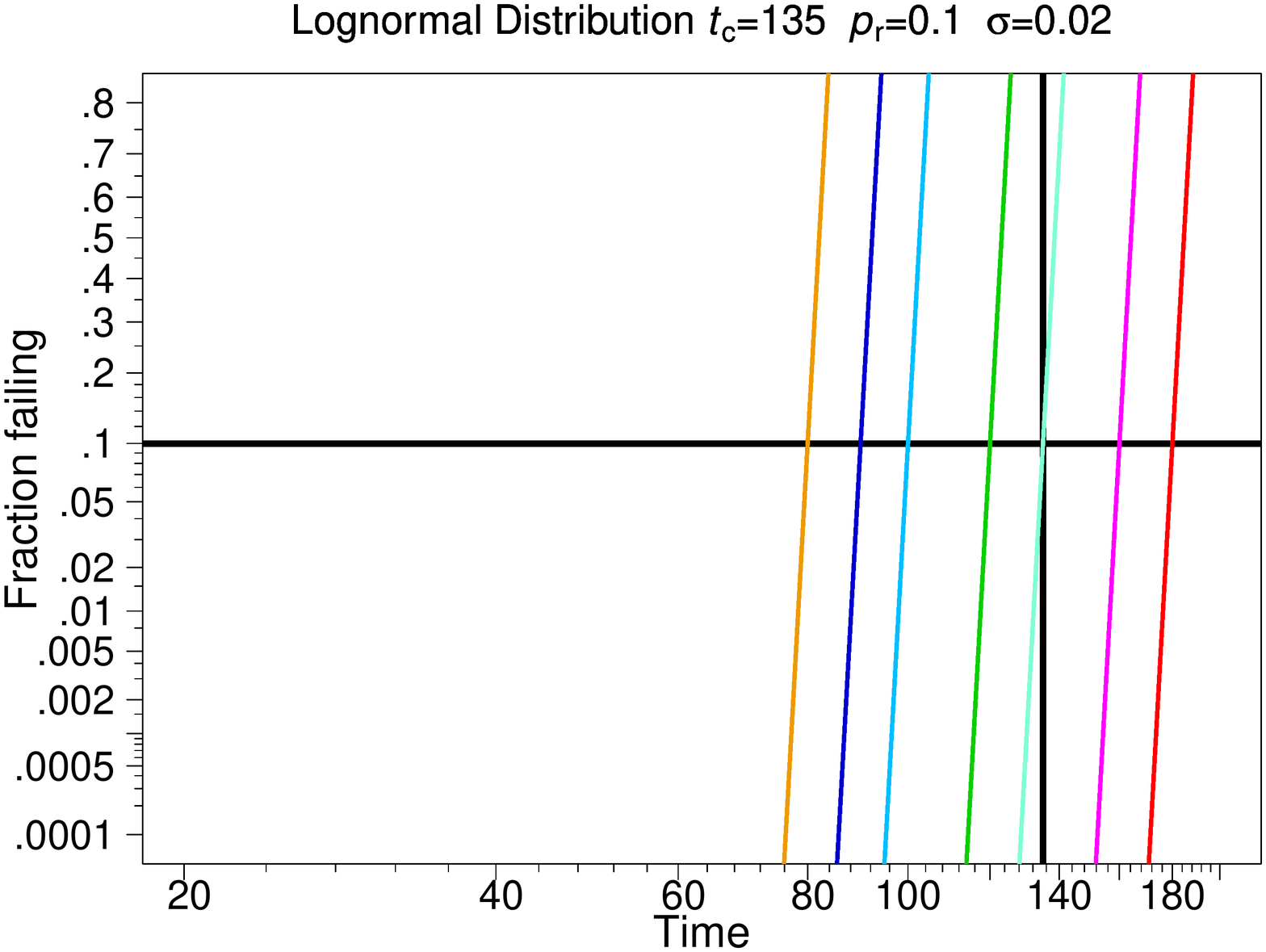}{3.5in}\\
\rsplidafiguresize{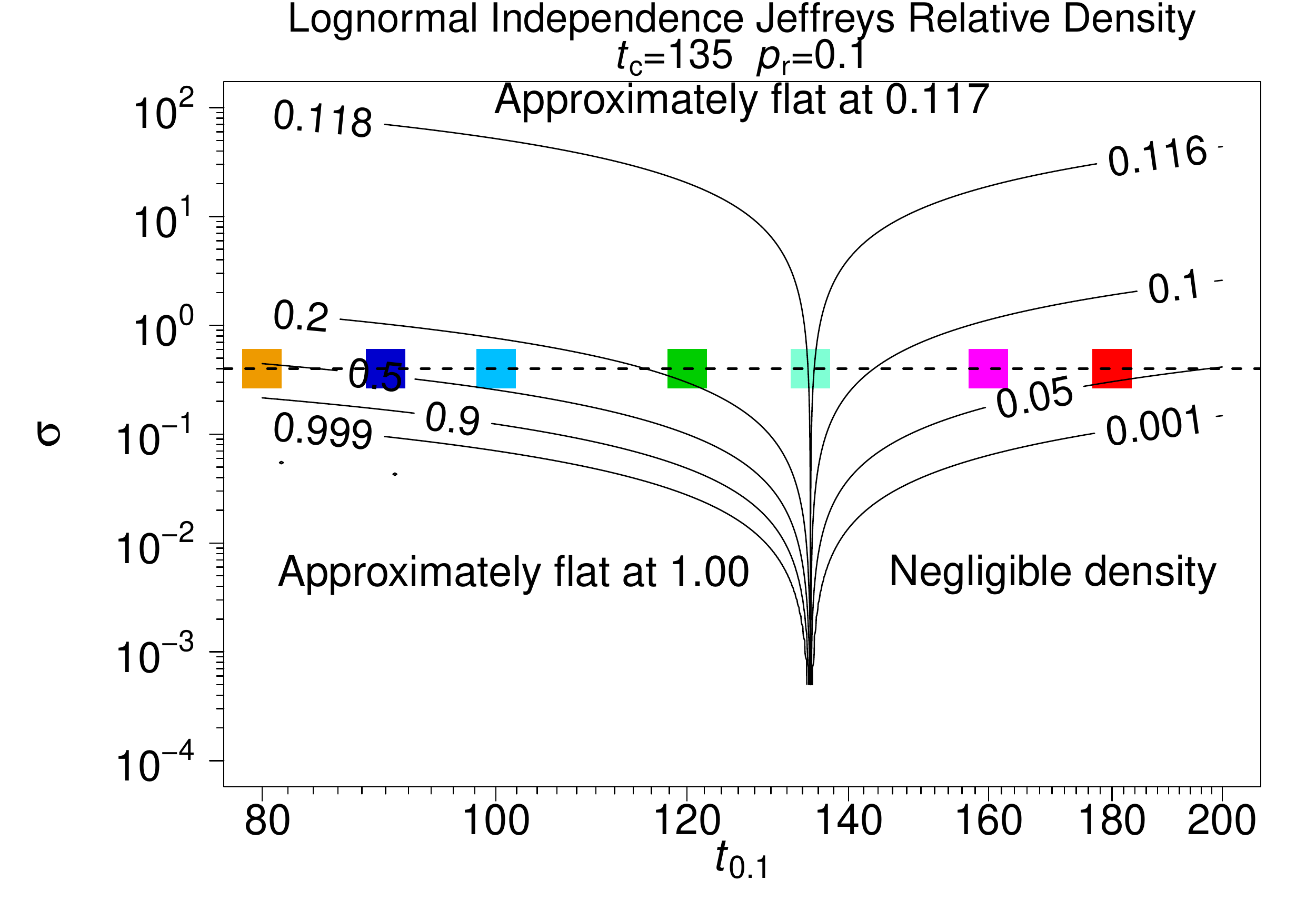}{3.5in}&
\rsplidafiguresize{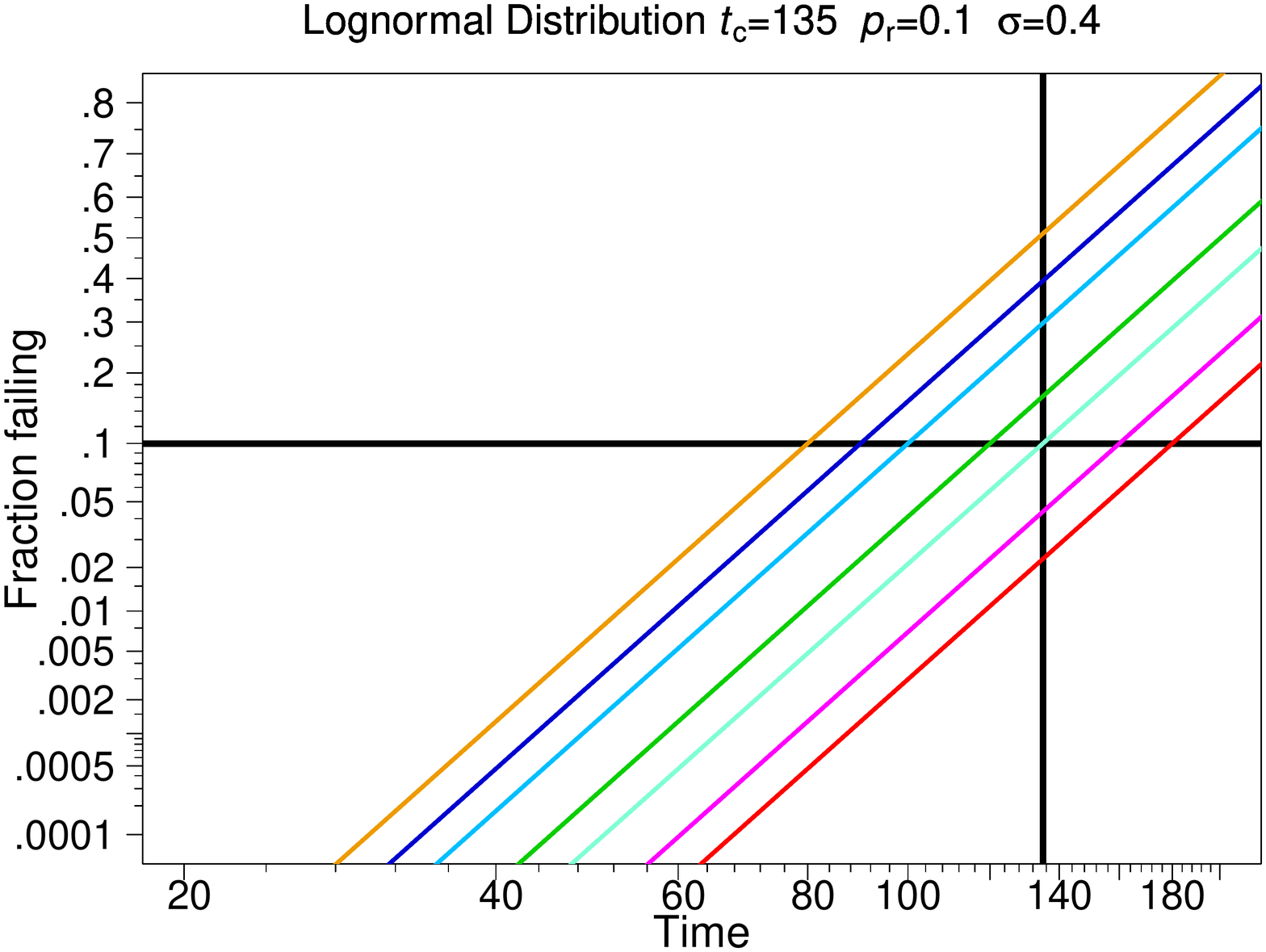}{3.5in}\\
\rsplidafiguresize{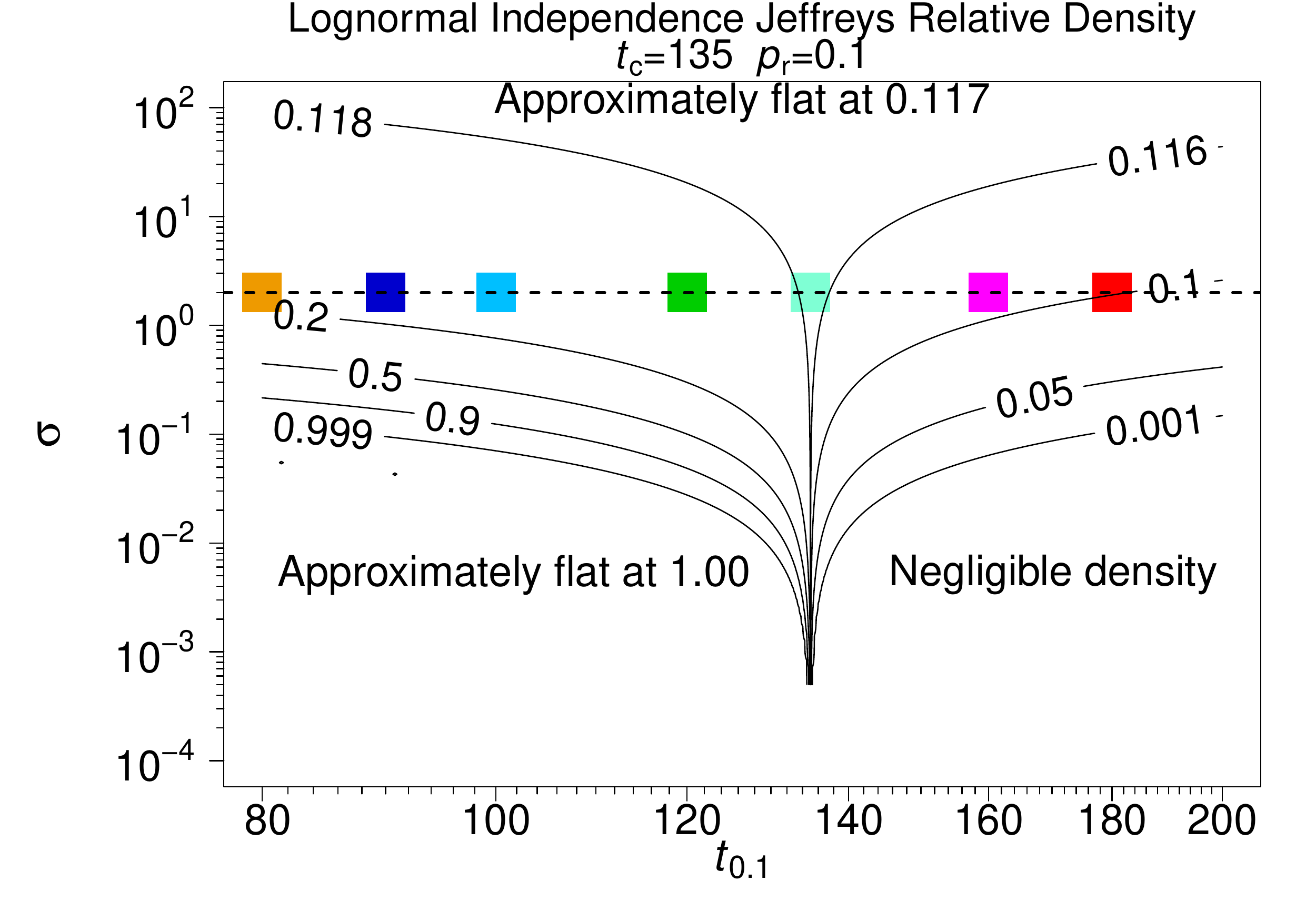}{3.5in}&
\rsplidafiguresize{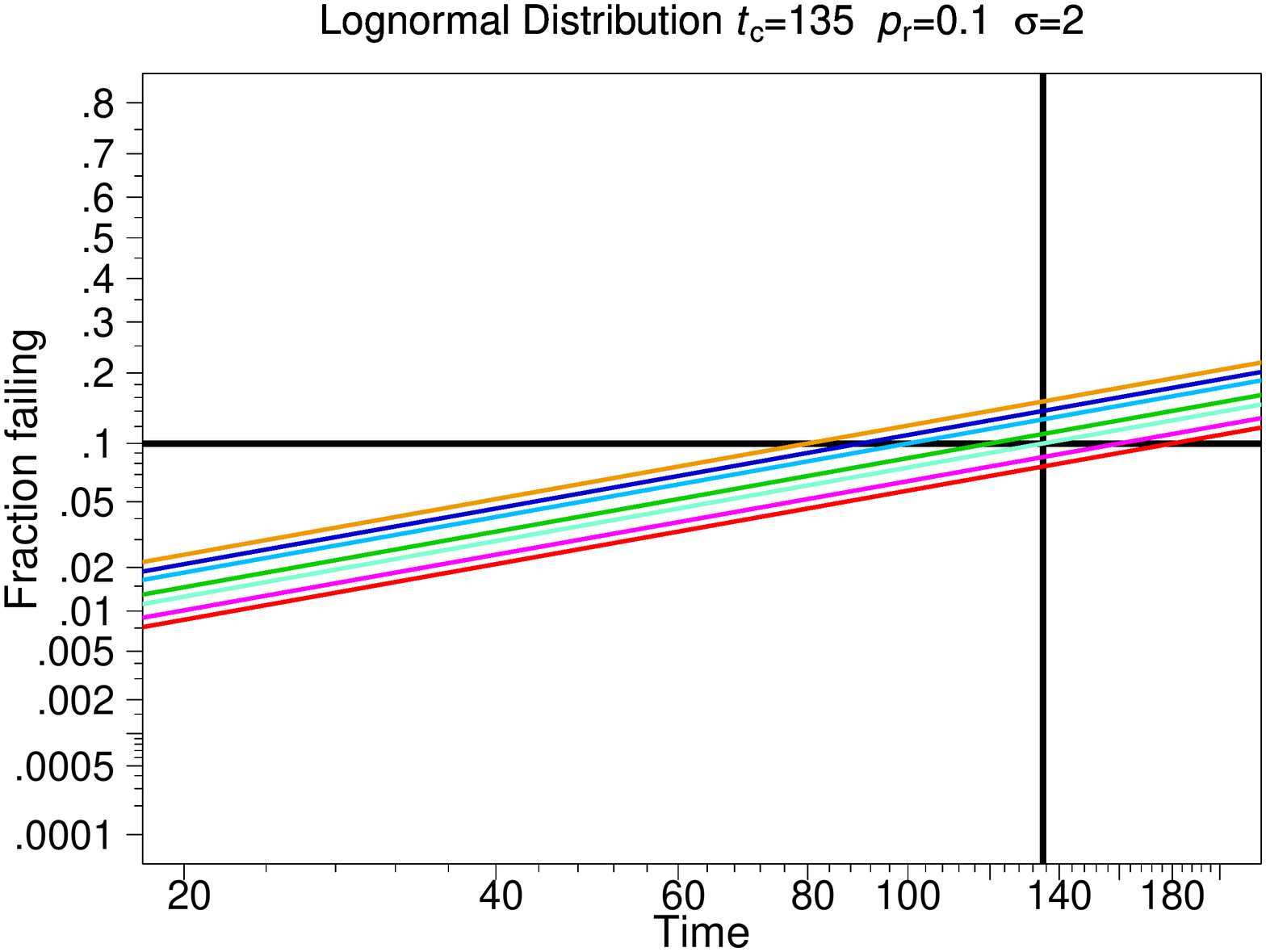}{3.5in}\\
\end{tabular}
\caption{IJ prior densities with points in the parameter space
  indicated by the squares (on the left) and Lognormal cdfs on
  lognormal probability scales for the different points (on the
  right) for $\sigma=$ 0.02 (top), 0.4 (middle), and 2.0 (bottom).
}
\label{S.figure:ijprior.explore}
\end{figure}

Every point in the $(t_{p_{r}}, \sigma)$ parameter space corresponds
to a particular lognormal cdf.  Figure~\ref{S.figure:ijprior.explore}
provides a visualization of the relationship for the particular case
of $p_{r}=0.10$. The horizontal lines in the
probability plots on the right are at $p_{r}=0.10$. The vertical
lines are at the censoring time $t_{c}=135$.  The horizontal
position of the colored squares in the density plots on the left
indicate the value of $t_{p_{r}}=$ 80, 90, 100, 120, 135, 160, and
180, controlling the horizontal location of the cdf in the plots on
the right. The vertical position of the squares indicate the value
of $\sigma=0.02, 0.4$, and $2.0$, which controls the slopes of the
cdfs on the lognormal scales on the right of Figure~\ref{S.figure:ijprior.explore}
(the slope of the lognormal
  cdf is $1/\sigma$ on the linear axes underlying lognormal
  probability scales, as described in
  \citet[][Section~6.2.3]{MeekerEscobarPascual2022}).

In the top-right probability plot in
Figure~\ref{S.figure:ijprior.explore}, consider, for example, the cdf
on the far right with $t_{p_{r}}=180$ and $\sigma=0.02$ (corresponding
to the red square on the far right in the top-left contour plot). Substituting
these parameter values, $p_{r}=0.10$, and $t_{c}=135$
into (\ref{S.equation:pr.t.le.tc}) gives
\begin{align*}
\Pr(T < t_{c}) &= \Phi_{\norm}\left [\frac{\log(135) -
        \log(180)}{0.02} + \Phi_{\norm}^{-1}(0.10)
\right ]= 1.298 \times 10^{-55}.
\end{align*}
If $t_{p_{r}}=135$ (corresponding to the cyan square in the top-left
contour plot), by definition of the distribution quantile, $\Pr(T <
t_{c})=0.10$ (for any value of $\sigma$) and if $t_{p_{r}}=120$ and
$\sigma=0.02$ (corresponding to the green square in the top-left
contour plot), $\Pr(T < t_{c})=0.999997$, showing the steepness of
the cliff.
%
%
%
%
%

\section{Implementation of and Experiences Using Stan with a CJ
  or an IJ Prior Distribution}
\label{S.section:experiences.using.stan.with.ij.prior}
Section~\ref{section:ij.prior} of the main paper gives expressions
for the CJ priors for both $y_{p}=\log(t_{p})$ and $\log(\sigma)$ as
well as the the IJ priors. These are easy to compute given the
algorithms to compute the FIM elements described in
Section~\ref{section:lls.fim} and could be used in conjunction with
a standard MCMC method like the Metropolis--Hastings algorithm. In
our work, we used two alternative methods to compute posterior
draws. The one based on {\tt rstan} \citep{rstan} was used for our
examples and the simulation.  Another that uses a simple rejection
method described in \citet{SmithGelfand1992} was used to check the
first method. The rest of this section describes some implementation
details and our experiences.

\subsection{Implementation Using Rstan}
\label{S.section:implementation.using.rstan}
We did not see a way to compute FIM elements within a Stan model. As
an alternative, we sent down (from R) vectors (length 200 was the default) of
values and then used a Stan-model function to compute the needed FIM
element values with linear interpolation. Expressions for the IJ
prior density were programmed directly in the model block of the
Stan model (as if they were part of the likelihood function).

These codes were
exercised extensively in preparation for the running of our simulation study. For
some factor-level combinations we noticed excessively long
run times (often with many treedepth exceedences) for some data
sets. The root cause was found to be extremely small stepsize values
obtained from the adaptation stage of the Stan NUTS sampler for an
IJ prior and a small number of failures (e.g., fewer than
ten). Because we never saw this behavior with flat priors (even with
only three failures), we suspect that the problems arose because of
our non-differentiable piecewise-linear approximation to the FIM
elements. We avoided the problems by disabling the NUTS adaptation and
specifying a stepsize. After some experimentation we found that, for
a given data set, the flat prior stepsize divided by 5 worked well
(i.e., fast sampling with few or no warnings). 

Section~\ref{S.section:ThousandTestThreeFail} looks closely at
estimation results for a data set with three failures from a sample
of size 1000. The likelihood for these data has an interesting
funnel shape with a sharp point at the top (Weibull) or bottom
(lognormal). When running the Stan NUTS sampler for a flat prior, divergent
transitions were sometimes observed, although they were not concentrated in the
tip.  Changing the default $\textrm{adapt}\_\textrm{delta}$ to 0.995
eliminated the divergent transitions.

\subsection{Implementation of Sampling from IJ/CJ Priors for Bayes
  Without Tears Plots}
\label{S.section:implementation.using.bayes.without.tears}
\citet{SmithGelfand1992} describe a simple accept/reject posterior
sampling method that accepts points from the prior with a
probability equal to the value of the relative likelihood at the
point. As illustrated in
Figures~\ref{figure:BearingCage.noninformative.informative}
and~\ref{figure:RocketMotor.noninformative.informative} (also see
the numerous plots in Section~\ref{S.section:careful.look.three.failures}),
comparing plots of prior points and likelihood contours with
a similar plot of posterior points and likelihood contours provides
insight into how the prior and likelihood combine to produce a
posterior distribution.

It is easy to compute values of the IJ prior
using the expressions in Section~\ref{section:ij.prior} used, for
example, to compute the contours in
Figure~\ref{figure:ijprior.density.examples}. However, we saw no simple way to
sample from these improper priors. Instead we used Stan to sample
from the prior density, in a manner similar to what we describe in
Section~\ref{S.section:implementation.using.rstan}, but with no data
contributing to the likelihood.
Because the Stan NUTS sampler cannot
be used to sample from an improper distribution, sampling was
constrained to a large rectangle (much larger than the boundaries of
the contour plots where the draws are plotted) so that the
resulting prior distributions are proper.

\subsection{Software for the Examples and Simulations}
\label{S.section:software.examples.simulation}
All of the computing for this project was done using R functions
that are available in the R Package RSplida. A windows version of
RSplida is available from
https://wqmeeker.stat.iastate.edu/RSplida.zip and the echapters
folder in RSplida
includes a file of commands that can be used to run the examples in
this paper.

\section{Lognormal Distribution Simulation Results and Conclusions}
\label{S.section:lognormal.simulation.results.conclusions}
\begin{figure}
\centering
\begin{tabular}{cc}
\rsplidafiguresize{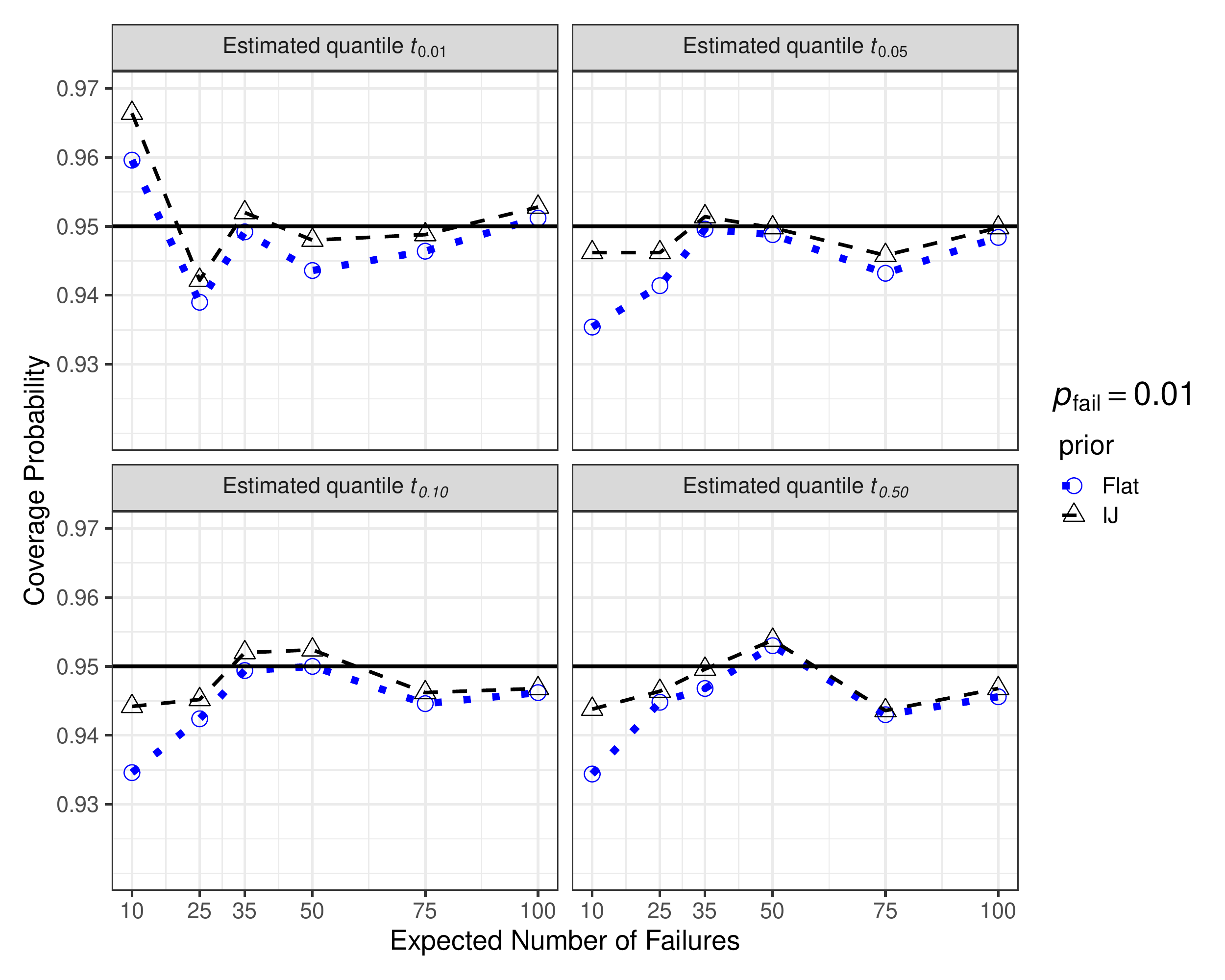}{6.0in}\\
\rsplidafiguresize{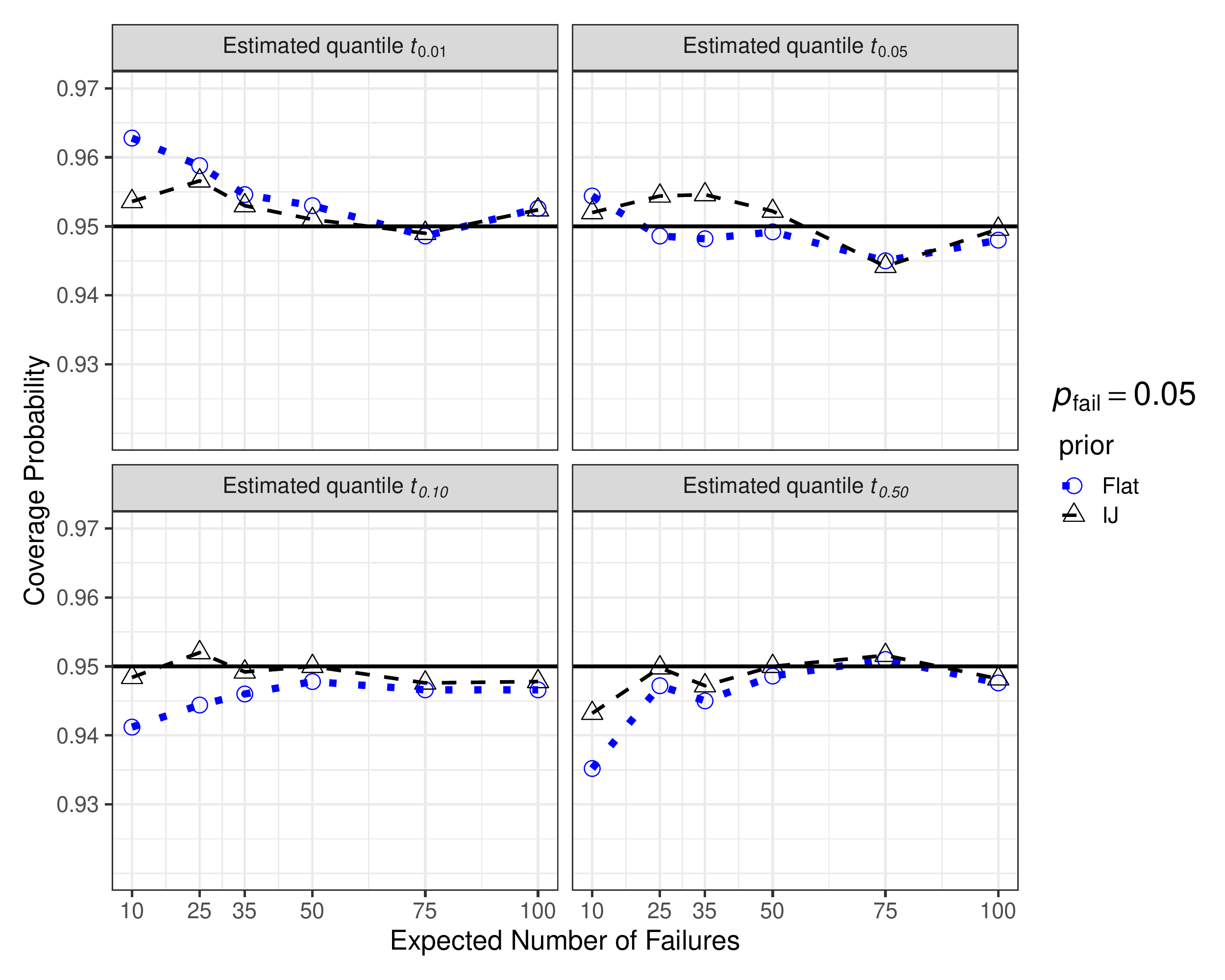}{6.0in}
\end{tabular}
\caption{Lognormal distribution two-sided estimated coverage probabilities for $p_{\fail}=0.01$
  on the top and $p_{\fail}=0.05$ on the bottom.}
\label{figure:lognormalSimulationTwoSidedResults.small.pfail}
\end{figure}

\begin{figure}
\centering
\begin{tabular}{cc}
\rsplidafiguresize{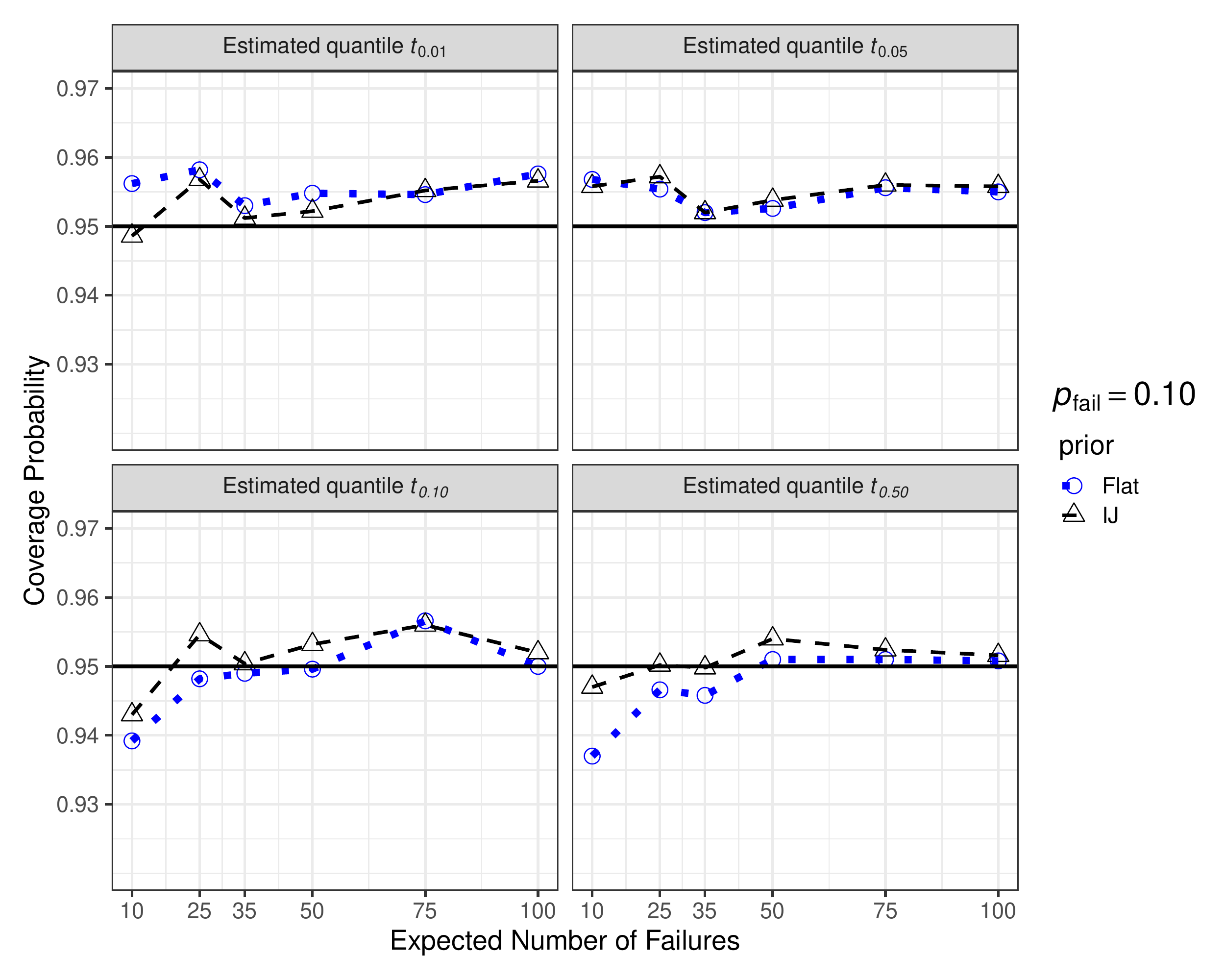}{6.0in}\\
\rsplidafiguresize{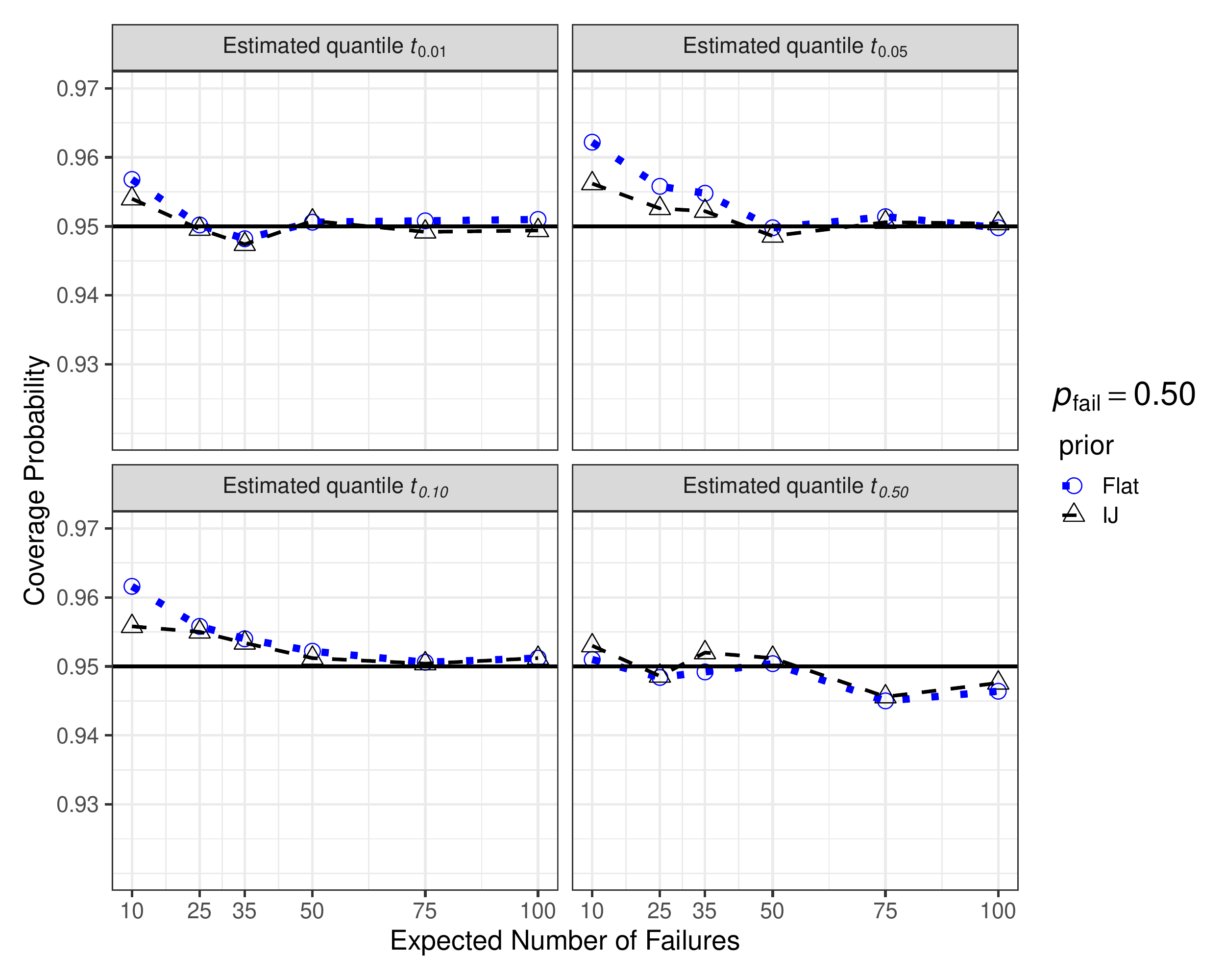}{6.0in}
\end{tabular}
\caption{Lognormal distribution two-sided estimated coverage probabilities for $p_{\fail}=0.10$
  on the top and $p_{\fail}=0.50$ on the bottom.}
\label{figure:lognormalSimulationTwoSidedResults.large.pfail}
\end{figure}

\begin{figure}
\centering
\begin{tabular}{cc}
\rsplidafiguresize{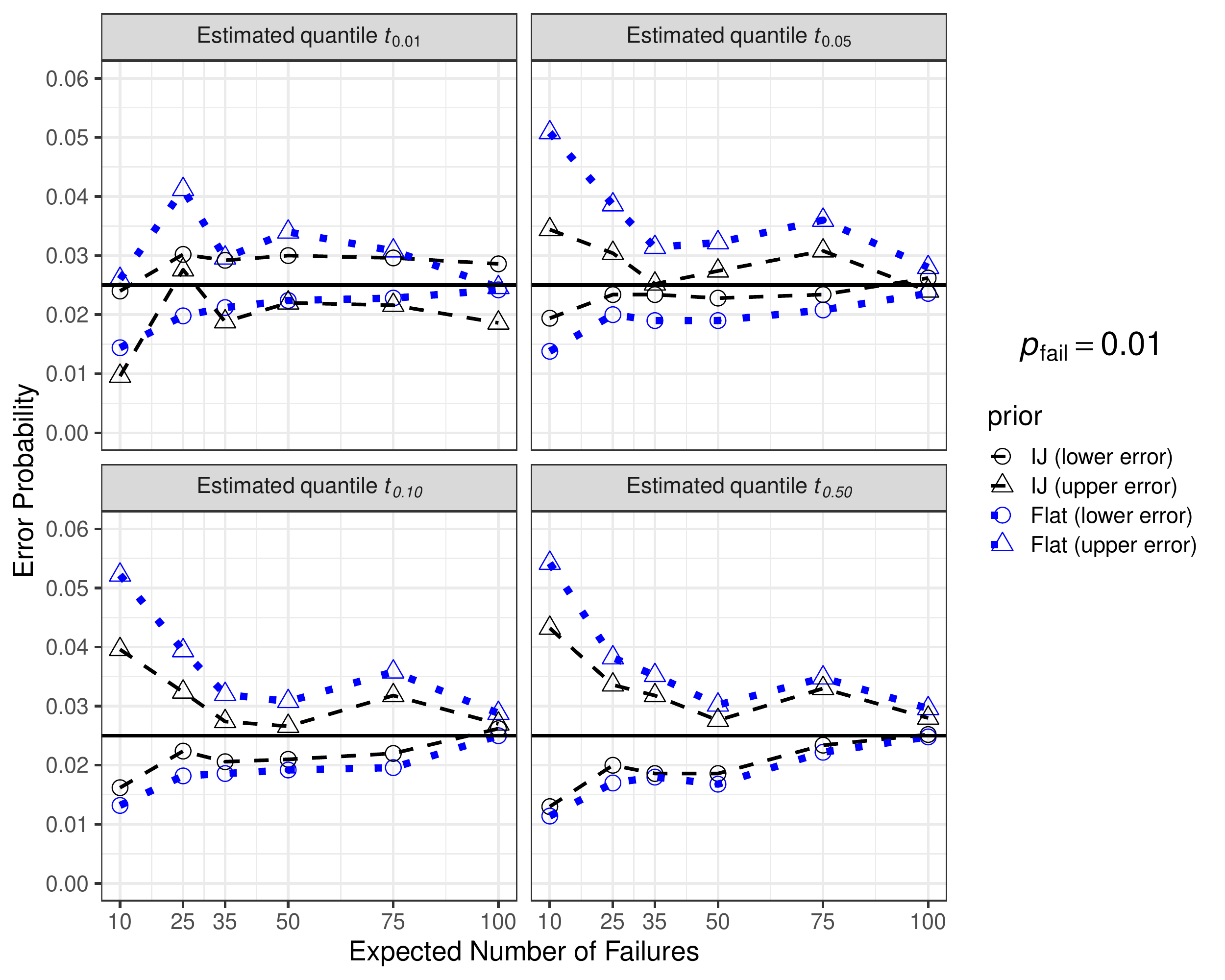}{6.0in}\\
\rsplidafiguresize{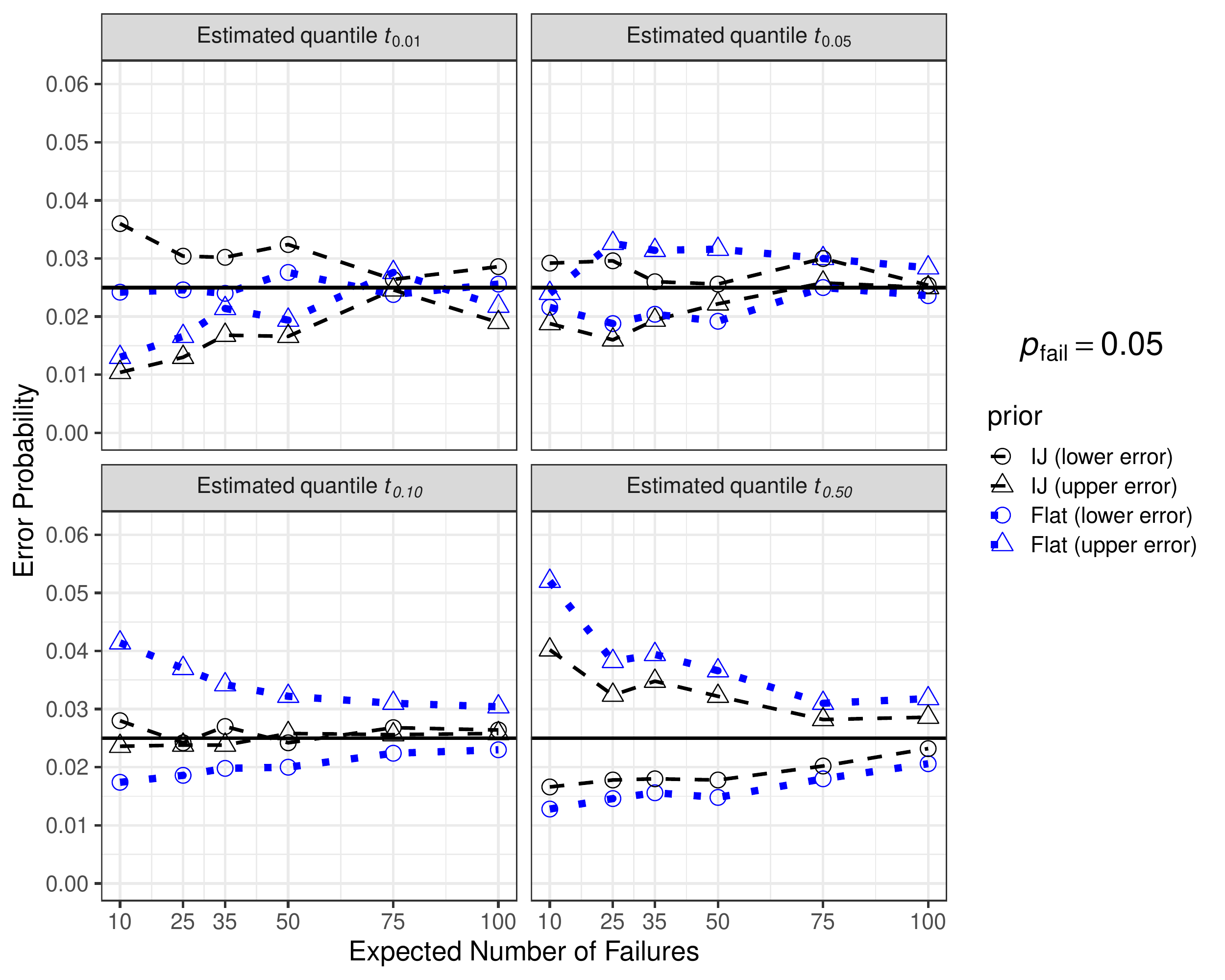}{6.0in}
\end{tabular}
\caption{Lognormal distribution one-sided
estimated error probabilities for $p_{\fail}=0.01$
  on the top and $p_{\fail}=0.05$ on the bottom.}
\label{figure:lognormalSimulationOneSidedResults.small.pfail}
\end{figure}

\begin{figure}
\centering
\begin{tabular}{cc}
\rsplidafiguresize{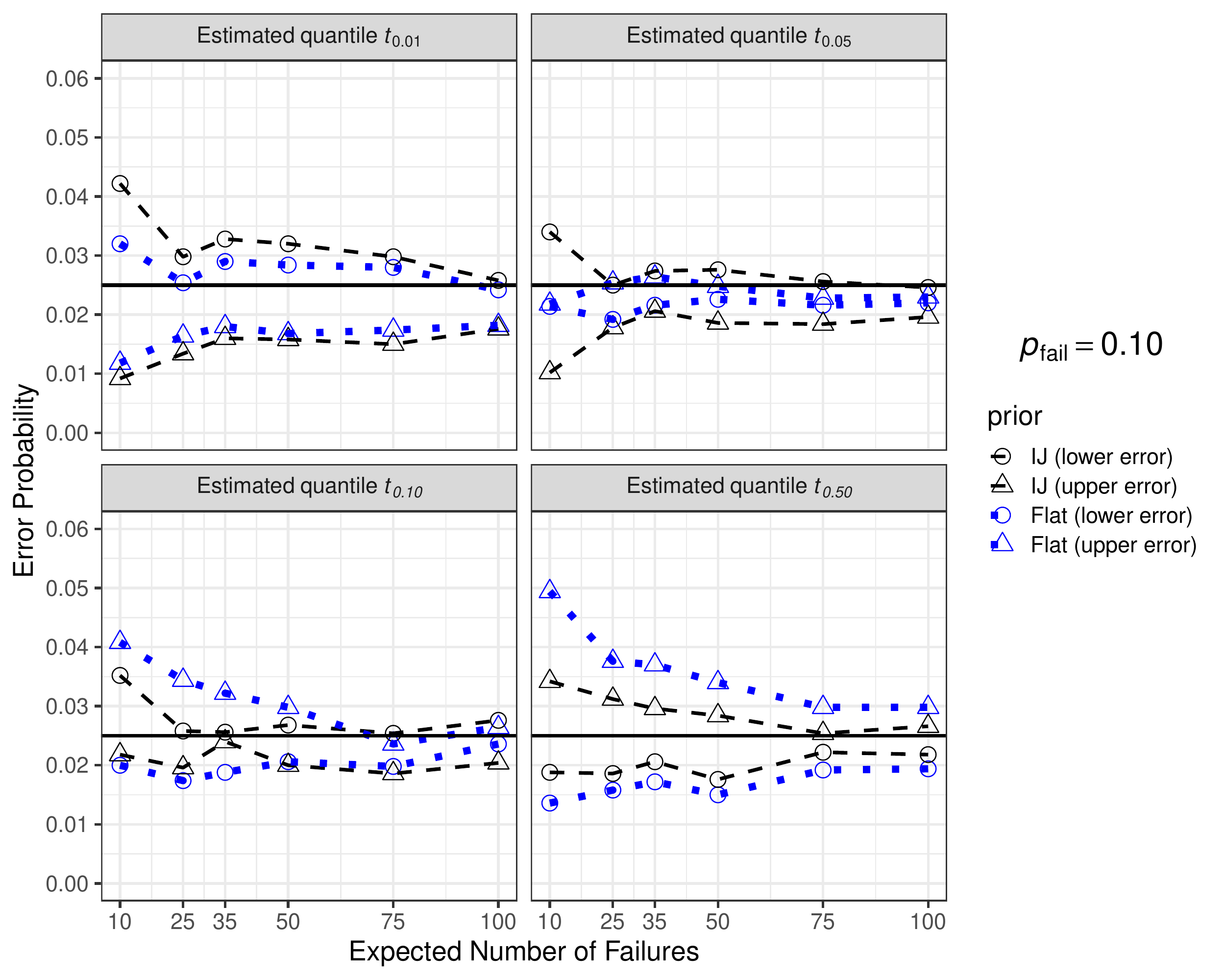}{6.0in}\\
\rsplidafiguresize{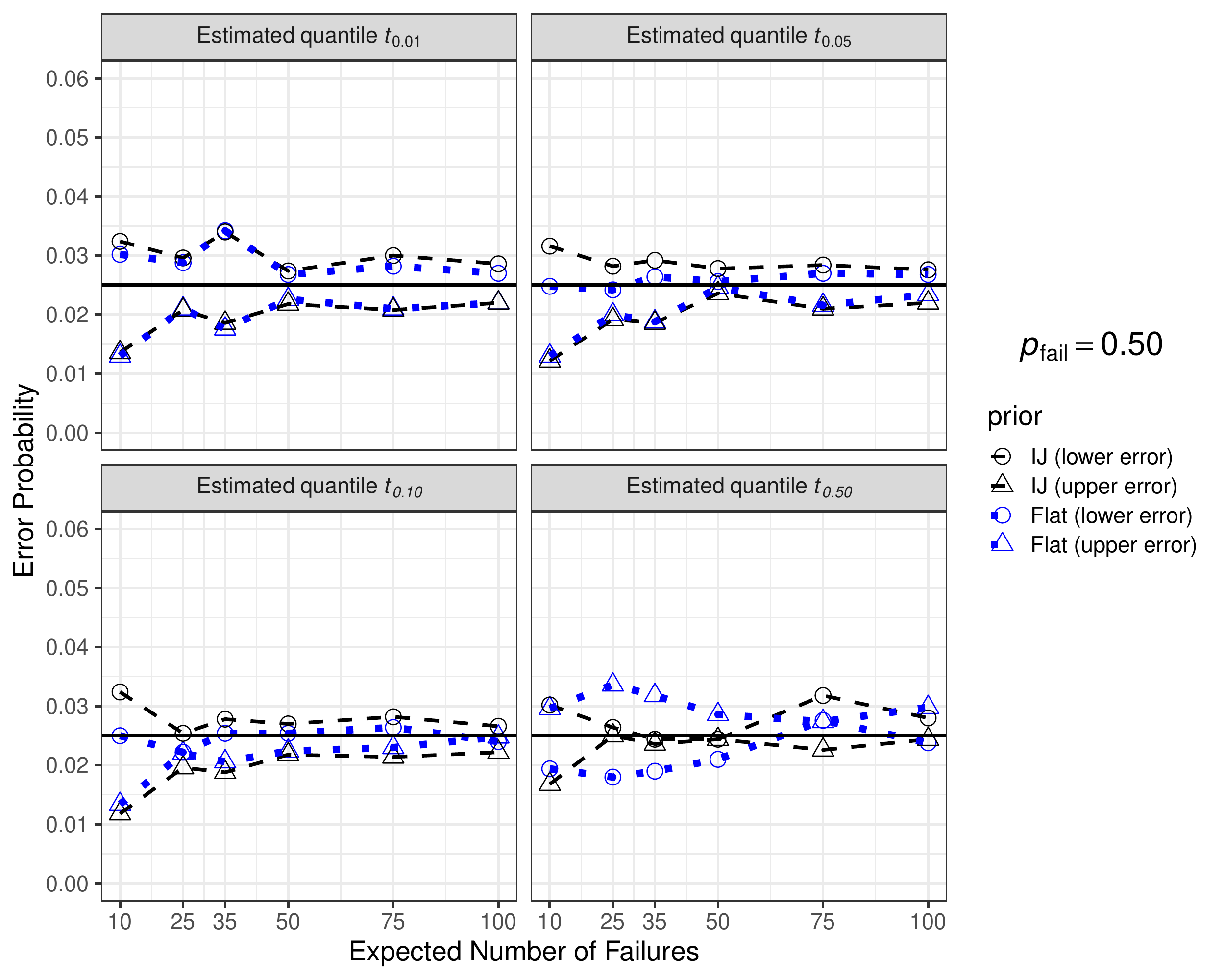}{6.0in}
\end{tabular}
\caption{Lognormal distribution  one-sided
estimated error probabilities for $p_{\fail}=0.10$
  on the top and $p_{\fail}=0.50$ on the bottom.}
\label{figure:lognormalSimulationOneSidedResults.large.pfail}
\end{figure}

Section~\ref{simulation.results.conclusions} of the main paper
presents the simulation results, evaluating the credible interval
coverage probabilities using different noninformative priors
for the Weibull distribution. Here we present similar results for
the lognormal distribution.
Figures~\ref{figure:lognormalSimulationTwoSidedResults.small.pfail}
($p_{\fail}=0.01$ and $p_{\fail}=0.05$) and~\ref{figure:lognormalSimulationTwoSidedResults.large.pfail} ($p_{\fail}=0.10$ and $p_{\fail}=0.50$) summarize
the lognormal distribution simulation results with two-sided coverage probabilities.
Figures~\ref{figure:lognormalSimulationOneSidedResults.small.pfail}
($p_{\fail}=0.01$ and $p_{\fail}=0.05$)
and~\ref{figure:lognormalSimulationOneSidedResults.large.pfail}
($p_{\fail}=0.10$ and $p_{\fail}=0.50$) summarize the lognormal
distribution simulation results with one-sided error probabilities.
Similar to the results in
Section~\ref{simulation.results.conclusions}, observations from
Figures~\ref{figure:lognormalSimulationTwoSidedResults.small.pfail}--\ref{figure:lognormalSimulationOneSidedResults.large.pfail} are
\begin{itemize}
\item
With a few exceptions the IJ priors, when compared to flat, tend to result in
coverage probabilities closer to the nominal values.
\item
Taking into account MC error, the coverage probabilities for the IJ
and flat priors are close to the 0.95 nominal credible level for
$\E(r) \geq 35$.
\end{itemize}

\section{A Careful Look at Examples with Three Failures}
\label{S.section:careful.look.three.failures}
This section looks carefully at Weibull and lognormal Bayesian
estimation results for two data sets with only three failures under
three different noninformative priors. The two data sets were chosen
based on the two extreme simulation factor-level combinations
described in Section~\ref{section:simulation.factor.levels} (sample
sizes 20 and 1000).  

In some examples, the Bayesian estimate of $F(t)$ does not appear to
agree well with the nonparametric estimates. With \typeI{}
censoring, to a very high degree of approximation, the Weibull ML estimate
at the censoring time is $\Fhat(t_{c}) \approx r/n$
\citep[as shown in][]{Escobar2010}.
Thus there is an \textit{invisible pseudo data point} at
$(t_{c}, r/n)$ and this point is indicated in the probability plots
in this section with the symbol $\msquare$. This invisible pseudo data point
has more influence than the other visible data points (smaller order
statistics have more variability).
When the Bayesian estimate of $F(t)$ does not agree well with
the points, this invisible pseudo data point will help explain why. We also
did experiments to assure that our conclusions in the section are
not sensitive to the location of the three failure times before the
censoring time (they are not).

The different point colors in the contour plots correspond to the
four different MCMC chains that were used in generating the points.
Recall that, as described in
Section~\ref{S.section:implementation.using.bayes.without.tears}
draws from an IJ/CJ priors are generated by using Stan.
Clusters of one color would indicate problems with the NUTS sampler, 
but we do not see any such problems.

\subsection{Example of Three Failures from a Sample of Size of Twenty}
\begin{figure}[!ht]
\begin{tabular}{cc}
\rsplidafiguresize{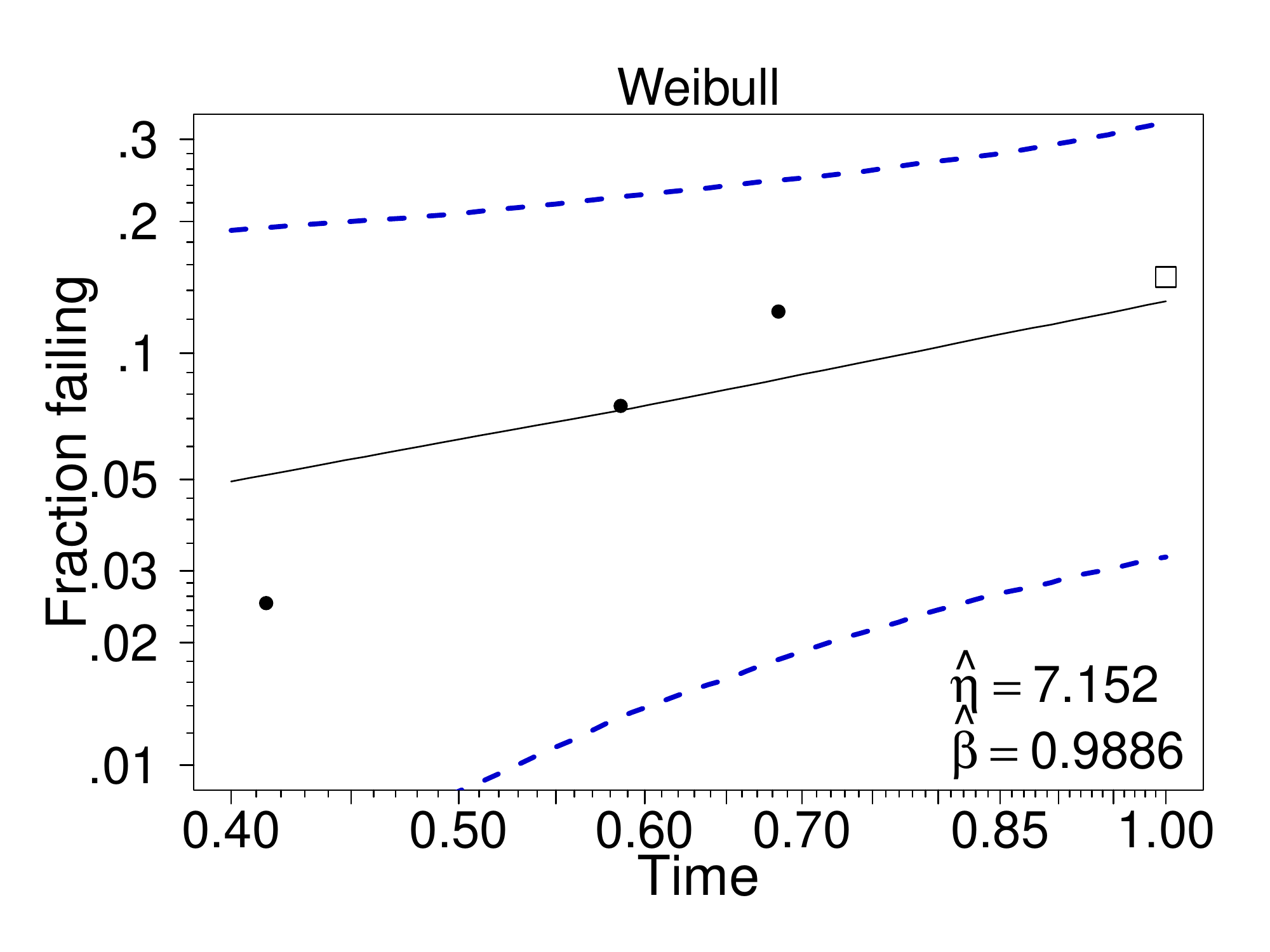}{3.5in}&
\rsplidafiguresize{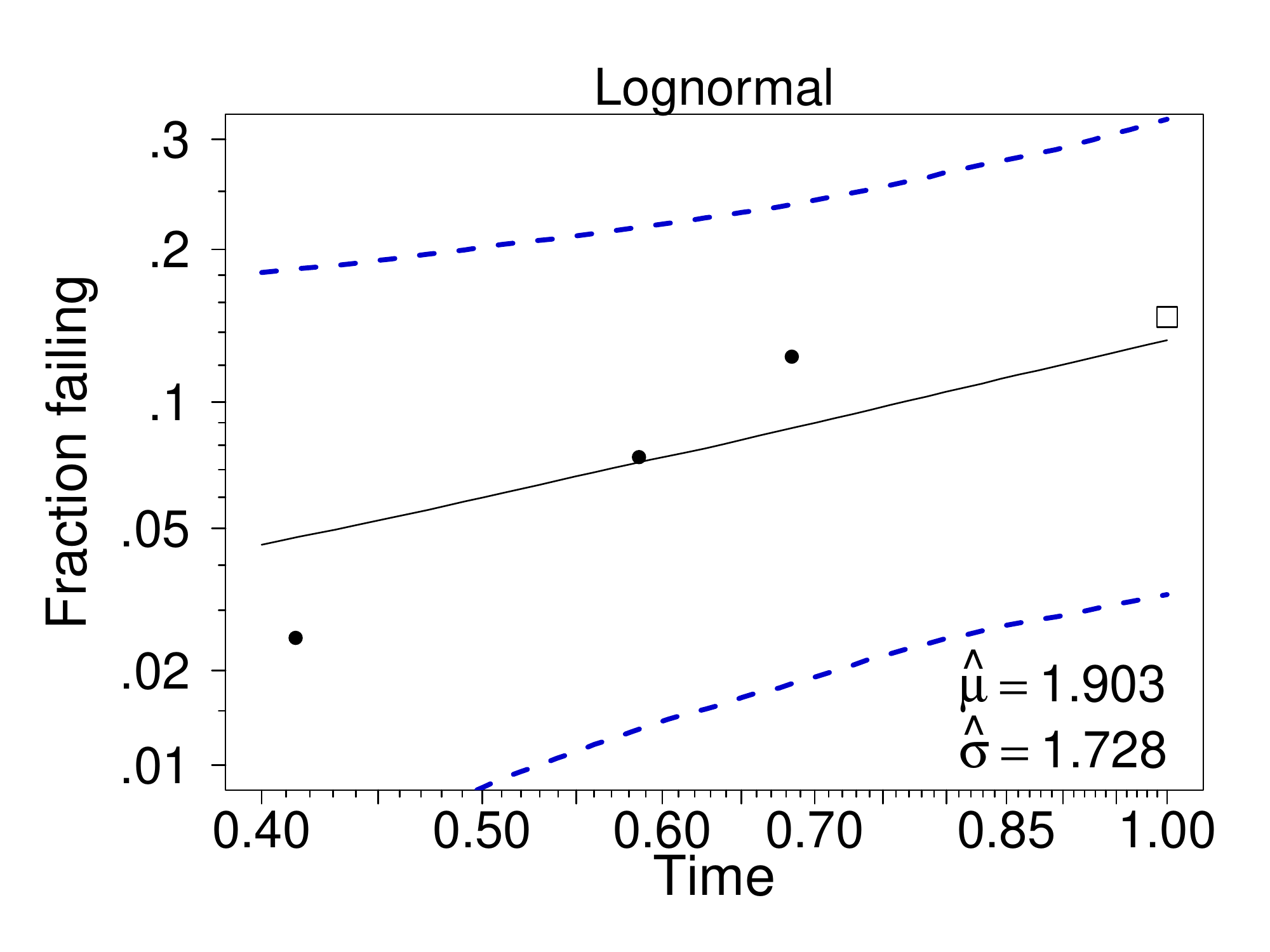}{3.5in}\\
\rsplidafiguresize{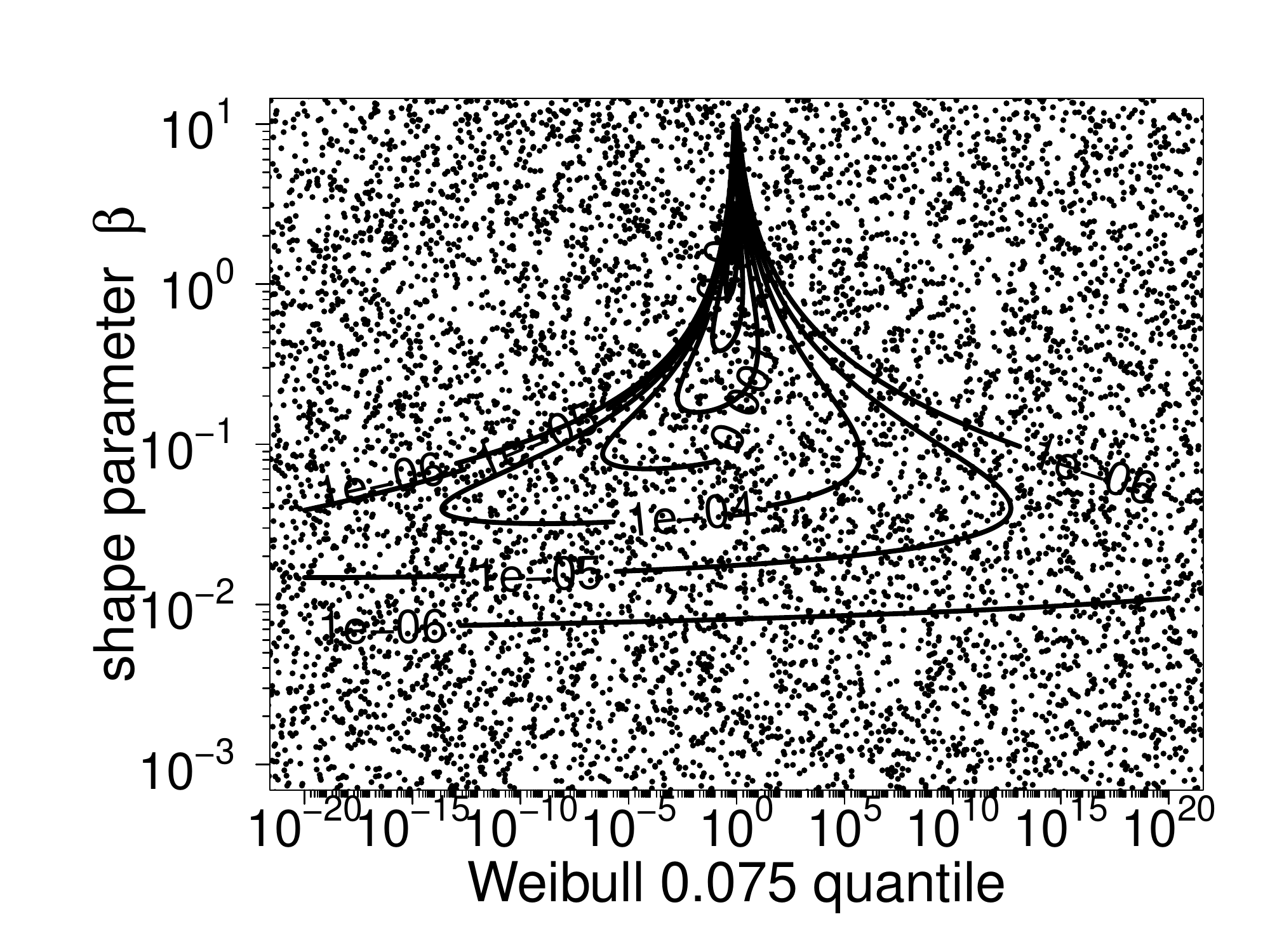}{3.5in}&
\rsplidafiguresize{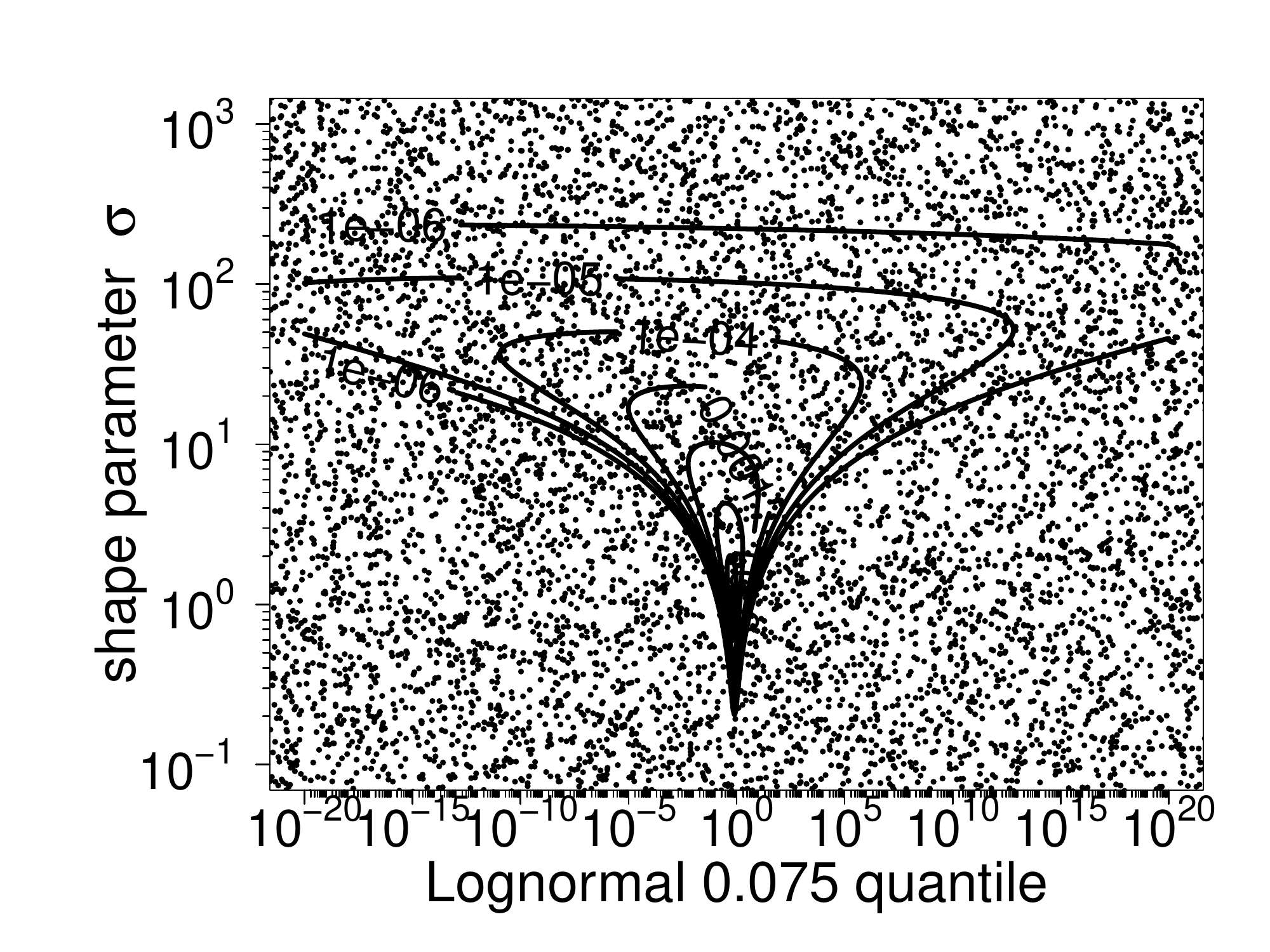}{3.5in}\\
\rsplidafiguresize{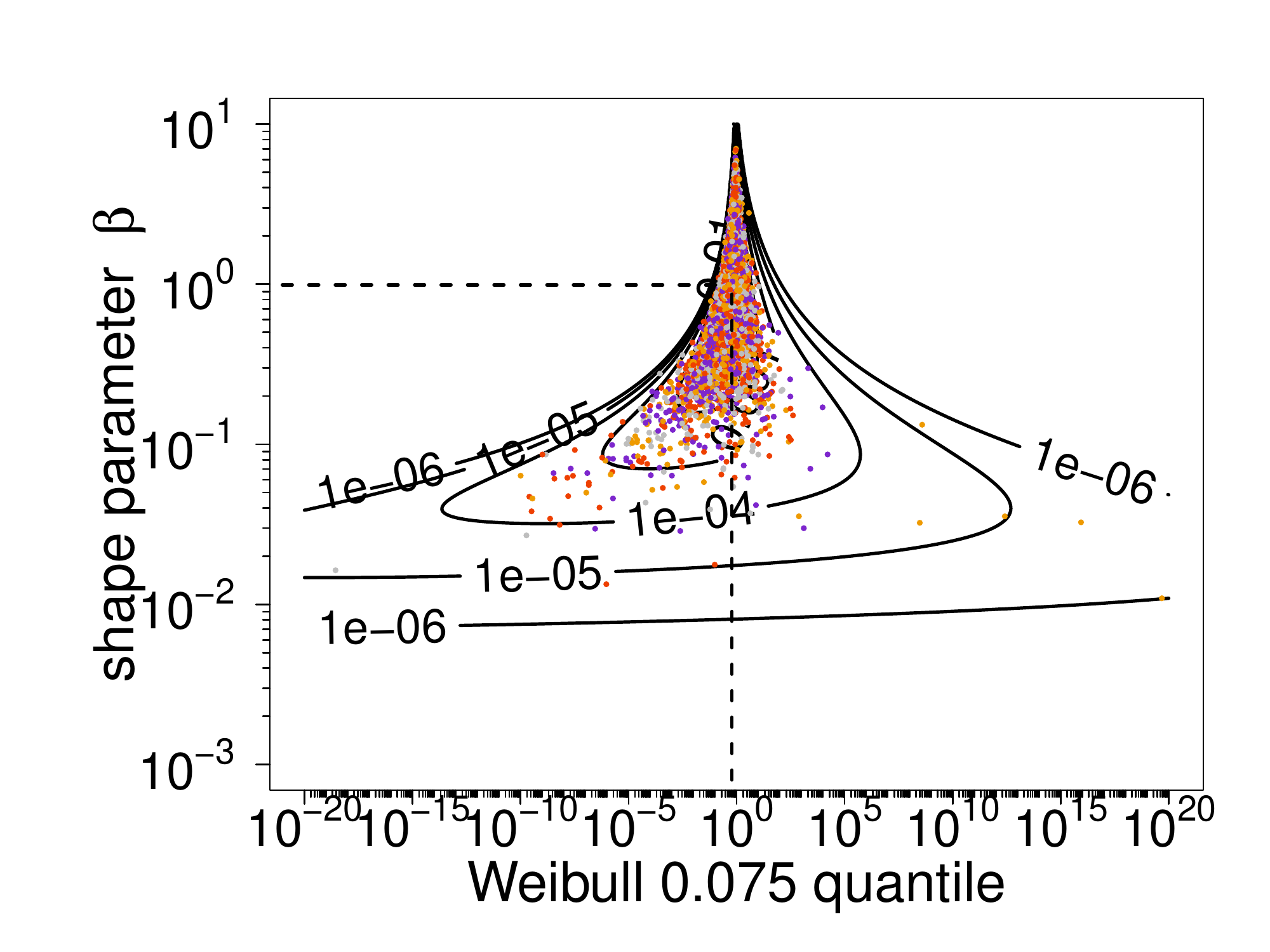}{3.5in}&
\rsplidafiguresize{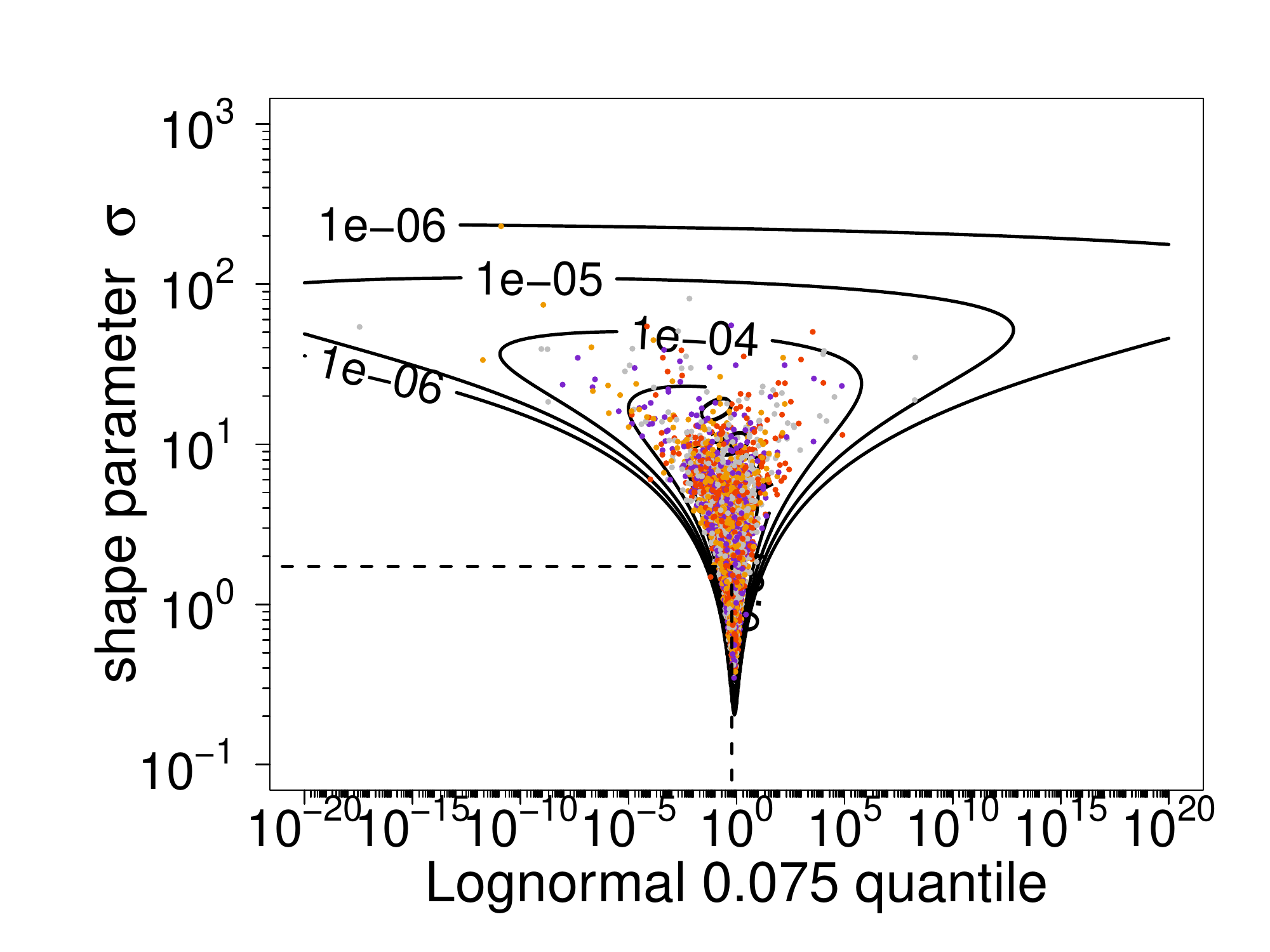}{3.5in}\\
\end{tabular}
\caption{Bayesian estimation results for data with $r=3$ and $n=20$
  using the Weibull distribution (on the left) and the lognormal
  distribution (on the right) showing an estimate of $F(t)$ on a
  probability plot (top), draws from the bounded joint prior and likelihood contours
  (middle), and posterior draws and likelihood contours (bottom) for
  a flat prior.}
\label{S.figure:small.sample.flat}
\end{figure}

\begin{figure}[!ht]
\begin{tabular}{cc}
\rsplidafiguresize{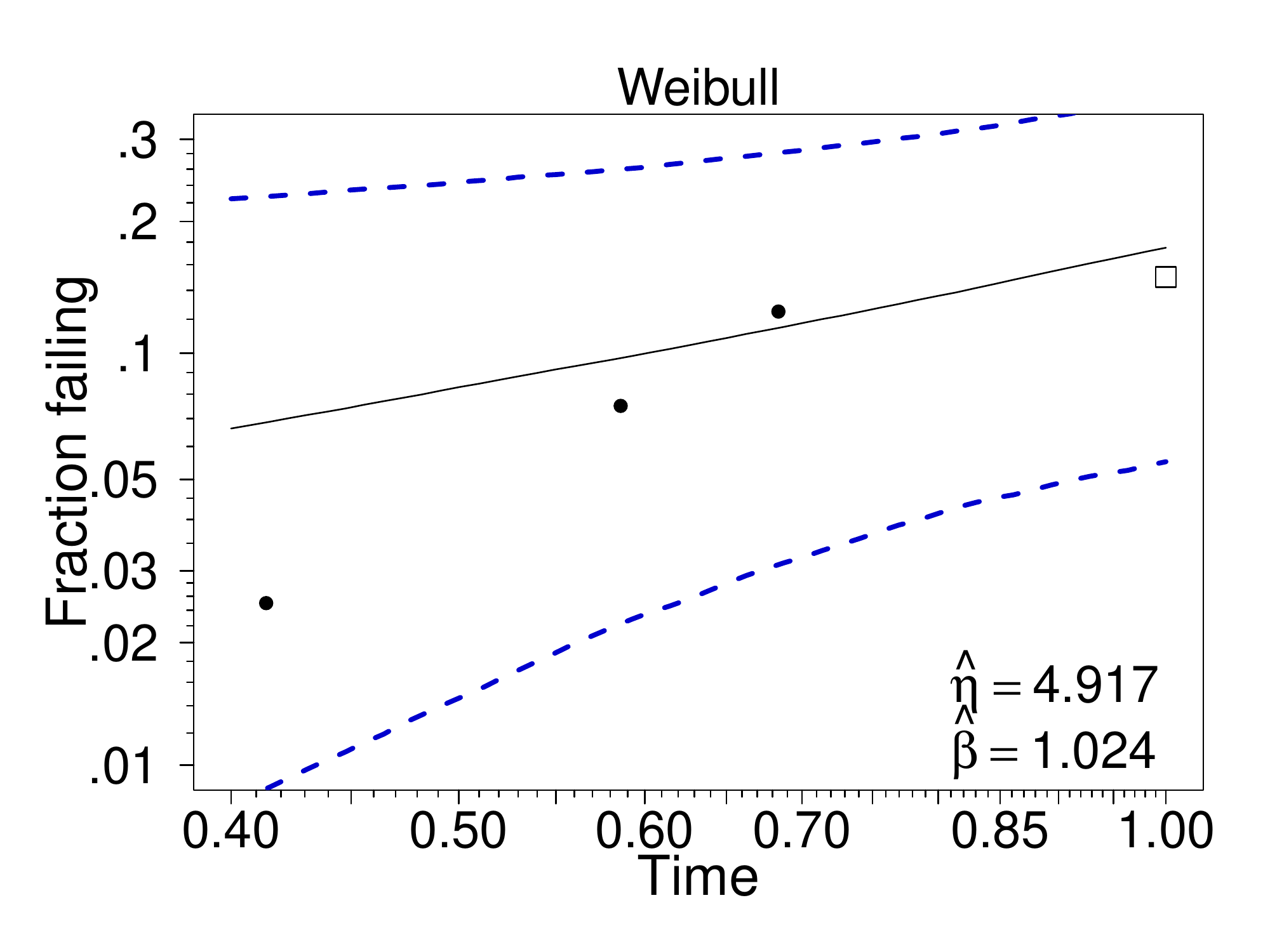}{3.5in}&
\rsplidafiguresize{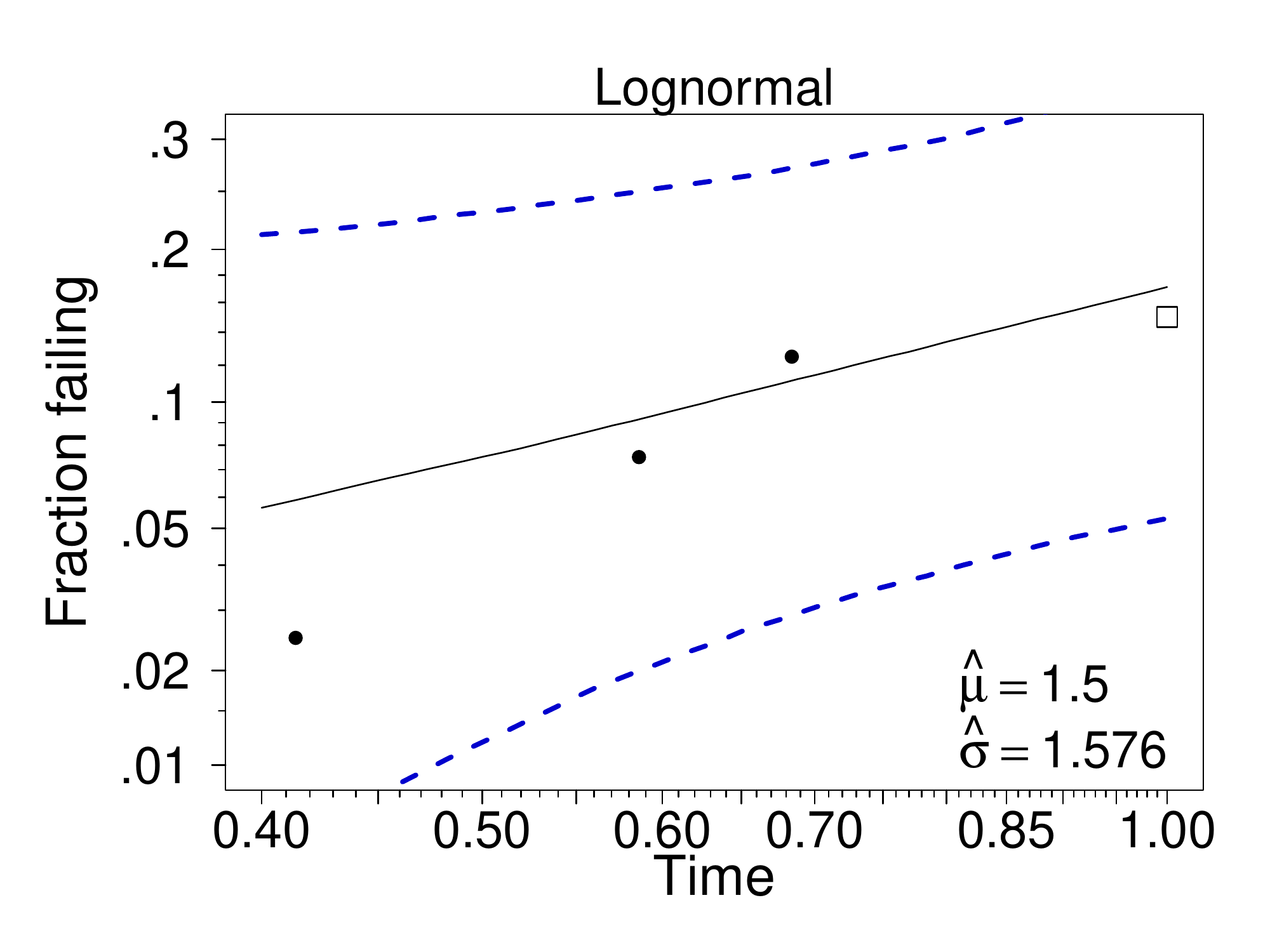}{3.5in}\\
\rsplidafiguresize{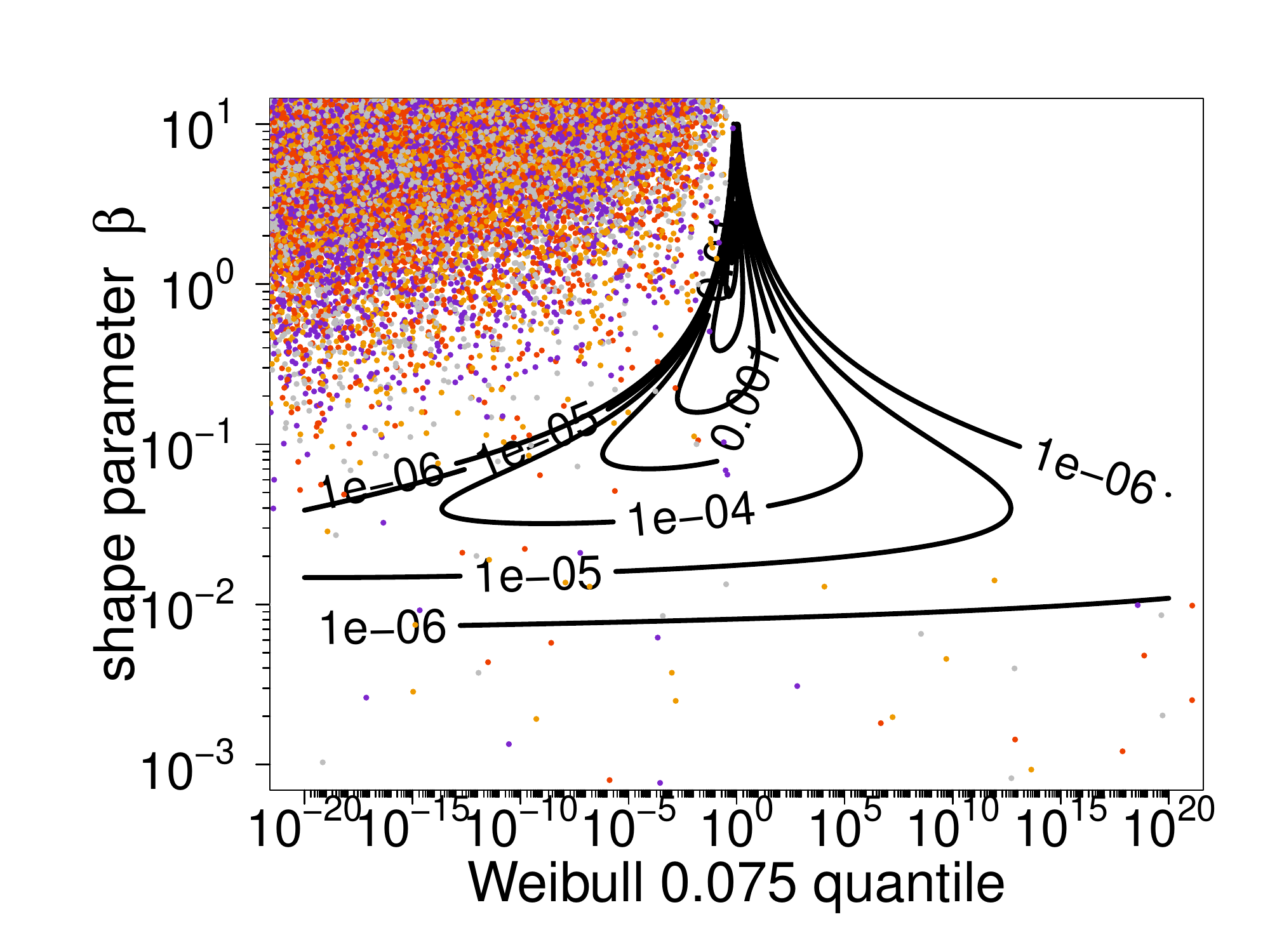}{3.5in}&
\rsplidafiguresize{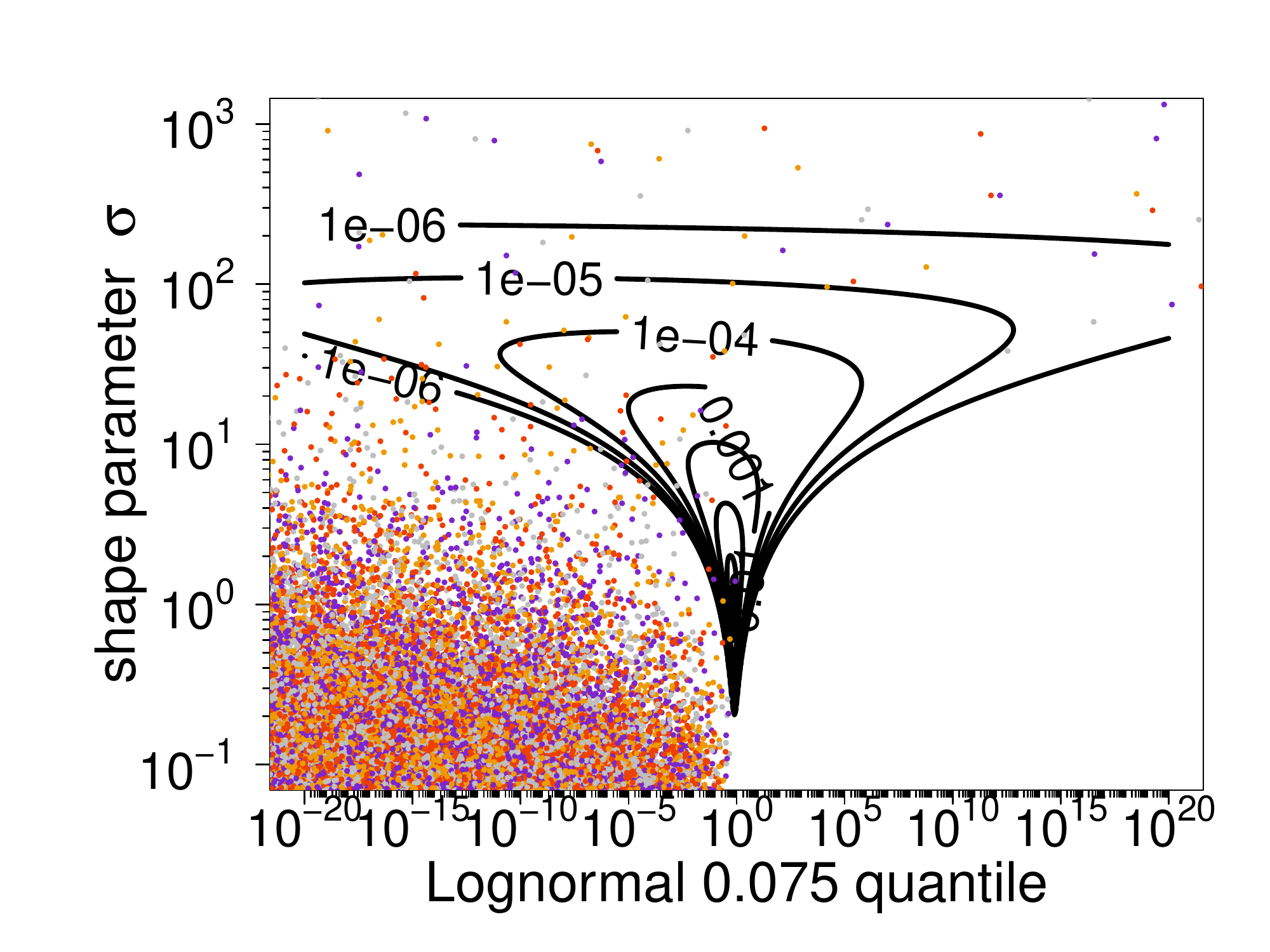}{3.5in}\\
\rsplidafiguresize{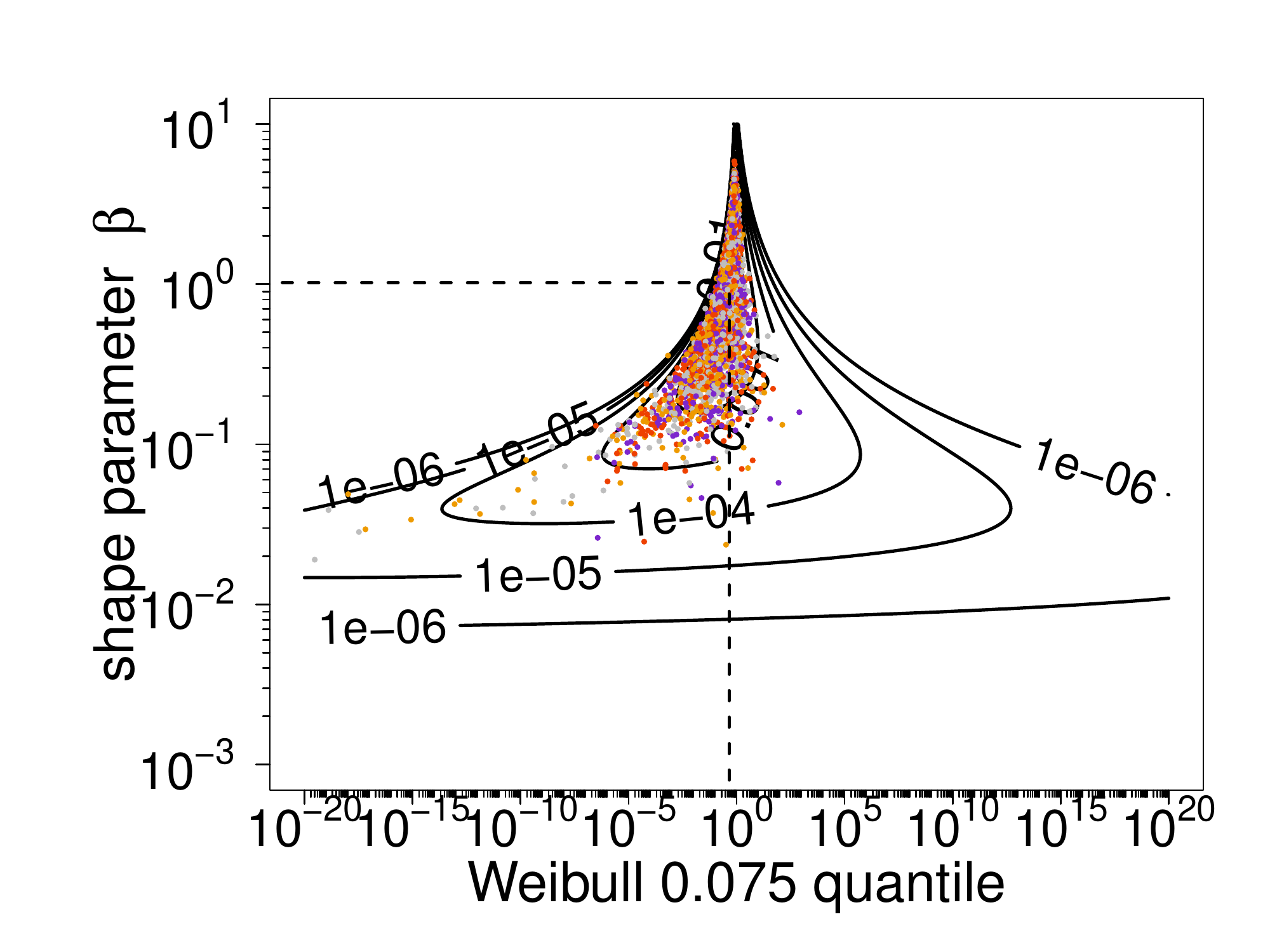}{3.5in}&
\rsplidafiguresize{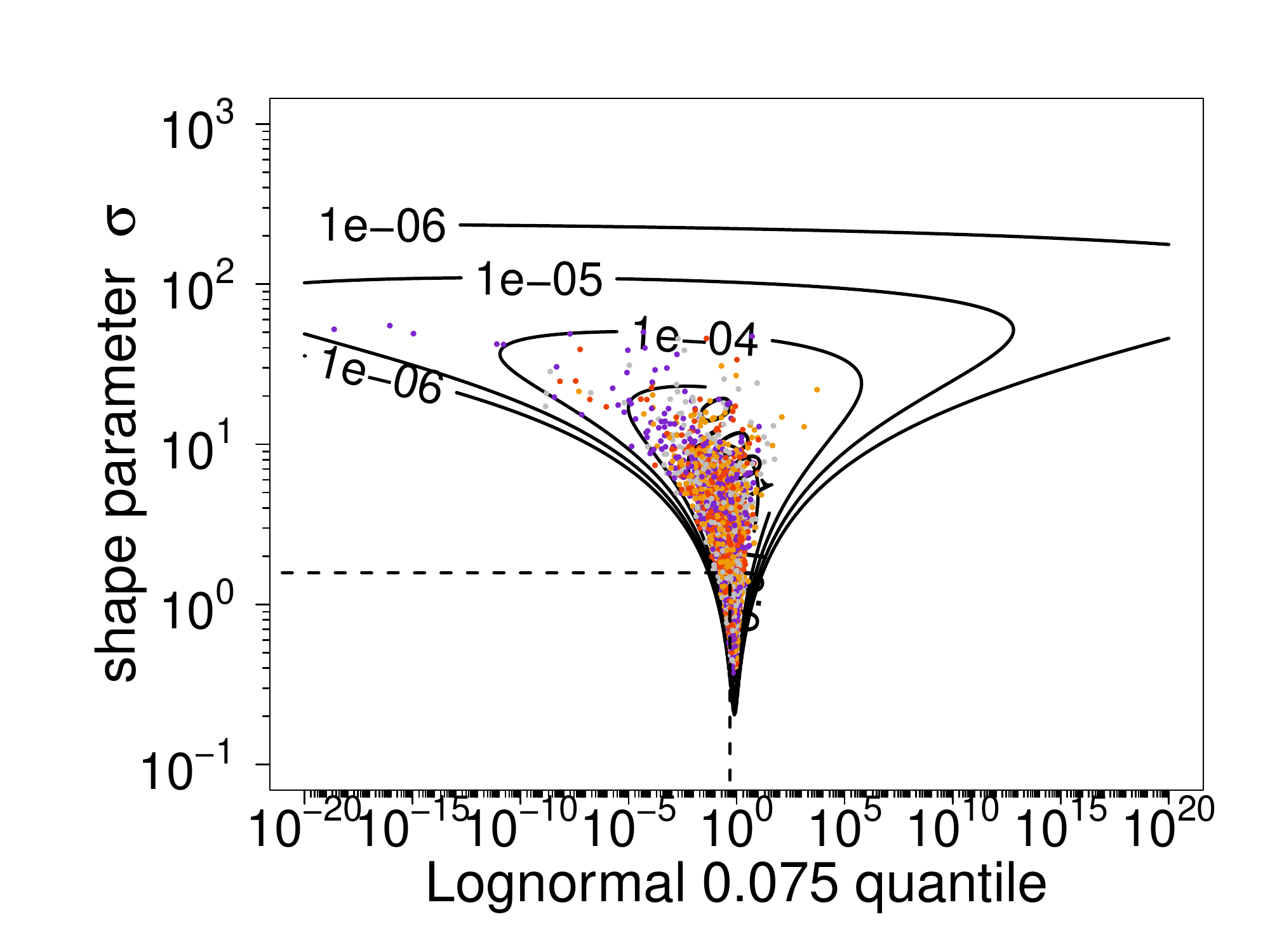}{3.5in}\\
\end{tabular}
\caption{Bayesian estimation results for data with $r=3$ and $n=20$
  using the Weibull distribution (on the left) and the lognormal
  distribution (on the right) showing an estimate of $F(t)$ on a
  probability plot (top), draws from the bounded joint prior  and likelihood contours
  (middle), and posterior draws and likelihood contours (bottom) for
  an IJ prior.  }
\label{S.figure:small.sample.IJ}
\end{figure}

The results in this section correspond to the factor-level
combination $\E\big( r \big)=10$ and $p_{\fail}=0.50$ resulting in a
sample of size $n=20$ and a censoring time of 1. To illustrate an
extreme case, a simulated sample resulting in three failures was
chosen. The failures were at times 0.414, 0.586, and
0.684. The value of $p_{r}$ for reparameterization was
chosen to be $(3/20)/2=0.075$, resulting in a likelihood that is
reasonably well behaved (except for the funnel shape with a sharp point).
Figures~\ref{S.figure:small.sample.flat} and \ref{S.figure:small.sample.IJ}
give results for flat and IJ priors, respectively. It is interesting to
see how the priors and likelihood functions combine to produce the
posterior distribution and we can see the difference between the
posterior resulting from the flat and IJ priors.

\subsection{Example of Three Failures from a Sample of Size of One
  Thousand} 
\label{S.section:ThousandTestThreeFail}
\begin{figure}[!ht]
\begin{tabular}{cc}
\rsplidafiguresize{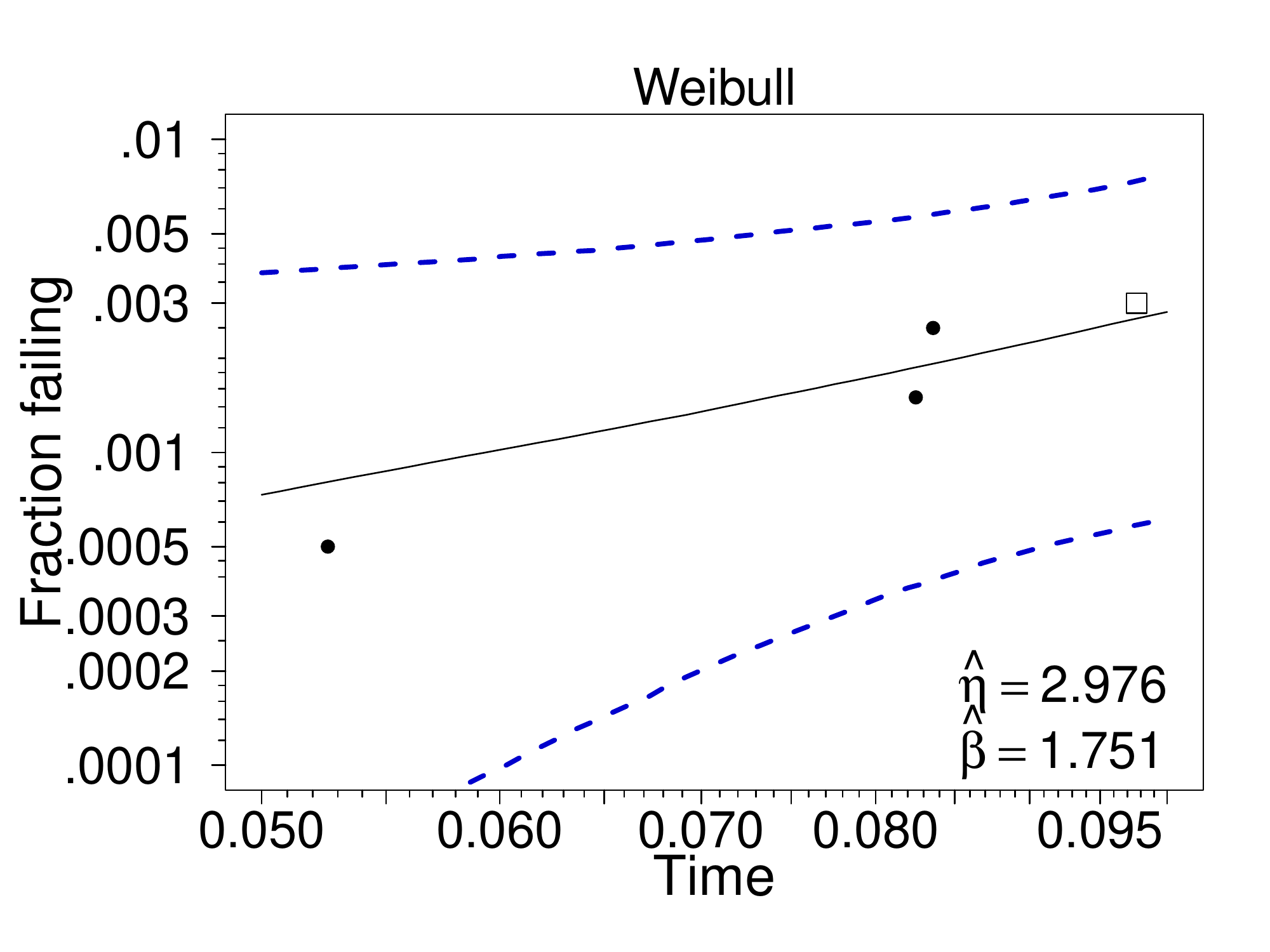}{3.5in}&
\rsplidafiguresize{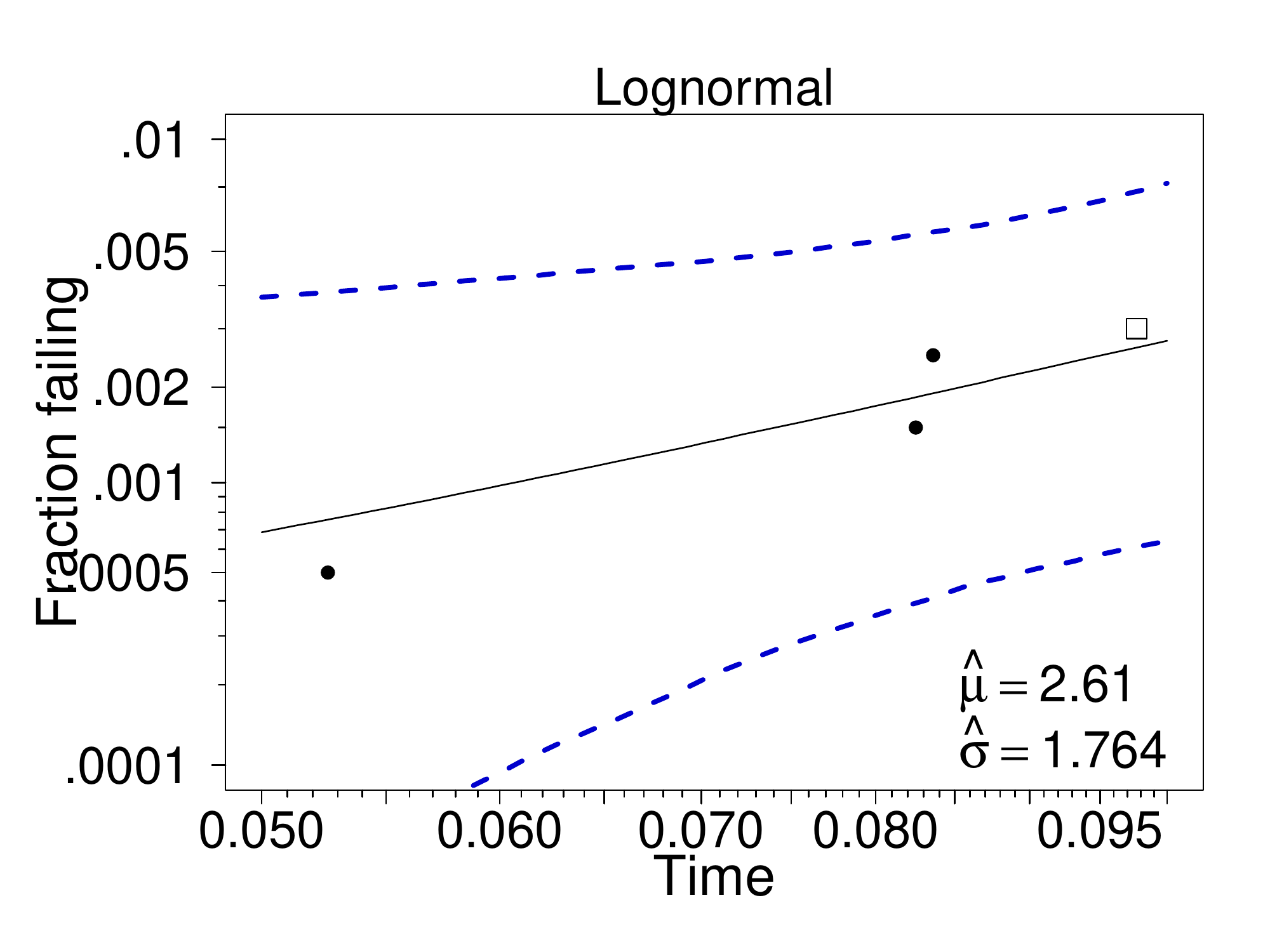}{3.5in}\\
\rsplidafiguresize{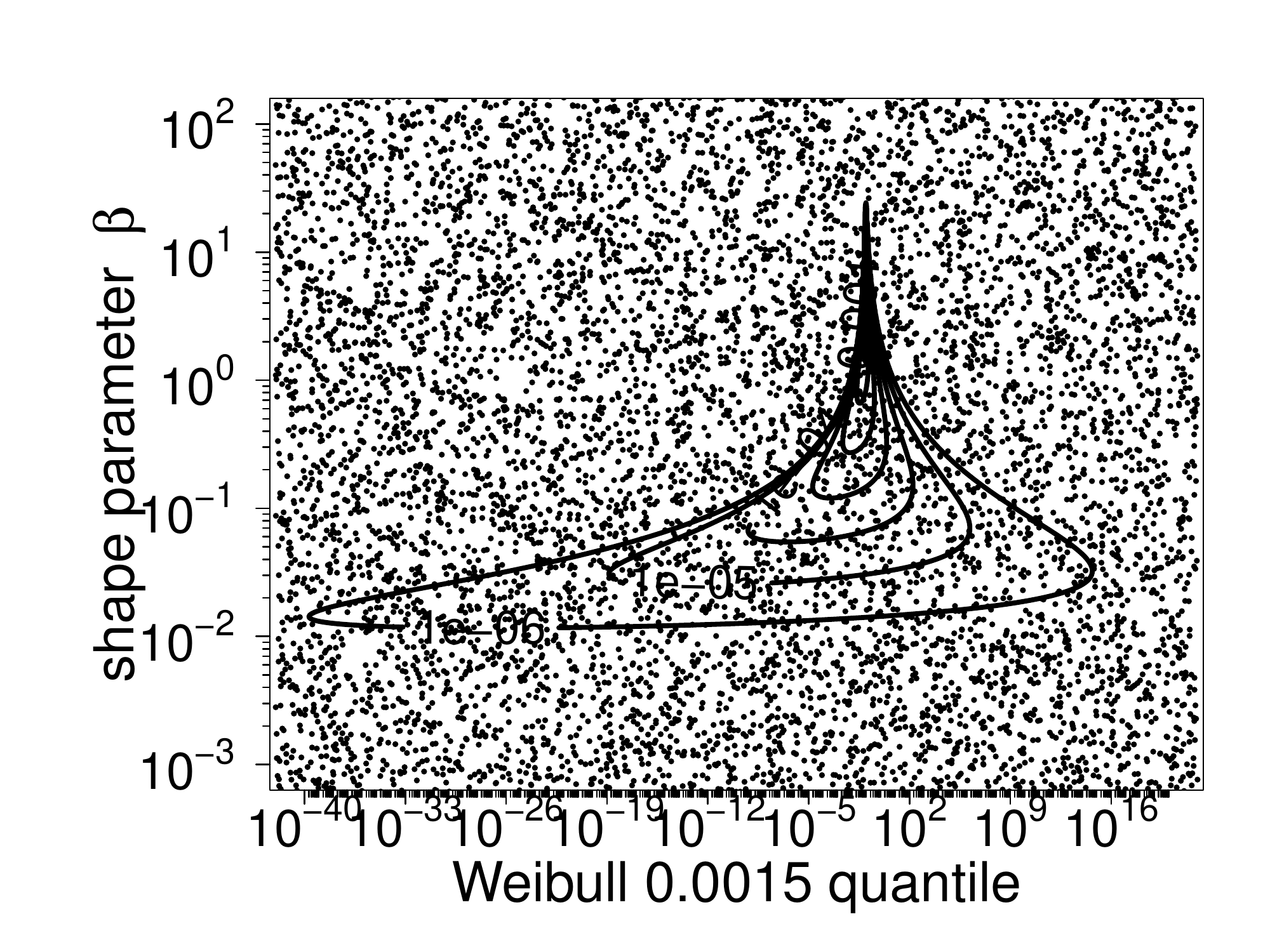}{3.5in}&
\rsplidafiguresize{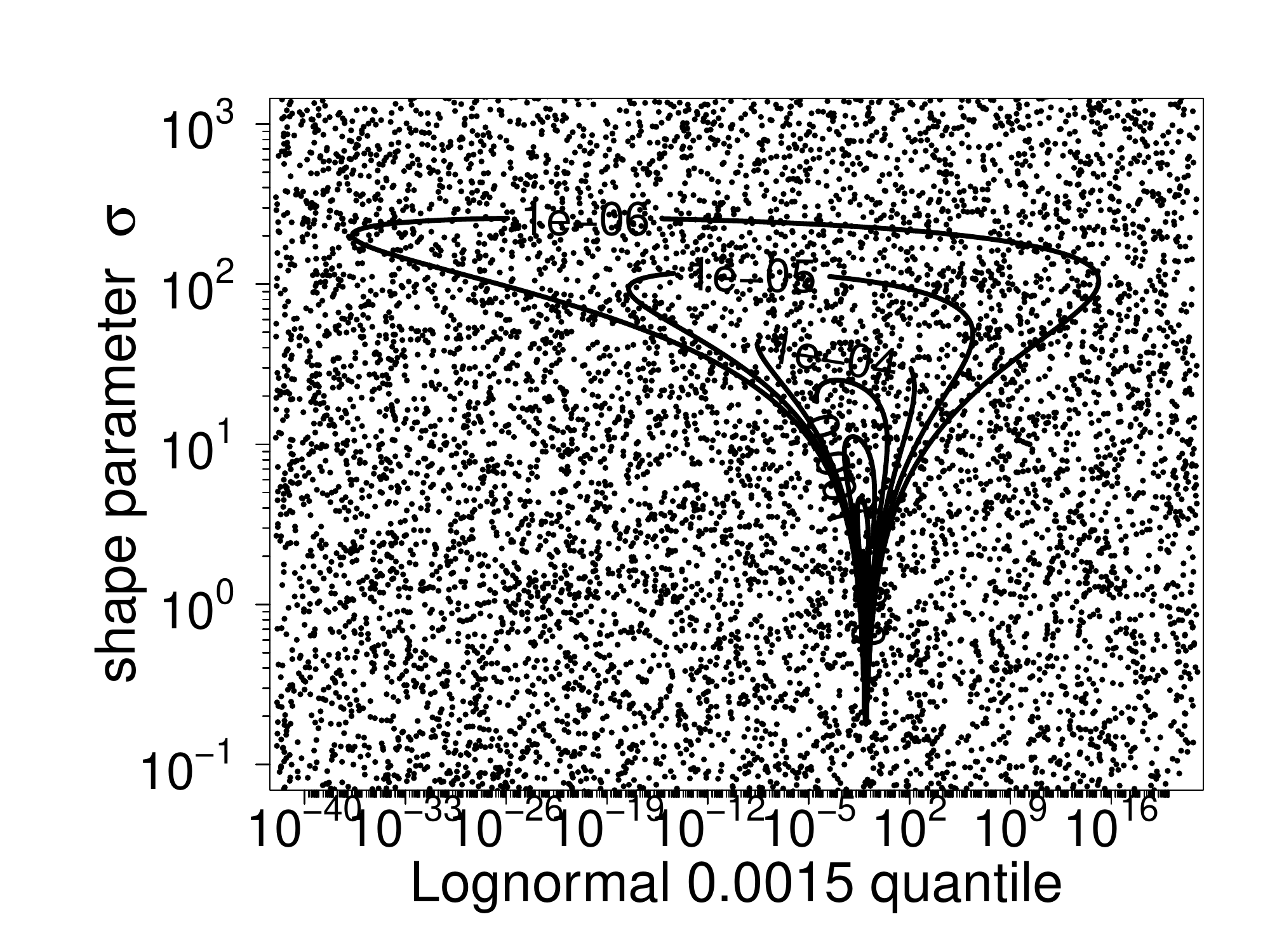}{3.5in}\\
\rsplidafiguresize{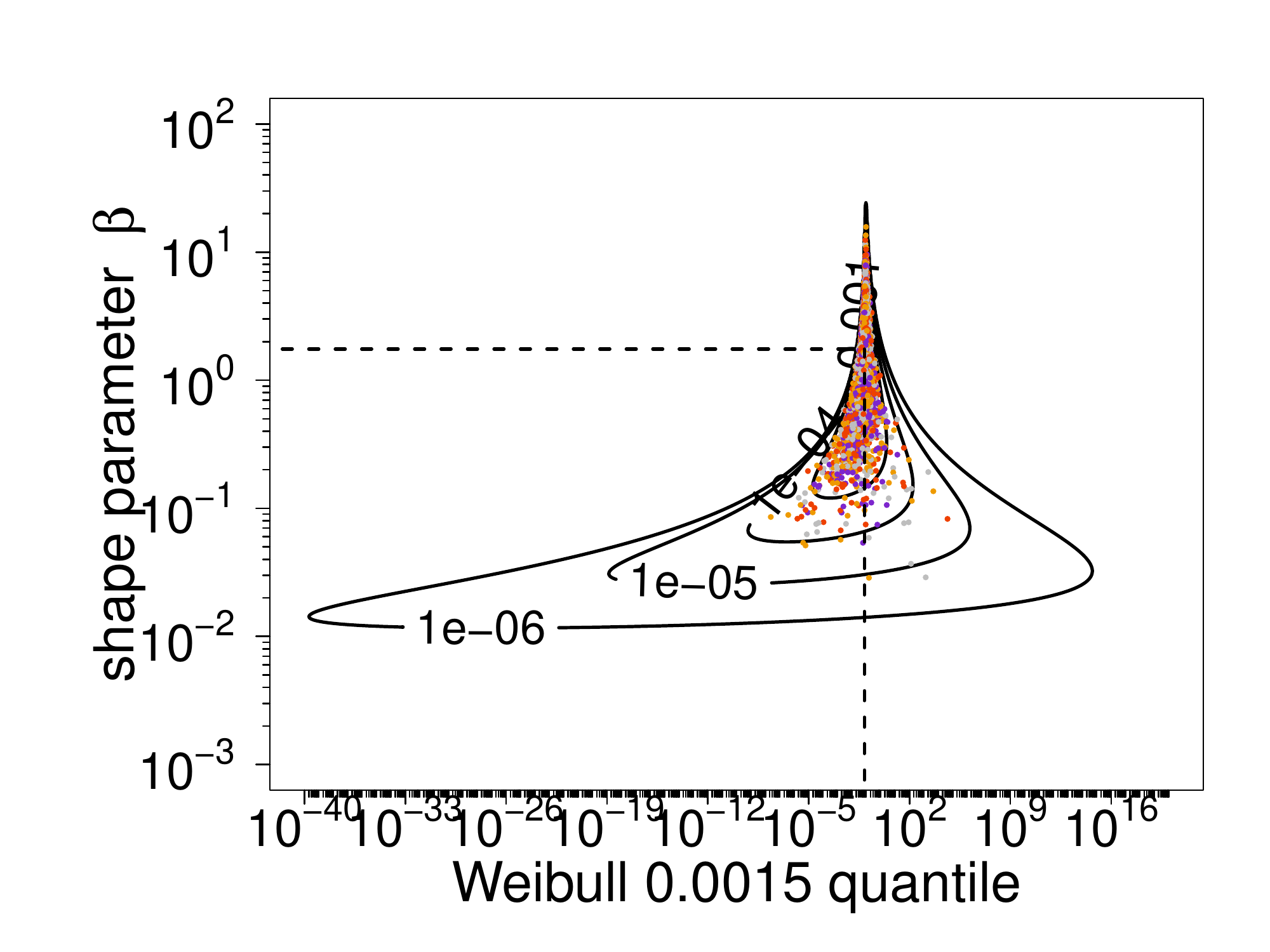}{3.5in}&
\rsplidafiguresize{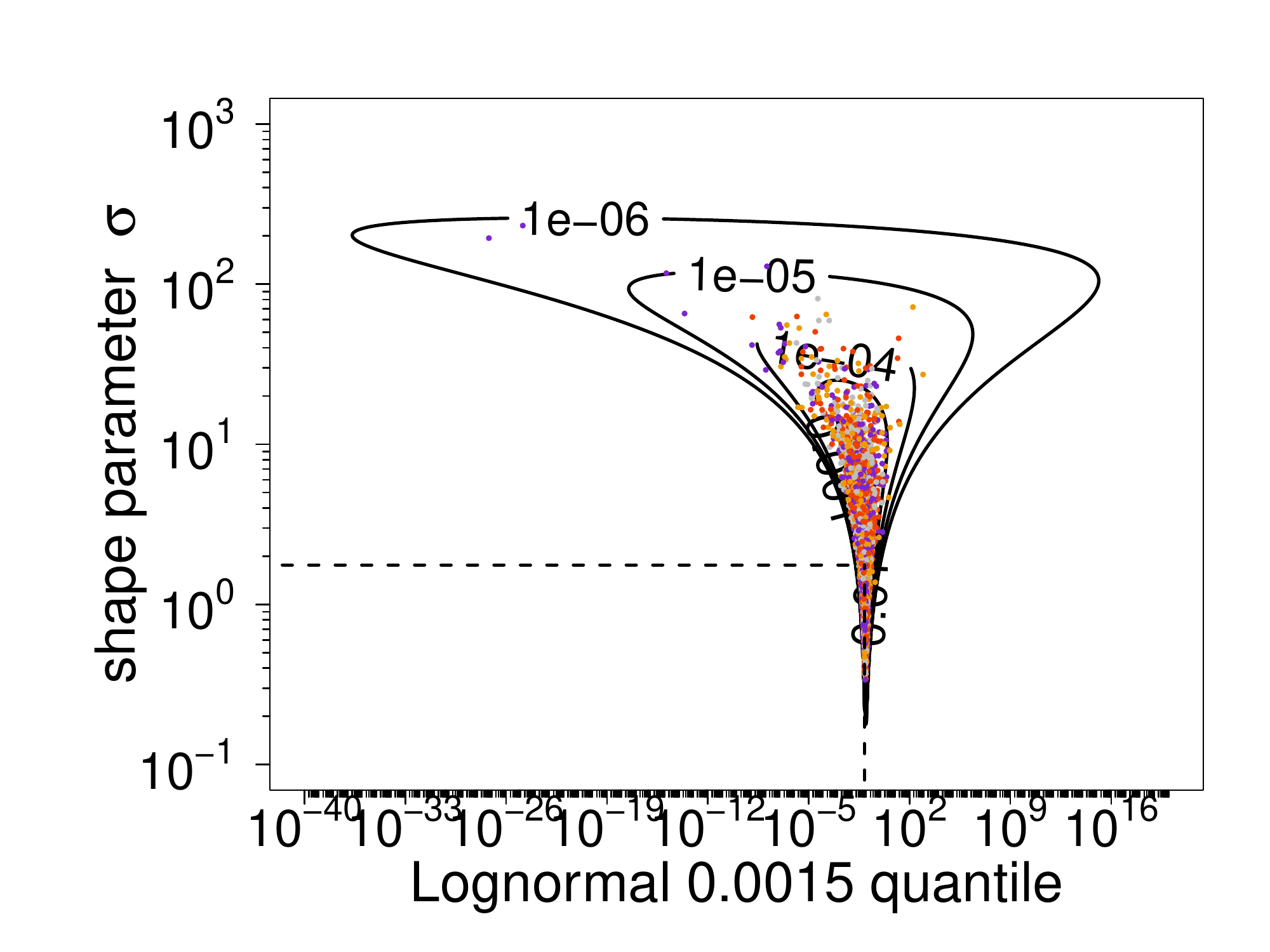}{3.5in}\\
\end{tabular}
\caption{Bayesian estimation results for data with $r=3$ and $n=1000$
  using the Weibull distribution (on the left) and the lognormal
  distribution (on the right) showing an estimate of $F(t)$ on a
  probability plot (top), draws from the bounded joint prior and likelihood contours
  (middle), and posterior draws and likelihood contours (bottom) for
  a flat prior.  }
\label{S.figure:large.sample.flat}
\end{figure}

\begin{figure}[!ht]
\begin{tabular}{cc}
\rsplidafiguresize{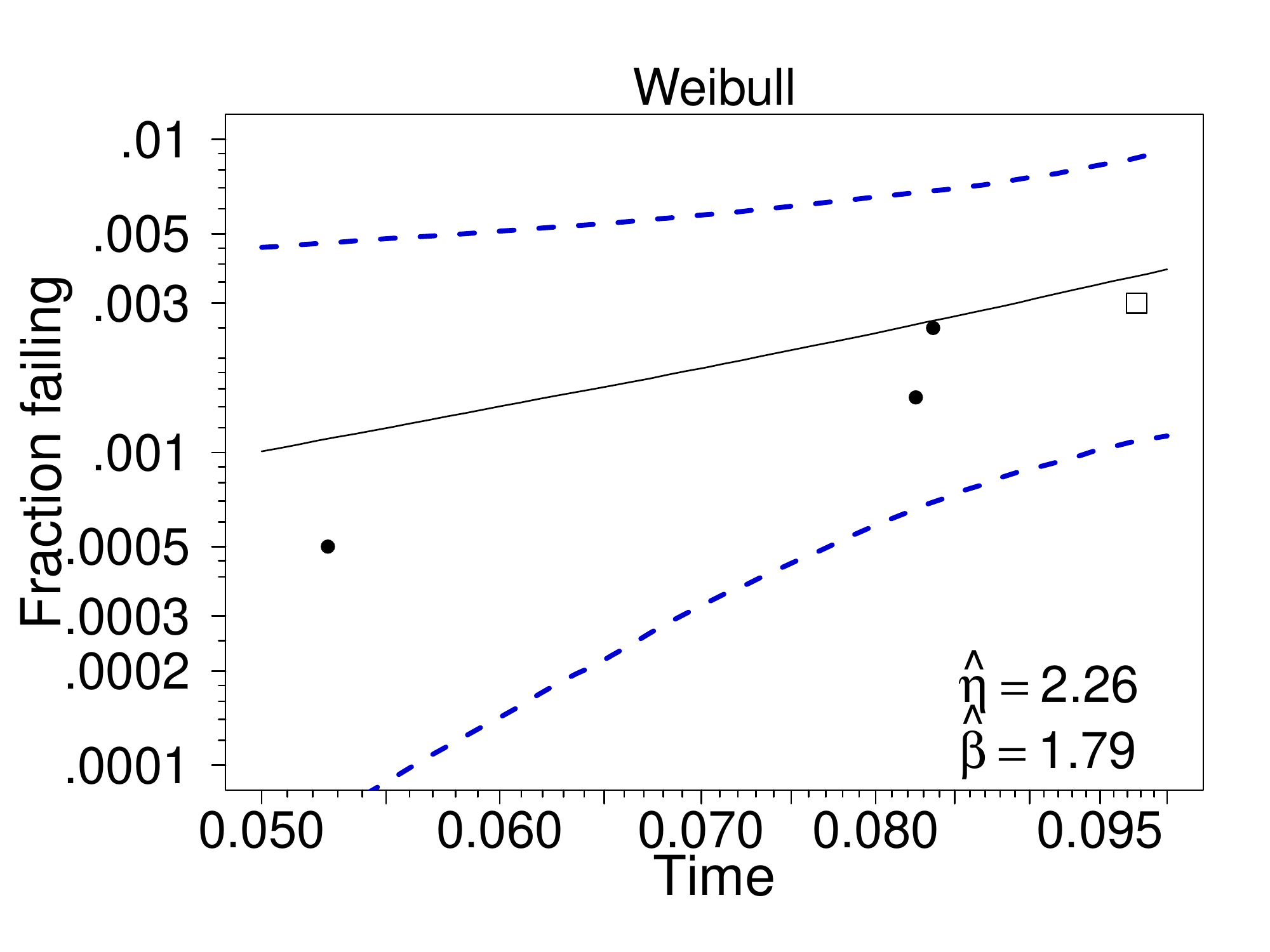}{3.5in}&
\rsplidafiguresize{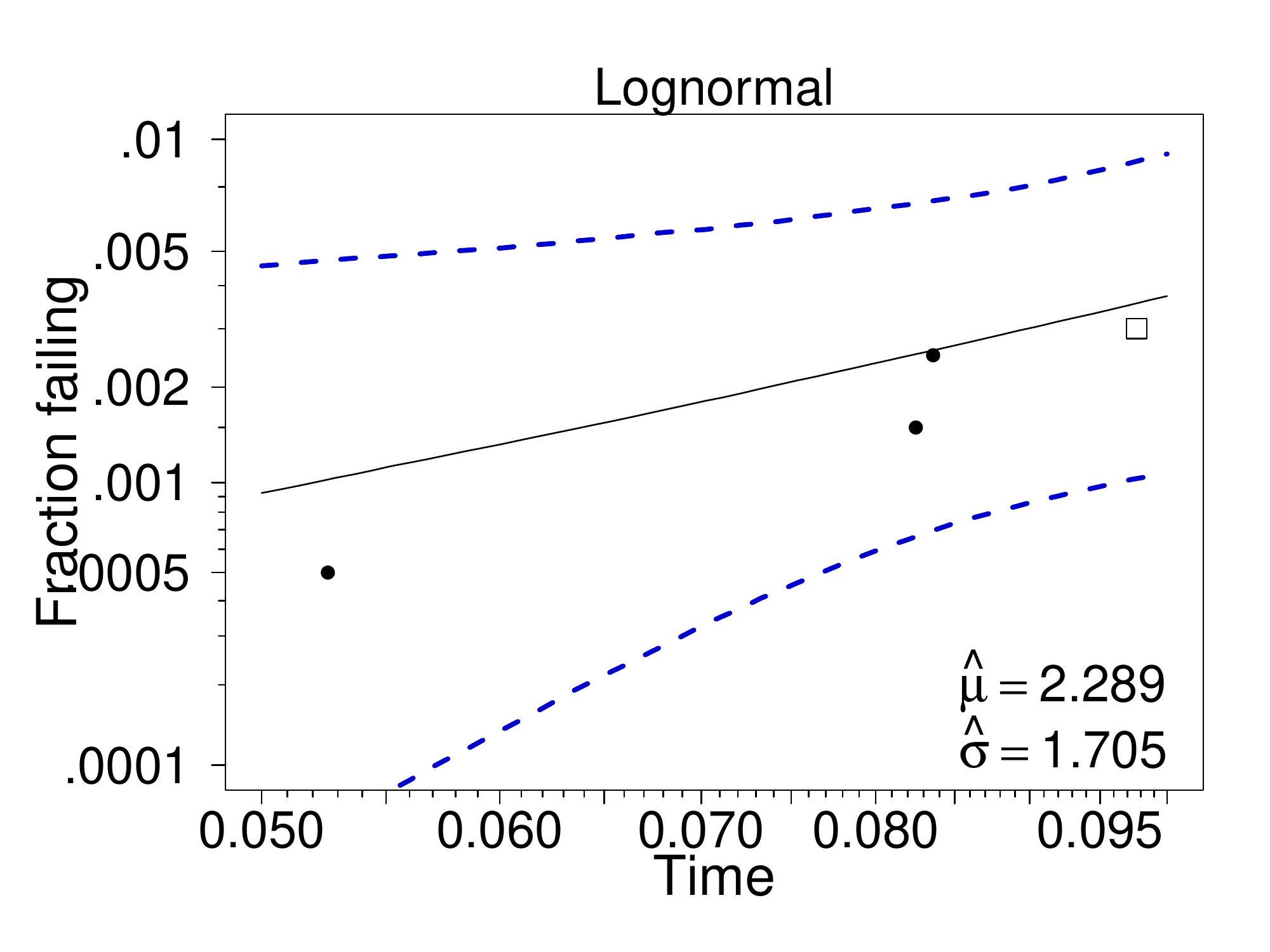}{3.5in}\\
\rsplidafiguresize{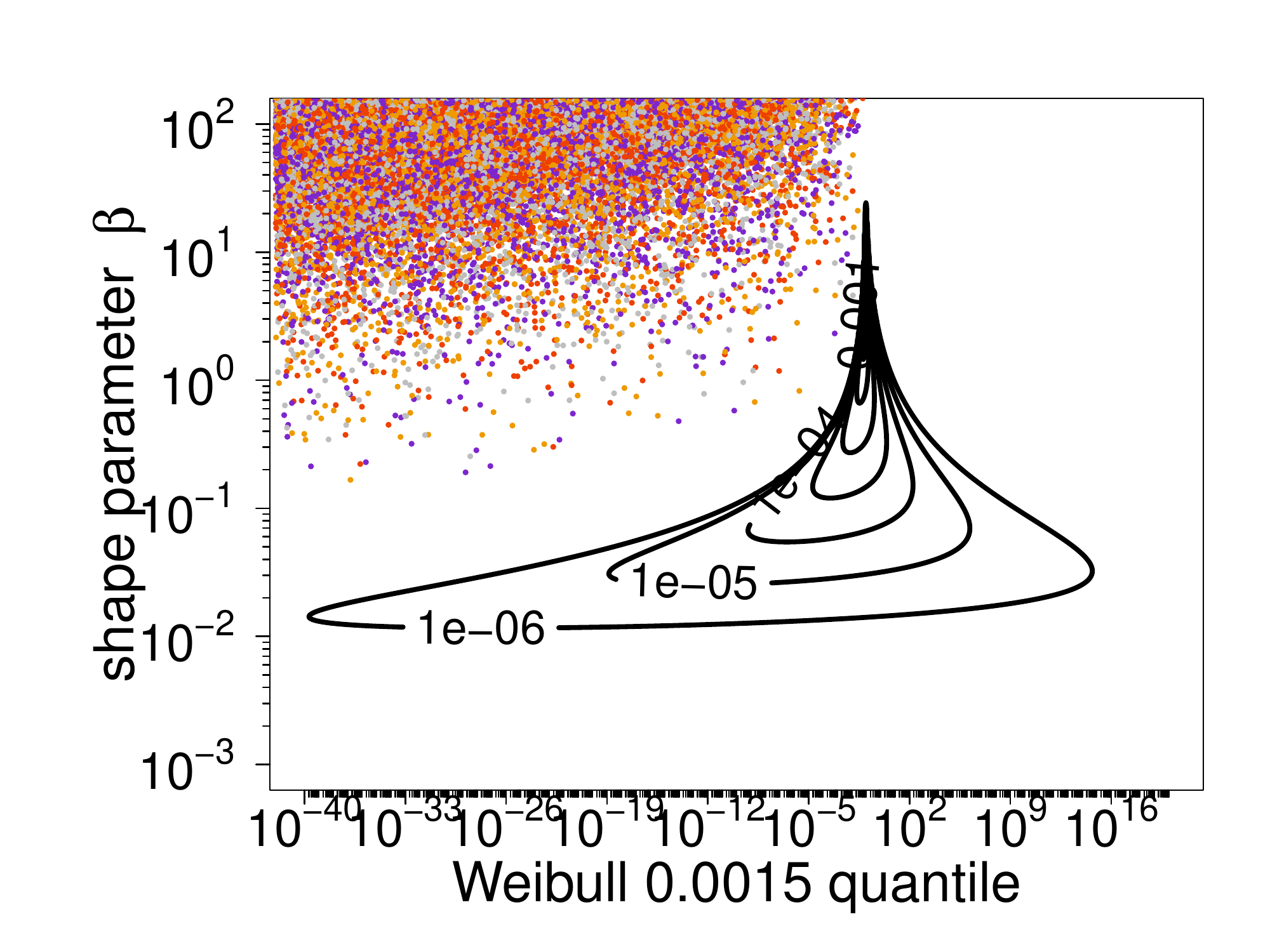}{3.5in}&
\rsplidafiguresize{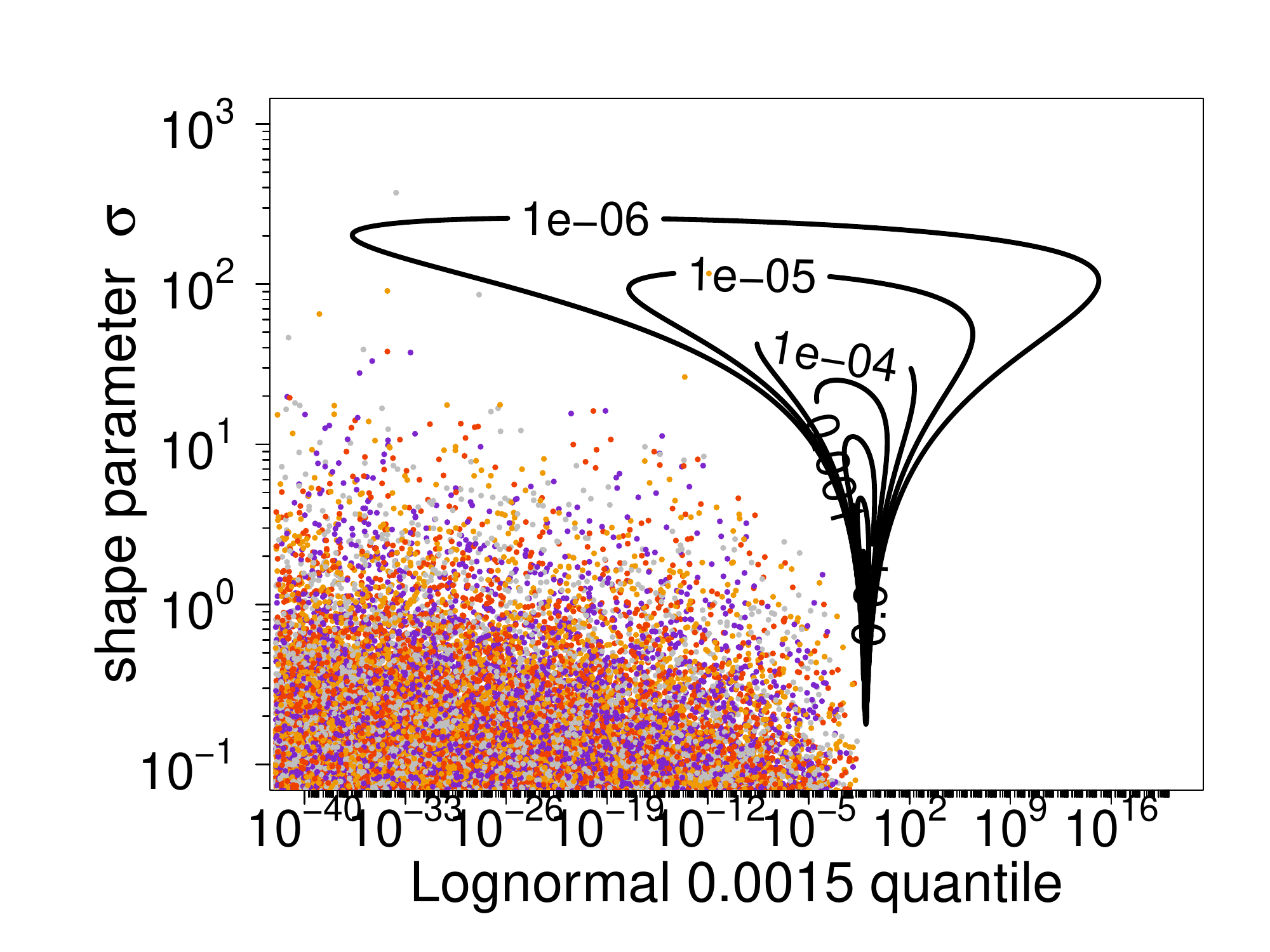}{3.5in}\\
\rsplidafiguresize{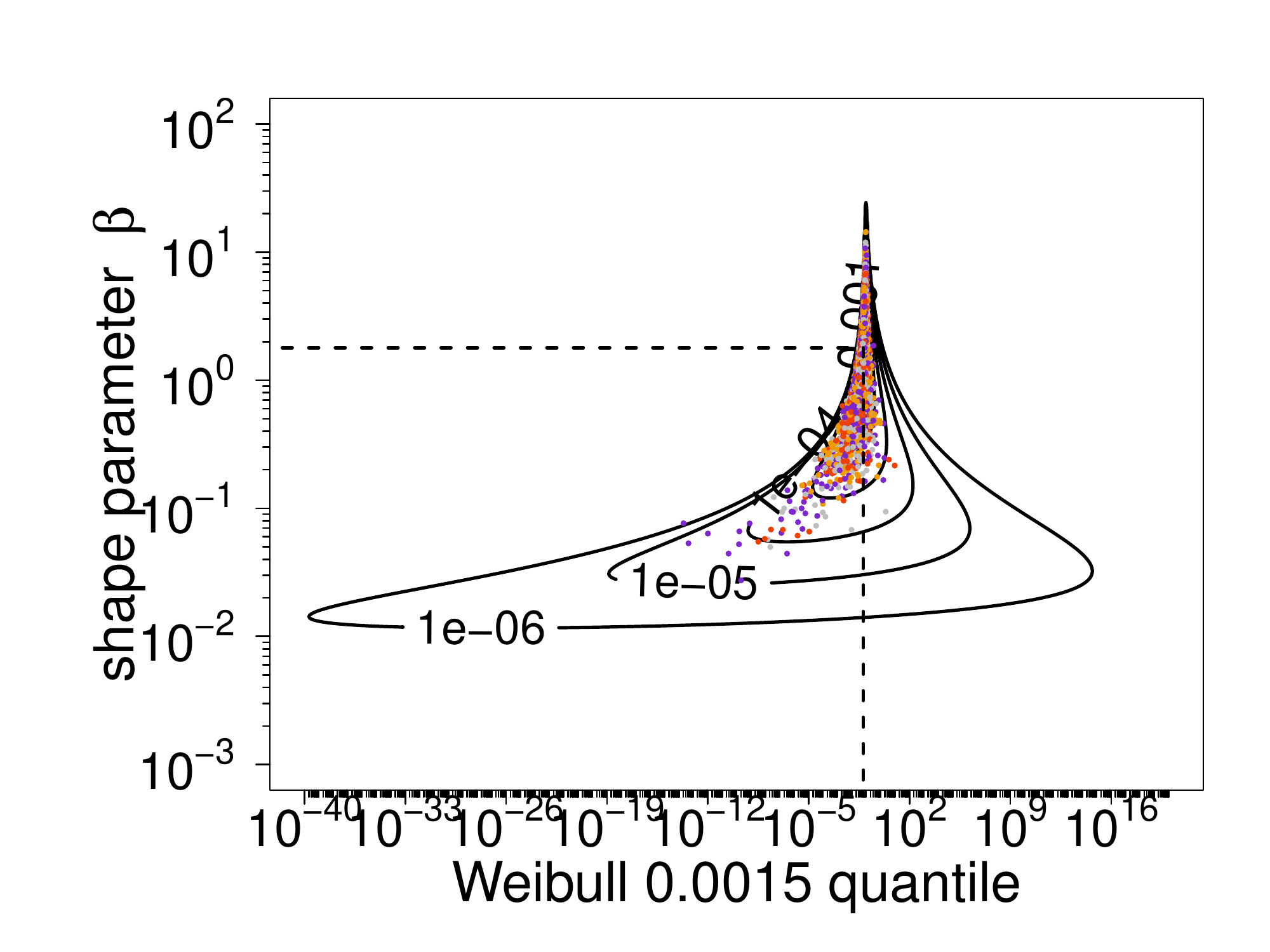}{3.5in}&
\rsplidafiguresize{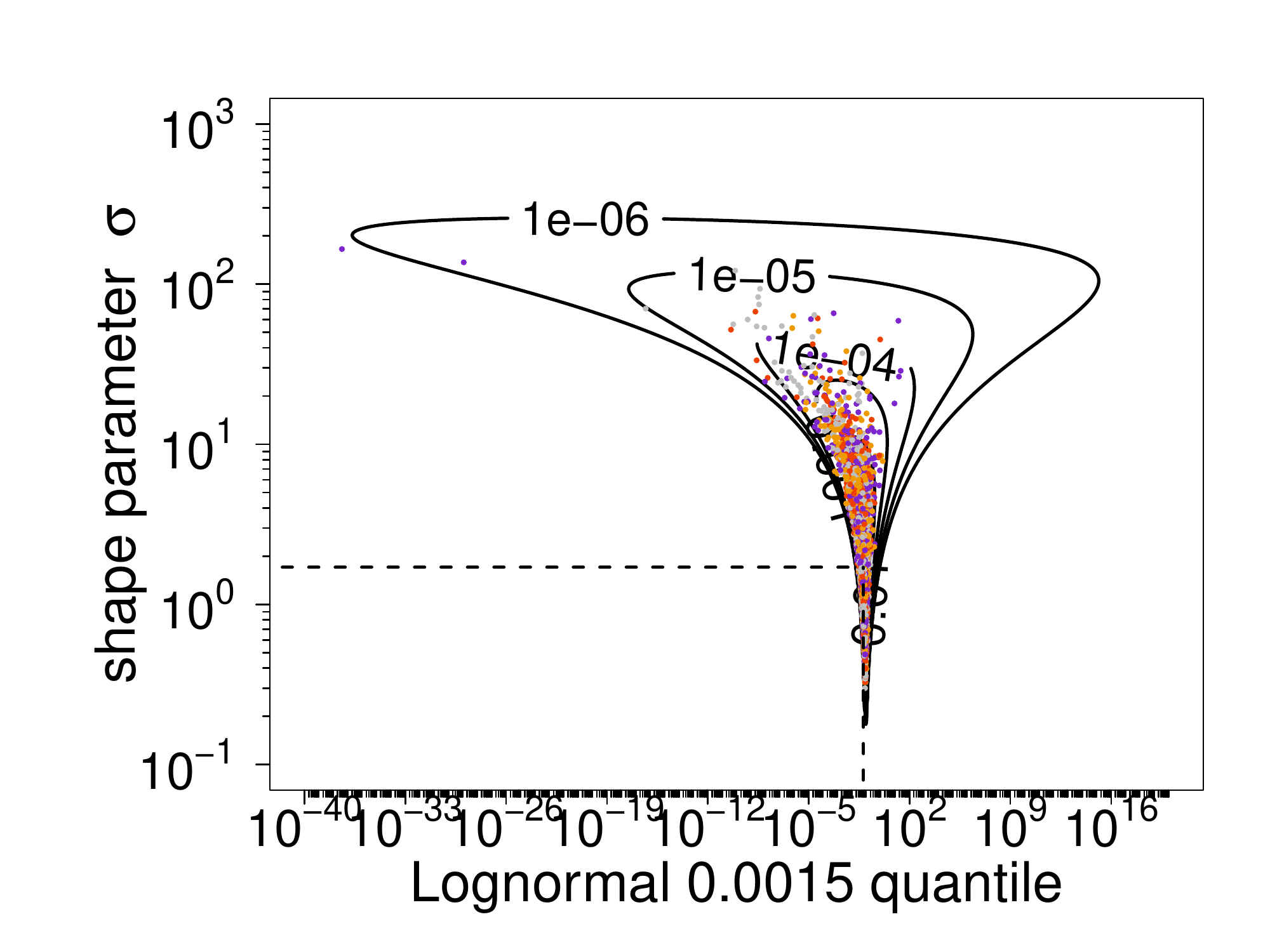}{3.5in}\\
\end{tabular}
\caption{Bayesian estimation results for data with $r=3$ and $n=1000$
  using the Weibull distribution (on the left) and the lognormal
  distribution (on the right) showing an estimate of $F(t)$ on a
  probability plot (top), draws from the bounded joint prior and likelihood contours
  (middle), and posterior draws and likelihood contours (bottom) for
  an IJ prior.  }
\label{S.figure:large.sample.IJ}
\end{figure}

The results in this section correspond to the
factor-level combination $\E\big( r \big)=10$ and $p_{\fail}=0.01$
resulting in a sample of size $n=1000$ and a censoring time of
0.0977. To illustrate an extreme, a simulated sample resulting in
three failures was chosen. The failures were at times 0.0526,
0.0825, and 0.0836. The value of $p_{r}$ for reparameterization was
chosen to be $(3/1000)/2=0.015$, resulting in a likelihood that is
reasonably well behaved (except for the funnel shape with a sharp point). 
Figures~\ref{S.figure:large.sample.flat} and \ref{S.figure:large.sample.IJ}
give Weibull (on the left) and lognormal (on the right) estimation
results for flat and IJ priors, respectively.

\section{Proofs of Limiting Results for Weakly Informative Prior Distributions}
\label{S.section:limiting.results.weakly.informative.prior.distributions}
As mentioned in
Section~\ref{section:weakly.informative.priors.lls.distributions}
the normal (lognormal) distribution with a large standard
deviation (log standard deviation) is often use to specify
weakly informative prior distributions. 

\subsection{Limit of a Normal Distribution as its Standard Deviation Increases}
As mentioned in
Section~\ref{section:weakly.informative.priors.lls.distributions} of
the paper, a normal distribution prior density with any mean will
approach a flat prior as the standard deviation of the normal
distribution increases. First we consider a $\NORM(\mu, \sigma)$
distribution truncated outside of $\mu \pm A$ for a value $A>0$
\begin{align}
\label{S.equation:norm.to.flat.limit}
\lim_{\sigma \to \infty}\,
\frac{\dfrac{1}{\sigma}\,\phi_{\norm}\left(\dfrac{x-\mu}{\sigma}\right)}
{
\Phi_{\norm}\left(\dfrac{A}{\sigma}\right) -
  \Phi_{\norm}\left(-\dfrac{A}{\sigma}\right)}=
\frac{1}{2A},
\end{align}
for any $\mu$ and $ \mu-A \le x \le \mu+A$. We use this truncated
distribution so that the density remains proper in the limit.

Although it is possible to use L'Hospital's Rule to compute the
limit in (\ref{S.equation:norm.to.flat.limit}) directly, the needed
derivatives are complicated, making the proof lengthy. Here we take
an alternative simpler path. The denominator in
(\ref{S.equation:norm.to.flat.limit}) can be written as
\begin{align*}
\Phi_{\norm}\left(\dfrac{A}{\sigma}\right) -
  \Phi_{\norm}\left(-\dfrac{A}{\sigma}\right)
  =\int_{-A/\sigma}^{A/\sigma }\phi_{\norm}(w)\, dw
\end{align*} 
for $A>0$. 
The mean value theorem for integrals says that 
\begin{align*}
\int_{-A/\sigma}^{A/\sigma} \phi_{\norm}(w)\, dw = \left[ -\frac{A}{\sigma}
  -\left(\frac{A}{\sigma} \right)\right] \phi_{\norm}(\zeta) = \frac{2A}{\sigma}
  \phi_{\norm}(\zeta),
\end{align*} 
where $-A/\sigma \leq \zeta \leq A/\sigma $. Then
\begin{align*}
\frac{\dfrac{1}{\sigma}\,\phi_{\norm}\left(\dfrac{x-\mu}{\sigma}\right)}
{\Phi_{\norm}\left(\dfrac{A}{\sigma}\right) -
  \Phi_{\norm}\left(-\dfrac{A}{\sigma}\right)}&=
\frac{\dfrac{1}{\sigma}\,\phi_{\norm}\left(\dfrac{x-\mu}{\sigma}\right)}
{ \dfrac{2A}{\sigma}
  \phi_{\norm}(\zeta)  }=
\frac{1}{2A}\,\frac{\phi_{\norm}\left(\dfrac{x-\mu}{\sigma}\right)}
{\phi_{\norm}(\zeta)  }.
\end{align*}
For large $\sigma$, both  $-A/\sigma$ and  $A/\sigma $ are
approximately zero, implying that  $\zeta$ is approximately
zero. Of course, $(x-\mu)/\sigma$ will also be  approximately
zero. Thus
\begin{align*}
\frac{1}{2A}\,\dfrac{\phi_{\norm}\left(\dfrac{x-\mu}{\sigma}\right)}
{\phi_{\norm}(\zeta)  } \approx \frac{1}{2A}\,\dfrac{\phi_{\norm}\left( 0 \right)}
{\phi_{\norm}(0)  } =\frac{1}{2A},
\end{align*}
giving the needed result.

\subsection{Limit of a Lognormal Distribution as its Log Standard Deviation Increases}
As mentioned in
Section~\ref{section:weakly.informative.priors.lls.distributions} of
the paper, a lognormal distribution
prior $f(t)$ with any log-mean will be proportional to $1/t$ as the
log standard deviation increases. We start by noting that, for any
values of $\mu$ and $t>0$, the
standard normal density has the limit
\begin{align*}
\lim_{\sigma \to \infty}\,
\phi_{\norm}\left(\dfrac{\log(t)-\mu}{\sigma}\right)
=\lim_{\sigma \to \infty}\,\frac{1}{\sqrt{2 \pi}}\exp\left[-\frac{1}{2}\left(\frac{\log(t)-\mu}{\sigma}\right)^{2}\right]
=\frac{1}{\sqrt{2 \pi}} > 0.
\end{align*}
This implies that, for any fixed large value of $\sigma$, the lognormal density
\begin{align*}
\frac{1}{\sigma t}\phi_{\norm}\left(\dfrac{\log(t)-\mu}{\sigma}\right)
 \approx \frac{1}{\sigma t \sqrt{2 \pi}},
\end{align*}
giving the needed result.

\section{Log-Truncated and Log-Reciprocal-Truncated Distributions}
\label{S.section:log.truncated.normal.distributions}
\subsection{Motivation for the Distributions}
\label{S.section:motivation.log.truncated.distributions}
As mentioned in
Section~\ref{section:motivation.for.partially.informative.prior} of the main
paper, when specifying an informative prior
distribution for a positive parameter (like a log-location-scale
distribution shape parameter $\sigma$ or $\beta=1/\sigma$ or a
log-location-scale distribution quantile $t_{p}$), a normal
distribution truncated below zero (denoted by $\TNORM$) is often
used. Although it is a slight abuse, to keep the notation simple and
standard, we employ $\mu$ and $\sigma$ to denote the parameters of
the $\TNORM$ distribution.  The pdf for the $\TNORM(\mu, \sigma)$
random variable $T>0$ is
   \begin{align}
\label{S.equation:tnorm.pdf}
    \dtnorm(t;\mu,\sigma)&=\left(\frac{1}{\sigma}\right)
\frac{\phi_{\norm}\left ( \dfrac{t-\mu}{\sigma}\right)}
     {1-\Phi_{\norm}\left(-\dfrac{\mu}{\sigma}\right)}, \,\,\, t>0.
   \end{align}
Informative (and perhaps weakly informative)
priors are specified in terms of the distributions for parameters
like $T=\sigma$, $T=\beta=1/\sigma$, or $T=t_{p}$.
Then expressions for the prior pdfs of the unconstrained parameters
$S=\log(\sigma)$, $S=\log(\beta)=\log(1/\sigma)$, or $S=\log(t_{p})$
are needed, as these transformed parameters are used in the MCMC sampling.

\subsection{Log-Truncated-Normal Distribution pdf}
\label{S.section:LTNORM}
\begin{figure}
\begin{tabular}{cc}
(a) & (b) \\[-3.2ex]
\rsplidafiguresizetwo{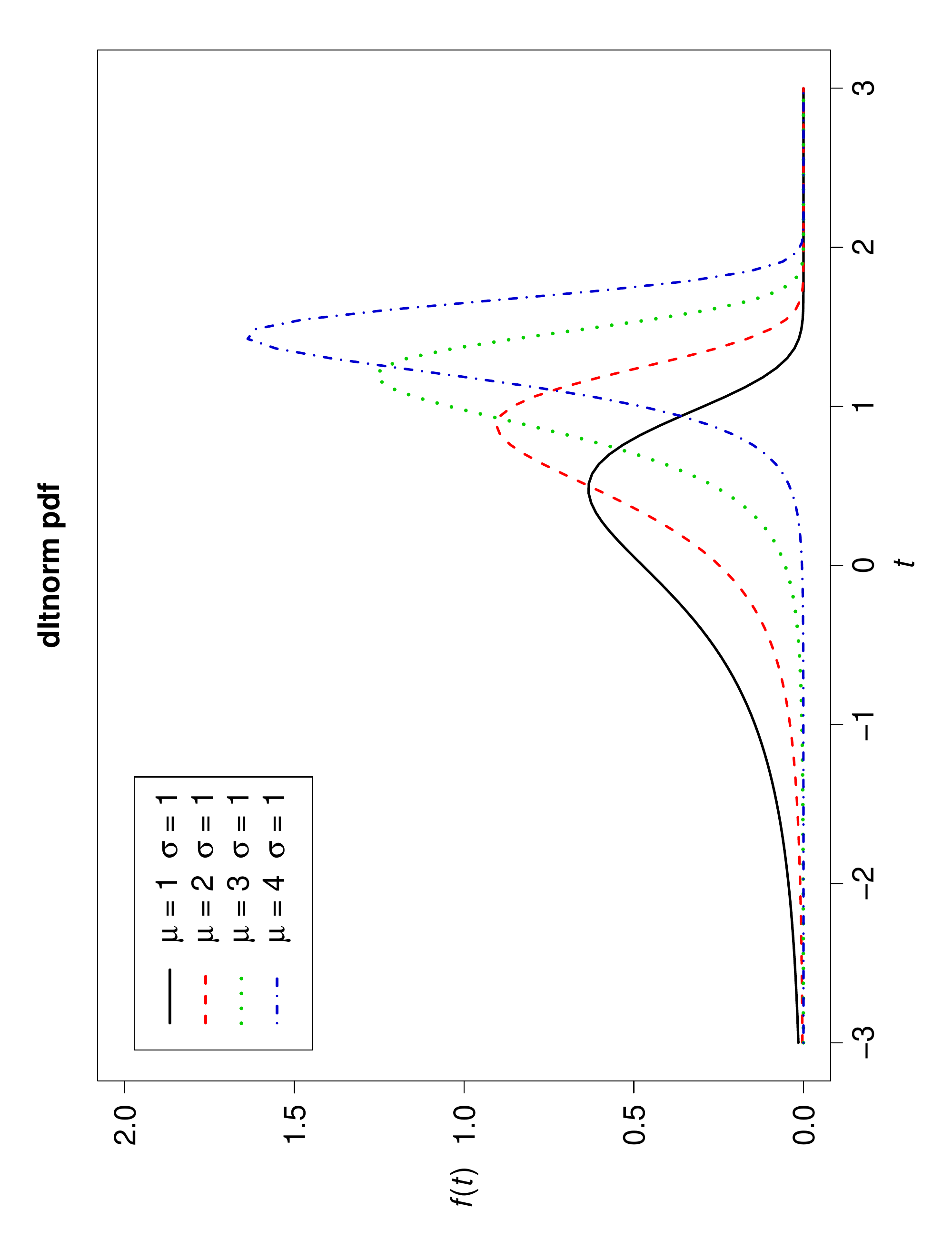}{2.65in}&
\rsplidafiguresizetwo{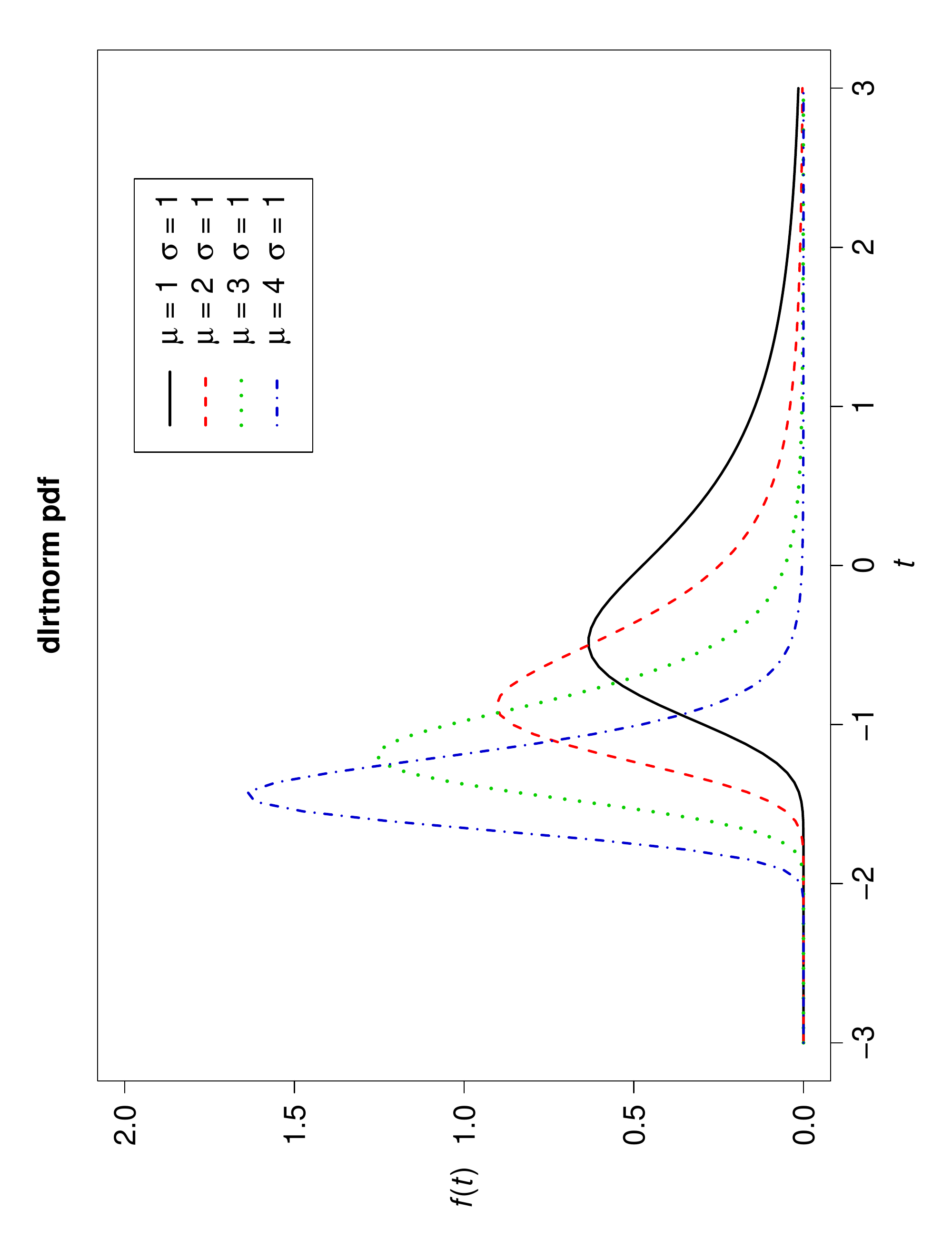}{2.65in}
\end{tabular}
\caption{pdfs for the log-truncated-normal distribution (a) and the
  log-reciprocal-truncated-normal distribution (b).}
\label{S.figure:transformtnorm}
\end{figure}

If $T \sim \TNORM(\mu, \sigma)$, the distribution of $S=\log(T)$ can
be obtained by using the standard random variable transformation
methods described, for example, in Chapter~2 of
\citet{CasellaBerger2002}. Specifically,
because $S$ is a monotone increasing function of $T$ and the inverse
of the transformation is $T=\exp(S)$, the pdf for $S$ is
   \begin{align}
\nonumber
    \dltnorm(s;\mu,\sigma)&=\dtnorm(\exp(s);\mu,\sigma) \times \exp(s)\\[1ex]
\label{S.equation:LTNORM.pdf}
    &= \left(\frac{1}{\sigma}\right)
\frac{\phi_{\norm}\left ( \dfrac{\exp(s)-\mu}{\sigma}\right)}
     {1-\Phi_{\norm}\left(-\dfrac{\mu}{\sigma}\right)} \times \exp(s),\\[1ex]
\nonumber
    &= \left(\frac{1}{\sigma}\right)
\frac{\phi_{\norm}\left (\dfrac{\mu-\exp(s)}{\sigma}\right)}
     {\Phi_{\norm}\left(\dfrac{\mu}{\sigma}\right)} \times \exp(s), \,\,\,-\infty<s<\infty.
   \end{align}
We call this the log-truncated-normal (LTNORM) distribution.
The simpler second expression is obtained by using the symmetry relationships
 $\Phi_{\norm}(z)=1-\Phi_{\norm}(-z)$ and $\phi_{\norm}(z)=\phi_{\norm}(-z)$.
Figure~\ref{S.figure:transformtnorm}(a) shows pdfs for the LTNORM distribution.

\subsection{Log-Reciprocal-Truncated-Normal Distribution pdf}
\label{S.section:LRTNORM}
Similar to Section~\ref{S.section:LTNORM}, if $T \sim \TNORM(\mu,
\sigma)$, then the distribution of $W=\log(1/T)=-\log(T)$ can, again, be
obtained by using the standard random variable transformation
methods. Specifically,
because $W$ is a monotone decreasing function of $T$ and the inverse
of the transformation is $T=\exp(-W)$, the pdf for $W$ is
   \begin{align}
\nonumber
\dlrtnorm(w;\mu,\sigma)&=\dtnorm(\exp(-w);\mu,\sigma) \times \exp(-w)\\[1ex]
\label{S.equation:LRTNORM.pdf}
    &= \left(\frac{1}{\sigma}\right)
\frac{\phi_{\norm}\left ( \dfrac{\exp(-w)-\mu}{\sigma}\right)}
     {1-\Phi_{\norm}\left(-\dfrac{\mu}{\sigma}\right)} \times
     \exp(-w),\\[1ex]
\nonumber
    &= \left(\frac{1}{\sigma}\right)
\frac{\phi_{\norm}\left ( \dfrac{\mu-\exp(-w)}{\sigma}\right)}
     {\Phi_{\norm}\left(\dfrac{\mu}{\sigma}\right)} \times \exp(-w) \,\,\,-\infty<w<\infty.
\end{align}
We call this the log-reciprocal-truncated-normal
(LRTNORM) distribution.
Figure~\ref{S.figure:transformtnorm}(b) shows pdfs for the LRTNORM distribution.

\subsection{Log-Truncated-Location-Scale-$t$
  and Log-Reciprocal-Truncated-Location-Scale-$t$ Distribution pdfs} 

As mentioned in
Section~\ref{section:motivation.for.partially.informative.prior}, a
useful generalization of the normal distribution is the
location-scale-$t$ (LST) distribution.  Similar to what is described
in Section~\ref{S.section:motivation.log.truncated.distributions}, if
priors are specified for the positive log-location-scale
distribution parameters (e.g., $\sigma$, $\beta=1/\sigma$, or
$t_{p}$) using a truncated LST distribution, pdfs of the
unconstrained parameters (e.g., $S=\log(\sigma)$,
$S=\log(\beta)=\log(1/\sigma)$, or $S=\log(t_{p})$) are needed. This
is because the pdfs of these transformed parameters are used to
specify priors in the MCMC sampling. We refer to these distributions
as LTLST and LRLST. Expressions for the LTLST and LRLST pdfs are
obtained in the same manner as in Sections~\ref{S.section:LTNORM}
and~\ref{S.section:LRTNORM} except that the standard normal pdf and
cdf are replaced by their LST counterparts and these depend on the
specified degrees of freedom parameter. In particular, the standard
LST pdf is the Student's $t$ pdf:
\begin{align*}
\phi_{\lst}(z;\dfreedom)&=\dfrac{\Gamma\left[(\dfreedom+1)/2\right]}
   {\Gamma(\dfreedom/2)\, \sqrt{\pi \dfreedom}}\,
   \frac{1}{\left(1+{z^{2}}/{\dfreedom}\right)^{(\dfreedom+1)/2}}, \,\,\,
   -\infty< z< \infty.
\end{align*}
The corresponding LST
cdf is $\Phi_{\lst}(z;\dfreedom)$. Then, following the same path
used in~(\ref{S.equation:LTNORM.pdf})
and~(\ref{S.equation:LRTNORM.pdf}) (without giving all of the steps), the LTLST
and LRLST pdfs are
   \begin{align*}
    \dltlst(s;\mu,\sigma,\dfreedom)&=\left(\frac{1}{\sigma}\right)
    \frac{\phi_{\lst}\left (\dfrac{\mu-\exp(s)}{\sigma};\dfreedom \right)}
         {\Phi_{\lst}\left(\dfrac{\mu}{\sigma};\dfreedom \right)} \times
         \exp(s), \,\,\,-\infty<s<\infty,
   \end{align*}
and
   \begin{align*}
\dlrtlst(w;\mu,\sigma,\dfreedom)&=\left(\frac{1}{\sigma}\right)
\frac{\phi_{\lst}\left ( \dfrac{\mu-\exp(-w)}{\sigma};\dfreedom \right)}
     {\Phi_{\lst}\left(\dfrac{\mu}{\sigma};\dfreedom \right)} \times \exp(-w) \,\,\,-\infty<w<\infty.
\end{align*}

\begingroup
\setstretch{0.9}
\bibliographystyle{chicago}
\addcontentsline{toc}{section}{\protect\numberline{}References}

\bibliography{main}

\endgroup

\end{document}